\pdfoutput=1
\documentclass[useAMS, usenatbib, a4paper,fleqn]{mn2e}
\usepackage{aas_macros ,multirow,graphicx, xspace, amsmath, url,acronym,deluxetable,microtype}
\usepackage{hyperref}
\usepackage{listings}
\usepackage{amssymb}
\usepackage{cleveref}
\usepackage[top=0.8in, bottom=0.8in, left=0.75in, right=0.75in]{geometry}

\acrodef{hst}[HST]{Hubble Space Telescope}
\acrodef{cfht}[CFHT]{Canada France Hawaii Telescope}
\acrodef{vlt}[VLT]{Very Large Telescope}
\acrodef{cos}[COS]{Cosmic Origins Spectrograph}
\acrodef{fos}[FOS]{Faint Object Spectrograph}
\acrodef{vimos}[VIMOS]{Visible Multi-Object Spectrograph}
\acrodef{deimos}[DEIMOS]{Deep Imaging Multi-Object Spectrograph}
\acrodef{gmos}[GMOS]{Gemini Multi-Object Spectrograph}
\acrodef{vvds}[VVDS]{\ac{vlt} \ac{vimos} Deep Survey}
\acrodef{gdds}[GDDS]{Gemini Deep Deep Survey}
\acrodef{idl}[IDL]{Interactive Data Language}
\acrodef{iraf}[IRAF]{Image Reduction and Analysis Facility}
\acrodef{gto}[GTO]{Guaranteed Observing Time}
\acrodef{go}[GO]{General Observer}
\acrodef{stsci}[STScI]{Space Telescope Science Institute}
\acrodef{eso}[ESO]{European Southern Observatory}
\acrodef{pi}[PI]{principal investigator}
\acrodef{mos}[MOS]{multi-object spectroscopy}
\acrodef{los}[LOS]{line-of-sight}
\acrodef{lss}[LSS]{large scale structure}
\acrodef{wmap}[WMAP]{Wilkinson Microwave Anisotropy Probe}
\acrodef{hipass}[HIPASS]{H~{\sc i} Parkes All Sky Survey}
\acrodef{whim}[WHIM]{warm-hot intergalactic medium}

\acrodef{ccd}[CCD]{charge-coupled device}
\acrodef{sdss}[SDSS]{Sloan Digital Sky Survey}
\acrodef{gimic}[GIMIC]{Galaxies-Intergalactic Medium Interaction Calculation}

\acrodef{uv}[UV]{ultra-violet}
\acrodef{nuv}[NUV]{near ultra-violet}
\acrodef{fuv}[FUV]{far ultra-violet}
\acrodef{igm}[IGM]{intergalactic medium}
\acrodef{cgm}[CGM]{circumgalactic medium}

\acrodef{psf}[PSF]{point spread function}
\acrodef{lsf}[LSF]{line spread function}
\acrodef{fwhm}[FWHM]{full width at half maximum}
\acrodef{qso}[QSO]{quasi-stellar object}
\acrodef{sn}[SN]{supernova}
\acrodef{agn}[AGN]{active galactic nuclei}
\acrodef{grb}[GRB]{gamma-ray burst}

\newcommand{\calcos}{{\sc calcos}\xspace} 
\newcommand{\sextractor}{{\sc sextractor}\xspace}
\newcommand{\esorex}{{\sc esorex}\xspace}
\newcommand{\vipgi}{{\sc vipgi}\xspace}
\newcommand{\vmmps}{{\sc vmmps}\xspace}
\newcommand{\gmmps}{{\sc gmmps}\xspace}
\newcommand{\vpfit}{{\sc vpfit}\xspace}
\newcommand{\camb}{{\sc camb}\xspace}

\crefname{equation}{equation}{equations}
\crefformat{equation}{equation~#2#1#3}
\Crefformat{equation}{Equation~#2#1#3}
\crefmultiformat{equation}{Equations~#2#1#3}{and~#2#1#3}{, #2#1#3}{, and~#2#1#3}

\crefname{figure}{figure}{figures}
\crefname{table}{table}{tables}
\crefname{section}{section}{sections}
\renewcommand{\arraystretch}{1.1} 

\newcommand{\hi}{H~{\sc i}\xspace}

\newcommand{\ha}{H$\alpha$\xspace}

\newcommand{\dfour}{D$4000$\xspace}

\newcommand{\nhi}{$N_{\rm HI}$\xspace}
\newcommand{\bhi}{$b_{\rm HI}$\xspace}
\newcommand{\mpc}{\,Mpc\xspace}
\newcommand{\hmpc}{\,$h^{-1}_{100}$Mpc\xspace}
\newcommand{\hhmpc}{\,$h^{-1}_{70}$Mpc\xspace}
\newcommand{\kpc}{\,kpc\xspace}
\newcommand{\cm}{\,cm$^{-2}$\xspace}
\newcommand{\xiag}{$\xi_{\rm ag}(r)$\xspace}
\newcommand{\xiaa}{$\xi_{\rm aa}(r)$\xspace}
\newcommand{\xigg}{$\xi_{\rm gg}(r)$\xspace}

\newcommand{\nabsrand}{$\alpha_{\rm abs}$\xspace}
\newcommand{\ngalrand}{$\alpha_{\rm gal}$\xspace}
\newcommand{\zgal}{$z_{\rm gal}$\xspace}
\newcommand{\zabs}{$z_{\rm abs}$\xspace}

\newcommand{\zrandgal}{$z_{\rm gal}^{\rm rand}$\xspace}

\newcommand{\rband}{$R$-band\xspace}
\newcommand{\SF}{`SF'\xspace}
\newcommand{\nSF}{`non-SF'\xspace}
\newcommand{\strong}{`strong'\xspace}
\newcommand{\weak}{`weak'\xspace}
\newcommand{\fn}{\footnotemark[9]}

\title[The IGM--galaxy cross-correlation at $z\lesssim 1$]{On the
  connection between the intergalactic medium and galaxies: The H~{\sc
    i}--galaxy cross-correlation at $z\lesssim 1$} \author[Nicolas Tejos et al.]{
\parbox[t]{\textwidth}{
\vspace{-1.0cm} 
Nicolas Tejos,$^{1}$\thanks{E-mail: \href{mailto:nicolas.tejos@durham.ac.uk}{nicolas.tejos@durham.ac.uk}}
Simon L. Morris,$^{1}$ 
Charles W. Finn,$^{1}$ 
Neil H. M. Crighton,$^{2}$
Jill Bechtold,$^3$
Buell T. Jannuzi,$^{3}$
Joop Schaye,$^{4}$
Tom Theuns,$^{1,5}$
Gabriel Altay,$^{6,1}$
Olivier Le F\'evre,$^{7}$ 
Emma Ryan-Weber$^{8}$ and
Romeel Dav\'e$^{9,10,11,3}$
}
\vspace*{6pt} \\ 
$^{1}$ Department of Physics, Durham University, South Road, Durham, DH1 3LE, UK\\ 
$^{2}$ Max-Planck-Institute for Astronomy, K{\"o}nigstuhl 17, D-69117, Heidelberg, Germany\\
$^{3}$ Department of Astronomy and Steward Observatory, University of Arizona, Tucson, AZ 85721, USA\\
$^{4}$ Leiden Observatory, Leiden University, PO Box 9513, NL-2300 RA Leiden, the Netherlands\\
$^{5}$ Department of Physics, University of Antwerp, Groenenborgerlaan 171, B-2020 Antwerpen, Belgium\\
$^{6}$ Center for Relativistic Astrophysics, School of Physics, Georgia Institute of Technology, 837 State Street, Atlanta, GA, USA\\
$^{7}$ Aix Marseille Universit\'e, CNRS, LAM, Laboratoire d'Astrophysique de Marseille, 38 rue F. Joliot-Curie, F-13388, Marseille, France\\
$^{8}$ Centre for Astrophysics and Supercomputing, Swinburne University of Technology, Hawthorn, VIC 3122, Australia\\
$^{9}$ University of the Western Cape, Bellville, Cape Town 7535, South Africa\\
$^{10}$ South African Astronomical Observatories, Observatory, Cape Town 7925, South Africa\\
$^{11}$ African Institute for Mathematical Sciences, Muizenberg, Cape Town 7945, South Africa
\vspace*{-0.5cm}}

\begin{document}
\date{Draft version}

\pagerange{\pageref{firstpage}--\pageref{lastpage}} \pubyear{2013}

\maketitle

\label{firstpage}

\begin{abstract}

We present a new optical spectroscopic survey of $1777$ `star-forming'
(\SF) and $366$ `non-star-forming' (\nSF) galaxies at redshifts $z\sim
0 - 1$ ($2143$ in total), $22$ \acs{agn} and $423$ stars, observed by
instruments such as \acs{deimos}, \acs{vimos} and \acs{gmos}, in $3$
fields containing $5$ \acp{qso} with \acs{hst} \ac{uv} spectroscopy.
We also present a new spectroscopic survey of $165$ \strong ($10^{14}
\le N_{\rm HI}\lesssim 10^{17}$ \cm), and $489$ \weak ($10^{13}
\lesssim N_{\rm HI} < 10^{14}$ \cm) intervening \hi (\lya) absorption
line systems at $z \lesssim 1$ ($654$ in total), observed in the
spectra of $8$ \acp{qso} at $z \sim 1$
by \acs{cos} and \acs{fos} on the \acs{hst}. Combining these new data
with previously published galaxy catalogs such as \acs{vvds} and
\acs{gdds}, we have gathered a sample of $654$ \hi~absorption systems
and $17509$ galaxies at transverse scales $\lesssim 50$ \mpc, suitable
for a two-point correlation function analysis. We present observational
results on the \hi--galaxy ($\xi_{\rm ag}$) and galaxy--galaxy
($\xi_{\rm gg}$) correlations at transverse scales $r_{\perp} \lesssim
10$ \mpc, and the \hi--\hi auto-correlation ($\xi_{\rm aa}$) at
transverse scales $r_{\perp} \lesssim 2$ \mpc. The two-point
correlation functions are measured both along and transverse to the
\acl{los}, $\xi(r_{\perp},r_{\parallel})$. We also infer the shape of
their corresponding `real-space' correlation functions, $\xi(r)$, from
the projected along the line-of-sight correlations, assuming power-laws
of the form $\xi(r)=(r/r_0)^{-\gamma}$. Comparing the results from
$\xi_{\rm ag}$, $\xi_{\rm gg}$ and $\xi_{\rm aa}$, we constrain the
\hi--galaxy statistical connection, as a function of both \hi column
density and galaxy star-formation activity. Our results are consistent
with the following conclusions: (i)~the bulk of \hi systems on
$\sim$~Mpc scales have little velocity dispersion ($\lesssim 120$ \kms)
with respect to the bulk of galaxies (i.e. no strong galaxy
outflow/inflow signal is detected); (ii)~the vast majority ($\sim
100\%$) of \strong \hi systems and \SF galaxies are distributed in the
same locations, together with $75 \pm 15\%$ of \nSF galaxies, all of
which typically reside in dark matter haloes of similar masses;
(iii)~$25 \pm 15\%$ of \nSF galaxies reside in galaxy clusters and are
not correlated with \strong \hi systems at scales $\lesssim 2$ \mpc;
and (iv)~$> 50\%$ of \weak \hi systems reside within galaxy voids
(hence not correlated with galaxies), and are confined in dark matter
haloes of masses smaller than those hosting \strong systems and/or
galaxies. We speculate that \hi systems within galaxy voids might still
be evolving in the linear regime even at scales $\lesssim 2$~\mpc.
\end{abstract}


\section{Introduction}

\acresetall

\subsection{Motivation}

The physics of the \ac{igm} and its connection with galaxies are key to
understanding the evolution of baryonic matter in the Universe. This is
because of the continuous interplay between the gas in the \ac{igm} and
galaxies: (i) galaxies are formed by the condensation and accretion of
primordial or enriched gas; and (ii) galaxies enrich their haloes and
the \ac{igm} via galactic winds and/or merger events.

Theoretical analyses---under a $\Lambda$ cold dark matter paradigm
($\Lambda$CDM)---suggest that: (i) the accretion happens in two major
modes: `hot' and `cold'
\citep[e.g.][]{Rees1977,White1978,White1991,Keres2005,VanDeVoort2011};
and (ii) galactic winds are mostly driven by \ac{sn} and/or \ac{agn}
feedback \citep[e.g.][]{Baugh2005, Bower2006, Lagos2008, Creasey2013}.

Models combining `N-body' dark matter simulations (collisionless,
dissipationless) with `semi-analytic' arguments \citep[e.g.][and
  references therein]{Baugh2006} have been successful in reproducing
basic statistical properties of luminous galaxies (e.g. luminosity
functions, clustering, star-formation histories, among
others). However, in order to provide predictions for the signatures of
`hot'/`cold' accretion and/or \ac{agn}/\ac{sn} feedback in the
\ac{igm}, a {\it full} hydrodynamical description is required.

In practice, hydrodynamical simulations still rely on unresolved
`sub-grid physics' to lower the computational cost
\citep[e.g.][]{Schaye2010, Scannapieco2012}, whose effects are not
fully understood. Therefore, observations of the \ac{igm} and galaxies
in the same volume are fundamental to testing these predictions and
helping to discern between different physical models
\citep[e.g.][]{Fumagalli2011,Oppenheimer2012,Stinson2012,Hummels2013,Ford2013,Rakic2013}.

Although the \ac{igm} is the main reservoir of baryons at all epochs
\citep[e.g.][]{Fukugita1998,Cen1999,Schaye2001,Dave2010,Shull2012}, its
extremely low densities make its observation difficult and
limited. Currently, the only feasible way to observe the \ac{igm} is
through intervening absorption line systems in the spectra of bright
background sources, limiting its characterization to being
one-dimensional. Still, an averaged three dimensional picture can be
obtained by combining multiple lines-of-sight (LOS) and galaxy surveys,
which is the approach adopted in this work (see \Cref{intro:strategy}).

The advent of the \ac{cos} on the \ac{hst} has revolutionized the study
of the \ac{igm} and its connection with galaxies at low-$z$ ($z\lesssim
1$). With a sensitivity $\sim 10$ times greater than that of its
predecessors, \ac{cos} has considerably increased the number of
\acp{qso} for which \ac{uv} spectroscopy is feasible. This capability
has been exploited for studies of the so-called \ac{cgm}, by
characterizing neutral hydrogen (\hi)\footnote{Note that at column
  densities $N_{\rm HI} \lesssim 10^{17}$ \cm the hydrogen gas is
  mostly ionized however.} and metal absorption systems in the vicinity
of known galaxies
\citep[e.g.][]{Tumlinson2011,Thom2012,Werk2013,Stocke2013,Keeney2013,Lehner2013}.

Studies of the \ac{cgm} implicitly assume a direct one-to-one
association between absorption systems and their closest observed
galaxy, which might not always hold because of incompleteness in the
galaxy surveys and projection effects. Given that metals are formed and
expelled by galaxies, a direct association between them seems sensible,
in accordance with predictions from low-$z$ simulations
\citep[e.g.][]{Oppenheimer2012}. However, the situation for neutral
hydrogen is more complicated, as \hi traces both enriched {\it and}
primordial material.\footnote{Note that whether truly primordial \hi
  clouds exist at low-$z$ is still to be observationally confirmed.}

The nature of the relationship between \hi and galaxies at low-$z$ has
been widely debated. Early studies have pointed out two distinct
scenarios for this connection: (i) a one-to-one physical association
because they both belong to the same dark matter haloes
\citep[e.g.][]{Mo1994b,Lanzetta1995,Chen1998}; and (ii) an indirect
association because they both trace the same underlying dark matter
distribution but not necessarily the same haloes
\citep[e.g.][]{Morris1991,Morris1993,Mo1994a,Stocke1995,Tripp1998}.
More recent studies have shown the presence of \hi absorption systems
within galaxy voids
\citep[e.g.][]{Grogin1998,Penton2002,Manning2002,Stocke1995,Tejos2012},
hinting at a third scenario: (iii) the presence of \hi absorption
systems that {\it are not} associated with galaxies \citep[although
  see][]{Wakker2009}.\footnote{Note that little can be said about low
  surface brightness galaxies, as current spectroscopic surveys are
  strongly biased against these, for obvious reasons \citep[although
    see][]{Ryan-Weber2006}.}

If we think of galaxies as peaks in the density distribution
\citep[e.g.][]{Press1974}, it is natural to expect high column density
\hi~systems to show a stronger correlation with galaxies than low
column density ones, owing to a density-\hi column density
proportionality \citep[e.g.][]{Schaye2001,
  Dave2010,Tepper-Garcia2012}. Similarly, we also expect the majority
of low column density \hi~systems to belong to dark matter haloes that
did not form galaxies. Thus, the relative importance of these three
scenarios should depend, to some extent, on the \hi column
density. \citet{Tejos2012} estimated that these three scenarios account
for $\sim 15\%$, $\sim 55\%$ and $\sim 30\%$ of the low-$z$ \hi systems
at column densities $N_{\rm HI} \gtrsim 10^{12.5}$ \cm, respectively,
indicating that the vast majority of \hi absorption line systems are
not physically associated with luminous galaxies \citep[see also][for a
  similar conclusion]{Prochaska2011b}.

\subsection{Study strategy}\label{intro:strategy}

In this paper we address the statistical connection between \hi and
galaxies at $z \lesssim 1$ through a clustering analysis
\citep[e.g.][]{Morris1993,Ryan-Weber2006,Wilman2007,Chen2009,Shone2010},
without considering metals. We focus only on hydrogen because it is the
best \ac{igm} tracer for a statistical study. Apart from the fact that
it traces both primordial and enriched material, it is also the most
abundant element in the Universe. Hence, current spectral sensitivities
allow us to find \hi inside and outside galaxy haloes, which is not the
case yet for metals at low-$z$ \citep[according to recent theoretical
  results; e.g.][]{Oppenheimer2012}.

Focusing on the second half of the history of the Universe ($z\lesssim
1$) has the advantage of allowing relatively complete galaxy surveys
even at faint luminosities ($\lesssim L^*$; elusive at higher
redshifts). Faint galaxies are important for statistical analyses as
they dominate the luminosity function, not just in number density, but
also in total luminosity and mass. Moreover, the combined effects of
structure formation, expansion of the Universe, and the reduced
ionization background, allow us to observe a considerable amount of \hi
systems and yet resolve the so-called \hi \lya-forest into individual
lines \citep[e.g.][]{Theuns1998a,Dave1999}. This makes it possible to
recover column densities and Doppler parameters through Voigt profile
fitting.

One major advantage of clustering over one-to-one association analyses
is that it does not impose arbitrary scales, allowing us to obtain
results for both small ($\lesssim 1$ \mpc) and large scales ($\gtrsim
1-10$ \mpc). In this way, we can make use of {\it all} the \hi and
galaxy data available, and not only those lying close to each
other. Results from the small scale association are important to
constraint the `sub-grid physics' adopted in current hydrodynamical
simulations. Conversely, results from the largest scales provide
information unaffected by these uncertain `sub-grid physics'
assumptions \citep[e.g.][]{Hummels2013,Ford2013,Rakic2013}. Moreover,
the physics and cosmic evolution of the diffuse \ac{igm} (traced by
\hi) obtained by cosmological hydrodynamical simulations
\citep[e.g.][]{Paschos2009,Dave2010} are in good agreement with
analytic predictions \citep[e.g.][]{Schaye2001}. Our results will be
able to test all of these predictions.

Another advantage to using a clustering analysis is that it properly
takes into account the selection functions of the surveys. Even at
scales $\lesssim 300$~\kpc~(the typical scale adopted for the
\ac{cgm}), a secure or unique \hi--galaxy one-to-one association is not
always possible. This is because \hi~and galaxies are clustered at
these scales and because surveys are never $100\%$ complete. Clustering
provides a proper statistical description, at the cost of losing
details on the physics of an individual \hi--galaxy pair. Thus, both
one-to-one associations and clustering results are complementary, and
needed, to fully understand the relationship between the \ac{igm} and
galaxies.

In this paper we present observational results for the \hi--galaxy
two-point correlation function at $z\lesssim 1$. Combining data from
\ac{uv} \ac{hst} spectroscopy of 8 \acp{qso} in 6 different fields,
with optical deep \ac{mos} surveys of galaxies around them, we have
gathered a sample of $669$ well identified intervening \hi absorption
systems and $17509$ galaxies at projected separations $\lesssim 50$
\mpc from the \ac{qso} \ac{los}. This dataset is the largest sample to
date for such an analysis.

Comparing the results from the \hi--galaxy cross-correlation with the
\hi--\hi and galaxy--galaxy auto-correlations, we provide constraints
on their statistical connection as a function of both \hi column
density and galaxy star-formation activity.

Our paper is structured as follows. Sections \ref{data:IGM} and
\ref{data:galaxy} describe the \ac{igm} and galaxy data used in this
work, respectively. The \ac{igm} sample is described in Section
\ref{abs:samples} while the galaxy sample is described in Section
\ref{gal:samples}. Section \ref{analysis} describes the formalisms used
to measure the \hi--galaxy cross-correlation and the \hi--\hi and
galaxy--galaxy auto-correlations. Our observational results are
presented in Section \ref{results} and discussed in Section
\ref{discussion}. A summary of the paper is presented in Section
\ref{summary}.

All distances are in co-moving coordinates assuming $H_0=70$ \kms
Mpc$^{-1}$, $\Omega_{\rm m}=0.3$, $\Omega_{\rm \Lambda}=0.7$, $k=0$,
unless otherwise stated, where $H_0$, $\Omega_{\rm m}$, $\Omega_{\rm
  \Lambda}$ and $k$ are the Hubble constant, mass energy density, `dark
energy' density and spatial curvature, respectively. Our chosen
cosmological parameters lie between the latest results from the
\acl{wmap} \citep{Komatsu2011} and the Planck satellite
\citep{Planck2013}.


\begin{table*}
  \begin{minipage}{0.9\textwidth}
    \centering
    \caption{Properties of the observed QSOs.}\label{tab:data:IGM}
    \begin{tabular}{@{}lccccccc@{}}
      \hline
      
      \multicolumn{1}{c}{QSO Name} & Field Name & R.A.          & Dec.         & $z_{\rm QSO}$ & \multicolumn{3}{c}{Magnitude}\\
                                   &            &  (hr min sec) & (deg min sec)&              & Visual (Band) & NUV (AB)& FUV (AB)\\ 
      \multicolumn{1}{c}{(1)}      & (2)        & (3)           & (4)          & (5)          & (6)           & (7)     & (8)     \\
      \hline
  
      Q0107-025A           &Q0107& 01 10 13.10   &  $-$02 19 52.0 & 0.96000 & 18.1 ($B$)& 18.1 & 19.3 \\
      Q0107-025B           &Q0107& 01 10 16.20   &  $-$02 18 50.0 & 0.95600 & 17.4 ($V$)& 17.5 & 18.6 \\
      Q0107-0232           &Q0107& 01 10 14.51   &  $-$02 16 57.5 & 0.72600 & 18.4 ($B$)& 18.9 & 20.1 \\
      J020930.7-043826     &J0209& 02 09 30.74   &  $-$04 38 26.3 & 1.12800 & 17.2 ($g$)& 17.5 & 18.5 \\
      J100535.24+013445.7  &J1005& 10 05 35.26   &  $+$01 34 45.6 & 1.08090 & 16.8 ($g$)& 17.4 & 18.6 \\ 
      J102218.99+013218.8  &J1022& 10 22 18.99   &  $+$01 32 18.8 & 0.78900 & 16.8 ($V$)& 17.2 & 18.1 \\   
      J135726.27+043541.4  &J1357& 13 57 26.27   &  $+$04 35 41.4 & 1.23176 & 17.2 ($g$)& 17.8 & 19.2 \\
      J221806.67+005223.6  &J2218& 22 18 06.69   &  $+$00 52 23.7 & 1.27327 & 17.8 ($V$)& 18.6 & 24.0\tablenotemark{a} \\

      \hline

    \end{tabular}
    \end{minipage}
    \begin{minipage}{0.9\textwidth}
    (1) Name of the QSO.
    (2) Name of the field.
    (3) Right ascension (J2000). 
    (4) Declination (J2000). 
    (5) Redshift of the QSO. 
    (6) Apparent visual magnitude; the band is given in parenthesis.
    (7) Apparent near-UV magnitude from GALEX.
    (8) Apparent far-UV magnitude from GALEX.\\
    $^{\rm a}$ The sudden decrease in flux is due to the presence of a Lyman Limit System.
    
  \end{minipage}
\end{table*}

\begin{table*} 
  \begin{minipage}{0.95\textwidth}
    \centering
    \caption{Summary of the QSO observations (HST spectroscopy).}\label{tab:data:HSTobs}
    \begin{tabular}{@{}lcccccccc@{}}
      \hline
      
      \multicolumn{1}{c}{QSO Name} & Instrument   & Grating & Wavelength & FWHM  & Dispersion & $\langle S/N \rangle$      & Exposure  &Program ID   \\
                                   &              &          & range (\AA)& (\AA) & (\AA/pixel)& (per pixel)                & time (h)      &           \\
      \multicolumn{1}{c}{(1)}      & (2)          & (3)      & (4)        & (5)   & (6)        & (7)                        & (8)           & (9)       \\ 
      \hline
      
      Q0107-025A           & COS  & G130M    & 1135--1460 & 0.07  & 0.01      & 9          & 7.8 & 11585      \\
                           & COS  & G160M    & 1460--1795 & 0.09  & 0.01      & 8          & 12.3& 11585      \\
                           & FOS  & G190H    & 1795--2310 & 1.39  & 0.36      & 28         & 7.5 & 5320, 6592 \\
                           & FOS  & G270H    & 2310--3277 & 1.97  & 0.51      & 32         & 2.4 & 6100       \\
      Q0107-025B           & COS  & G130M    & 1135--1460 & 0.07  & 0.01      & 9          & 5.9 & 11585      \\
                           & COS  & G160M    & 1460--1795 & 0.09  & 0.01      & 7          & 5.9 & 11585      \\
                           & FOS  & G190H    & 1795--2310 & 1.39  & 0.36      & 28         & 1.8 & 5320, 6592 \\
                           & FOS  & G270H    & 2310--3277 & 1.97  & 0.51      & 32         & 1.8 & 6100       \\
      Q0107-0232           & COS  & G160M    & 1434\tablenotemark{a}--1795 & 0.09 & 0.01& 7& 23.2& 11585      \\
                           & FOS  & G190H    & 1795--2310 & 1.39  & 0.36      & 18         & 9.1 & 11585      \\
      J020930.7-043826     & COS  & G130M    & 1277\tablenotemark{a}--1460 &0.07 &0.01  & 12         & 3.9& 12264      \\
                           & COS  & G160M    & 1460--1795 &0.09   & 0.01       & 10         & 7.8& 12264      \\
                           & COS  & G230L    & 1795--3084 &0.79   & 0.39       & 12         & 4.0& 12264      \\
      J100535.24+013445.7  & COS  & G130M    & 1135--1460 &0.07   & 0.01       &  9         & 3.9& 12264      \\
                           & COS  & G160M    & 1460--1795 &0.09   & 0.01       &  9         & 6.2& 12264      \\
      J102218.99+013218.8  & COS  & G130M    & 1135--1460 &0.07   & 0.01       &  6         & 0.6& 11598      \\
                           & COS  & G160M    & 1460--1795 &0.09   & 0.01       &  5         & 0.8& 11598      \\
      J135726.27+043541.4  & COS  & G130M    & 1135--1460 &0.07   & 0.01       & 9         & 3.9& 12264      \\
                           & COS  & G160M    & 1460--1795 &0.09   & 0.01       & 7          & 7.8& 12264      \\
                           & COS  & G230L    & 1795--3145 &0.79   & 0.39       & 11         & 4.0& 12264      \\
      J221806.67+005223.6  & COS  & G230L    & 2097\tablenotemark{b}--3084 &0.79 &0.39  & 10         & 5.6& 12264      \\

      \hline

    \end{tabular}
    \end{minipage}
  \begin{minipage}{0.95\textwidth}
    (1) Name of the QSO.
    (2) Instrument.
    (3) Grating. 
    (4) Wavelength range used for a given setting. 
    (5) Full-width at half maximum of the line spread function of the spectrograph. 
    (6) Dispersion.
    (7) Average signal-to-noise ratio per pixel over the given wavelength range.
    (8) Exposure time of the observations.
    (9) HST program ID of the observations.\\
    $^{\rm a}$ Due to the presence of a Lyman Limit System blocking shorter wavelengths.\\
    $^{\rm b}$ Due to poor signal-to-noise data at shorter wavelengths.\\
    
  \end{minipage}
\end{table*}

\section{Intergalactic medium data}\label{data:IGM}

We used \ac{hst} spectroscopy of 8 \acp{qso} to characterize the
diffuse \ac{igm} through the observations of intervening \hi~absorption
line systems. We used data from \ac{cos} \citep{Green2012} taken under
\ac{hst} programs \ac{go} 12264 (PI: Morris), \ac{go} 11585 (PI:
Crighton) and \ac{go} 11598 (PI: Tumlinson); and data from the \ac{fos}
\citep{Keyes1995} taken under \ac{hst} programs \ac{go} 5320 (PI:
Foltz), \ac{go} 6100 (PI: Foltz) and \ac{go} 6592 (PI: Foltz).

Data from program \ac{go} 12264 were taken to study the statistical
relationship between \hi~absorption line systems and galaxies at
redshift $z\lesssim 1$. We selected four \acp{qso} at $z_{\rm QSO} \sim
1$ (namely J020930.7-043826, J100535.24+013445.7, J135726.27+043541.4
and J221806.67+005223.6) lying in fields of view that were already
surveyed for their galaxy content by the \ac{vvds}
\citep{LeFevre2005,LeFevre2013} and the \ac{gdds}
\citep{Abraham2004}. Data from programs \ac{go} 5320, \ac{go} 6100,
\ac{go} 6592 and \ac{go} 11585 contain spectroscopy of three \acp{qso}
(namely Q0107-025A, Q0107-025B and Q0107-0232) whose \acp{los} are
separated by $\sim 0.4-1$~\mpc. This triple \ac{qso} field is ideal for
measuring the characteristic sizes of the \hi~absorption systems but it
can also be used to address the connection between \hi~systems and
galaxies \citep[e.g.][]{Crighton2010}. Data from program \ac{go} 11598
were originally taken to investigate the properties of the \ac{cgm} by
targeting \acp{qso} whose \ac{los} lie within $\lesssim 150$~\kpc~of a
known galaxy. For this paper we used one \ac{qso} observed under
program \ac{go} 11598 (namely J102218.99+013218.8), for which we have
conducted our own galaxy survey around its \ac{los} (see
\Cref{data:galaxy}).  Given that this \ac{los} contains only one
pre-selected galaxy, this selection will not affect our results on the
\ac{igm}--galaxy statistical connection.

\Cref{tab:data:IGM} summarizes our \ac{qso} sample while
\Cref{tab:data:HSTobs} gives details on their HST observations.

\subsection{Data reduction}

\subsubsection{COS data}
Individual exposures from \ac{cos} were downloaded from the \ac{stsci}
archive and reduced using \calcos v2.18.5 in combination with Python
routines developed by the authors.\footnote{Available at
  \url{https://github.com/cwfinn/COS/}} A full description of the
reduction process will be presented in Finn et al. (2013, in prep.),
here we present a summary.

Individual files corresponding to single central wavelength setting,
stripe and FP-POS (i.e. \verb|x1d| files) were obtained directly from
\calcos. The source extraction was performed using a box of 25 pixels
wide along the spatial direction for all G130M exposures, and 20 pixels
for all G160M and G230L exposures. The background extraction was
performed using boxes encompassing as much of the background signal as
possible, whilst avoiding regions close to the detector edges. We set
the background smoothing length in \calcos to $1$ pixel and performed
our own background smoothing procedure masking out portions of the
spectra affected by strong geocoronal emission lines (namely the
\hi~\lya and \ion{O}{1}~$\lambda\lambda 1302,1306$) and pixels with bad
data quality
flags\footnote{\url{http://www.stsci.edu/hst/cos/pipeline/cos_dq_flags}}. We
interpolated across the gaps to get the background level in these
excluded regions. The background smoothing lengths were set to 1000
pixels for the \ac{fuv}A stripes, 500 pixels for the \ac{fuv}B stripes
and 100 pixels for all \ac{nuv} stripes, along the dispersion
direction.

\begin{figure*}
  \begin{minipage}{1\textwidth}
    \includegraphics[width=1\textwidth]{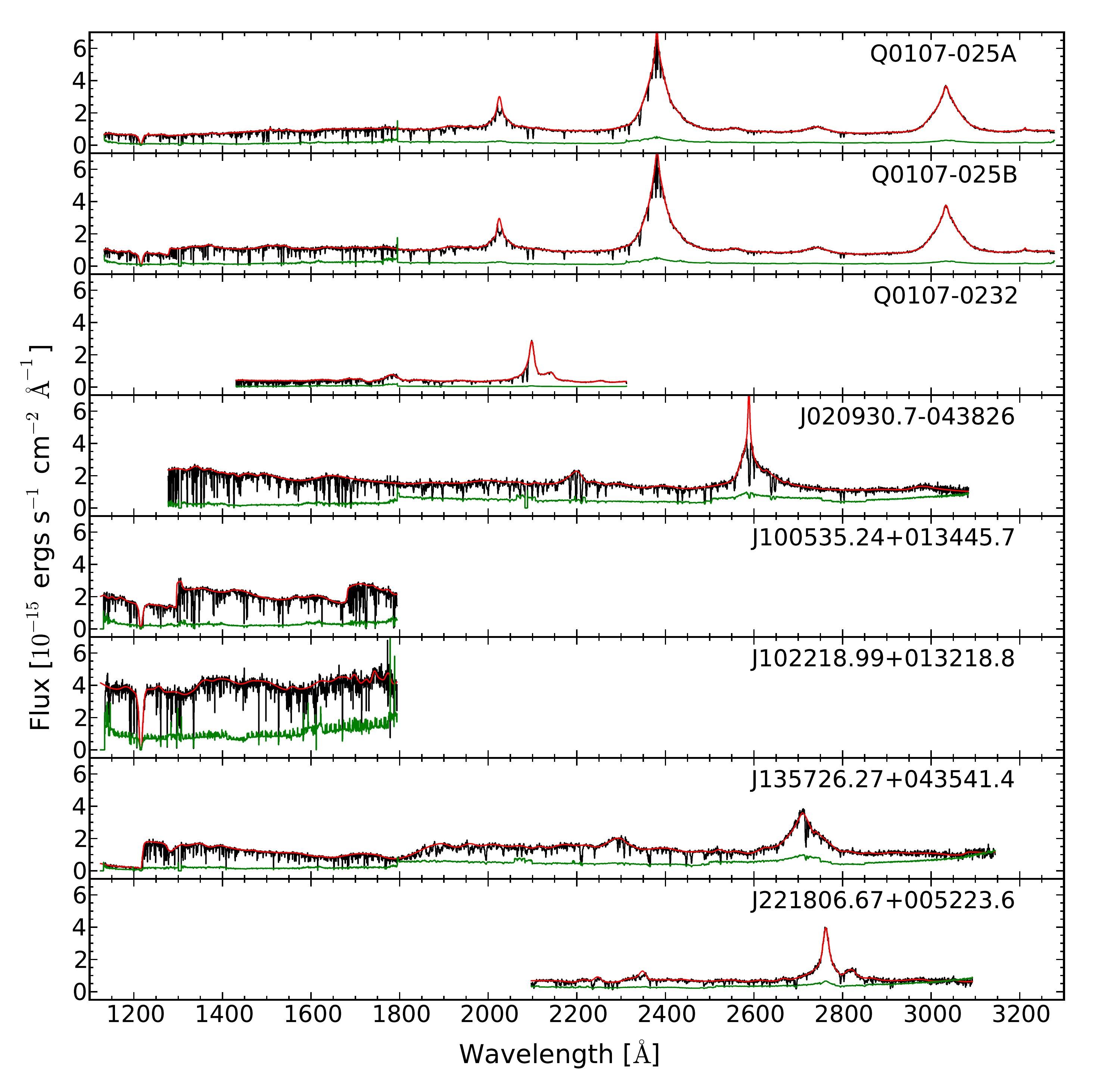}
  \end{minipage}
  
  \caption{Observed spectra of our sample of QSOs: flux (black lines),
    uncertainty (multiplied by a factor of $5$ for clarity; green
    lines) and continuum fit (red lines). Wavelengths
    $\lambda<1795$~\AA~and $\lambda \ge 1795$~\AA~correspond to data
    from the FUV and NUV channels respectively (see
    Table~\ref{tab:data:HSTobs}). The FUV spectra have been re-binned
    to match the resolution of the NUV spectra for
    clarity.}\label{fig:all_quasars}

\end{figure*}

The error array was calculated in the same way as in \calcos, but using
our new background estimation. Each spectrum was then flux calibrated
using sensitivity curves provided by \ac{stsci}.

Co-alignment was performed by cross-correlating regions centred on
strong Galactic absorption features (namely, \ion{C}{2} $\lambda 1334$,
\ion{Al}{2} $\lambda 1670$, \ion{Si}{2} $\lambda 1260$, \ion{Si}{2}
$\lambda 1526$ and \ion{Mg}{2} $\lambda\lambda 2796, 2803$ \AA). For
each grating we pick the central wavelength setting and FP-POS position
with the most accurately determined wavelength solutions from
\ac{stsci} as a reference. These are FP-POS $=3$ for all gratings,
central wavelengths of $1309$ and $1600$~\AA~for the G130M and G160M
gratings respectively, and $2950$~\AA~(using only the `B' stripe) for
the G230L grating. All other settings for each grating are then
cross-correlated on these ones, assuming the reference and comparison
settings both contain one of the absorption features
specified. Wavelengths offsets are then applied to the comparison
settings to match the reference ones. These offsets typically amount to
a resolution element or less. For those settings that could not be
aligned on any of the Galactic features specified, we manually searched
for other strong absorption lines on which to perform the
cross-correlation. Strong absorption lines were always found. We then
scaled the fluxes of the comparison setting such that its median flux
value matches that of either the reference or the already calibrated
setting in the overlapped region.

At this point we changed some pixel values according to their quality
flags: flux and error values assigned to pixels with bad data quality
flags were set to zero, while pixels with warnings had their exposure
times reduced by a factor of two. We then re-scaled the wavelength
binning of each exposure to have a constant spacing equal to the
dispersion for the grating, using nearest-neighbour interpolation. The
combined wavelength binning therefore consists of three wavelength
scales, one for the G130M grating ($\lambda < 1460$~\AA), one for the
G160M grating ($1460 \leq \lambda < 1795$~\AA) and one for the G230L
grating ($\lambda \ge 1795$\AA).

The co-addition was then performed via modified exposure time
weighting. Finally, the combined \ac{fuv} and \ac{nuv} spectra were
re-binned to ensure Nyquist sampling (two pixels per resolution
element). Both are binned onto a linear wavelength scale with spacing
equal to 0.0395~\AA~for the \ac{fuv}, and a spacing equal to
0.436~\AA~for the \ac{nuv}.

\subsubsection{FOS data}
Individual exposures from \ac{fos} were downloaded from the \ac{stsci}
archive and reduced using the standard \textsc{calfos}
pipeline. Wavelength corrections given by \citet{Petry2006} were
applied to each individual exposure. As described by
\citeauthor{Petry2006}, these corrections were determined using a
wavelength calibration exposure taken contemporaneously with the G190H
grating science exposures, and were verified using Galactic \ion{Al}{2}
and \ion{Al}{3} absorption features. The shortest wavelength region of
the \ac{fos} G190H settings overlap with the longest wavelength
\ac{cos} settings, and we confirmed that the wavelength scales in these
overlapping regions were consistent between the two instruments. Then
we combined all individual exposures together, resampling to a common
wavelength scale of $0.51$ \AA\ per pixel.

\subsection{Continuum fitting}
We fit the continuum of each \ac{qso} in a semi-automatized and
iterative manner: (i) we first divide each spectrum in multiple chunks,
typically of $12$ \AA\ at wavelengths shorter than that of the
\hi~\lya~emission from the \acp{qso} (at larger wavelengths we used
much longer intervals but these are not relevant for the present work);
(ii) we then fit straight line segments through the set of points given
by the central wavelength and the median flux values for each chunk;
(iii) we then removed pixels with flux values falling $3 \times$ their
uncertainty below the fit value; (iv) we repeat steps (ii) and (iii)
until a converged solution is reached; (v) we fit a cubic spline
through the final set of median points to get a smooth continuum. The
success of this method strongly depends on the presence of emission
lines, and on number and positions of the chosen wavelength
chunks. Therefore, we visually inspect the solution and improve it by
adding and/or removing points accordingly, making sure that the
distribution of flux values above the continuum fit is consistent with
a Gaussian tail. We checked that the use of these subjective steps does
not affect the final results significantly (see \Cref{abs:check}).

In \Cref{fig:all_quasars} we show our \ac{qso} spectra (black lines)
with their corresponding uncertainties (green lines) and continuum fit
(red lines). We refer the reader to Finn et al. (2013, in prep.) for
further details on the continuum fitting process (including the
continuum fit associated to the peaks of the broad emission lines).


\section{Galaxy data}\label{data:galaxy}
\begin{table*}
  \begin{minipage}{0.8\textwidth}
    \centering
      \caption{Summary of galaxy observations
      (spectroscopy).\tablenotemark{a}}\label{tab:data:Galobs}
    \begin{tabular}{@{}ccccccc@{}}
      \hline
      
      Field Name & Instrument    & Gratting & Wavelength &  Dispersion & Exposure   &Reference   \\
                 &               &          & range (\AA)&  (\AA/pixel)& time (h)   &            \\
      (1)        & (2)           & (3)      & (4)        & (5)         &(6)         & (7)        \\ 
      
      \hline

      Q0107  & DEIMOS      & 1200l/mm   &6400-9100     &  0.3     &0.99     &  This paper\\ 
             & GMOS        & R400       &5000-9000     &  0.7     &0.90     &  This paper\\
             & VIMOS       & LR\_red    &5500-9500     &  7.3     &0.96     &  This paper\\
             & CFHT-MOS    & O300       &5000-9000     &  3.5     &0.83     &  \citet{Morris2006}\\
      J0209  & GMOS        &R150\_G5306 &5500-9200     &  3.4     &21       &  \ac{gdds}\\
      J1005  & VIMOS       & LR\_red    &5500-9500     &  7.3     &0.96   & This paper \\
             & VIMOS       & LR\_red    &5500-9500     &  7.3     &0.83   & \ac{vvds} \\
      J1022  & VIMOS       & LR\_red    &5500-9500     &  7.3     &0.96   &  This paper\\
      J1357  & VIMOS       & LR\_red    &5500-9500     &  7.3     &0.83   &  \ac{vvds}\\
      J2218  & VIMOS       & LR\_red    &5500-9500     &  7.3     &0.96   &  This paper\\
             & VIMOS       & LR\_red    &5500-9500     &  7.3     &0.83   &  \ac{vvds}\\
      
      \hline
      
    \end{tabular}
    \end{minipage}
  \begin{minipage}{0.8\textwidth}
    (1) Name of the field.
    (2) Instrument.
    (3) Gratting.
    (4) Wavelength range. 
    (5) Dispersion.
    (6) Exposure time of the observations.
    (7) Reference of the observations.\\
    $^{a}$ Redshift uncertainties for each instrument setup are described in \Cref{data:galaxy}.
  \end{minipage}
\end{table*}

Our chosen \acp{qso} are at $z_{\rm QSO} \sim 0.7-1.3$, so we aim to
target galaxies at $z\lesssim 1$, corresponding to the last $\sim 7$
Gyr of cosmic evolution. The majority of these \acp{qso} lie in fields
already surveyed for their galaxy content. We used archival galaxy data
from: the \ac{vvds} \citep{LeFevre2005,LeFevre2013}, \ac{gdds} \citep{Abraham2004}
and the \ac{cfht} \ac{mos} survey published by
\citet{Morris2006}. Despite the existence of some galaxy data around
our \ac{qso} fields we have also performed our own galaxy surveys using
\ac{mos} to increase the survey completeness\footnote{Note that the
  largest of these surveys, the \ac{vvds}, has a completeness of only
  about $20-25\%$.}. We acquired new galaxy data from different
ground-based \ac{mos}, namely: the \acf{vimos} \citep{LeFevre2003} on
the \ac{vlt} under programs 086.A-0970 (PI: Crighton) and 087.A-0857
(PI: Tejos); the \ac{deimos} \citep{Faber2003} on Keck under program
A290D (PIs: Bechtold and Jannuzi); and the \ac{gmos} \citep{Davies1997}
on Gemini under program GS-2008B-Q-50 (PI:
Crighton). \Cref{tab:data:Galobs} summarizes the observations taken to
construct our galaxy samples.

The following sections provide detailed descriptions of the
observations, data reduction, selection functions and construction of
our new galaxy samples. We also give information on the subsamples of
the previously published galaxy surveys used in this work.

\subsection{VIMOS data}\label{data:vimos}
\subsubsection{Instrument setting} 
We used the low-resolution (LR) grism with 1.0 arcsecond slits ($R
\equiv \lambda/\Delta \lambda \approx 200$) due to its high multiplex
factor in the dispersion direction (up to $4$). As we needed to target
galaxies up to the \acp{qso} redshifts ($z_{\rm{QSO}} \sim 0.7-1.3$),
we used that grism in combination with the OS-red filter giving
coverage between $5500 - 9500$ \AA.

\subsubsection{Target selection, mask design and pointings} 
We used \rband~pre-imaging to observe objects around our \ac{qso}
fields and \sextractor v2.5 \citep{Bertin1996} to identify them and
assign \rband~magnitudes, using zero points given by ESO. For fields
J1005, J1022 and J2218 we added a constant shift of $\sim 0.38$
magnitudes to match those reported by the \ac{vvds} survey in objects
observed by both surveys (see \Cref{vimos:vvds_comparison} and
\Cref{fig:VIMOS_VVDS_comparison}). No correction was added to the Q0107
field. For objects in fields J1005, J1022 and J2218 we targeted objects
at $R<23.5$, giving priority to those with $R<22.5$. For objects in
field Q0107 we targeted objects at $R<23$, giving priority to those
with $R<22$. We did not impose any morphological star/galaxy separation
criteria, given that unresolved galaxies will look like point sources
(see \Cref{gal:class_star}).
The masks were designed using the \vmmps \citep{Bottini2005} using the
`Normal Optimization' method (random) to provide a simple selection
function. We targeted typically $\sim 70-80$ objects per mask per
quadrant, equivalent to $\sim 210-320$ objects per pointing. We used
three pointings of one mask each, shifted by $\sim 2.5$ arcminutes
centred around the \ac{qso}.

\subsubsection{Data reduction for field Q0107} 
The spectroscopic data were taken in 2011 and the reduction was
performed using \vipgi \citep{Scodeggio2005} using standard
parameters. We took three exposures per pointing of $1155$ s, followed
by lamps. The images were bias corrected and combined using a median
filter. Wavelength calibration was performed using the lamp exposures,
and further corrected using five skylines at $5892$, $6300$, $7859$,
$8347$ and $8771$ \AA~\citep{Osterbrock1996,Osterbrock1997}.  Finally,
the slits were spectrophotometrically calibrated using standard star
spectra \citep{Oke1990,Hamuy1992,Hamuy1994} taken at dates similar to
our observations. The extraction of the one-dimensional (1D) spectra
was performed by collapsing objects along the spatial axis, following
the optimal weighting algorithm presented in \citet{Horne1986}. Our
wavelength solutions per slit show a quadratic mean $rms \lesssim 1$
\AA~in more than $75\%$ of the slits and a $rms \lesssim 2$ \AA~in all
the cases. We consider these as good solutions, given that the pixel
size for the low resolution mode is $\sim 7$ \AA. These data were taken
before the recent update of the \ac{vimos} \acp{ccd} on August 2010,
and so fringing effects considerably affected the quality of the data
at $\gtrsim 7500$ \AA. We attempted to correct for this with no
success.

\subsubsection{Data reduction for fields J1005, J1022 and J2218} 
The spectroscopic data were taken on 2011 and the reduction was
performed using \esorex v.3.9.6.  All three pointings of fields J1005
and J1022 were observed, while only `pointing 3' of J2218 was
observed. Due to a problem with focus, data from `quadrant 3' of
`pointing 1' and `pointing 3' of field J1022 were not usable. `Pointing
2' (middle one) of fields J1005 and J1022 were observed twice to
empirically asses the redshift uncertainty (see
\Cref{vimos:redshift}). We took three exposures per pointing of 1155 s
followed by lamps. The reduction was performed using a
\verb|peakdetection| parameter (threshold for preliminary peak
detection in counts) of $500$ when possible, and decreasing it when
needed to minimize the number of slits lost (we typically lost $\sim 1$
slit per quadrant). We also set the \verb|cosmics| parameter to `True'
(cleaning cosmic ray events) and stacked our 3 images using the
median. Wavelength calibration was further improved using four skylines
at $5577.34$, $6300.30$, $8827.10$ and $9375.36$
\AA~\citep{Osterbrock1996,Osterbrock1997} with the \verb|skyalign|
parameter set to 1 (1st order polynomial fit to the expected
positions). The slits were spectrophotometrically calibrated using
standard star spectra \citep{Oke1990,Hamuy1992,Hamuy1994} taken at
dates similar to our observations. The extraction of the
one-dimensional (1D) spectra was performed by collapsing the objects
along the spatial axis, following the optimal weighting algorithm
presented in \citet{Horne1986}. Our wavelength solutions per slit show
a quadratic mean $rms \lesssim 1$ \AA~in more than $90\%$ of the cases,
which we considered as satisfactory for a pixel size of $\sim 7$
\AA. These data were taken after a recent update to the \ac{vimos}
\acp{ccd} on August 2010, and so no important fringing effects were
present.

\subsubsection{Redshift determination}\label{vimos:redshift}
Redshifts for our new galaxy survey were measured by cross-correlating
galaxy, star, and \ac{qso} templates with each observed spectrum. We
used templates from the
\ac{sdss}\footnote{\url{http://www.sdss.org/dr7/algorithms/spectemplates/}}
degraded to the lower resolution of our \ac{vimos} observations. Galaxy
templates were redshifted from $z=0$ to $z=2$ using intervals of
$\Delta z = 0.001$. The \ac{qso} template was redshifted between $z=0$
to $z=4$ using larger intervals of $\Delta z=0.01$. Star templates were
shifted $\pm 0.005$ around $z=0$ using intervals of $\Delta z=0.0001$
to help improve the redshift measurements and quantify the redshift
uncertainty (see below). We improved the redshift solution by fitting a
parabola to the $3$ redshift points with the largest cross-correlation
values around each local maximum. This technique gives comparable
redshift solutions (within the expected errors) to that obtained by
decreasing the redshift intervals by a factor $\sim 10$, but at a much
lower computational cost. Before computing the cross-correlations, we
masked out regions at the very edges of the wavelength coverage
($<5710$ and $>9265$ \AA) and those associated with strong sky
emission/absorption features (between $5870-5910$, $6275-6325$ and
$7550-7720$ \AA). For the Q0107 field we additionally masked out the
red part at $> 7550$ \AA~because of fringing problems. We visually
inspected each 1-dimensional and 2-dimensional spectrum and looked for
the `best' redshift solution (see below).

\subsubsection{Redshift reliability}\label{gal:reliability} 
For each targeted object we manually assigned a redshift reliability
flag. We used a very simple scheme based on three labels: `a'
(`secure'), `b' (`possible') and `c' (`uncertain'). As a general rule,
spectra assigned with `a' flags have at least $3$ well identified
spectral features (either in emission or absorption) or $2$ well
identified emission lines; spectra assigned with `c' flag are those
which do not show clear spectral features either due to a low
signal-to-noise ratio or because of an intrinsic lack of such lines
observed at the \ac{vimos} resolution (e.g. some possible A, F and G
type stars appear in this category); spectra assigned with `b' flags
are those that lie in between the two aforementioned categories.

\subsubsection{Uncertainty of the semi-automatized process} 
The process includes subjective steps (determining the `best' template
and redshift, and assigning a redshift reliability). This uncertainty
was estimated by comparing two sets of redshifts obtained independently
by three of the authors (N.T. versus S.L.M. and N.T. versus N.H.M.C.)
in two subsamples of the data. We found discrepancies in $\lesssim 5\%$
of the cases, the vast majority of which were for redshifts labelled as
`b'.

\begin{figure}
  \includegraphics[width=0.5\textwidth]{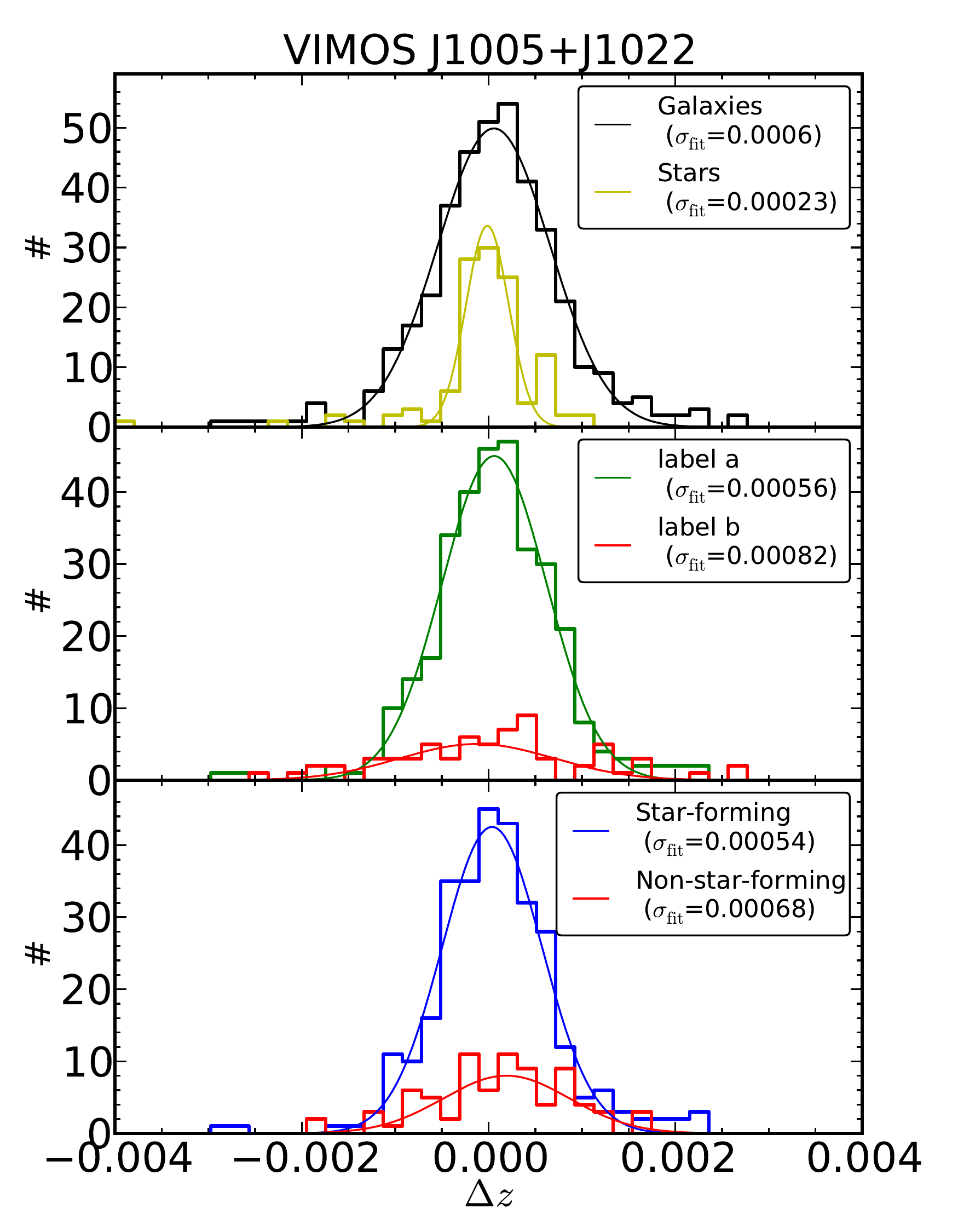}
  \caption{Histograms of the measured redshift difference between two
    independent observations of a same object in fields J1005 and
    J1022. Top panel: all identified galaxies (black lines) and stars
    (yellow lines). Middle panel: galaxies with `secure' redshifts
    (label `a'; green lines) and with `possible' redshifts (label `b';
    red lines). Bottom panel: galaxies classified as `star-forming'
    (blue lines) and as `non-star-forming' (red lines; see
    \Cref{gal:sclass}). Best Gaussian fits to the histograms and
    standard deviation values are also shown.}

  \label{fig:VIMOS_autocomparison}
\end{figure}

\subsubsection{Further redshift calibration for fields J1005, J1022 and J2218}\label{vimos:redshift_calib}
Even though the wavelength calibration from the \esorex reduction was
generally satisfactory, we found a $\sim 1$ pixel systematic
discrepancy between the obtained and expected wavelength for some
skylines in localized areas of the spectrum (particularly towards the
red end). This effect was most noticeable in quadrant 3, where the
redshift difference between objects observed twice showed a
distribution displaced from zero by $\sim 0.001$ ($\sim 1$ pixel). A
careful inspection revealed that the other quadrants also showed a
similar but less strong effect ($\lesssim$ 0.5 pixel). We corrected for
this effect using the redshift solution of the stars. For a given
quadrant we looked at the mean redshift of the stars and applied a
systematic shift of that amount to all the objects in that
quadrant. This correction placed the mean redshift of stars at zero,
and therefore corrected the redshift of all objects accordingly.

\subsubsection{Redshift statistical uncertainty for fields J1005, J1022 and
  J2218} In order to assess the redshift uncertainty for these fields,
we measured a redshift difference between two independent observations
of the same object. These objects were observed twice, and come mainly
from our `pointing 2' in fields J1005 and J1022, but there is also a
minor contribution ($\lesssim 10\%$) of objects that were observed
twice using different pointings. \Cref{fig:VIMOS_autocomparison} shows
the observed redshift differences for all galaxies and stars (top
panel); galaxies with `secure' and `possible' redshifts (middle panel);
and galaxies classified as `star-forming' (\SF) or `non-star-forming'
(\nSF) based on the presence of current, or recent, star formation (see
\Cref{gal:sclass}; bottom panel). All histograms are centred around
zero and do not show evident systematic biases. The redshift difference
of all galaxies show a standard deviation of $\approx 0.0006$. A
somewhat smaller standard deviation is observed for galaxies with
`secure' redshifts and/or those classified as \SF~(note that there is a
large overlap between these two samples), and consequently a somewhat
larger standard deviation is observed for galaxies with `possible'
redshift and/or classified as \nSF. This behaviour is of course
expected, as it is simpler to measure redshifts for galaxies with
strong emission lines (for which the peak in the cross-correlation
analysis is also better constrained) than for galaxies with only
absorption features (at a similar signal-to-noise ratio). From this
analysis we take $\approx 0.0006/\sqrt{2} = 0.0004$ as the
representative redshift uncertainty of our \ac{vimos} galaxy survey in
these fields. This uncertainty corresponds to $\approx 120-60$ \kms~at
redshift $z=0-1$. This uncertainty is $\sim 2$ times smaller than that
claimed for the \ac{vvds} survey \citep{LeFevre2005,LeFevre2013}.

\begin{figure}
  \includegraphics[width=0.47\textwidth]{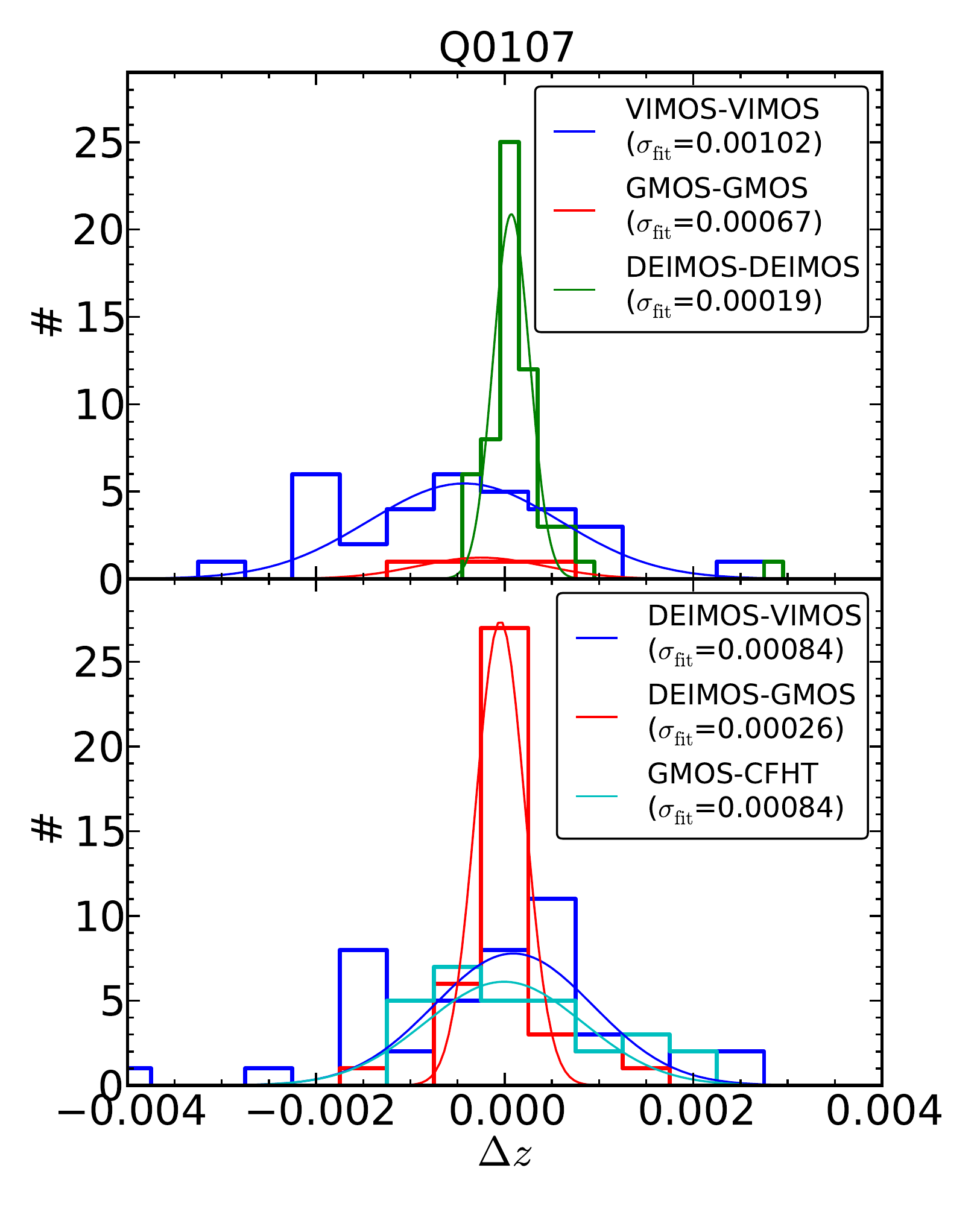}
  \caption{Histograms of the measured redshift difference between two
    independent observations of the same object in field Q0107. Top
    panel shows it for galaxies observed twice by the same instrument:
    \ac{vimos}-\ac{vimos} (blue lines), \ac{gmos}-\ac{gmos} (red lines)
    and \ac{deimos}-\ac{deimos} (green lines). Bottom panel shows it
    for objects observed twice by different instruments, after shifting
    to match the \ac{deimos} mean: \ac{deimos}-\ac{vimos} (blue lines),
    \ac{deimos}-\ac{gmos} (red lines) and \ac{gmos}-\ac{cfht} (cyan
    lines). Best Gaussian fits to the histograms and standard deviation
    values are also shown.}

  \label{fig:redshift_autocomparison_Q0107}
\end{figure}

\subsubsection{Further redshift calibration for field Q0107}\label{}
We did not see systematic differences between quadrants, as was seen
for fields J1005, J1022 and J2218. \ac{vimos} observations of the Q0107
field were reduced differently, and the data come mainly from the blue
part of the spectrum. Therefore, such an effect might not be present
or, if present, might be more difficult to detect. However, we did find
a systematic shift between the redshifts measured from \ac{vimos}
compared to those measured from \ac{deimos}. Given the much higher
resolution of \ac{deimos}, we used its frame as reference for all our
Q0107 observations. Thus, we corrected the Q0107 \ac{vimos} redshifts
to match the \ac{deimos} frame. This correction was $\sim 0.0008$
($\lesssim 1$ \ac{vimos} pixel) and the result is shown in the bottom
panel of \Cref{fig:redshift_autocomparison_Q0107} (blue lines).

\subsubsection{Redshift statistical uncertainty for field Q0107} 
In order to assess the redshift uncertainty, we used objects that were
observed twice in the Q0107 field. We found a distribution of redshift
differences centred at $\sim 0$ with a standard deviation of $\approx
0.001$ (see top panel of \Cref{fig:redshift_autocomparison_Q0107}),
corresponding to a single \ac{vimos} uncertainty of $\approx
0.001/\sqrt{2} \approx 0.0007$.  Another way to estimate the \ac{vimos}
uncertainty in the Q0107 field is by looking at the redshift difference
for objects that were observed twice, once by \ac{vimos} and another
time by \ac{deimos} (44 in total; see bottom panel of
\Cref{fig:redshift_autocomparison_Q0107}). In this case, the
distribution shows a standard deviation of $\approx \sqrt{0.00084}$,
corresponding to a single \ac{vimos} uncertainty of
$\sqrt{0.00084^2-0.00013^2} \approx 0.0008$, given that the uncertainty
of a \ac{deimos} single measurement is $\approx 0.00013$ (see
below). So, we take a value of $\approx 0.00075$ as the representative
redshift uncertainty of a single \ac{vimos} observation in the Q0107
field. This uncertainty corresponds to $\approx 220-110$ \kms~at
redshift $z=0-1$. This uncertainty is larger than that of fields J1005,
J1022 and J2218, consistent with the poorer quality detector being
used.


\subsection{DEIMOS data}\label{data:deimos}
\subsubsection{Instrument setting} 
We patterned our \ac{deimos} observations to resemble the Deep
Extragalactic Evolutionary Probe 2 (DEEP2) `1 hour' survey
\citep{Coil2004}. We used the $1200$ line mm$^{-1}$ grating with a 1.0
arcsecond slit giving a resolution of $R\sim 5000$ over the wavelength
range $6400-9100$ \AA.

\subsubsection{Target selection}

We used $B$, $R$ and $I$ bands~pre-imaging to select objects around our
Q0107 field. We used \sextractor v2.5 \citep{Bertin1996} to identify
them and assign $B$, $R$ and $I$ magnitudes to them. We used color cuts
as in \citet[][see also \citealt{Newman2012}]{Coil2004} to target
galaxies\footnote{Note that \citet{Coil2004} presented $B-R \le 0.5$
  but should have been $B-R \ge 0.5$, which is what we used.}:

\begin{equation}
\begin{split}
B - R & \le  2.35 (R-I) - 0.45 \ \rm{or} \\
R - I & \ge  1.15  \ \rm{or}  \\
B − R & \ge  0.5 \ \rm{.}
\end{split}
\end{equation}

\noindent We also gave priority to objects within 1 arcminute of the
Q0107-025A \ac{los}. We targeted objects up to $R=24.5$ magnitudes, but
we assigned higher priorities to the brightest ones. In an attempt to
be efficient, we also imposed a star/galaxy morphological criteria of
\verb|CLASS_STAR|$< 0.97$ (although see
\Cref{gal:class_star})\footnote{The parameter CLASS\_STAR assigns a
  value of $1$ to objects that morphologically look like stars, and a
  value of $0$ to objects that look like galaxies. Values in between
  $1$ and $0$ are assigned for less certain objects
  \citep{Bertin1996}.}.

\subsubsection{Data reduction} 
The observations were taken in 2007 and 2008. The reduction was
performed using the DEEP2 DEIMOS Data
Pipeline\footnote{\url{http://astro.berkeley.edu/~cooper/deep/spec2d/}}
\citep{Newman2012}, from which galaxy redshifts were also obtained.

\subsubsection{Redshift reliability} 
The redshift reliability for \ac{deimos} data was originally based on
four subjective categories: (0) `still needs work', (1) `not good
enough', (2) `possible', (3) `good' and (4) `excellent'. In order to
have a unified scheme we matched those \ac{deimos} labels with our
previously defined \ac{vimos} ones (see \Cref{vimos:redshift}) as
follows: \ac{deimos} label 4 is matched to label `a'
(\{4\}$\rightarrow$ \{`a'\}); \ac{deimos} labels 3 and 2 are matched to
label `b' (\{3,2\}$\rightarrow$ \{`b'\}); and \ac{deimos} labels 1 and
0 are matched to label `c' (\{1,0\}$\rightarrow$ \{`c'\}).

\subsubsection{Redshift statistical uncertainty for field Q0107} 
In order to assess the redshift uncertainty, we used objects that were
observed twice in the Q0107 field. We found a distribution of redshift
differences centred at $\sim 0$ with a standard deviation of $\approx
0.00019$ (see top panel of \Cref{fig:redshift_autocomparison_Q0107}),
corresponding to a single \ac{deimos} uncertainty of $\approx
0.00019/\sqrt{2} \approx 0.00013$. So, we take a value of $\approx
0.00013$ as the representative redshift uncertainty of a single
\ac{deimos} observation in the Q0107 field. This uncertainty
corresponds to $\approx 40-20$ \kms~at redshift $z=0-1$.


\subsection{GMOS data}\label{data:gmos}

\subsubsection{Instrument setting} 
We used the R400 grating centred on a wavelength of 7000~\AA\ with a
1.5 arcseconds slit giving a resolution of $R=639$.

\subsubsection{Target selection, mask design and pointings} 
We used \rband~pre-imaging to select objects around our Q0107 field. We
used \sextractor v2.5 \citep{Bertin1996} to identify objects and assign
them \rband magnitudes. The masks were designed using
\gmmps\footnote{\url{http://www.gemini.edu/?q=node/10458}}. Top
priority was given to objects with $R < 22$, followed by those with
$22\le R< 23$ and last priority to those with $23 \le R < 24$.  We
typically targeted $\sim 40$ objects per mask. Six masks were taken,
three around \ac{qso} C, two around \ac{qso} B, and one around \ac{qso}
A, where many objects had already been targeted in previous
observations.

\subsubsection{Data reduction} 
The observations were taken in 2008
. Three 1080~s offset science exposures were taken for each mask,
dithered along the slit to cover the gaps in the \ac{ccd}
detectors. Arcs were taken contemporaneously to the science
exposures. We used the Gemini \ac{iraf} package to reduce the spectra.
A flat-field lamp exposure was divided into each bias-subtracted
science exposure to remove small-scale variations across the \acp{ccd},
and the fringing pattern seen at red wavelengths. The dithered images
(both arcs and science) were then combined into a single exposure. The
spectrum for each mask was wavelength-calibrated by identifying known
arc lines and fitting a polynomial to match pixel positions to
wavelengths.  Finally the wavelength-calibrated 2-d spectra were
extracted to produced 1-d spectra. The typical $rms$ scatter of the
known arc line positions around the polynomial fit ranged from 0.5 to
1.0 \AA, depending on how many arc lines were available to fit (bluer
wavelength ranges tended to have fewer arc lines). A 0.75~\AA~$rms$
scatter corresponds to a velocity error of 38~\kms at 6000~\AA.

\subsubsection{Redshift determination and reliability}\label{gmos:redshift}

We determined redshifts by using the same method to that of the
\ac{vimos} spectra: plausible redshifts were identified as peaks in the
cross-correlation measured between the \ac{gmos} spectra and spectral
templates (see \Cref{vimos:redshift} for further details). Redshifts
reliabilities were also assigned following the definitions in our
\ac{vimos} sample.

\subsubsection{Further redshift calibration} 
We found a systematic shift of the redshifts measured from \ac{gmos}
with respect to those measured from \ac{deimos} for the 40 objects
observed by these two instruments. Given the much higher resolution of
\ac{deimos} we used its frame as reference for our Q0107
observations. Thus, we corrected all \ac{gmos} redshifts to match the
\ac{deimos} frame. This correction was $\sim 0.0004$ or $\sim 80$~\kms
($\lesssim 1$ \ac{gmos} pixel) and the result is shown in the bottom
panel of \Cref{fig:redshift_autocomparison_Q0107} (red lines).

\subsubsection{Redshift statistical uncertainty for field Q0107} 

There were only 3 objects that were observed twice using \ac{gmos} (see
top panel of \Cref{fig:redshift_autocomparison_Q0107}), and so we did
not take the uncertainty from such an small sample. Instead, we use
objects observed by both \ac{gmos} and \ac{deimos} to estimate the
\ac{gmos} redshift uncertainty. The distribution of redshift
differences for objects with both \ac{gmos} and \ac{deimos} spectra
(see bottom panel of \Cref{fig:redshift_autocomparison_Q0107}) shows a
standard deviation of $\approx 0.00027$. Given that the uncertainty of
\ac{deimos} alone is $\approx 0.00013$ we estimate the \ac{gmos}
uncertainty to be $\approx \sqrt{0.00027^2 -0.00013^2 }\approx
0.00024$. This uncertainty corresponds to $\approx 70-35$ \kms~at
redshift $z=0-1$.

\subsection{CFHT MOS data}\label{data:cfht}
We used the \ac{cfht} galaxy survey of the Q0107 field presented by
\citet{Morris2006}. There are $61$ galaxies in this sample, $29$ of
which were also observed by our \ac{gmos} survey. We use only redshift
information from this sample without assigning a particular template or
redshift label. We refer the reader to \citet{Morris2006} for details
on the data reduction and construction of the galaxy sample.

\subsection{VVDS}\label{data:vvds}
Three of the \acp{qso} presented in this paper (namely:
J100535.24+013445.7, J135726.27+043541.4 and J221806.67+005223.6) were
chosen because they lie in fields already surveyed for galaxies by the
\ac{vvds} survey \citep{LeFevre2005,LeFevre2013}. For our purposes, we
use a subsample of the whole \ac{vvds} survey, selecting only galaxies
in those fields. We refer the reader to \citet{LeFevre2005} and
\citet{LeFevre2013} for details on the data reduction and construction
of these galaxy catalogs.

\subsubsection{Redshift reliability} 
The redshift reliability for \ac{vvds} data was originally based on six
categories: (0) `no redshift', (1) `50\% confidence'; (2) `75\%
confidence'; (3) `95\% confidence'; (4) `100\% confidence'; (8) `single
emission line'; and (9) `single isolated emission line'
\citep{LeFevre2005,LeFevre2013}. They expanded this classification
system for secondary targets (objects which are present by chance in
the slits) by the use of the prefix `2'. Similarly the prefix `1' means
`primary \ac{qso} target', while the prefix `21' means `secondary
\ac{qso} target'. In order to have a unified scheme we matched those
\ac{vvds} labels with our previously defined \ac{vimos} ones (see
\Cref{vimos:redshift}) as follows: \ac{vvds} label 4, 3 and their
corresponding extensions are matched to label `a'
(\{4,14,24,214,3,13,23,213\}$\rightarrow$ \{`a'\}); \ac{vvds} labels 2,
9 and their corresponding extensions are matched to label `b'
(\{2,12,22,212,9,19,29,219\}$\rightarrow$ \{`b'\}); and \ac{vvds}
labels 1, 0 and their corresponding extensions are matched to label `c'
(\{1,11,21,211,0,10,20,210\}$\rightarrow$ \{`c'\}).

\begin{figure*}
  \begin{minipage}{0.4\textwidth}
  \includegraphics[width=1\textwidth]{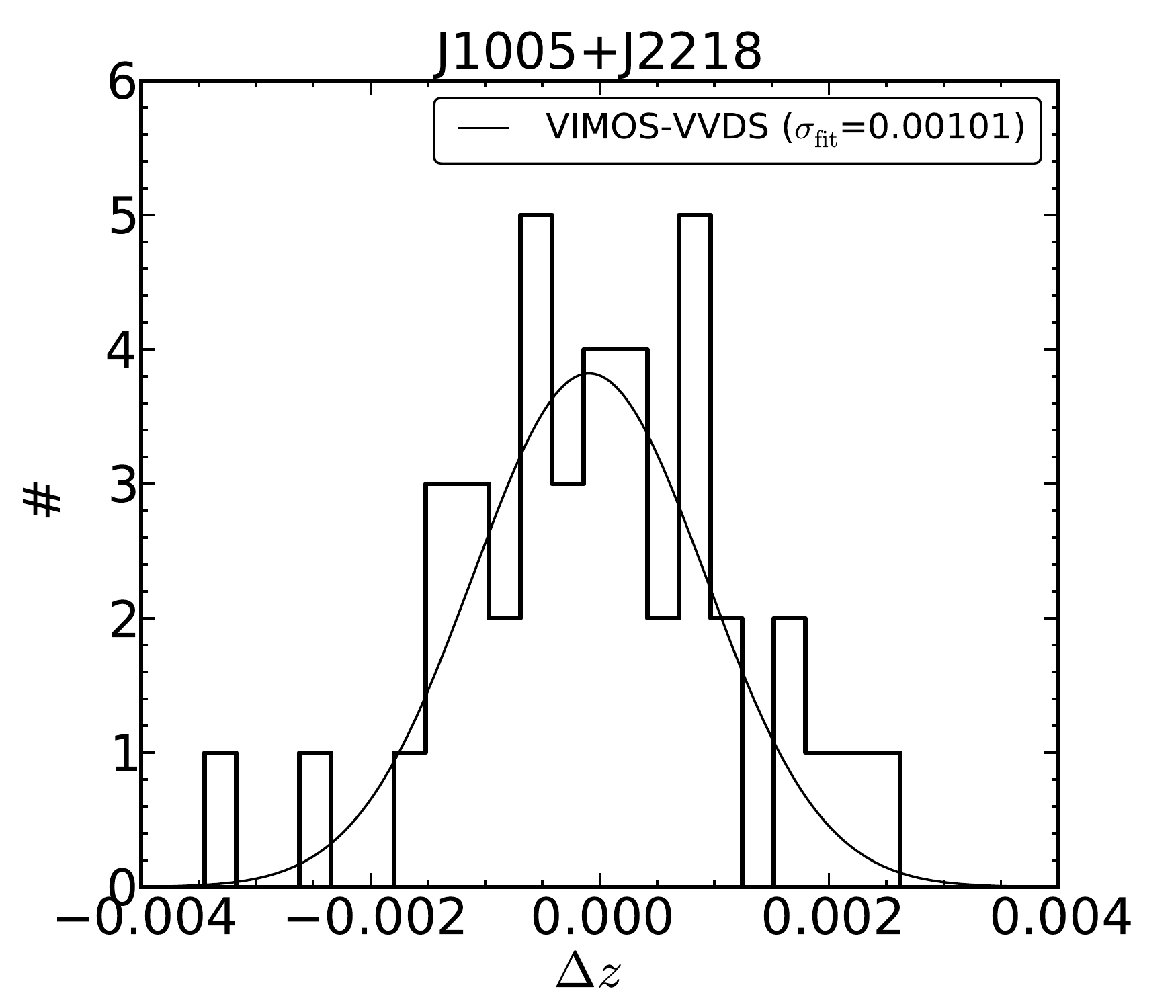}
  \end{minipage}
  \begin{minipage}{0.4\textwidth}
  \includegraphics[width=1\textwidth]{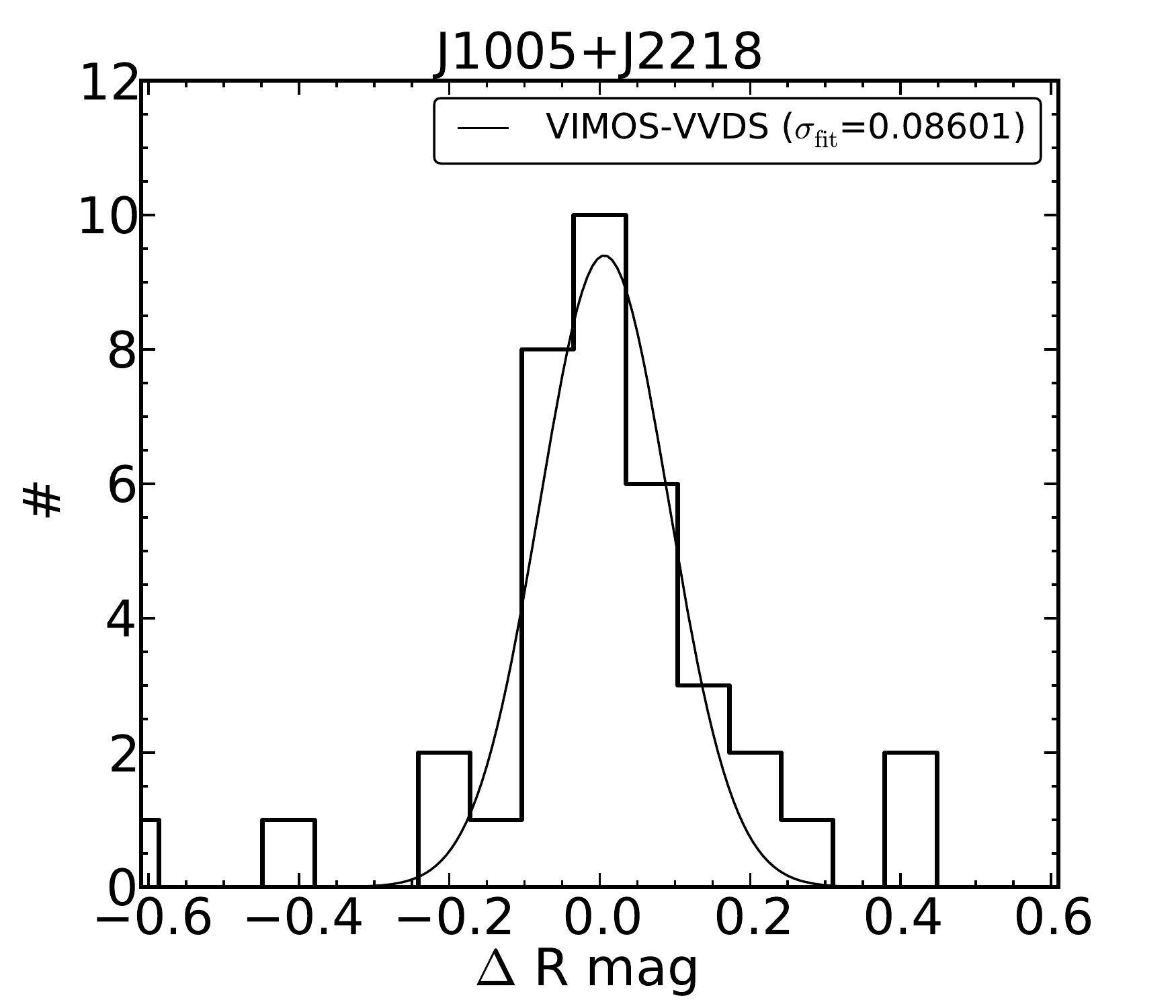}
  \end{minipage}
    
  \caption{Difference in redshift (left panel) and \rband magnitude
    (right panel) measurements for galaxies in common between our VIMOS
    sample and the VVDS survey in fields J1005 and J2218. Best Gaussian
    fits to the histograms and standard deviation values are also
    shown. We see a good agreement in both redshift and magnitude
    measurements between the two surveys. The redshift difference
    distribution has a mean of $\lesssim 0.0001$ and a standard
    deviation of $\sigma_{\Delta z} \approx 0.001$, while the magnitude
    difference distribution has a mean of $\approx 0.006$ with a
    standard deviation of $\sigma_{\Delta {\rm R}} \approx 0.09$
    magnitudes. See \Cref{vimos:vvds_comparison} for further
    details.} \label{fig:VIMOS_VVDS_comparison}

\end{figure*}

\subsubsection{Consistency check between our VIMOS and VVDS sample}\label{vimos:vvds_comparison}

We performed a consistency check by comparing the redshifts and \rband
magnitudes obtained for galaxies in common between our \ac{vimos}
sample and the \ac{vvds} survey in fields J1005 and J2218 (the only
ones with such overlap). We found a good agreement in redshift
measurements between the two surveys, with a mean of the distribution
being $\approx 0.0003$ and a standard deviation of $\sigma_{\Delta z}
\approx 0.001$. This standard deviation is consistent with the
quadratic sum of the typical \ac{vvds} uncertainty ($\sim
0.0013/\sqrt{2}$) and our \ac{vimos} one ($\sim 0.0006/\sqrt{2}$), as
$\sim \sqrt{0.0006^2+0.0013^2}/\sqrt{2} \approx 0.001$.  In order to
place all galaxies in a single consistent frame we shifted the
\ac{vvds} redshifts by $0.0003$. The left panel of
\Cref{fig:VIMOS_VVDS_comparison} shows the distribution of these
redshift differences after applying the correction.

The right panel of \Cref{fig:VIMOS_VVDS_comparison} shows the
distribution of \rband~magnitude differences. We also see a good
agreement in the magnitude difference distribution (by construction,
see \Cref{data:vimos}), with a mean of $0.006$ and a standard deviation
of $\sigma_{\Delta z} \approx 0.09$. We note that this standard
deviation is greater than $\sqrt{2} \times$ the typical magnitude
uncertainty as given by \sextractor of $\sim 0.02$. Thus, we caution
the reader that our reported \rband~magnitude uncertainties might be
underestimated by a factor of $\sim 3$.


\subsection{GDDS}\label{data:gdds}
One of the \acp{qso} presented in this paper (namely: J020930.7-043826)
was chosen because it lies in a field already surveyed for galaxies by
the \ac{gdds} survey. For our purposes we use a subsample of the whole
\ac{gdds} survey selecting only galaxies in this field. We refer the
reader to \citet{Abraham2004} for details on data reduction and
construction of this galaxy catalog.

\subsubsection{Redshift reliability} 
The redshift reliability for \ac{gdds} data was originally based on
five subjective categories: (0) `educated guess', (1) `very insecure';
(2) `reasonably secure' (two or more spectral features); (3) `secure'
(two or more spectral features and continuum); (4) `unquestionably
correct'; (8) `single emission line' (assumed to be \ion{O}{2}); and
(9) `single emission line' \citep{Abraham2004}. In order to have a
unified scheme we matched those \ac{gdds} labels with our previously
defined \ac{vimos} ones (see \Cref{vimos:redshift}) as follows:
\ac{gdds} label 4 and 3 are matched to label `a' (\{4,3\}$\rightarrow$
\{`a'\}); \ac{gdds} labels 2, 8 and 9 are matched to label `b'
(\{2,8,9\}$\rightarrow$ \{`b'\}); and \ac{gdds} labels 1 and 0 are
matched to label `c' (\{1,0\}$\rightarrow$ \{`c'\}).



\section{IGM samples}\label{abs:samples}

\subsection{Absorption line search}\label{abs:search}
The search of absorption line systems in the continuum normalized
\ac{qso} spectra was performed manually (eyeballing), based on a
iterative process described as follows: (i) we first searched for all
possible features (\hi and metal lines) at redshift $z=0$ and $z=z_{\rm
  QSO}$, and labelled them accordingly. (ii) We then searched for strong
\hi~absorption systems, from $z=z_{\rm QSO}$ until $z=0$, showing at
least $2$ transitions (e.g. \lya~and \lyb or \lyb and \lyc, and so
on). This last condition allowed us to identify (strong) \hi~systems at
redshifts greater than $z> 0.477$ even for spectra without \ac{nuv}
coverage ($\lambda > 1795$ \AA). (iii) When a \hi~system is found, we
labelled all the Lyman series transitions accordingly and looked for
possible metal transitions at the same redshift. (iv) We then performed
a search for `high-ionization' doublets (namely: \ion{Ne}{8},
\ion{O}{6}, \ion{N}{5}, \ion{C}{4} and \ion{Si}{4}), from $z=z_{\rm
  QSO}$ until $z=0$, independently of the presence of \hi. (v) We
assumed the remaining unidentified features to be \hi~\lya~and repeated
step (iii), unless there is evidence indicating otherwise (e.g. no
detection of the \lyb transition when the spectral coverage and
signal-to-noise would allow it). For all of the identified transitions
we set initial guesses in number of velocity components, column
densities and Doppler parameters, for a subsequent Voigt profile
fitting.

This algorithm allowed us to identify the majority but not all the
absorption line systems observed in our \ac{qso} spectral sample. The
remaining unidentified features are typically very narrow and
inconsistent with being \hi (assuming a minimum temperature of the
diffuse \ac{igm} of $T\sim 10^{4}$ K, implies a $b_{\rm HI}\sim 10$
\kms; e.g. \citealt{Dave2010}), so we are confident that our \hi~sample
is fairly complete.

\begin{figure*}
  \begin{minipage}{0.32\textwidth}
    \centering
  \includegraphics[width=1\textwidth]{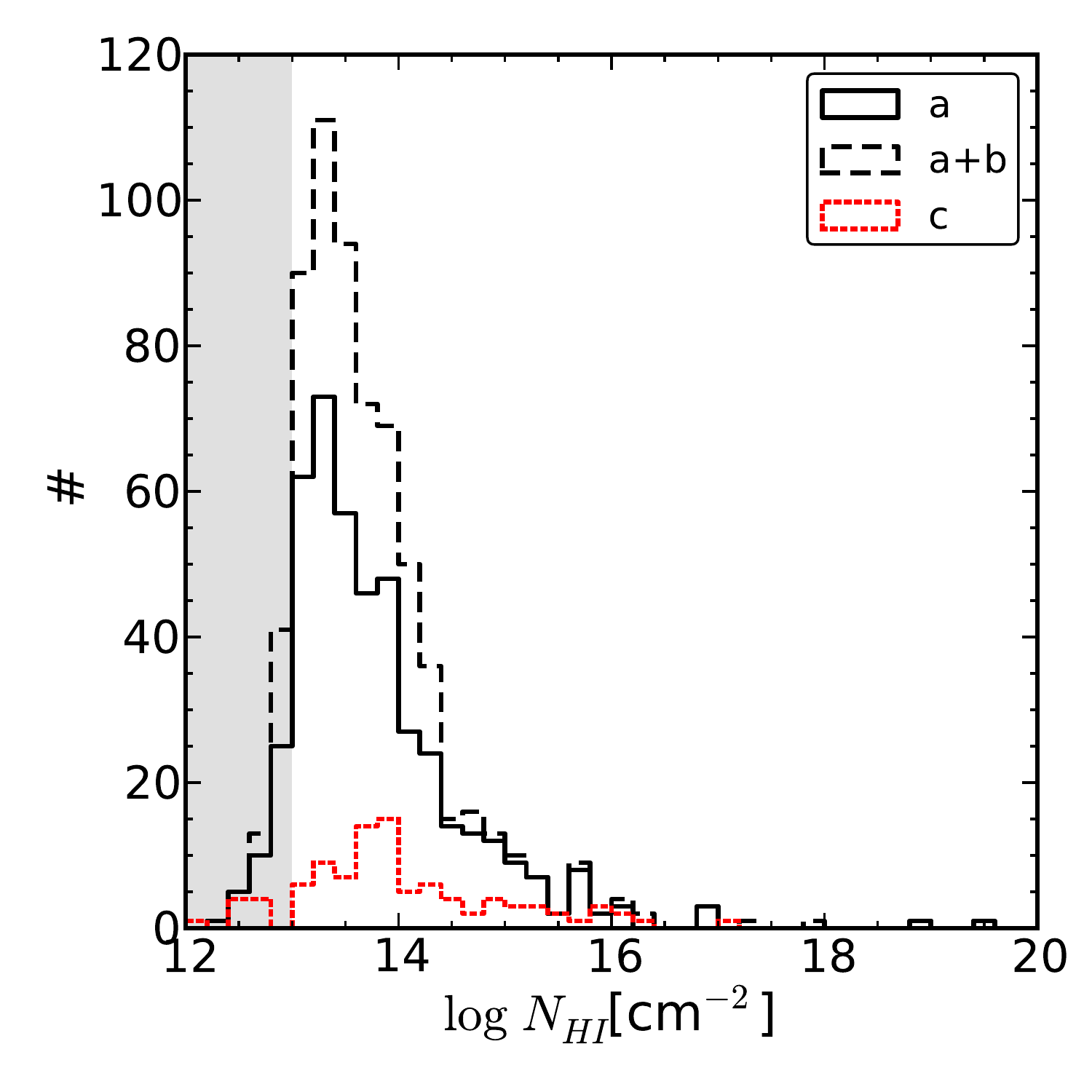}
  \end{minipage}
  \begin{minipage}{0.32\textwidth}
    \centering
  \includegraphics[width=1\textwidth]{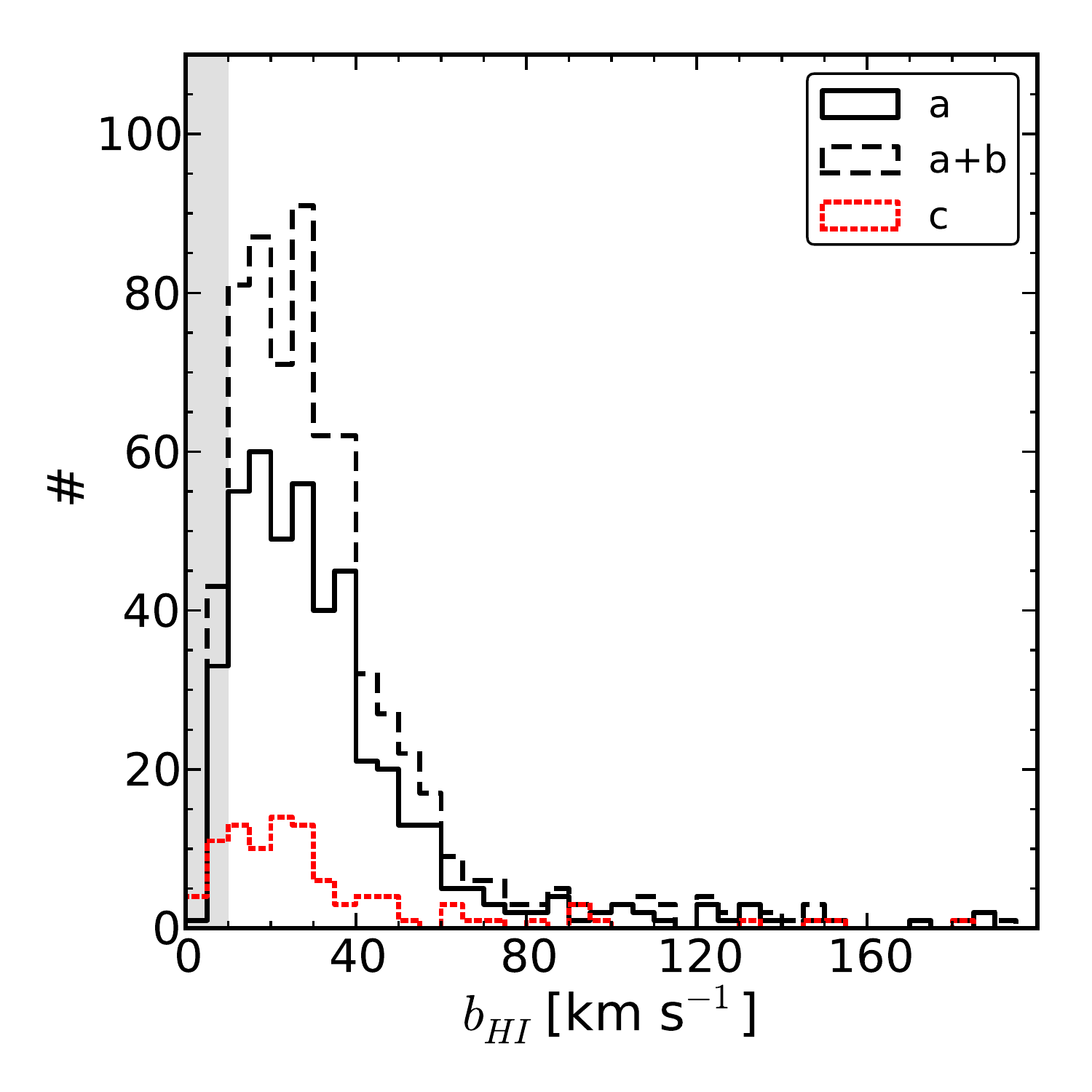}
  \end{minipage}
  \begin{minipage}{0.32\textwidth}
    \centering
  \includegraphics[width=1\textwidth]{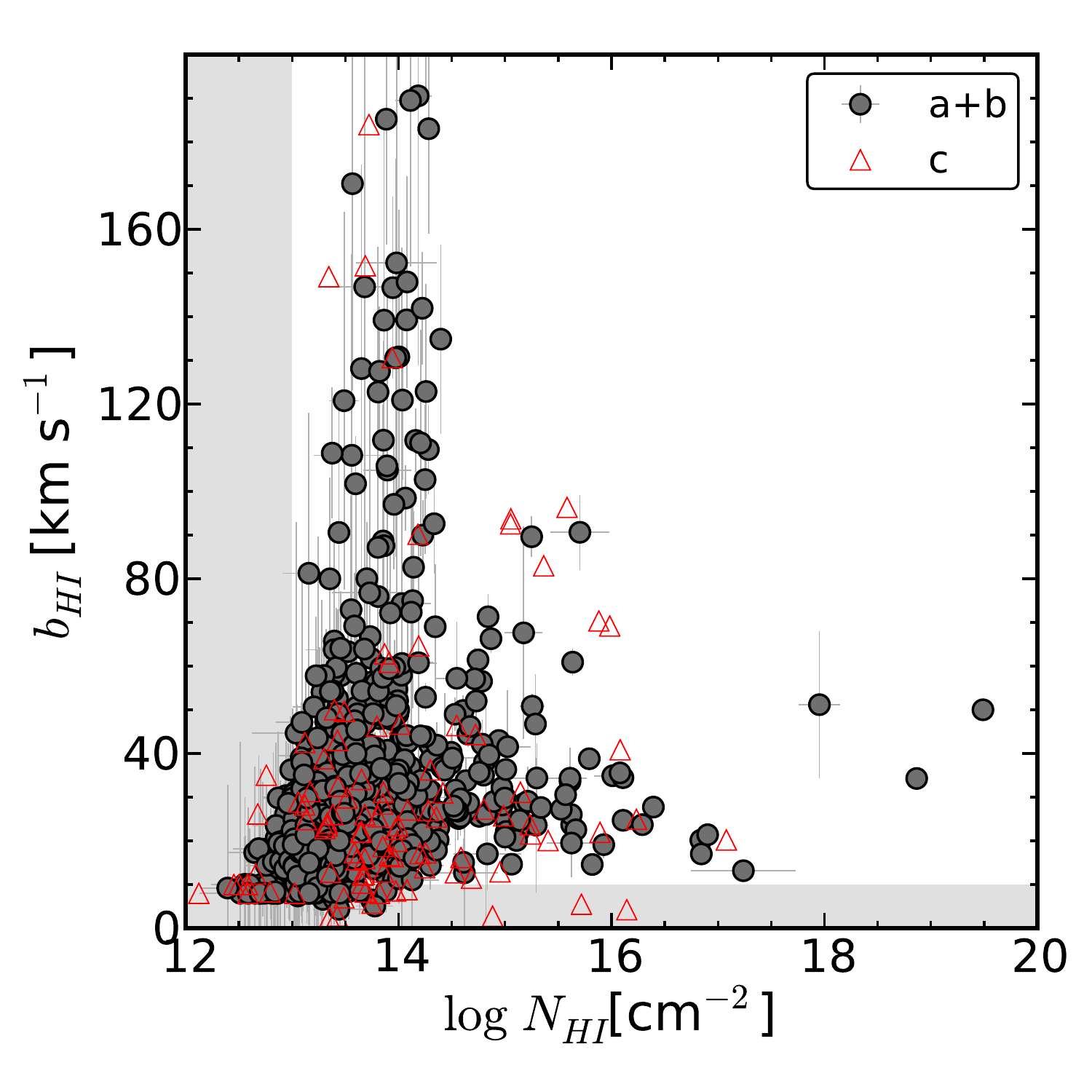}
  \end{minipage}
  
  \caption{The first two panels show the observed \hi column density
    (\nhi; left panel) and Doppler parameter (\bhi; middle panel)
    distributions for `secure' systems (`a' label; black solid lines),
    `secure' plus `probable' systems (`a$+$b' labels; dashed black
    lines), and `uncertain' systems (`c' label; dotted red lines; see
    \Cref{abs:reliability} for definitions of these labels). The right
    panel shows the distribution of Doppler parameters as a function of
    column density for `secure' plus `probable' systems (`a$+$b'
    labels; grey circles), and `uncertain' systems (`c' label; red open
    triangles; uncertainties not shown). Grey shaded areas show regions
    with low completeness levels. For further details see
    \Cref{abs:completeness}.}\label{fig:logn_b_dist}

\end{figure*}

\subsection{Voigt profile fitting}
We fit Voigt profiles to the identified absorption line systems using
\vpfit\footnote{\url{http://www.ast.cam.ac.uk/~rfc/vpfit.html}}. We
accounted for the non-Gaussian \ac{cos} \ac{lsf}, by interpolating
between the closest \ac{cos} \ac{lsf} tables provided by
\ac{stsci}\footnote{\url{http://www.stsci.edu/hst/cos/performance/spectral_resolution}}
at a given wavelength. We used the guesses provided by the absorption
line search (see \Cref{abs:search}) as the initial input of \vpfit, and
modified them when needed to reach satisfactory solutions. For
intervening absorption systems we kept solutions having the least
number of velocity components needed to minimize the reduced
$\chi^2$.\footnote{Our typical reduced $\chi^2$ values are on the order
  $\lesssim 1.2$.} For fitting \hi~systems, we used at least two
spectral regions associated to their Lyman series transitions when the
spectral coverage allowed it. This means that for \hi~systems showing
only \lya transition, we also included their associated \lyb regions
(even though they do not show evident absorption) when available. This
last step provides confident upper limits to the column density of
these systems. For strong \hi systems we used regions associated to as
many Lyman series transitions as possible, but excluding spectral
regions of poor signal-to-noise ($S/N \lesssim 1$). We refer the reader
to Finn et al. (2013, in prep.)  for further details on the Voigt
profile fitting process.

In the following we will present only results for \hi~systems; a
catalog of metal systems will be published elsewhere.

\subsection{Absorption line reliability}\label{abs:reliability}

For each \hi~absorption system we assigned a reliability flag. We used
a scheme based on three labels:

\begin{itemize}
\item {\it Secure (`a'):} systems at redshifts that allow the detection
  of either \lya~and \lyb or \lyb~and \lyc transitions in a given
  spectrum, whose $\log N_{\rm HI}$ values are greater than $30\times$
  their uncertainties as quoted by \vpfit.\smallskip

\item {\it Probable (`b'):} systems at redshifts that only allow the
  detection of the \lya transition in a given spectrum, whose $\log
  N_{\rm HI}$ values are greater than $30\times$ their uncertainties as
  quoted by \vpfit.\smallskip

\item {\it Uncertain (`c'):} systems at any redshift, whose $\log
  N_{\rm HI}$ values are smaller than $30\times$ their uncertainties as
  quoted by \vpfit. Systems in this category will be excluded from the
  correlation analyses presented in this paper.\smallskip
\end{itemize}

\subsection{Consistency check of subjective steps}\label{abs:check}

The whole process of finding and characterizing \ac{igm} absorption
lines involves subjective steps. We checked that this fact does not
affect our final results by comparing redshift, column density and
Doppler parameter values for \hi~systems obtained
independently---including the continuum fitting---by two of the authors
(N.T. versus C.W.F.) in the J020930.7-043826 \ac{qso} spectrum. We
found values consistent with one another at the $1\sigma$ level in
$\sim 90\%$ of cases for $\log N_{\rm HI}$ and $b_{\rm HI}$, and in
$100\%$ of cases for redshifts. The vast majority of discrepancies were
driven by weak absorption systems close to the level of detectability,
for which the differences in the continuum fitting are more important.

\subsection{\nhi and \bhi distributions and completeness}\label{abs:completeness}

In \Cref{fig:logn_b_dist} we show the observed \hi column density
(\nhi; left panel) and Doppler parameter (\bhi; middle panel)
distributions for `secure' systems (`a' label; black solid lines),
`secure' plus `probable' systems (`a$+$b' labels; dashed black lines),
and `uncertain' systems (`c' label; dotted red lines; see
\Cref{abs:reliability}). We see sudden decreases in the number of
systems at $N_{\rm HI} \lesssim 10^{13}$ \cm and $b_{\rm HI} \lesssim
10$ \kms, which indicate the observational completeness limits of our
sample and/or our selection (shown as grey shaded areas in
\Cref{fig:logn_b_dist}).

Theoretical results point out that the \hi column density distribution
is well described by a power law of the form $f(N_{\rm HI}) \propto
N_{\rm HI}^{-\beta}$ with $\beta \sim - 1.7-1.8$, extending
significantly below $\sim 10^{13}$ \cm
\citep[e.g.][]{Theuns1998a,Paschos2009,Dave2010,Tepper-Garcia2012}. This
has been observationally confirmed from higher signal-to-noise ratio
data ($S/N \sim 20-40$) at least down to $N_{\rm HI} \sim 10^{12.3}$
\cm \citep{Williger2010}. Our current \nhi completeness limit is
therefore not physical, and driven by the signal-to-noise ratio of our
sample. Indeed, using the results from \citet{Keeney2012}, the expected
minimum rest-frame equivalent width for \hi lines detected in the
\ac{fuv}-\ac{cos}---in which the majority of weak lines are
detected---at the $3\sigma$ confidence level, for our typical
signal-to-noise ratio ($S/N \sim 10$; see \Cref{tab:data:HSTobs}), is
$\sim 40$ m\AA. This limit corresponds to $N_{\rm HI} \sim 10^{13}$~\cm
for a typical Doppler parameter of $b_{\rm HI} \sim 30$~\kms, which is
consistent with what we observe.

The same theoretical results point out that the \hi Doppler parameter
distribution for the diffuse \ac{igm} peaks at $\sim 20-40$ \kms, with
almost negligible contribution of lines with $b_{\rm HI} < 10$ \kms
\citep{Paschos2009,Dave2010,Tepper-Garcia2012}. Given that the
\ac{fuv}-\ac{cos} data have spectral resolutions of about $\sim 16$
\kms, these samples should include the vast majority of real \hi
systems at $N_{\rm HI} \gtrsim 10^{13}$ \cm. On the other hand, the
\ac{nuv}-\ac{cos} and \ac{fos} data (see \Cref{tab:data:HSTobs}) have
spectral resolutions of about $\sim 100$ \kms, which introduces some
unresolved lines. Unresolved blended systems also add some unphysically
broad lines in all our data. This observational effect explains, in
part, the tail at large \bhi (see middle panel of
\Cref{fig:logn_b_dist}). We note that very broad lines can also be
explained by physical mechanisms, such as temperature, turbulence,
Jeans smoothing and Hubble flow broadenings
\citep[e.g.][]{Rutledge1998,Hui1999,Theuns2000,Dave2010,Tepper-Garcia2012}. There
are a total of $58/766 \sim 8\%$ of systems with $b_{\rm HI} \ge 80$
\kms. Such a small fraction does not affect our results significantly.

We also note that the typical \bhi uncertainties are of the order of
$\sim 10$ \kms, and so scatter of a similar amount is expected in the
\bhi distributions. This explains the presence of lines with $b_{\rm
  HI} \lesssim 10$ \kms, all of which are consistent with $10$ \kms
within the errors. However, as we do not use the actual \bhi values in
any further analysis, this uncertainty does not affect our results.

The right panel of \Cref{fig:logn_b_dist} shows the distribution of
\bhi as a function of $\log N_{\rm HI}$ for `secure' plus `probable'
systems (`a$+$b' labels; grey circles), and `uncertain' systems (`c'
label; red open triangles; uncertainties not shown). We see that there
are not strong correlations between these values, apart from the
presence of the upper and lower \bhi envelopes. The upper envelope is
consistent with an observational effect, as higher \nhi values are
required to observe lines with larger \bhi, for a fixed signal-to-noise
ratio \citep[e.g.][]{Paschos2009,Williger2010}. The lower envelope is
consistent with a physical effect, driven by the temperature-density
relation of the diffuse \ac{igm}: \hi systems with larger \nhi probe,
on average, denser regions for which the temperature---a component of
the \bhi---is also, on average, larger
\citep[e.g.][]{Hui1997,Theuns1999, Schaye1999,
  Paschos2009,Dave2010,Tepper-Garcia2012}. A proper analysis of these
two effects is beyond the scope of this paper.

\begin{table}
  \centering
  \begin{minipage}{0.37\textwidth}
    \centering
    \caption{Summary of the \hi~survey used in this
      paper.\tablenotemark{a}}\label{tab:abs_survey_summary}
    \begin{tabular}{@{}lcccc@{}}
    \hline
               & Secure& Probable& Uncertain& Total\\
               & (`a')  & (`b')     & (`c')     &      \\
    \hline
    \multicolumn{5}{c}{Q0107-025A}\\
    \hline
    \hi        &   76   &  29       &  15        & 120  \\
    \ \ Strong &   26   &  1        &  10        & 37  \\
    \ \ Weak   &   50   &  28       &  5         & 83  \\
    \hline
    \multicolumn{5}{c}{Q0107-025B}\\
    \hline
    \hi        &   45   &  6        &  16       &  67 \\
    \ \ Strong &   22   &  1        &  2        &  25 \\
    \ \ Weak   &   23   &  5        &  14       &  42 \\
    \hline
    \multicolumn{5}{c}{Q0107-0232}\\
    \hline
    \hi        &   26   &  20      &  4       &   50\\
    \ \ Strong &   19   &  6       &  0       &   25\\
    \ \ Weak   &   7    &  14      &  4       &   25\\
    \hline
    \multicolumn{5}{c}{J020930.7-043826}\\
    \hline
    \hi        &   74   &  60      &  22       &   156\\
    \ \ Strong &   17   &  10      &  6        &   33\\
    \ \ Weak   &   57   &  50      &  16       &   123\\
    \hline
    \multicolumn{5}{c}{J100535.24+013445.7}\\
    \hline
    \hi        &   70   &  61      &  8       &   139\\
    \ \ Strong &   9    &  8       &  5       &   22\\
    \ \ Weak   &   61   &  53      &  3       &   117\\
    \hline
    \multicolumn{5}{c}{J102218.99+013218.8}\\
    \hline
    \hi        &   50   &  10      &  6       &   66\\
    \ \ Strong &   5    &  5       &  0       &   10\\
    \ \ Weak   &   45   &  5       &  6       &   56\\
    \hline
    \multicolumn{5}{c}{J135726.27+043541.4}\\
    \hline
    \hi        &   86   &  46      &  10      &   142\\
    \ \ Strong &   23   &  9       &  4       &   36\\
    \ \ Weak   &   63   &  37      &  6       &   106\\
    \hline
    \multicolumn{5}{c}{J221806.67+005223.6}\\
    \hline
    \hi        &   5   &  12      &  9       &   26\\
    \ \ Strong &   5   &  8       &  9       &   22\\
    \ \ Weak   &   0   &  4       &  0       &   4\\
    \hline
    \multicolumn{5}{c}{Total}\\
    \hline
    \hi        &   453  &  216     &  97       &  766 \\
    \ \ Strong &   126  &  47      &  37       &  210 \\
    \ \ Weak   &   327  &  169     &  60       &  556 \\

    \hline                                     
    \end{tabular}
  \end{minipage}
  \begin{minipage}{0.37\textwidth}
    $^{a}$ See \Cref{abs:reliability} and \Cref{abs:logn} for definitions.\\
\end{minipage}
  
\end{table}

\subsection{Column density classification}\label{abs:logn}

One of our goals is to test whether the cross-correlation between \hi
absorption systems and galaxies depends on \hi column density. To do
so, we split our \hi sample into subcategories based on a column
density limit. We define \strong systems as those with column densities
$N_{\rm HI} \ge 10^{14}$ \cm, and \weak systems as those with $N_{\rm
  HI} < 10^{14}$ \cm. The transition column density of $10^{14}$ \cm is
somewhat arbitrary but was chosen such that: (i) the \hi--galaxy
cross-correlation for \strong systems and the galaxy--galaxy
auto-correlation have similar amplitudes; and (ii) the \strong systems
sample is large enough to measure the cross-correlation at relatively
high significance. A larger column density limit (e.g. $\sim
10^{15-16}$ \cm) does indeed give a stronger \hi--galaxy clustering
amplitude, but it also increases the noise of the measurement.

We note that there might not necessarily be a physical mechanism
providing a sharp lower \nhi limit for the \hi--galaxy
association. However, recent theoretical results
\citep[e.g.][]{Dave2010} suggest that there might still be a physical
meaning for such a column density limit. We will discuss more on this
issue in \Cref{discussion:logn}.


\begin{figure*}
  \begin{minipage}{1\textwidth}
    \includegraphics[width=1\textwidth]{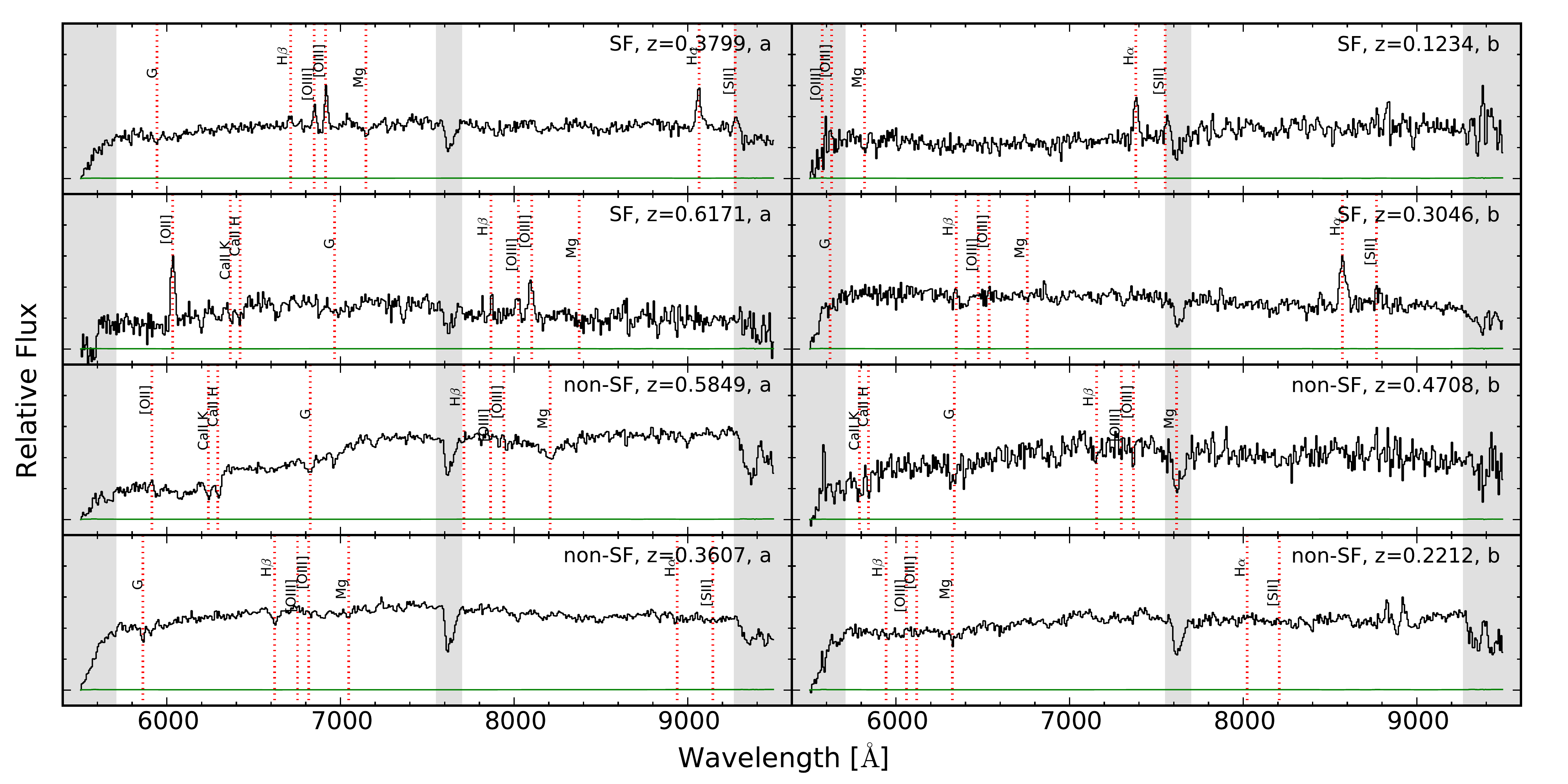}
  \end{minipage}
  
  \caption{Examples of galaxy spectra taken with \ac{vimos} (black
    lines) and their uncertainties (green lines). The left panels show
    spectra with `secure' redshifts (`a' labels) while the right panels
    show spectra with `possible' redshifts (`b' labels). The top four
    panels show examples of \SF galaxies while the bottom four panels
    show examples of \nSF galaxies. Grey shaded areas show regions
    affected by poor sensitivity (edges) or by telluric absorption
    (middle) excluded from the redshift determination process. Red
    dotted lines show the position of some spectral features for each
    galaxy spectrum. }\label{fig:example_gals}

\end{figure*}

\subsection{Summary}\label{abs:summary}
Our \ac{igm} data are composed of \ac{hst} data from the \ac{cos} and
\ac{fos} instruments taken on $8$ different \acp{qso} (see
\Cref{tab:data:IGM,tab:data:HSTobs}). We have split our \hi absorption
line system sample into \strong and \weak based on a column density
limit of $10^{14}$ \cm. Our survey is composed by a total of $\sim 669$
well identified (i.e. `a' or `b') \hi systems with $N_{\rm} \gtrsim
10^{13}$ \cm. \Cref{tab:abs_survey_summary} shows a summary of our \hi
survey. Tables \ref{tab:HI_Q0107A} to \ref{tab:HI_J2218} present the
survey in detail.


\section{Galaxy samples}\label{gal:samples}
In this section we describe our galaxy samples. In the following, we
will refer to our new galaxy surveys in terms of the instrument used
(either \ac{vimos}, \ac{deimos} and \ac{gmos}), to distinguish them
from the \ac{vvds} or \ac{gdds} surveys.

\subsection{Spectral type classification}\label{gal:sclass}

One of our goals is to test whether the cross-correlation between
\hi~absorption systems and galaxies depends on the galaxy spectral type
\citep[either absorption or emission line dominated;
  e.g.][]{Chen2009}. To do so, we need to classify our galaxy sample
accordingly.


We took a conservative approach by considering only two galaxy
subsamples: those which have not undergone important star formation
activity over their past $\sim 1$ Gyr and those which have. In terms
  of their spectral properties the former type has to show a strong
  \dfour~break and no significant emission lines (including \ha~and
        [\ion{O}{2}]). The latter type are the complementary galaxies,
        i.e. those with measurable emission lines. We henceforth name
        these subsamples as `non-star-forming' (\nSF) and
        `star-forming' (\SF) galaxies respectively, deliberately
        avoiding the misleading terminology of `early' and `late'
        types. Summarizing,

\begin{itemize}

\item {\it Non-star-forming galaxies (\nSF):} those galaxies which show
  no measurable star formation activity over the past $\gtrsim 1$ Gyr
  (e.g. Early, Bulge, Elliptical, Red Luminous Galaxy and S0
  templates).\smallskip

\item {\it Star-forming galaxies (\SF):} those galaxies which show
  evidence of current or recent ($\lesssim 1$ Gyr) star formation
  activity (e.g. Late, Sa, Sb, Sc, SBa, SBb, SBc and Starburst
  templates).

\end{itemize}

We note that we are not classifying galaxies on morphology, even though
the template names might suggest that. Our classification is based
solely on the presence or absence of spectral features associated with
star-formation activity. As an example, \Cref{fig:example_gals} shows
$8$ galaxies with a variety of signal-to-noise ratios, redshifts,
redshift reliabilities, and spectral classifications.

This template matching scheme was used only for our \ac{vimos} and
\ac{gmos} galaxies because in both the redshifts were determined using
template matches. For the rest of our data we used different
approaches, described in the following sections.

\subsubsection{DEIMOS data}
The \ac{deimos} reduction pipeline provides three weights from a
principal component analysis: $w_1$ (`absorption-like'), $w_2$
(`emission-like') and $w_3$ (`star-like'). Thus, for \ac{deimos} data
we use these weights to define star-forming and non-star-forming
galaxies as follows: if $\max(f w_1,w_2)=f w_1$ we assigned that object
to be a \nSF~galaxy; if $\max(f w_1,w_2)=w_2$ we assigned that object
to be a \SF~galaxy; and if $z<0.005$ we assigned that object to be a
`star' (this last condition takes precedence over the previous
ones). We used $f=0.2$ to be conservative in the definition of
\nSF~galaxies. This value also minimizes the `uncertain-identification
rate' in field Q0107 (see below). We did not use the information
provided by $w_3$ because we found $7$ objects with $z>0.005$
(galaxies) showing $\max(w_1,w_2,w_3)=w_3$, probably because of their
low signal-to-noise spectra.

\subsubsection{CFHT data}
In the case of the \ac{cfht} survey, we did not perform a spectral type
split, and so we will only use these galaxies for results involving the
whole galaxy population. We note that there is a large overlap between
our \ac{gmos} and the \ac{cfht} samples and that the \ac{cfht} sample
is comparatively small ($61$ galaxies). Thus, this choice does not
compromise our analysis.

\subsubsection{VVDS data}\label{gal:sclass:vvds}
In the case of the \ac{vvds} survey we used a color cut to split the
sample into red and blue galaxies. We chose this approach because the
current \ac{vvds} survey does not provide spectral classification for
galaxies in the fields used in this work. We used a single color limit
of $B-R = 2.15$ (no `k-correction' applied\footnote{If we knew the
  spectral type of the galaxies we would not have required the color
  split in the first place.}) to split our sample. Thus, galaxies with
$B-R<2.15$ were assigned to our \SF sample, whereas those with $B-R \ge
2.15$ were assigned to our \nSF sample. We chose this limit as it gives
the same proportion of \nSF/\SF galaxies as in the rest of our
sample. Objects with no $B-R$ color measurement were left out of this
classification, and so these will only contribute to the results
involving the whole galaxy population.

\begin{table}
  \begin{minipage}{0.45\textwidth}
    \centering
    \caption{Summary of the galaxy surveys used in this
      paper.}\label{tab:gal_survey_summary}
    \begin{tabular}{@{}lccccc@{}}
      \hline
      & Secure & Possible & Uncertain & Undefined & Total\\
      & (`a')  & (`b')    & (`c')   & (`n') & \\
      \hline
    \multicolumn{6}{c}{Our new survey}\\
    \hline    
    Galaxies   & 1634  & 509       & 0       & 0     & 2143\\
    \ \ \SF    & 1336  & 441       & 0       & 0     & 1777\\
    \ \ \nSF   & 298   & 68        & 0       & 0     & 366\\
    Stars      & 451   & 42        & 0       & 0     & 493 \\
    AGN        & 2     & 20        & 0       & 0     & 22\\
    Unknown    & 0     & 0         & 893     & 0     & 893\\
    
    \hline
    \multicolumn{6}{c}{GGDS survey\tablenotemark{a}}\\
    \hline
    Galaxies   & 41  & 12       & 0    & 0     & 53\\
    \ \ \SF    & 32  & 11       & 0    & 0     & 43\\
    \ \ \nSF   & 9   & 1        & 0    & 0     & 10\\
    Stars      & 1   & 0        & 0    & 0     & 1 \\
    AGN        & 1   & 0        & 0    & 0     & 1\\
    Unknown    & 0   & 0        & 5    & 0     & 5\\
    \hline
    \multicolumn{6}{c}{VVDS survey\tablenotemark{b}}\\
    \hline
    Galaxies   & 9458   & 7903     & 0     & 0    & 17361\\
    \ \ \SF    & 3766   & 3179     & 0    & 0     & 6945\\
    \ \ \nSF   & 789    & 639      & 0    & 0     & 1428\\
    Stars      & 1      & 2        & 0     & 0    & 3 \\
    AGN        & 138    & 131      & 0     & 0    & 269\\
    Unknown    & 0      & 0        & 8394  & 0    & 8394\\
     
    \hline
    \multicolumn{6}{c}{CFHT survey}\\
    \hline
    Galaxies   & 0   & 0        & 0    & 31     & 31\\
    \hline
    \multicolumn{6}{c}{Total}\\
    \hline
    
    Galaxies   & 11133 & 8424   & 0       & 31    & 19588\\
    \ \ \SF    & 5134  & 3631   & 0       & 0     & 8765\\
    \ \ \nSF   & 1096  & 708    & 0       & 0     & 1804\\
    Stars      & 453   & 44     & 0       & 0     & 497\\
    AGN        & 141   & 151    & 0       & 0     & 292\\
    Unknown    & 0     & 0      & 9292    & 0     & 9292\\
    \hline                                     
    \end{tabular}
  \end{minipage}
  \begin{minipage}{0.45\textwidth}
    $^{\rm a}$ Only objects in field J0209.\\
    $^{\rm b}$ Only objects in fields J1005, J1357 and J2218.
  \end{minipage}
\end{table}

\subsubsection{GDDS data}
The \ac{gdds} survey provides spectral classification based on three
binary digits, each one referring to `young' (`100'),
`intermediate-age' (`010') and `old' (`001') stellar populations
\citep{Abraham2004}. The \ac{gdds} spectral classification also allowed
for objects dominated by one or more types, so `101' could mean that
the object has strong \dfour break and yet some strong emission
lines. In order to match \ac{gdds} galaxies to our spectral
classification we proceeded in the following way. Galaxies classified
as `old' were matched to our \nSF~sample (\{`001'\}$\rightarrow$
\{\nSF\}); and galaxies classified as not being `old' where matched to
our \SF~sample (\{$\not=$`001'\}$\rightarrow$ \{\SF\}).

\subsubsection{Uncertainty in the spectral classification scheme} 
We quantified the uncertainty in this spectral classification by
looking at the `uncertain-classification rate', i.e. the fraction of
(duplicate) galaxies that were not consistently classified as either
\SF~or \nSF~over the total number of (duplicate) galaxies. For fields
J1005, J1022 and J2218 this uncertainly-classification rate corresponds
to $11/667 \sim 2\%$. None of these uncertainly-classified galaxies
show redshift differences $\gtrsim 0.005$ (catastrophic). For the Q0107
field this uncertain-classification rate corresponds to $25/280\sim
9\%$. From these, $4/25$ show redshift differences $\gtrsim 0.005$, all
of which are galaxies labelled as `b' (`possible'); and $19/25$ were
driven by observations using different instruments. The higher
uncertain-identification rate for Q0107 is therefore mostly driven by
the inhomogeneity of our samples.

For fields J1005 and J2218 we also checked whether the color cut limit
used to split the \ac{vvds} sample (see \Cref{gal:sclass:vvds}) gives
consistency with the actual spectral classification of our \ac{vimos}
sample, for common objects observed by these two surveys. In this case,
the uncertain-classification rate corresponds to $2/40 \sim 5\%$, all
of which were conservative in the sense that the \ac{vvds}
classification (uncertain) was \SF whereas the \ac{vimos} one
(reliable) was \nSF.


\subsection{Treatment of duplicates}\label{gal:duplicates}
For objects observed with different instruments and/or showing
different redshift confidences, we combined their redshift information
considering the following priorities:

\begin{itemize}
  \item {\it Redshift label priority:} we gave primary priority to
    redshifts labelled as `a', `b' and `c', in that order.\smallskip

  \item {\it Instrument priority:} we gave secondary priority to
    redshifts measured with \ac{deimos}, \ac{gmos}, \ac{vimos} and
    \ac{cfht}, in that order. We based this choice on spectral
    resolution.
  
\end{itemize}

We therefore chose the redshift given by the highest priority and took
the average when $2$ or more observations had equivalent
priorities. The spectral classification of uncertainly-classified
objects (i.e., being classified as both \SF~{\it and} \nSF) was set to
be \SF, ensuring a conservative \nSF~classification.

\begin{figure*}
  \begin{minipage}{0.49\textwidth}
  \includegraphics[width=1.05\textwidth]{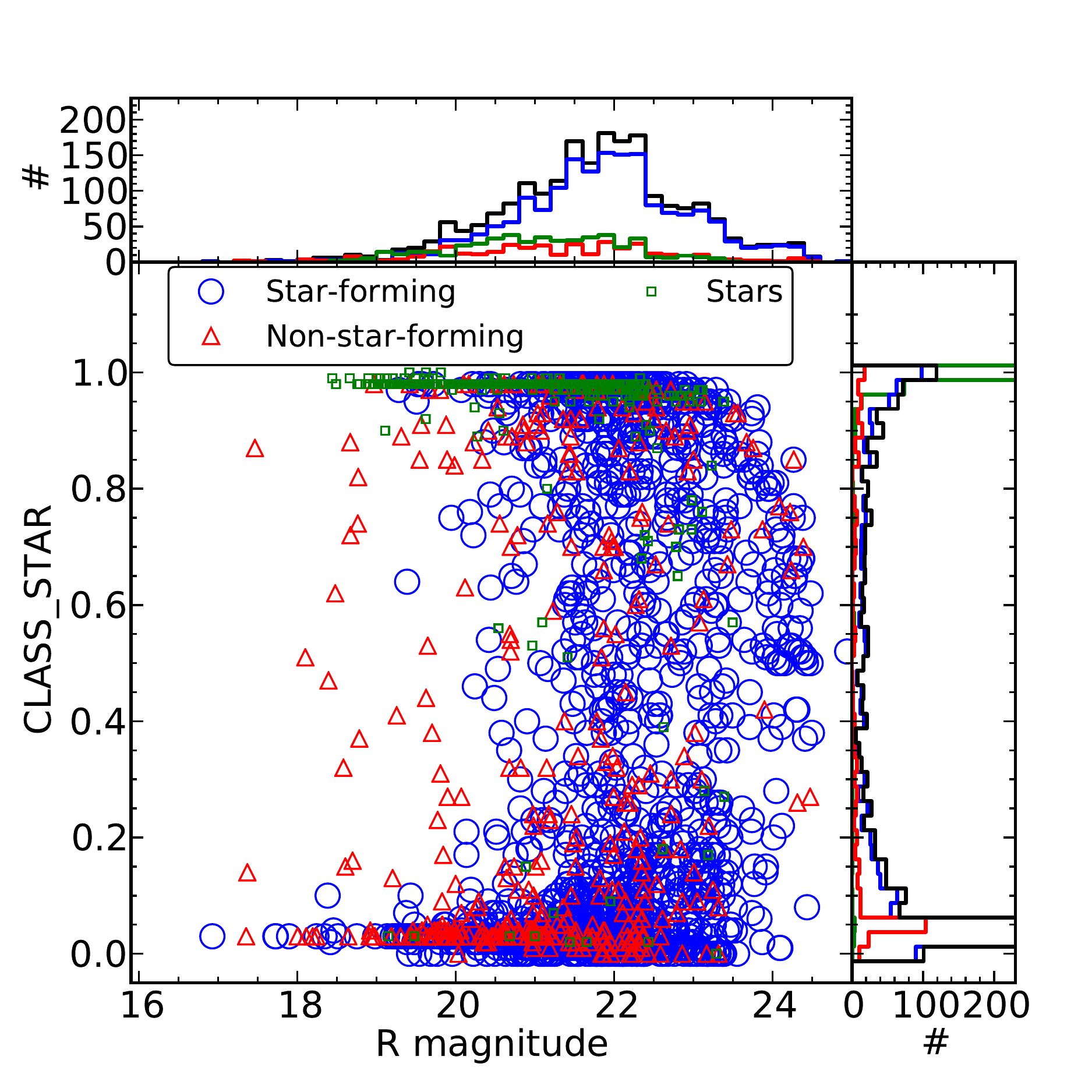}
  \end{minipage}
  \begin{minipage}{0.49\textwidth}
  \includegraphics[width=1.05\textwidth]{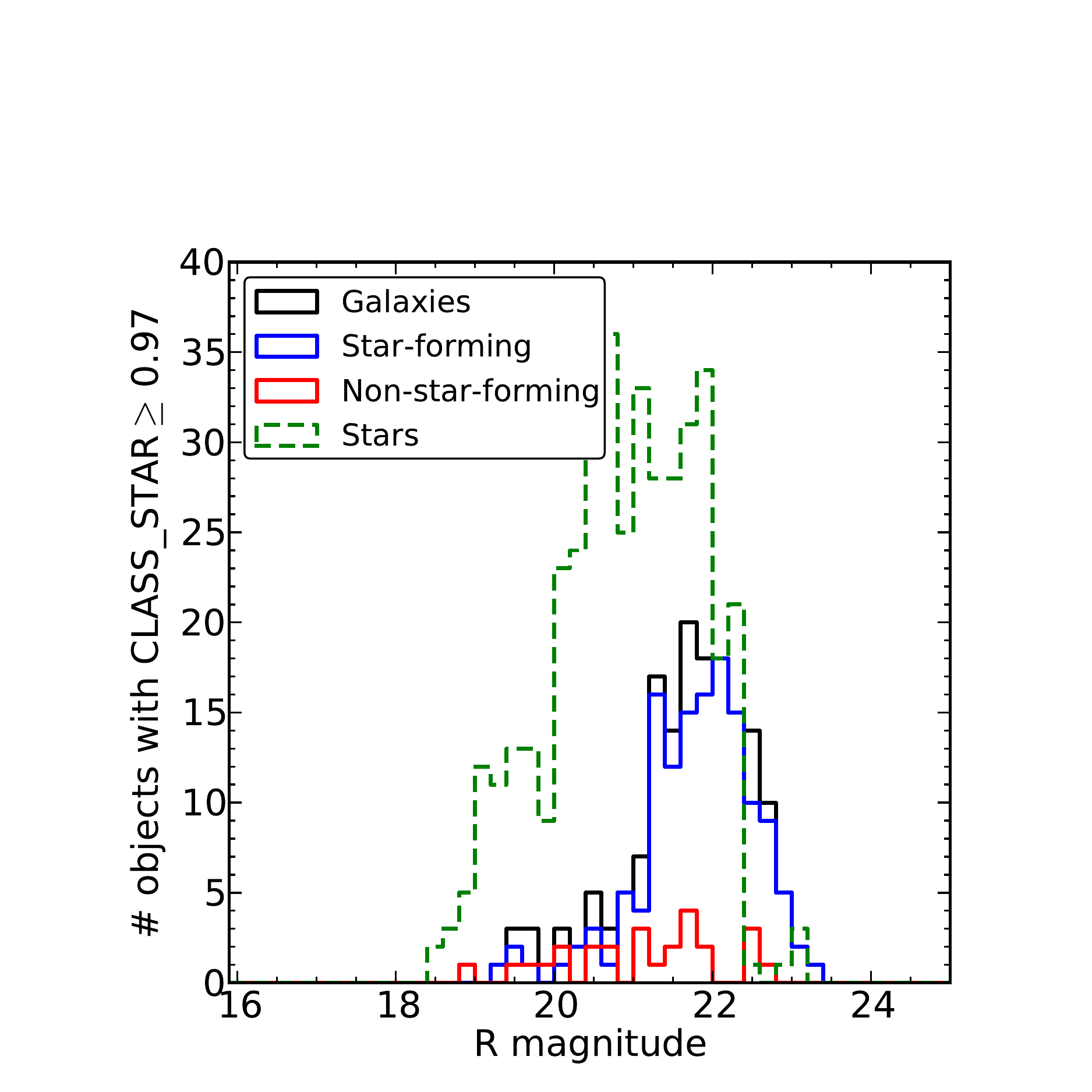}
  \end{minipage}
    
  \caption{Left: \sextractor CLASS\_STAR as a function of \rband
    magnitude for objects with spectroscopic redshifts: \SF galaxies
    (big blue open circles), \nSF galaxies (small red open triangles)
    and stars (small green squares. Histograms are shown around the
    main panel truncated at $230$ counts. The sudden decreases of
    objects at $R\sim 22$ and $R\sim23$ are due to our target selection
    (see \Cref{data:galaxy}). Right: Histogram of objects with
    CLASS\_STAR $\ge 0.97$: all galaxies (solid black), \SF~galaxies
    (solid blue), \nSF~galaxies (solid red) and stars (dashed
    green). We see a significant number of unresolved galaxies at
    $R\gtrsim 21$ mag (see \Cref{gal:class_star} for further
    discussion).} \label{fig:class_star}

\end{figure*}

\subsection{Star/galaxy morphological separation}\label{gal:class_star}

Our \ac{deimos} observations deliberately avoided star-like
(unresolved) objects, based on the \verb|CLASS_STAR|\fn parameter
provided by \sextractor (\Cref{data:deimos}). We found that this
selection misses a number of faint, unresolved galaxies and so it might
introduce an undesirable bias selection \citep[see
  also][]{Prochaska2011a}. This motivated our subsequent \ac{vimos} and
\ac{gmos} selection, for which no morphological criteria were imposed
(see \Cref{data:vimos} and \Cref{data:gmos}). Here we summarize our
findings regarding this issue.

\begin{figure*}
  \begin{minipage}{0.49\textwidth}
  \includegraphics[width=1\textwidth]{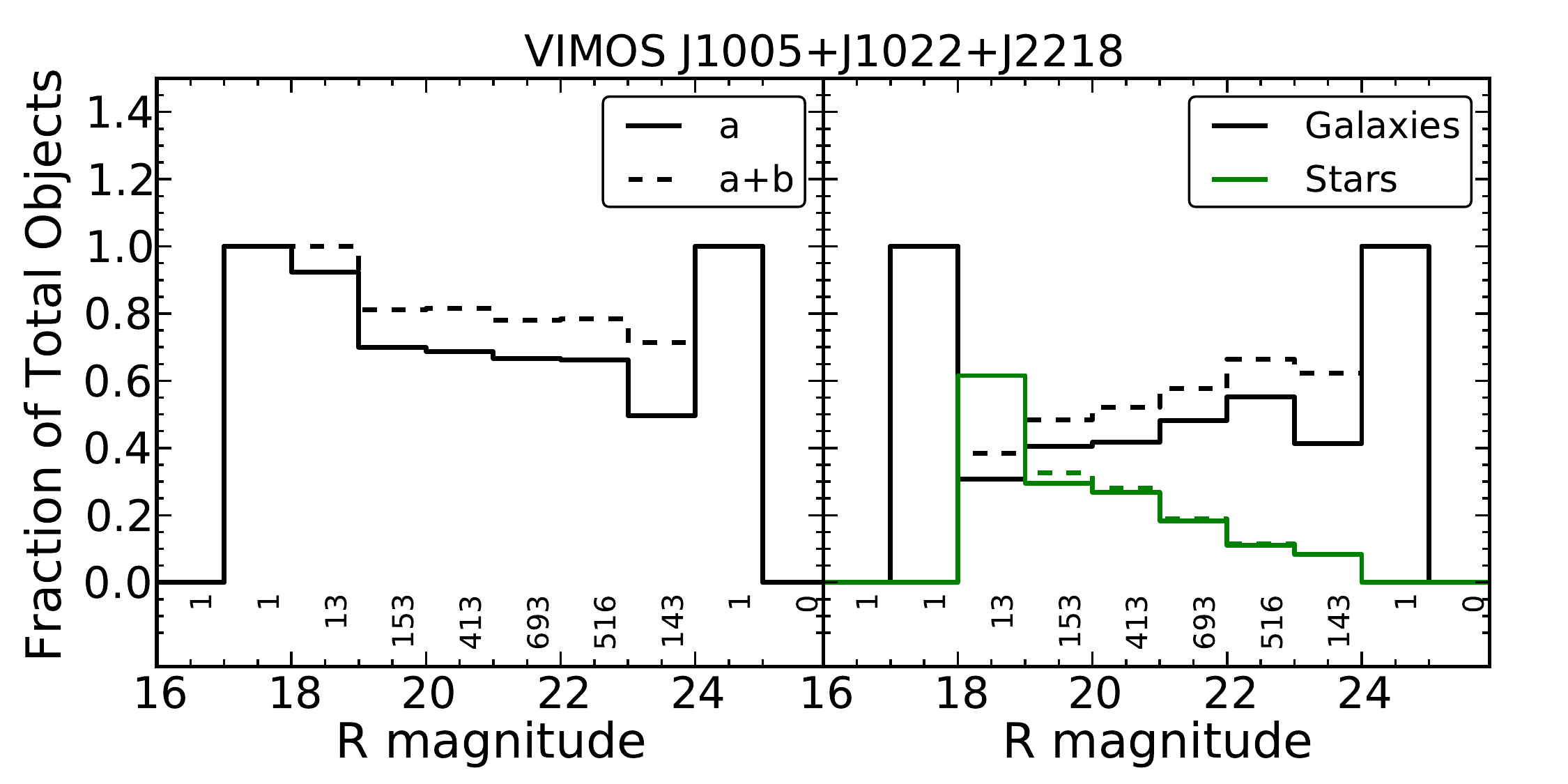}
  \end{minipage}
  \begin{minipage}{0.49\textwidth}
  \includegraphics[width=1\textwidth]{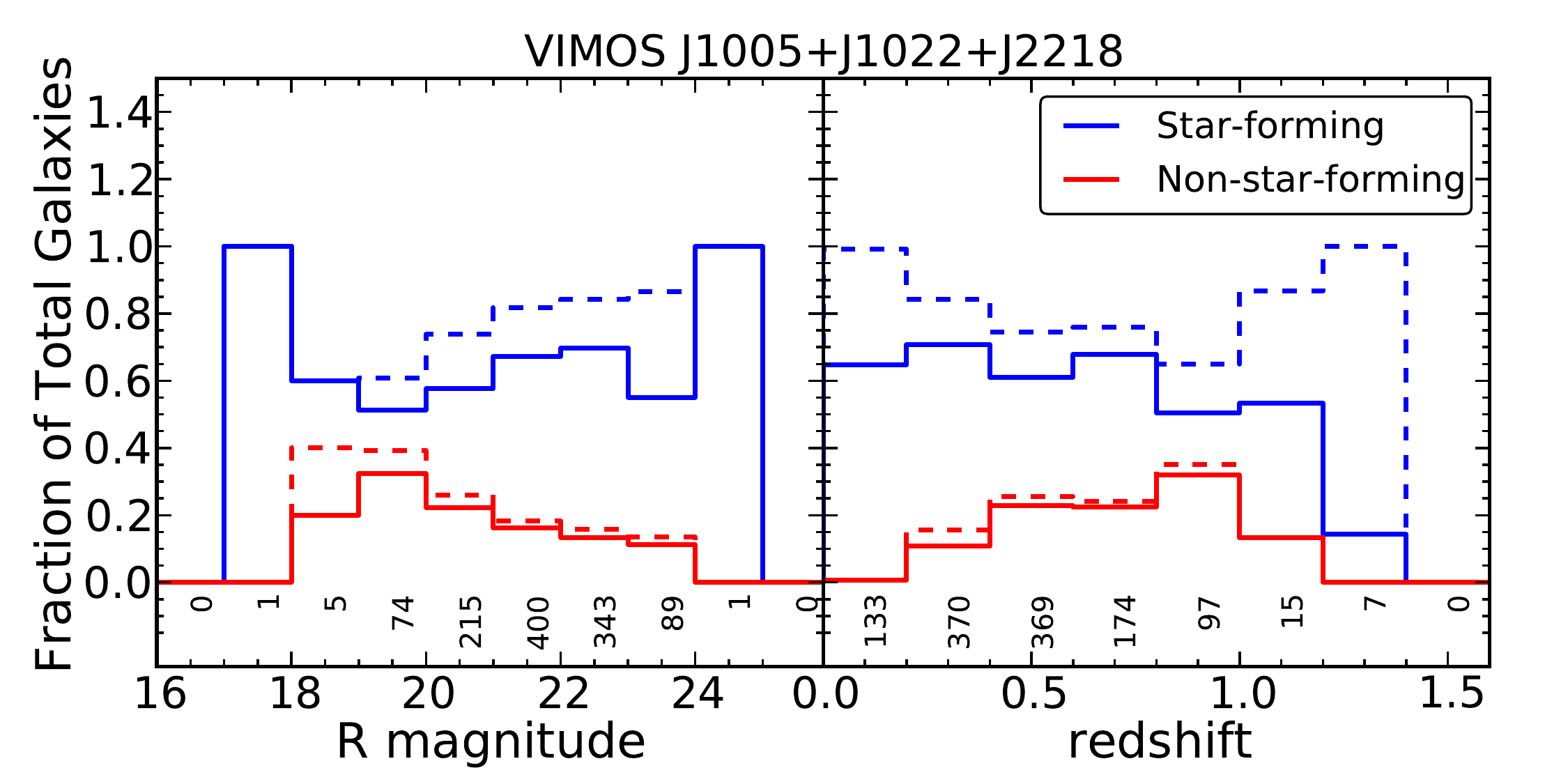}
  \end{minipage}
  
  \begin{minipage}{0.49\textwidth}
  \includegraphics[width=1\textwidth]{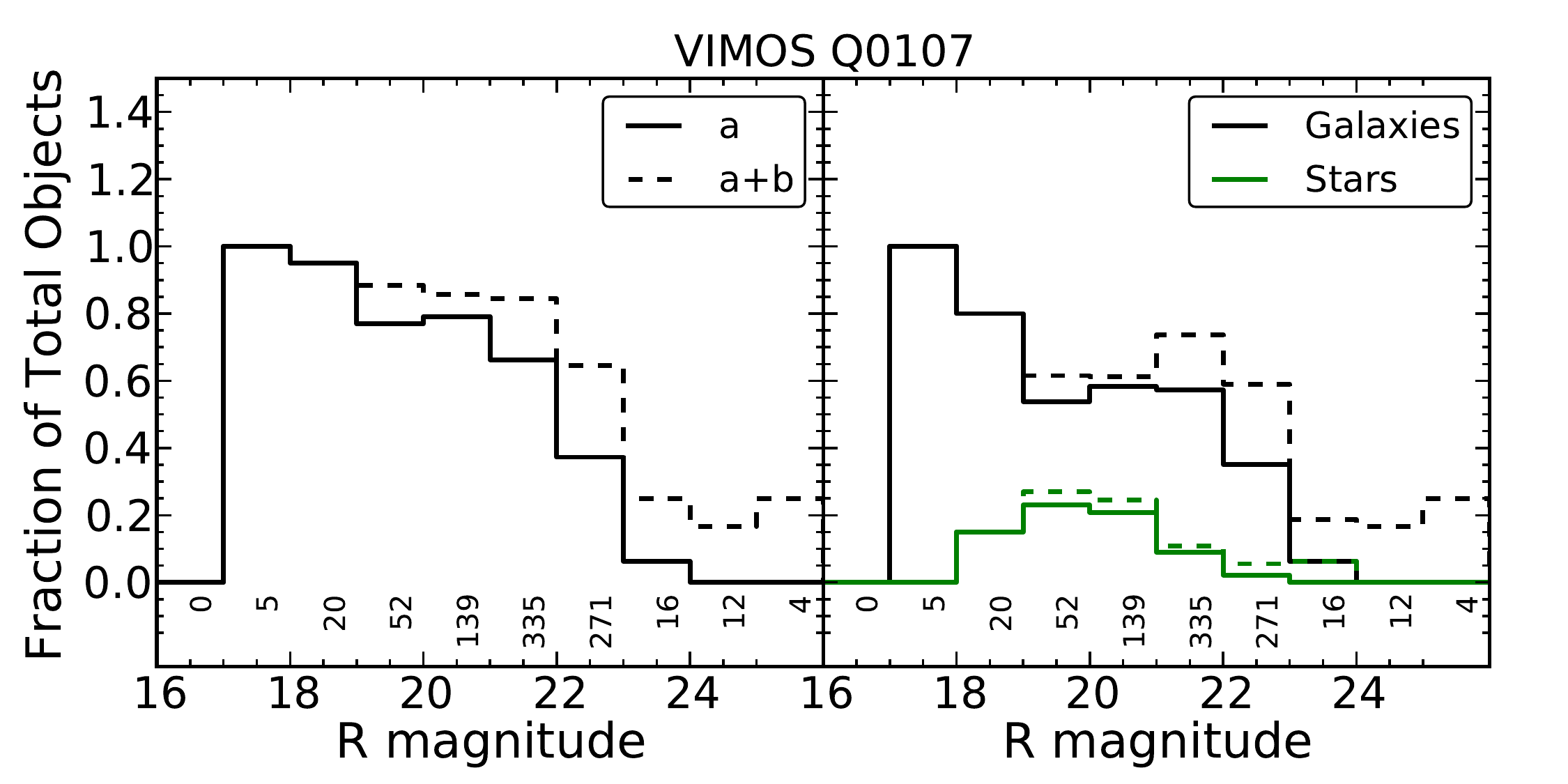}
  \end{minipage}
  \begin{minipage}{0.49\textwidth}
  \includegraphics[width=1\textwidth]{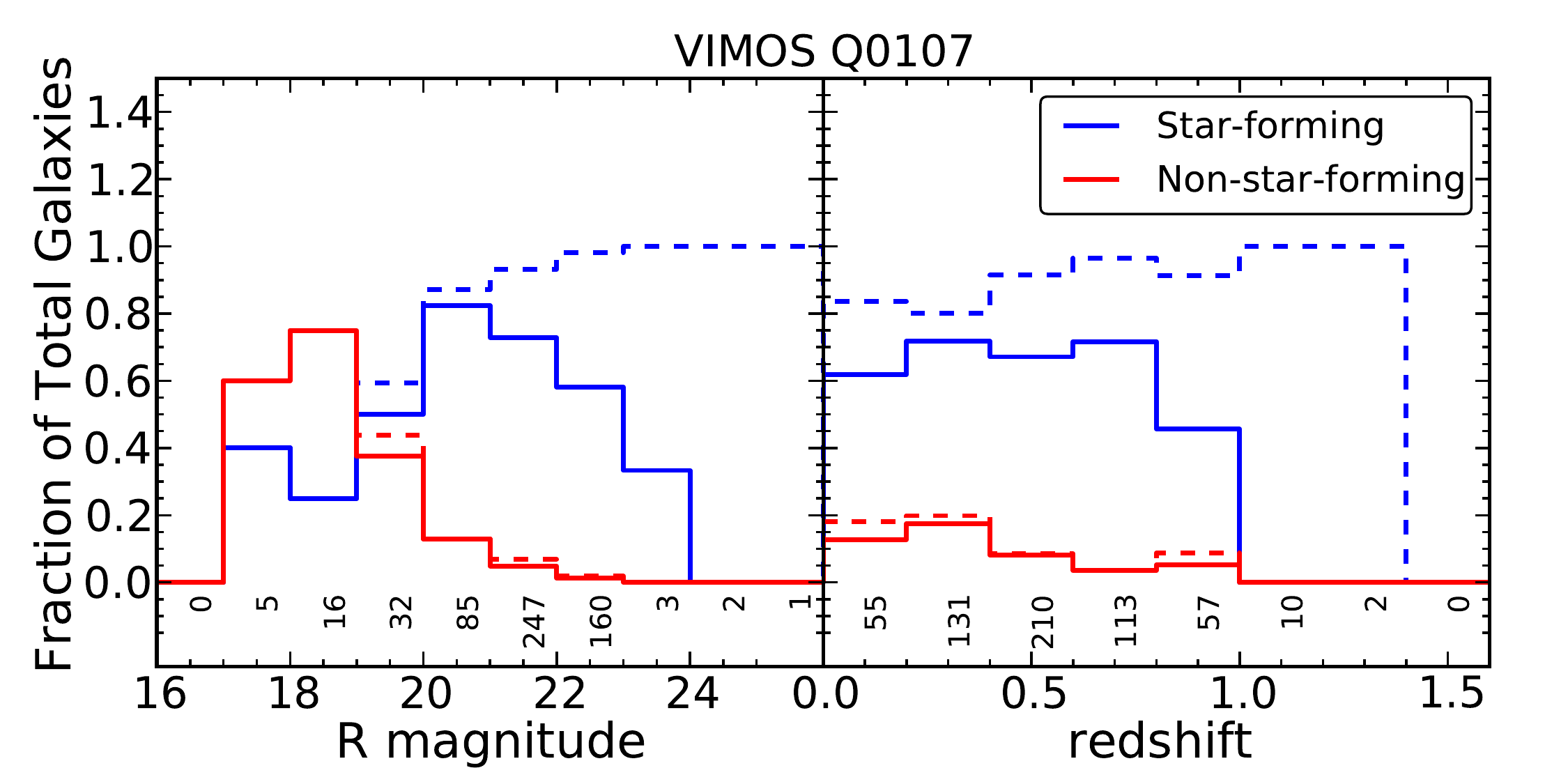}
  \end{minipage}

  \begin{minipage}{0.49\textwidth}
  \includegraphics[width=1\textwidth]{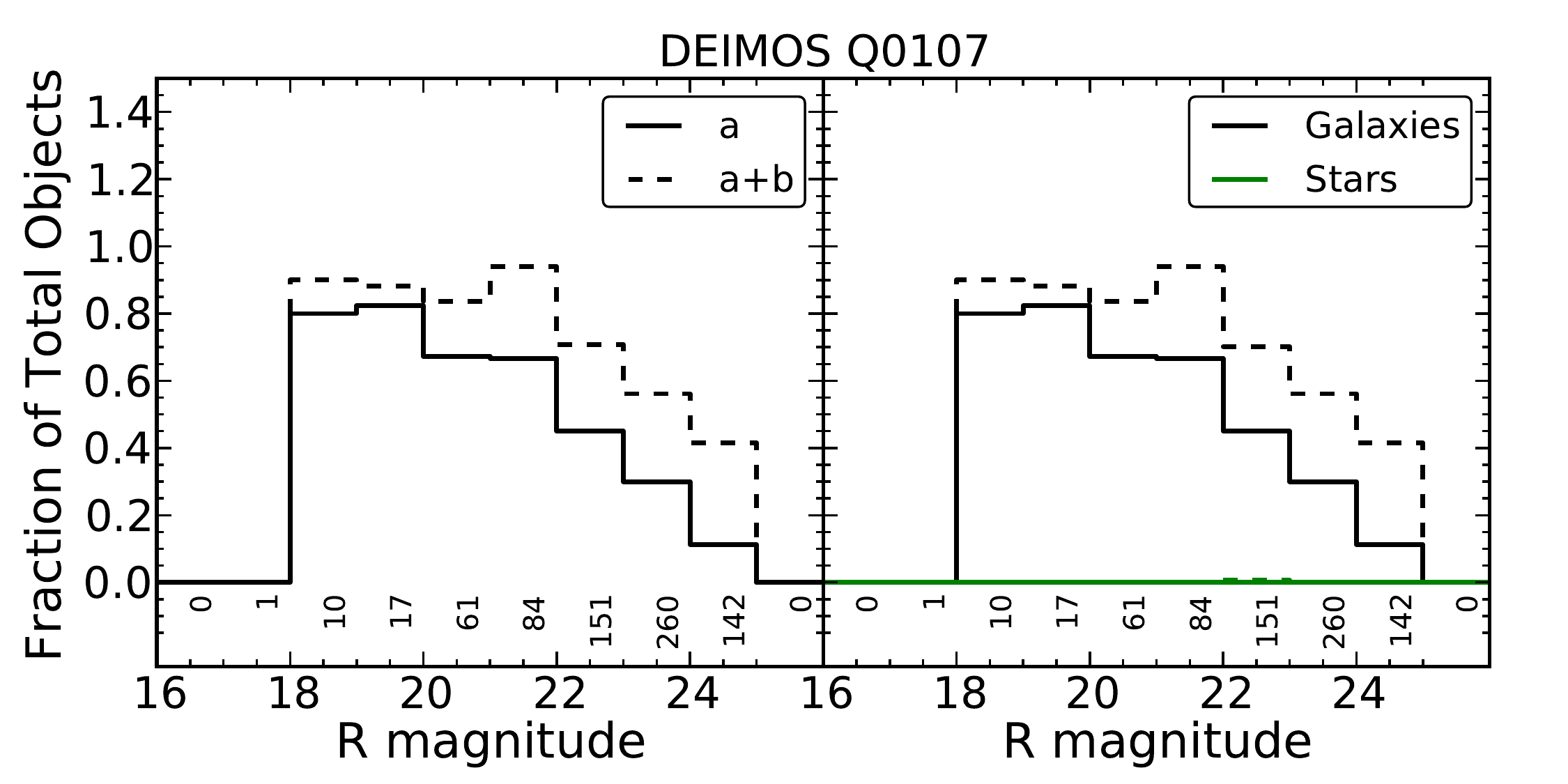}
  \end{minipage}
  \begin{minipage}{0.49\textwidth}
  \includegraphics[width=1\textwidth]{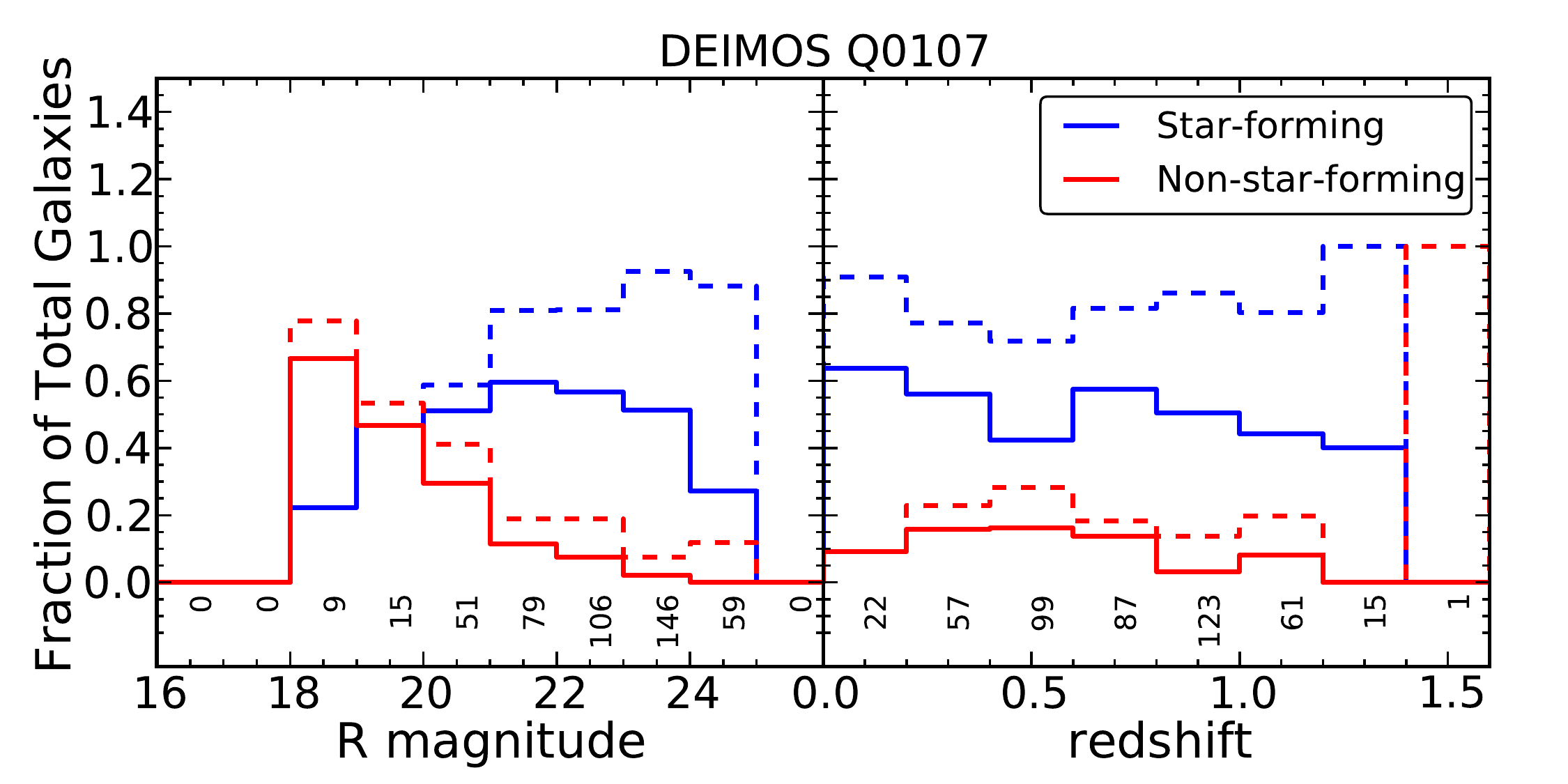}
  \end{minipage}

  \begin{minipage}{0.49\textwidth}
  \includegraphics[width=1\textwidth]{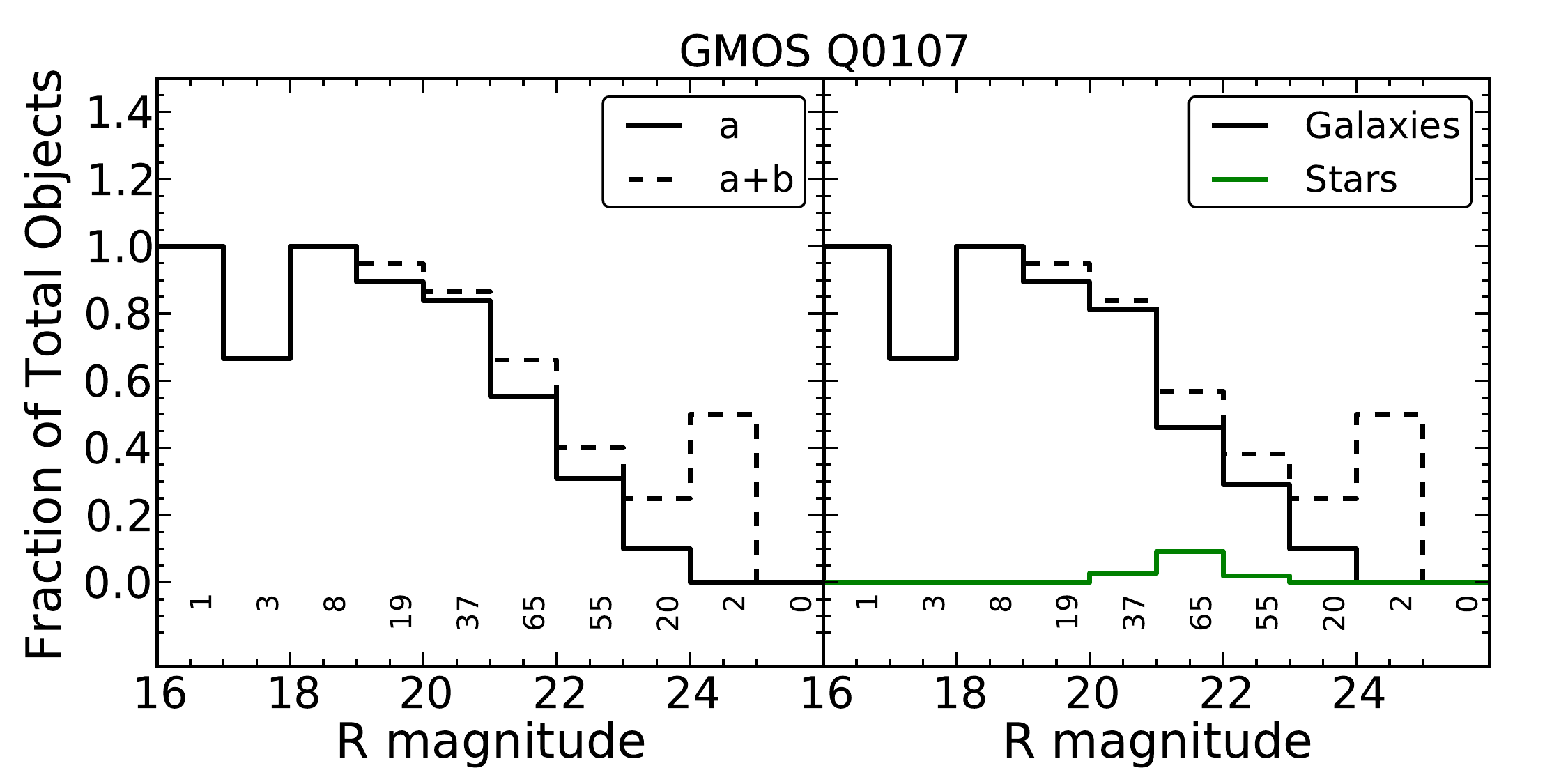}
  \end{minipage}
  \begin{minipage}{0.49\textwidth}
  \includegraphics[width=1\textwidth]{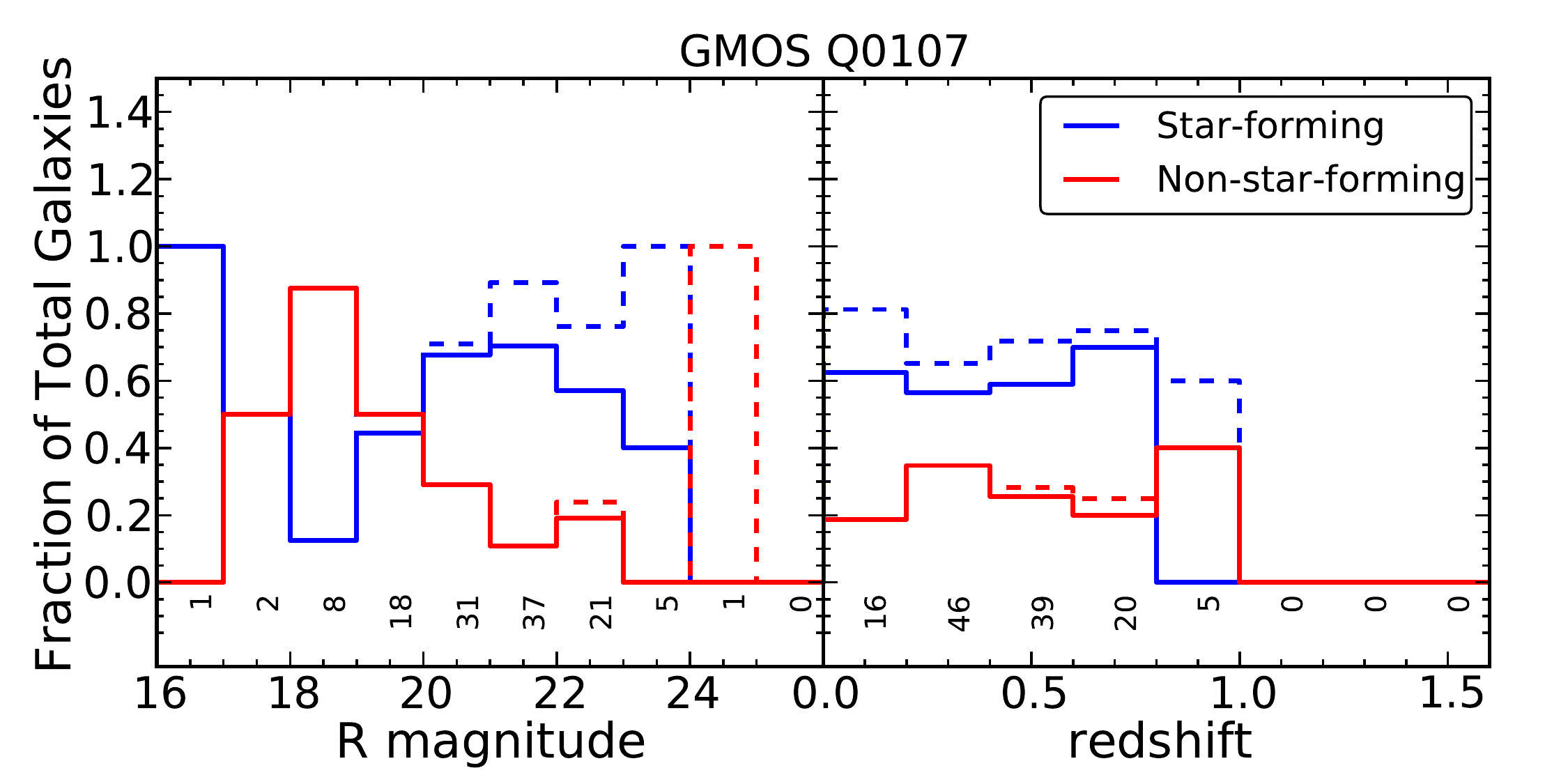}
  \end{minipage}
    
  \caption{Success rate of assigning redshifts for our new galaxy
    surveys. From top to bottom: \ac{vimos} (J1005, J1022 and J2218);
    \ac{vimos} (Q0107); \ac{deimos} (Q0107); and \ac{gmos} (Q0107). The
    first and second-column panels show the fraction of targeted
    objects with assigned redshift and the fraction of those that were
    identified as galaxies (black lines) and stars (green lines), as
    function of apparent \rband magnitude, respectively. The third and
    fourth-column panels show the fraction of galaxies that were
    classified as `star-forming' (blue lines) and/or `non-star-forming'
    (red lines), as a function of \rband magnitude and redshift,
    respectively. All these fractions are shown for both objects with
    high (`a' label; solid lines) and any (`a+b' label; dashed lines)
    redshift confidence. The number of objects corresponding to a
    fraction of $1$ (total) are labeled at the bottom of each bin. See
    \Cref{gal:completeness} for further
    discussion.}\label{fig:completeness}

\end{figure*}

The left panel of \Cref{fig:class_star} shows \verb|CLASS_STAR| values
as a function of \rband~magnitude for objects with spectroscopic
redshifts: \SF~galaxies (big blue open circles), \nSF~galaxies (small
red open triangles) and stars (small green squares). The sudden
decrease of objects at $R\sim 22.5$, $R\sim23.5$ and $R\sim24.5$
magnitudes are due to our target selection (see
\Cref{data:galaxy}). The fraction of \nSF~with respect to \SF~galaxies
is higher at brighter magnitudes (see \Cref{gal:completeness}). We see
a bimodal distribution of objects having \verb|CLASS_STAR| $\sim 0$
(resolved) and \verb|CLASS_STAR| $\sim 1$ (unresolved). The vast
majority of resolved objects are galaxies but some stars also fall in
this category due to the non-uniform \ac{psf} that varies across the
imaging field of view. On the other hand, the vast majority of bright
unresolved objects are stars, but a significant fraction of faint ones
are galaxies. The right panel of \Cref{fig:class_star} shows a
histogram of objects with \verb|CLASS_STAR| $\ge 0.97$ as a function of
\rband~magnitude. Such objects are typically excluded from galaxy
spectroscopic surveys. We find unresolved galaxies over a wide range of
magnitudes, but more importantly at $R\gtrsim 21$. At $R\gtrsim 22$
unresolved galaxies dominate over stars, and so a \verb|CLASS_STAR| $<
0.97$ criteria indeed introduces an undesirable selection bias. Even at
magnitudes brighter than $R\sim 21$, where the fraction of unresolved
galaxies is small, this morphological bias is still undesirable for
galaxy-absorber direct association studies. In our survey, $2$($7$) out
of $33$($82$) $R \le 21$ ($R \le 24$) unresolved galaxies lie at $\le
300$ \kpc~(physical) from a \ac{qso} \ac{los} which might have been
left out based on a morphological selection. As mentioned, our
\ac{deimos} survey is indeed affected by this selection effect, but our
\ac{vimos} and \ac{gmos} surveys are not, which allowed us to overcome
this potential problem in all our fields, including Q0107.

Neither the \ac{vvds} nor the \ac{gdds} data are affected in this
way. The \ac{vvds} survey targeted objects based only on magnitude
limits, while the \ac{gdds} survey used photometric redshifts to avoid
low-$z$ galaxies, with no morphological criteria imposed.


\begin{figure*}
  \begin{minipage}{0.42\textwidth}
    \includegraphics[width=1\textwidth]{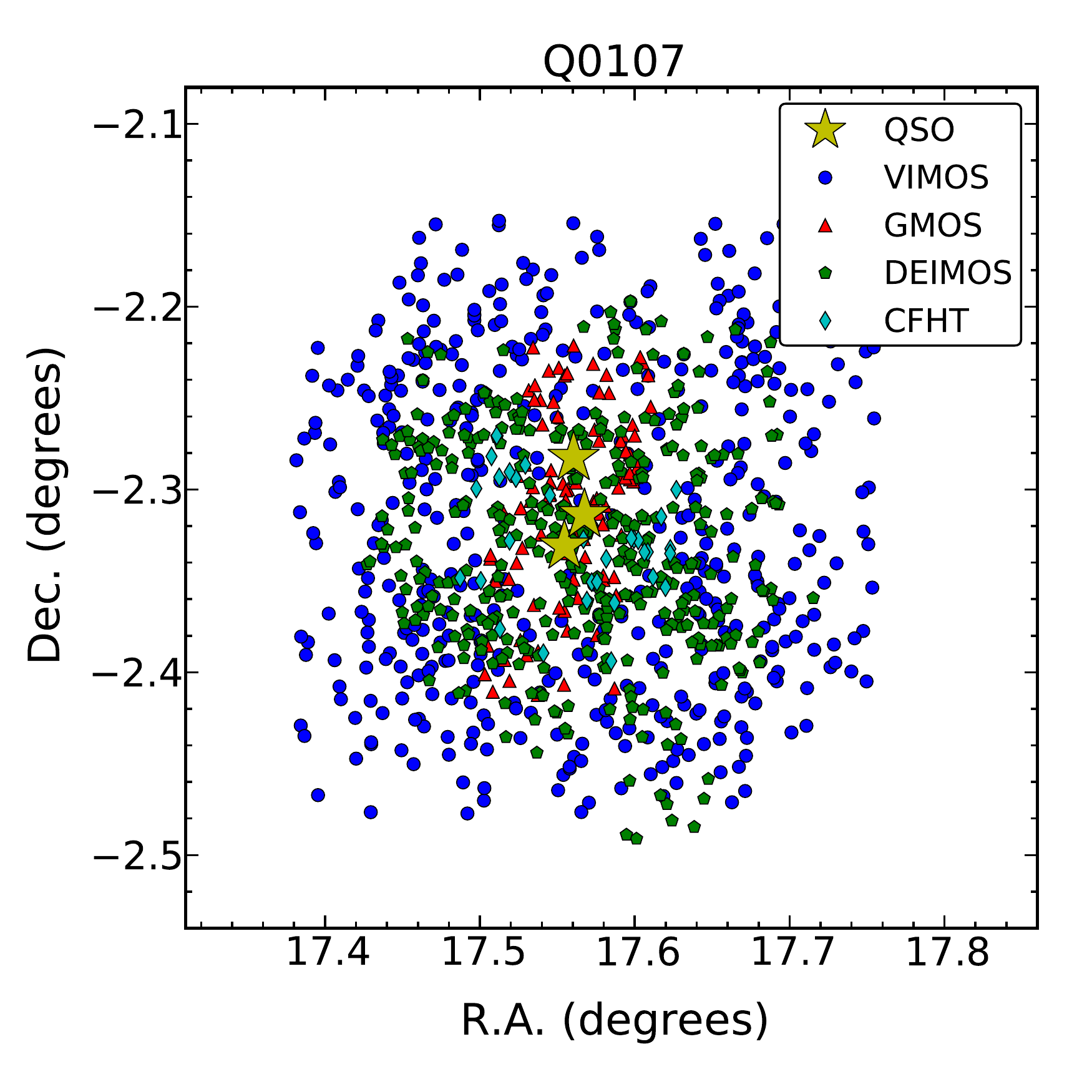}
  \end{minipage}
  \begin{minipage}{0.42\textwidth}
    \includegraphics[width=1\textwidth]{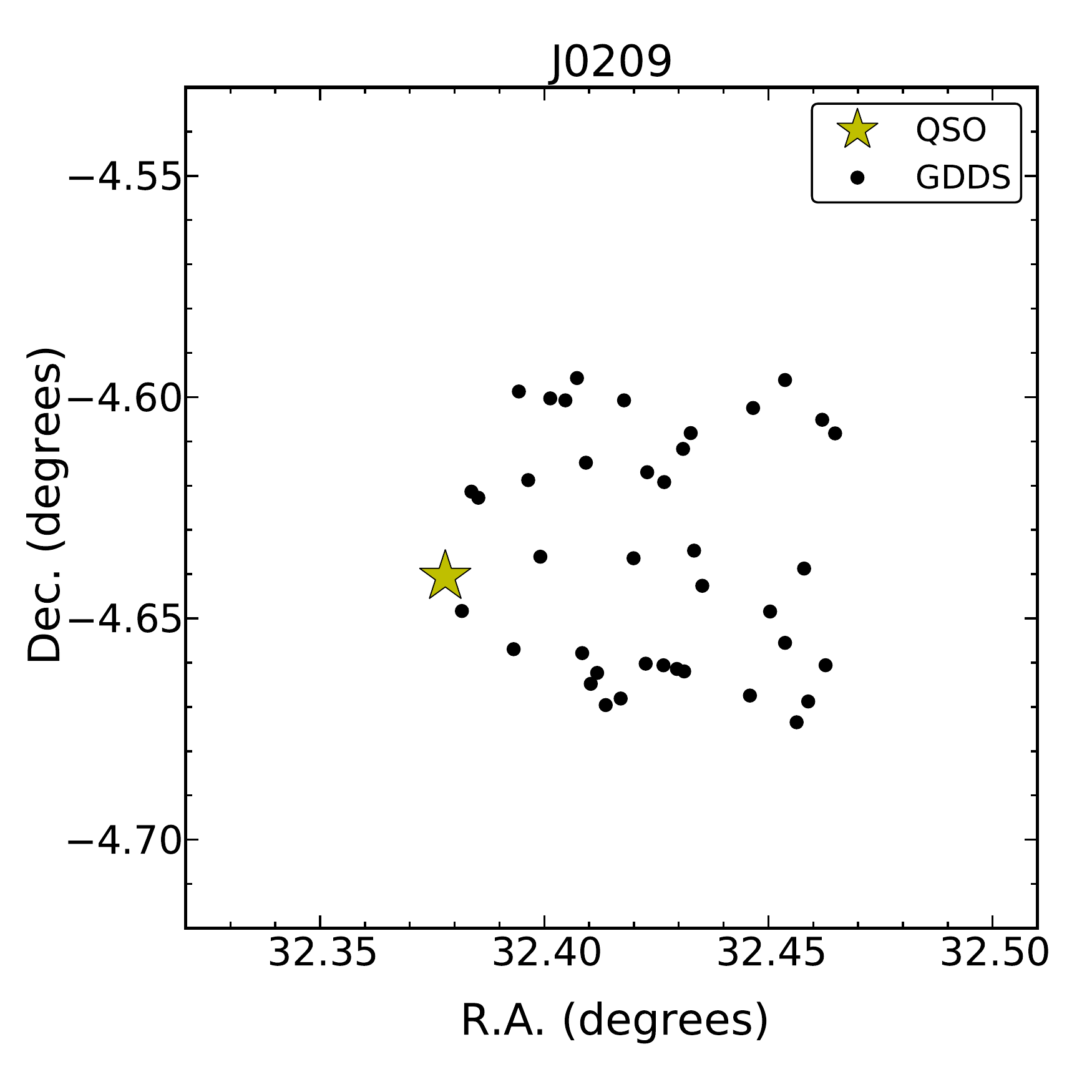}
  \end{minipage}
  \begin{minipage}{0.42\textwidth}
    \includegraphics[width=1\textwidth]{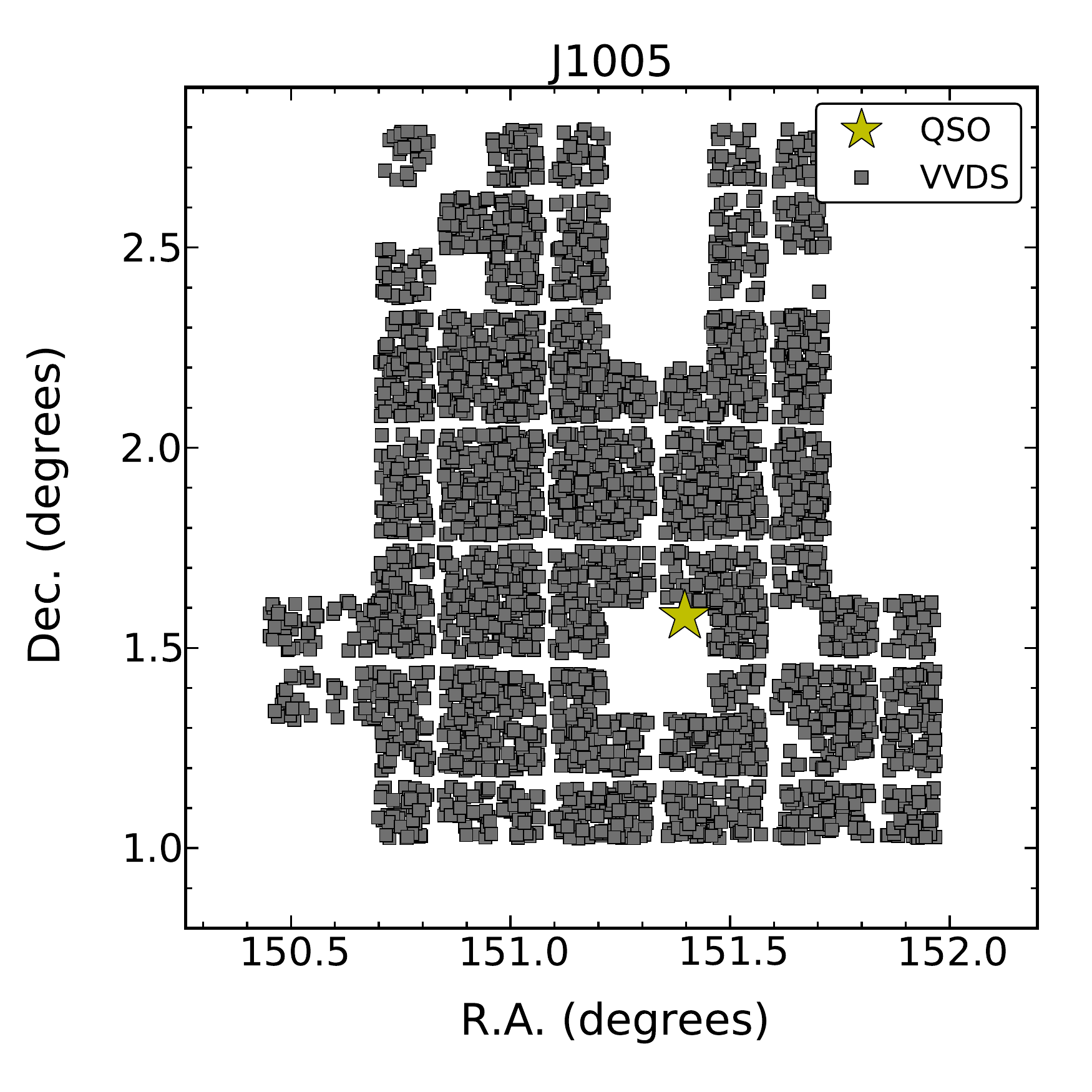}
  \end{minipage}
  \begin{minipage}{0.42\textwidth}
  \includegraphics[width=1\textwidth]{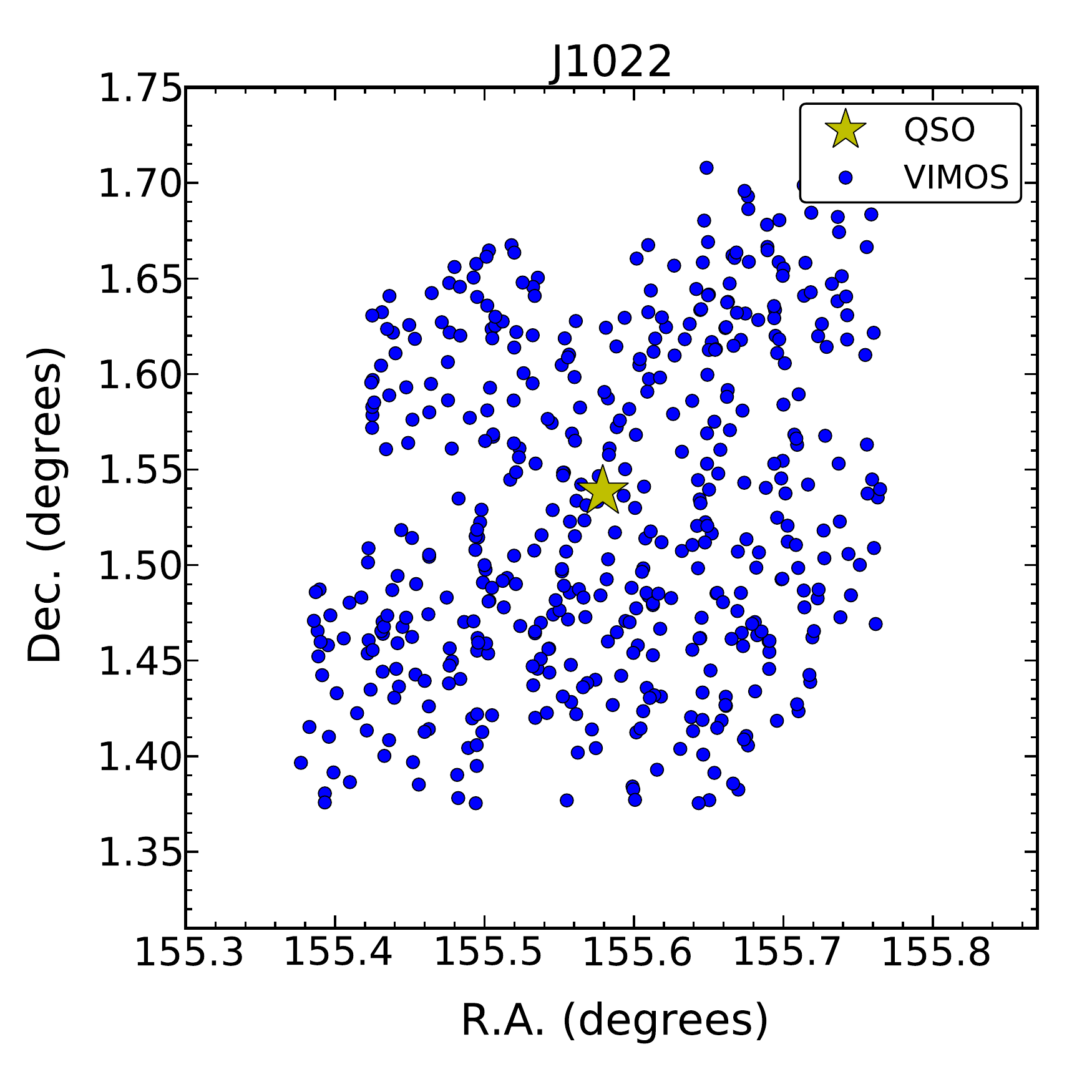}
  \end{minipage}
  \begin{minipage}{0.42\textwidth}
    \includegraphics[width=1\textwidth]{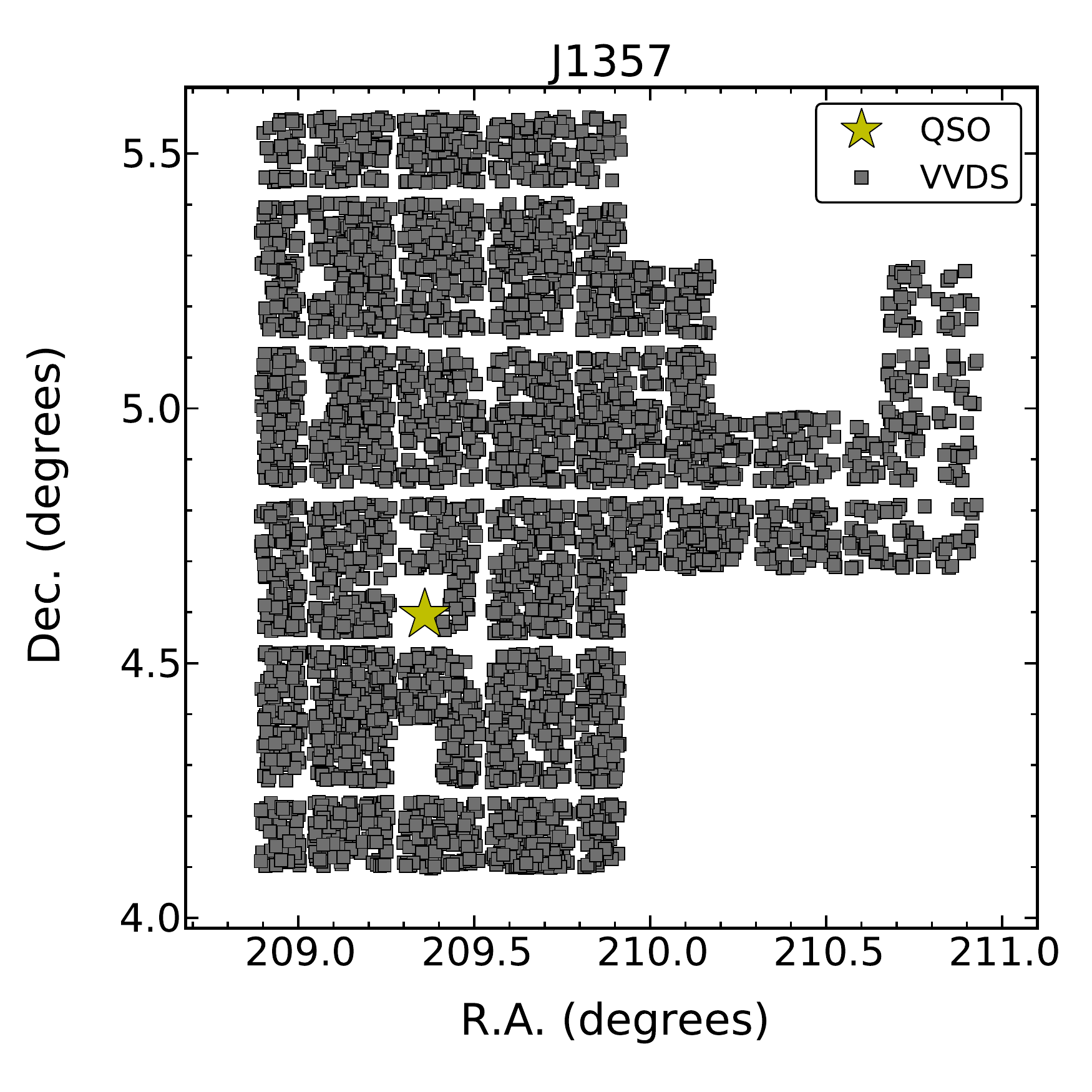}
  \end{minipage}
  \begin{minipage}{0.42\textwidth}
    \includegraphics[width=1\textwidth]{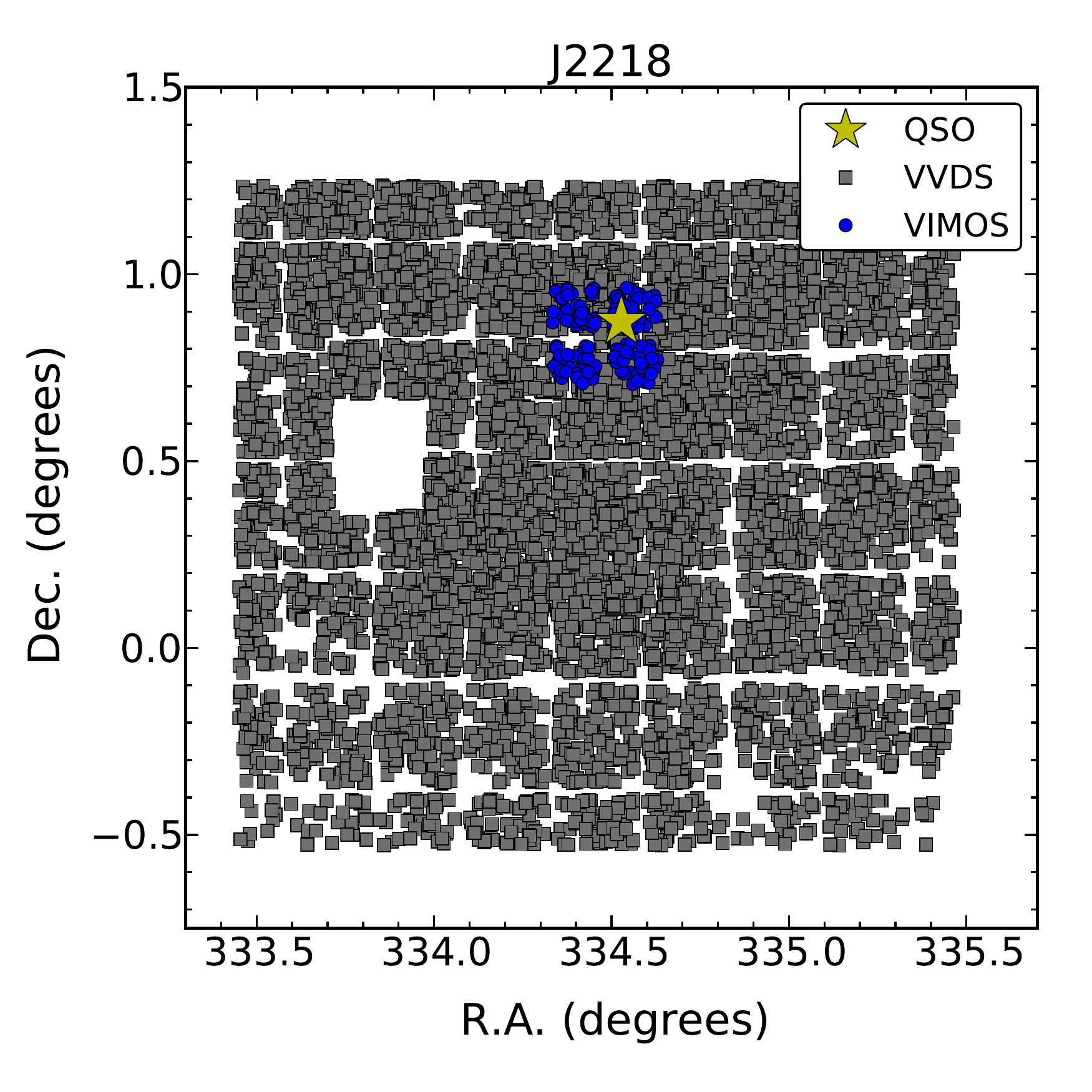}
  \end{minipage}
      
  \caption{Distribution on the sky of galaxies and background \acp{qso}
    (yellow stars) for each field. Blue circles, red triangles and
    green pentagons correspond to our new \ac{vimos}, \ac{gmos} and
    \ac{deimos} galaxies respectively; while black circles, grey
    squares and cyan diamonds correspond to \ac{gdds}, \ac{vvds} and
    \ac{cfht} \ac{mos} galaxies respectively.}\label{fig:gal_sky_dist}

\end{figure*}

\subsection{Completeness}\label{gal:completeness}

The completeness of a survey is defined as the fraction of detected
objects with respect to the total number of objects that could be
observed given the selection criteria. In the case of our galaxy survey
the completeness can be decomposed in: (i) the fraction of objects with
successful redshift determination with respect to the total number of
targeted objects; (ii) the fraction of targeted objects with respect to
the total number of objects detected by \sextractor; and (iii) the
fraction of objects detected by \sextractor with respect to the total
number of objects that could be observed. In the following we will
focus only on the first of these terms for our new galaxy data. For the
completeness of \ac{vvds}, \ac{gdds} and \ac{cfht} surveys we refer the
reader to \citet{LeFevre2005}, \citet{LeFevre2013}, \citet{Abraham2004}
and \citet{Morris2006}.

In \Cref{fig:completeness} we show the success rate of assigning
redshifts as a function of \rband~apparent magnitude, for all objects
(first-column panels) and for galaxies and/or stars (second-column
panels). We present them separately for each of our new galaxy surveys
because of their different selection functions. From top to bottom:
\ac{vimos} (J1005, J1022 and J2218), \ac{vimos} (Q0107), \ac{deimos}
(Q0107), and \ac{gmos} (Q0107). All of these fractions are computed for
objects whose redshifts have been measured at high (label `a', solid
lines) and/or any confidence (label `a+b', dashed lines). We see that
our surveys have a $\sim 70-90 \%$ success rate for objects with
$R\lesssim 22$ magnitudes, and a $\lesssim 40\%$ success rate for
objects with $22 \lesssim R \lesssim 24$, except for our \ac{vimos}
survey of fields J1005, J1022 and J2218, which shows a $\sim 70-90 \%$
success rate even for faint objects. As mentioned in \Cref{data:galaxy}
our \ac{vimos}, \ac{gmos} and \ac{deimos} surveys were limited at
$R=23-23.5$, $R=24$, $R=24.5$ respectively, and so the small
contribution of objects fainter than those limits correspond to
untargeted objects that happened to lie within the slits. These objects
correspond to a very small fraction of the total, and so we left them
in. The higher success rate for brighter objects is expected given the
higher signal-to-noise ratio of those spectra. For objects brighter
than $R\sim 22$ magnitudes, the fraction of identified galaxies is
$\gtrsim 50\%$, and the fraction of identified stars varies: from $\sim
0\%$ in our \ac{deimos} survey (by construction; see
\Cref{data:deimos}), $\lesssim 10\%$ in our \ac{gmos} survey, to $\sim
20-10 \%$ in our \ac{vimos} surveys. The fraction of identified
galaxies and stars at fainter magnitudes is $\lesssim 50\%$ and
$\lesssim 10\%$ respectively.

\Cref{fig:completeness} also shows how the galaxy completeness depends
on our galaxy spectral type classification (see \Cref{gal:sclass}). The
third and fourth-column panels show the fraction of galaxies classified
as \SF~(blue lines) and \nSF~(red lines) over the total number of
galaxies as a function of \rband~magnitude and redshift
respectively. Excluding magnitude bins with $<10$ galaxies, we see that
the fraction of \nSF~galaxies decreases with \rband~apparent
luminosity, consistent with the higher signal-to-noise ratio spectra
for the brighter objects. The fraction of \SF~galaxies shows a flatter
behavior because the redshift determination depends more on the
signal-to-noise of the emission lines than the signal-to-noise of the
continuum. The fraction of \nSF~galaxies dominates over \SF~ones at
$R\lesssim19$ (see also left panel \Cref{fig:class_star}), with a
contribution of $\sim 50-70\%$, although these bins have typically
$<20$ objects. At fainter magnitudes ($R\gtrsim20$), \SF~galaxies
dominate over \nSF~ones with a contribution of $\sim 60-90\%$.  Despite
these magnitude trends, we see that our galaxy sample is dominated by
the \SF type over the whole redshift range (except for the one galaxy
observed at $z>1.4$ in the \ac{deimos} survey), as might have been
expected from our conservative spectral classification
(\Cref{gal:sclass}). \SF~(\nSF) galaxies account for $\sim
60-80\%$~($\sim 20-30\%$) of the total galaxy fraction at $z\lesssim
1$, with a mild decrease (increase) with redshift. This redshift trend
is most apparent in our \ac{vimos} survey of fields J1005, J1022 and
J2218, which we explain as follows. The \dfour~\AA~break becomes
visible at $5500$ \AA~for redshifts $\sim 0.4$ and moves towards
wavelength ranges of higher spectral quality ($\sim 6000-7500$~\AA) at
$z\sim 0.7-0.9$. Simultaneously, \ha and [\ion{O}{3}] emission lines
are shifted towards poor quality spectral ranges ($\gtrsim 8000$ \AA;
due to the presence of sky emission lines) at $z\sim 0.2$ and $z\sim
0.6$, and are out of range at $z \gtrsim 0.4$ and $z \gtrsim 0.8$
respectively. At $z\gtrsim 1$ the only emission line available is
[\ion{O}{2}] which explains the rise in the fraction of low redshift
confidence (`b' labels) \SF~galaxies.

\subsection{Summary}
Our galaxy data is composed of a heterogeneous sample obtained from $4$
different instruments (see \Cref{tab:data:Galobs}), taken around $8$
different \ac{qso} \ac{los} in $6$ different fields (see
\Cref{fig:gal_sky_dist} and \Cref{tab:data:IGM}). For fields with
observations from more than one instrument, we have made sure that the
redshift frames are all consistent. We have also split the galaxies
into `star-forming' (\SF) and `non-star-forming' (\nSF), based on
either spectral type (for those lying close to the \ac{qso} \ac{los},
i.e., \ac{vimos}, \ac{deimos}, \ac{gmos} and \ac{gdds} samples) or
color (\ac{vvds} sample). \Cref{tab:gal_survey_summary} shows a summary
of our galaxy survey. Tables \ref{tab:gals_Q0107} to
\ref{tab:gals_J2218} present our new galaxy survey in detail. We refer
the reader to \citet{LeFevre2005}, \citet{LeFevre2013},
\citet{Abraham2004} and \citet{Morris2006} for retrieving the
\ac{vvds}, \ac{gdds} and \ac{cfht} data respectively.

Our final dataset comprises $19588$ ($11133$) galaxies with good
(excellent) spectroscopic redshifts at $z\lesssim 1$ around \ac{qso}
\ac{los} with $669$ ($453$) good (excellent) \hi~absorption line
systems. This is currently the largest sample suitable for a
statistical analysis on the \ac{igm}--galaxy connection to date.



\section{Correlation analysis}\label{analysis}
The main goal of this paper is to address the connection between the
\ac{igm} traced by \hi~absorption systems and galaxies in a statistical
manner. To do so, we focus on a two-point correlation analysis rather
than attempting to associate individual \hi~systems with individual
galaxies.

The two-point correlation function, $\xi(r)$, is defined as the
probability excess of finding a pair of objects at a distance $r$ with
respect to the expectation from a randomly distributed
sample.\footnote{Assuming isotropy, $\xi$ is a function of distance
  only.} Combining the results from the \hi--galaxy cross-correlation
with those from the \hi--\hi~and galaxy--galaxy auto-correlations for
different subsamples of \hi~systems and galaxies, we aim to get further
insights into the relationship between the \ac{igm} and galaxies.

\begin{figure*}
  \begin{minipage}{1\textwidth}
    \flushleft
    \includegraphics[width=\textwidth]{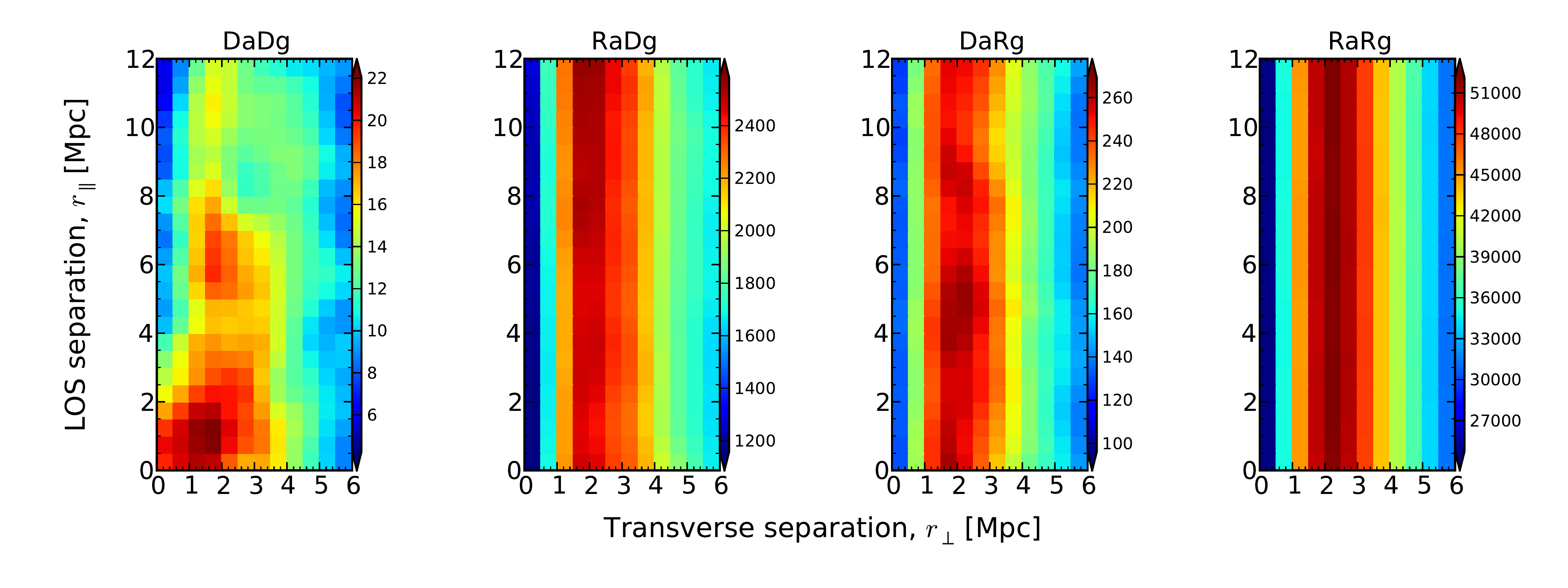}
  \end{minipage}
  \caption{Total number of cross-pairs between \hi~absorption systems
    and galaxies as a function of separations along ($r_{\parallel}$;
    $y$-axes) and transverse to the line-of-sight ($r_{\perp}$;
    $x$-axes). From left to right: $D_{\rm a}D_{\rm g}$ is the number
    of observed `data-data' absorber--galaxy pairs; $R_{\rm a}R_{\rm
      g}$ is the number of `random-random' absorber--galaxy pairs; and
    $D_{\rm a}R_{\rm a}$ and $R_{\rm a}D_{\rm g}$ are the number of
    `data-random' and `random-data' absorber--galaxy pairs
    respectively. We used an arbitrary binning of $0.5$ \mpc~with a
    Gaussian smoothing of standard deviation of $0.5$ \mpc~along both
    directions.} \label{fig:counts_ag}

\end{figure*}

\subsection{Two-dimensional correlation measurements}\label{correlation:2d}

In order to measure these spatial correlation functions we converted
all \hi~systems and galaxy positions given in (RA, DEC, $z$)
coordinates into a Cartesian co-moving system $(X,Y,Z)$. We first
calculated the radial co-moving distance to an object at redshift $z$
as,

\begin{equation}
R(z) = \frac{c}{H_0} \int_0^z \! \frac{1}{\sqrt{\Omega_{\rm
      m}(1+z')^3+\Omega_{\Lambda}}} \, {\rm d} z' \ \rm{.}
\label{eq:Rz}
\end{equation}

\noindent where $c$ is the speed of light, $H_0$ is the Hubble constant
and, $\Omega_{\rm m}$ and $\Omega_{\rm \Lambda}$ are the mass and `dark
energy' density parameters, respectively. Let (RA$_{0}$, DEC$_{0}$) be
the central coordinates of a given independent field. We then
transformed (RA,DEC,$z$) to $(X,Y,Z)$ as follows:

\begin{equation}
\begin{split}
X &\equiv R(z)\cos{(\Delta \delta)} \cos{(\Delta \alpha)}\\ Y &\equiv
R(z)\cos{(\Delta \delta)} \sin{(\Delta \alpha)}\\ Z &\equiv
R(z)\sin{(\Delta \delta)} \ \rm{,}
\end{split}
\end{equation}

\noindent where $\Delta \delta \equiv ({\rm DEC} - {\rm DEC}_{0})$ and
$\Delta \alpha \equiv ({\rm RA} - {\rm RA}_{0}) \cos({\rm DEC}_{0})$,
both in radians. Note that all our fields are far away from the poles
and each of them has small angular coverage (`pencil beam' surveys),
making this transformation accurate.  For fields with only one \ac{qso}
we chose $({\rm RA}_{0}, {\rm DEC}_{0}) = ({\rm RA}_{\rm QSO}, {\rm
  DEC}_{\rm QSO})$, while for the triple \ac{qso} field we took the
average position as the central one.

Given that peculiar velocities add an extra component to the redshifts
(in addition to cosmological expansion), our $(X,Y,Z)$ will be affected
differently, producing distortions even for actually true isotropic
signals. This is because the $X$ coordinate is parallel to the
\ac{los}, while the $Y$ and $Z$ coordinates are perpendicular to
it. Let $R(z)$ be the radial co-moving distance at redshift $z$
(\Cref{eq:Rz}) and $\Delta \theta$ a small ($\ll1$) angular separation
in radians. The transverse co-moving separation can be then
approximated by $\approx R(z) \Delta \theta$, implying that our $X$
coordinate will be affected a factor of $\approx 1/\Delta\theta$ times
that of the $Y$ and $Z$ coordinates for a fixed redshift difference. As
an example, a redshift difference of $\Delta z=0.0007$ at $z=0.5$
($\approx 140$ \kms) will roughly correspond to a radial co-moving
difference of $\approx 2$ \mpc, while only to a $\lesssim 0.02$
\mpc~difference in the transverse direction for co-moving separations
$\lesssim 20$ \mpc. We therefore measured the auto- and
cross-correlations both along and transverse to the \ac{los},
$\xi(r_{\perp},r_{\parallel})$, independently. In terms of our
Cartesian coordinates we have that $r_{\parallel,ij} \equiv |X_i- X_j|$
and $r_{\perp,ij} \equiv \sqrt{|Y_i - Y_j|^2 + |Z_i- Z_j|^2}$, are the
along the \ac{los} and transverse to the \ac{los} distances between two
objects at positions $(X_i,Y_i,Z_i)$ and $(X_j,Y_j,Z_j)$
respectively. Deviations from an isotropic signal in our
$(r_{\perp},r_{\parallel})$ coordinates can then be attributed to
redshift uncertainties and peculiar velocities, including \ac{lss} bulk
motions between the objects in the sample.\\

We used the \citet{Landy1993} estimator to calculate the galaxy--galaxy
auto-correlation as,

\begin{equation}
\xi^{\rm LS}_{\rm gg}(r_{\perp},r_{\parallel}) = \frac{D_{\rm g}D_{\rm
    g}/n_{\rm gg}^{\rm DD} - 2 D_{\rm g}R_{\rm g}/n_{\rm gg}^{\rm
    DR}}{R_{\rm g}R_{\rm g}/n_{\rm gg}^{\rm RR}} + 1 \ \rm{,}
\label{eq:LSgg}
\end{equation}

\noindent where $D_{\rm g}D_{\rm g}$ is the number of observed
`data-data' galaxy--galaxy pairs, $R_{\rm g}R_{\rm g}$ is the number of
`random-random' galaxy--galaxy pairs and $D_{\rm g}R_{\rm g}$ is the
number of `data-random' galaxy--galaxy pairs, all of which are measured
at the given $(r_{\perp},r_{\parallel})$ scales; and $n_{\rm gg}^{\rm
  DD}$, $n_{\rm gg}^{\rm DR}$ and $n_{\rm gg}^{\rm RR}$ are the
normalization factors for each respective pair count. Let $N_{\rm
  gal}^{\rm real}$ and $N_{\rm gal}^{\rm rand} \equiv \alpha_{\rm gal}
N_{\rm gal}^{\rm real}$ be the total number of real and random galaxies
respectively, then

\begin{equation}
\begin{split}
n^{\rm DD}_{\rm gg} & = N_{\rm gal}^{\rm real}(N_{\rm gal}^{\rm
  real}-1)/2\\ n^{\rm DR}_{\rm gg} & = \alpha_{\rm gal} (N_{\rm
  gal}^{\rm real})^2 \\ n^{\rm RR}_{\rm gg} & = \alpha_{\rm gal} N_{\rm
  gal}^{\rm real}(\alpha_{\rm gal} N_{\rm gal}^{\rm real}-1)/2\rm{.}
\end{split}
\end{equation}

The \hi--\hi~auto-correlation, $\xi^{\rm LS}_{\rm aa}$, was calculated
in a similar fashion as $\xi^{\rm LS}_{\rm gg}$,

\begin{equation}
\xi^{\rm LS}_{\rm aa}(r_{\perp},r_{\parallel}) = \frac{D_{\rm a}D_{\rm
    a}/n_{\rm aa}^{\rm DD} - 2 D_{\rm a}R_{\rm a}/n_{\rm aa}^{\rm
    DR}}{R_{\rm a}R_{\rm a}/n_{\rm aa}^{\rm RR}} + 1 \ \rm{,}
\label{eq:LSaa}
\end{equation}

\noindent where $D_{\rm a}D_{\rm a}$ is the number of observed
`data-data' absorber-absorber pairs, $R_{\rm a}R_{\rm a}$ is the number
of `random-random' absorber-absorber pairs and $D_{\rm a}R_{\rm a}$ is
the number of `data-random' absorber-absorber pairs, all of which
measured at the given $(r_{\perp},r_{\parallel})$ scales; and $n_{\rm
  aa}^{\rm DD}$, $n_{\rm aa}^{\rm DR}$ and $n_{\rm aa}^{\rm RR}$ are
the normalization factors for each respective pair count. Let $N_{\rm
  abs}^{\rm real}$ and $N_{\rm abs}^{\rm rand} \equiv \alpha_{\rm abs}
N_{\rm abs}^{\rm real}$ be the total number of real and random
\hi~systems respectively, then

\begin{equation}
\begin{split}
n^{\rm DD}_{\rm aa} & = N_{\rm abs}^{\rm real}(N_{\rm abs}^{\rm
  real}-1)/2\\ n^{\rm DR}_{\rm aa} & = \alpha_{\rm abs} (N_{\rm
  abs}^{\rm real})^2 \\ n^{\rm RR}_{\rm aa} & = \alpha_{\rm abs} N_{\rm
  abs}^{\rm real}(\alpha_{\rm abs} N_{\rm abs}^{\rm real}-1)/2\rm{.}
\end{split}
\end{equation}

The \hi--galaxy cross-correlation, $\xi^{\rm LS}_{\rm ag}$, was
calculated using a generalized version of the \citet{Landy1993}
estimator,

\begin{equation}
\xi^{\rm LS}_{\rm ag}(r_{\perp},r_{\parallel}) = \frac{D_{\rm a}D_{\rm
    g}/n_{\rm ag}^{\rm DD} - D_{\rm a}R_{\rm g}/n_{\rm ag}^{\rm DR} -
  R_{\rm a}D_{\rm g}/n_{\rm ag}^{\rm RD}}{R_{\rm a}R_{\rm g}/n_{\rm
    ag}^{\rm RR}} + 1
\label{eq:LSag}
\end{equation}

\noindent \citep[e.g.][]{Adelberger2003}, where $D_{\rm a}D_{\rm g}$ is
the number of observed `data-data' absorber--galaxy pairs, $R_{\rm
  a}R_{\rm g}$ is the number of `random-random' absorber--galaxy pairs,
and $D_{\rm a}R_{\rm a}$ and $R_{\rm a}D_{\rm g}$ are the number of
`data-random' and `random-data' absorber--galaxy pairs respectively,
all of which are measured at the given $(r_{\perp},r_{\parallel})$
scales. Following previous conventions the normalization factors in
this case are,

\begin{equation}\begin{split}
n^{\rm DD}_{\rm ag} & = N_{\rm abs}^{\rm real} N_{\rm gal}^{\rm
  real}\\ n^{\rm DR}_{\rm ag} & = \alpha_{\rm gal} N_{\rm abs}^{\rm
  real} N_{\rm gal}^{\rm real} \\ n^{\rm RD}_{\rm ag} & = \alpha_{\rm
  abs} N_{\rm abs}^{\rm real} N_{\rm gal}^{\rm real} \\ n^{\rm RR}_{\rm
  ag} & = \alpha_{\rm gal} \alpha_{\rm abs} N_{\rm abs}^{\rm real}
N_{\rm gal}^{\rm real} \ \rm{.}
\end{split}\end{equation}

This approach makes the random samples a crucial component of the
analysis. A detailed description of the random generator algorithms is
presented in \Cref{random}.\\

\citet{Landy1993} showed that $\xi^{\rm LS}$ minimizes the observed
variance and so is preferable over other proposed estimators
\citep[e.g.][]{Sharp1979,Hewett1982,Davis1983,Hamilton1993}. Given the
limited nature of any survey, all estimators are biased towards lower
correlation amplitudes. This is because the mean densities of our two
populations are estimated from the survey itself. In order for us to
measure a positive correlation on a certain scale, the measured $\xi$
needs to be negative at another. This leads to an observed correlation
amplitude which is lower than the underlying real one, $\xi^{\rm
  real}$, assumed to be positive. This is a well known bias commonly
referred to as the `integral constraint'. \citet{Landy1993} showed that
$\xi^{\rm LS}$ and $\xi^{\rm real}$ are related as

\begin{equation}
1 + \xi^{\rm LS} = \frac{\ \ 1+\xi^{\rm real}}{1+\xi_V} \ \rm{,}
\end{equation}

\noindent where $\xi_V$ is the `integral constraint' (scalar) defined
as

\begin{equation}
\xi_V \equiv \int_V \, G(r)\xi^{\rm real}(r) \, {\rm d}^2 V \ \rm{.}
\end{equation}

\noindent Here $G(r)$ is a normalized geometric window function
(positive) which gives the probability of having two volume elements
separated by a distance $r$ in the survey. In the case of our auto- and
cross-correlations, $G$ is given by $G_{\rm gg} \approx R_{\rm g}R_{\rm
  g}/n^{\rm gg}_{\rm RR}$, $G_{\rm aa} \approx R_{\rm a}R_{\rm
  a}/n^{\rm aa}_{\rm RR}$ and $G_{\rm ag} \approx R_{\rm a}R_{\rm
  g}/n^{\rm ag}_{\rm RR}$. Although we cannot know a priori the
amplitude of $\xi^{\rm real}$, we made a small correction using

\begin{equation}
\xi = (1+\tilde{\xi}_V)(1+\xi^{\rm LS}) - 1 \ \rm{,}
\end{equation}

\noindent where $\tilde{\xi}_V \equiv \int_V \, G(r) \xi^{\rm LS}(r) \,
          {\rm d}^2 V$, which still helps because of the discrete
          nature of all our cross-pair counts (including the
          randoms).\\

The computation of $\xi_{\rm gg}^{\rm LS}$, $\xi_{\rm aa}^{\rm LS}$ and
$\xi_{\rm ag}^{\rm LS}$ was performed {\it after} summing all the
cross-pairs from our $N_{\rm f}=6$ independent fields,

\begin{equation}
D_{\rm g}D_{\rm g}(r_{\perp},r_{\parallel}) = \sum\limits_i^{N_{\rm f}}
D_{\rm g}D_{\rm g}(r_{\perp},r_{\parallel})_{i} \ \rm{,}
\label{eq:DDsum}
\end{equation}

\noindent where $(D_{\rm g}D_{\rm g})_{i}$ is the number of `data-data'
galaxy--galaxy pairs in the $i$-th field, and so on for the rest of the
cross-pair counts. In contrast to measuring $\xi^{\rm LS}$ for each
independent field and then taking a weighted average, our adopted
approach reduces the `shot noise'.

Another way to reduce the `shot noise' is by using large bin sizes for
counting the cross-pairs, but this will limit the spatial resolution of
our $\xi$~measurements. Therefore, we have chosen to compute the
cross-pairs at scales $r_{\perp}<10$ \mpc using a linear grid of $0.5$
\mpc~in both $(r_{\perp},r_{\parallel})$ coordinates and apply a
Gaussian filter of $0.5$ \mpc~standard deviation (in both directions)
to smooth the final counts distribution obtained from \Cref{eq:DDsum}
{\it before} applying \Cref{eq:LSgg,eq:LSaa,eq:LSag}. We treated the
edges of the grid as if they were mirrors for the smoothing. As an
example, \Cref{fig:counts_ag} shows the number of cross-pairs between
\hi~absorption systems and galaxies for our `Full Sample' (defined in
\Cref{results}) using our adopted binning and smoothing.

An isotropic smoothing is desirable to avoid introducing artificial
distortions, especially at the smallest scales. The use of a smoothing
filter is justified by assuming that the underlying matter distribution
that gives rise to \hi~absorption systems and galaxies (and hence to
the data-data cross-pairs) is also smooth. Our approach offers a
compromise between reducing the `shot noise' while keeping a relatively
small bin size. We caution though, that if the geometry of \hi~clouds
does contain sharp edges at scales smaller than our adopted binning or
smoothing length, then we would not be able to detect such a feature.

\subsection{Random samples}\label{random}

One of the crucial steps for a correlation analysis is the construction
of the random samples. In order to cancel out any possible bias we
preserved the sensitivity function of the real survey in our random
samples. A detailed description of the random generator algorithms for
\hi~absorption systems and galaxies is presented in the following
sections.

\begin{figure*}
  \begin{minipage}{0.49\textwidth}
    \includegraphics[width=1\textwidth]{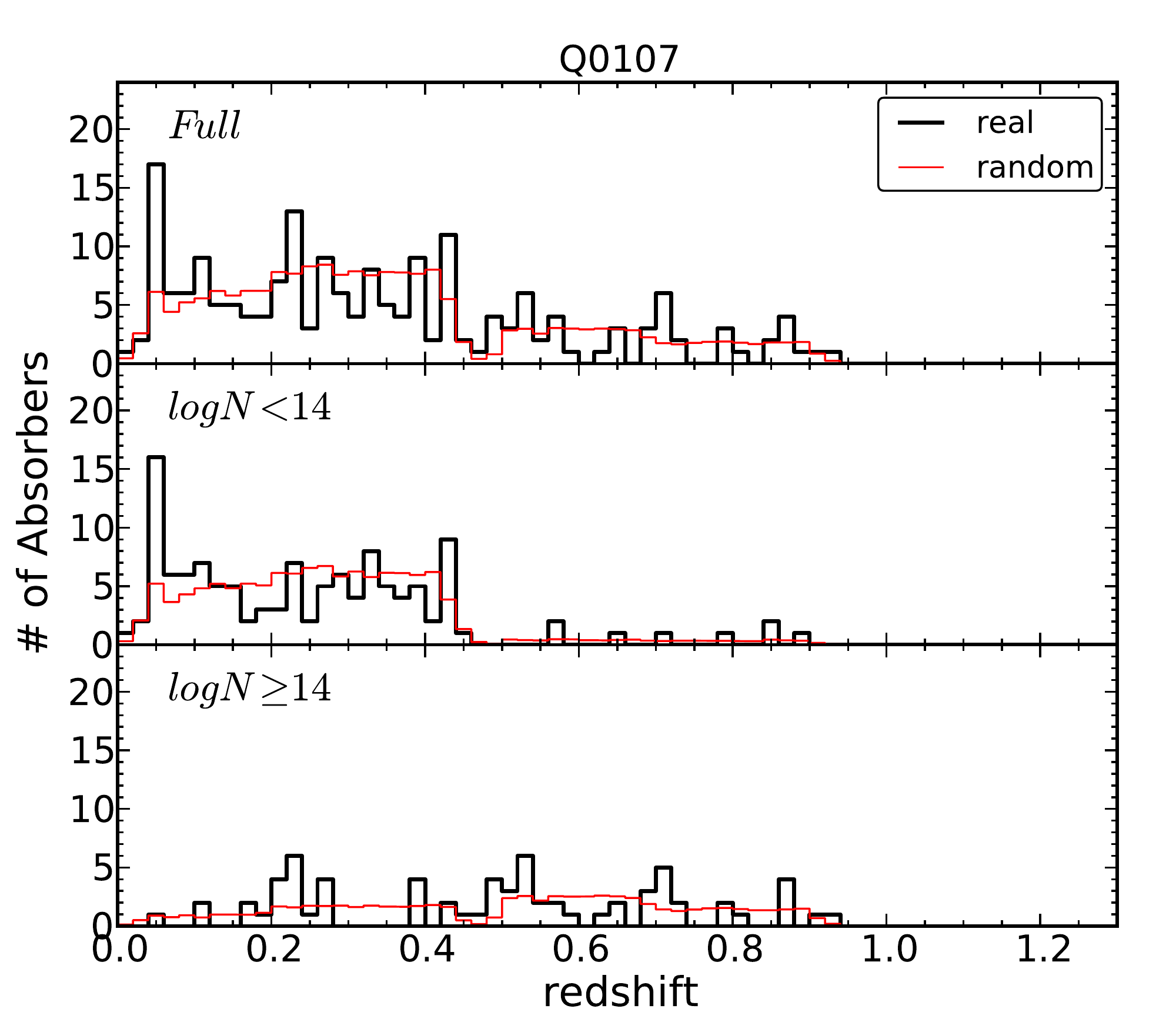}
  \end{minipage}
  \begin{minipage}{0.49\textwidth}
    \includegraphics[width=1\textwidth]{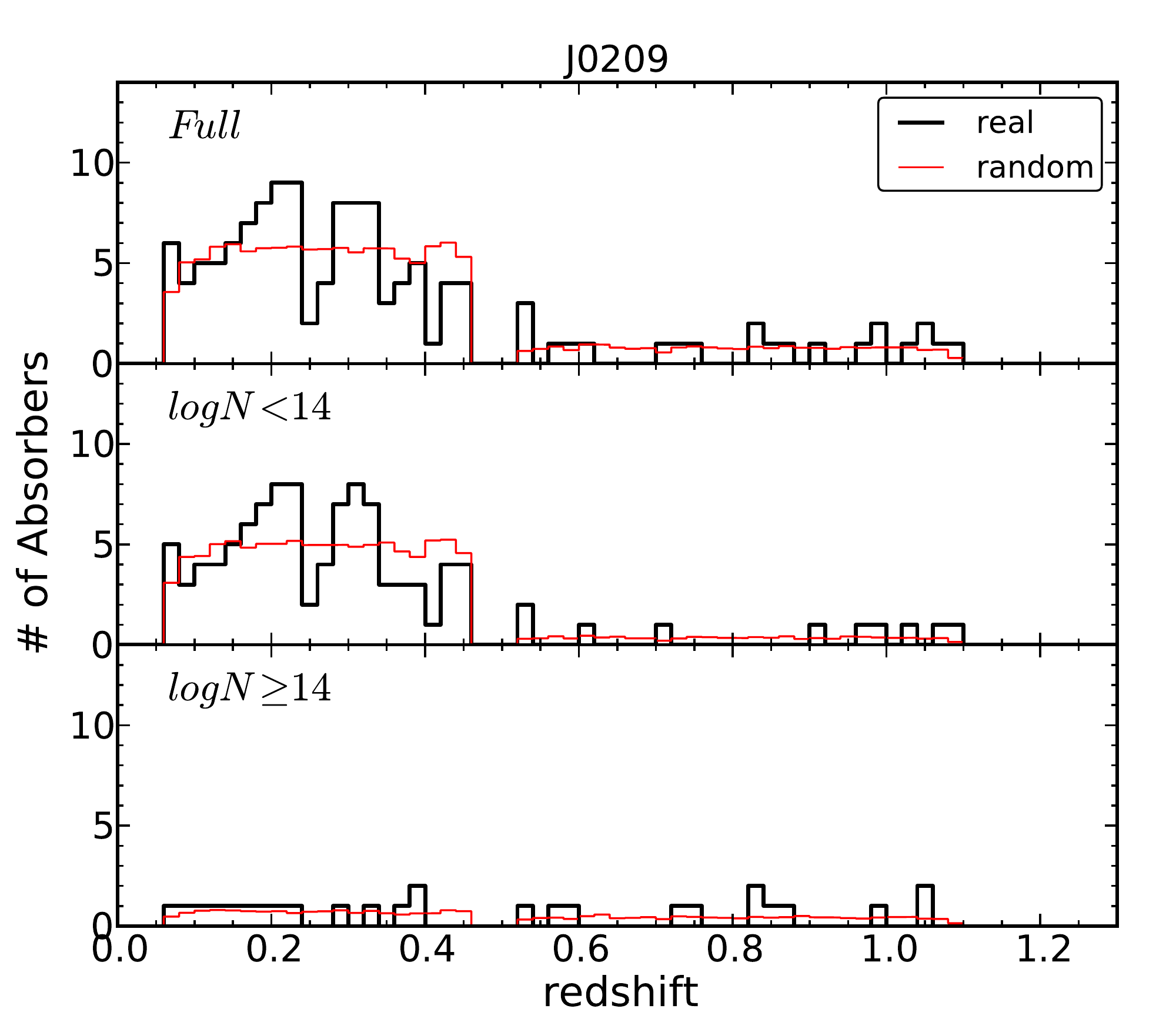}
  \end{minipage}
  \begin{minipage}{0.49\textwidth}
    \includegraphics[width=1\textwidth]{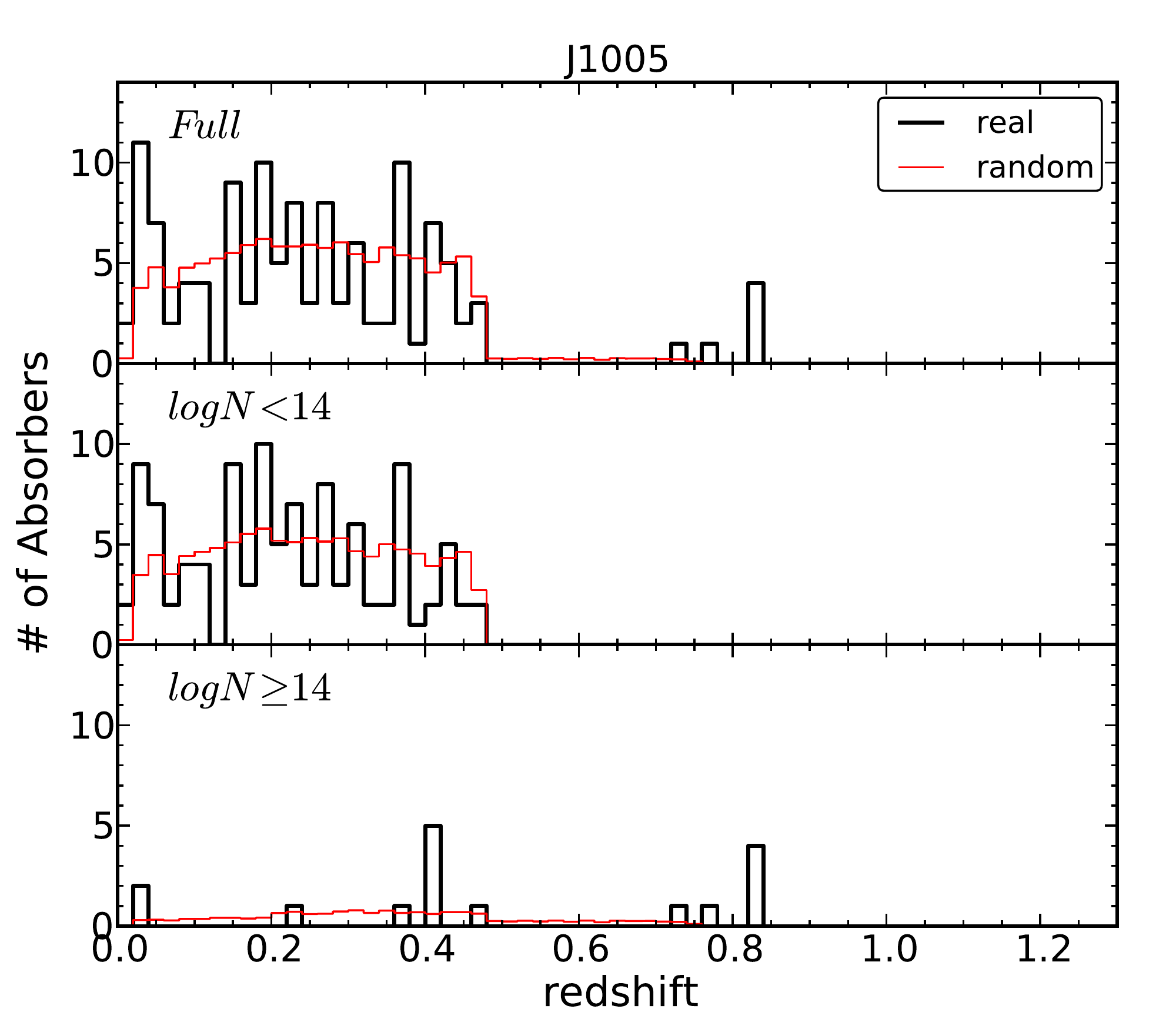}
  \end{minipage}
  \begin{minipage}{0.49\textwidth}
  \includegraphics[width=1\textwidth]{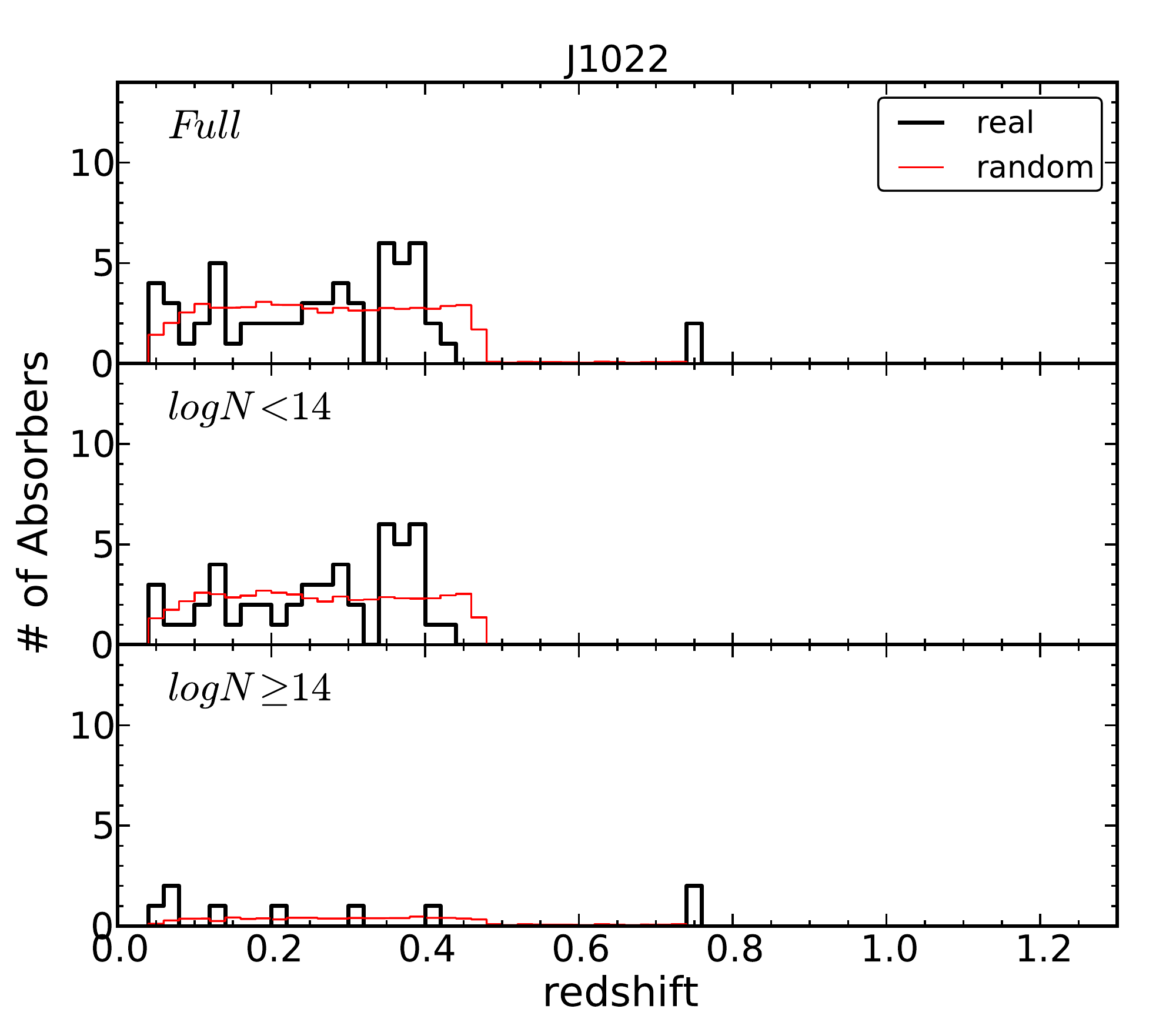}
  \end{minipage}
  \begin{minipage}{0.49\textwidth}
    \includegraphics[width=1\textwidth]{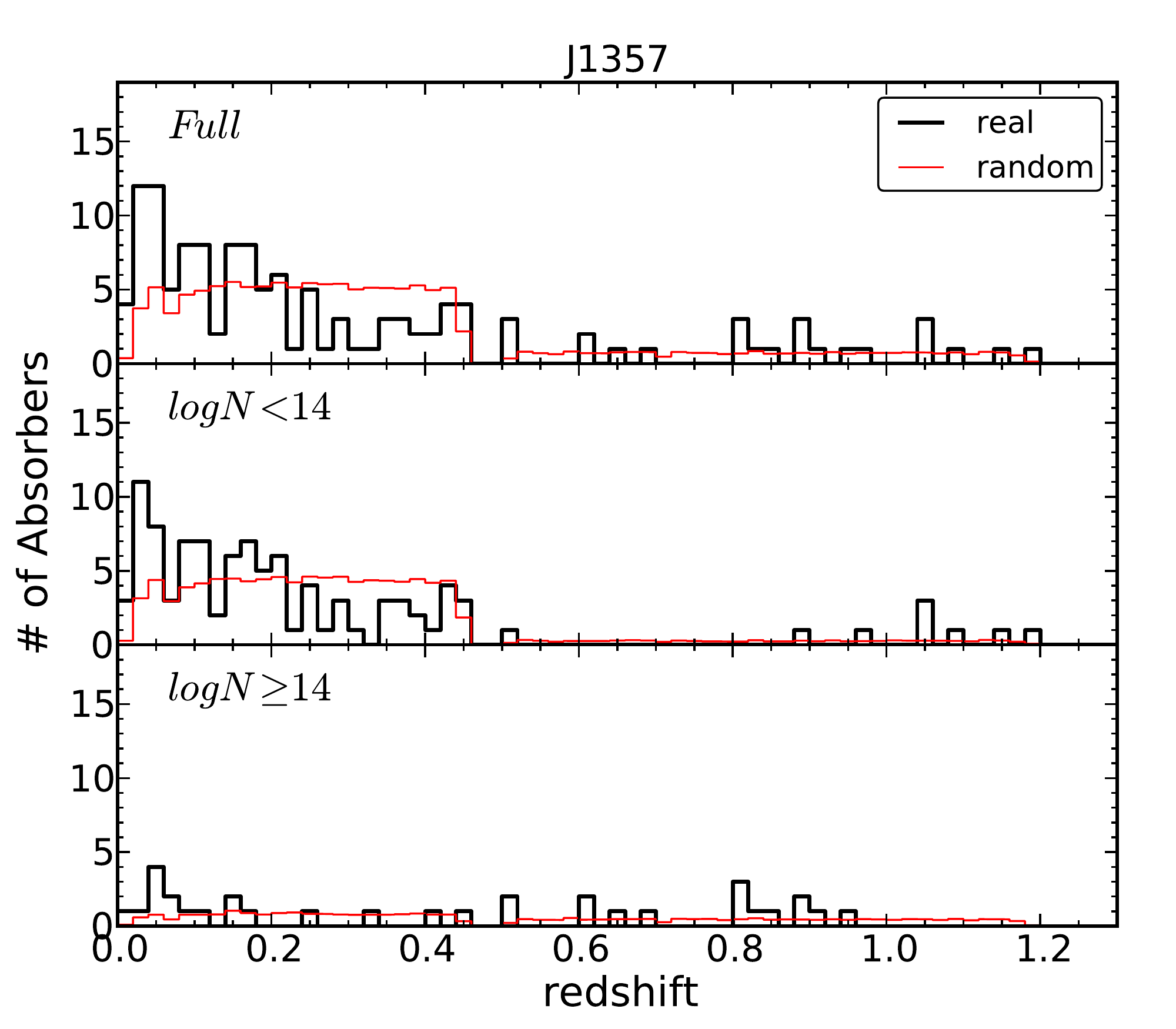}
  \end{minipage}
  \begin{minipage}{0.49\textwidth}
    \includegraphics[width=1\textwidth]{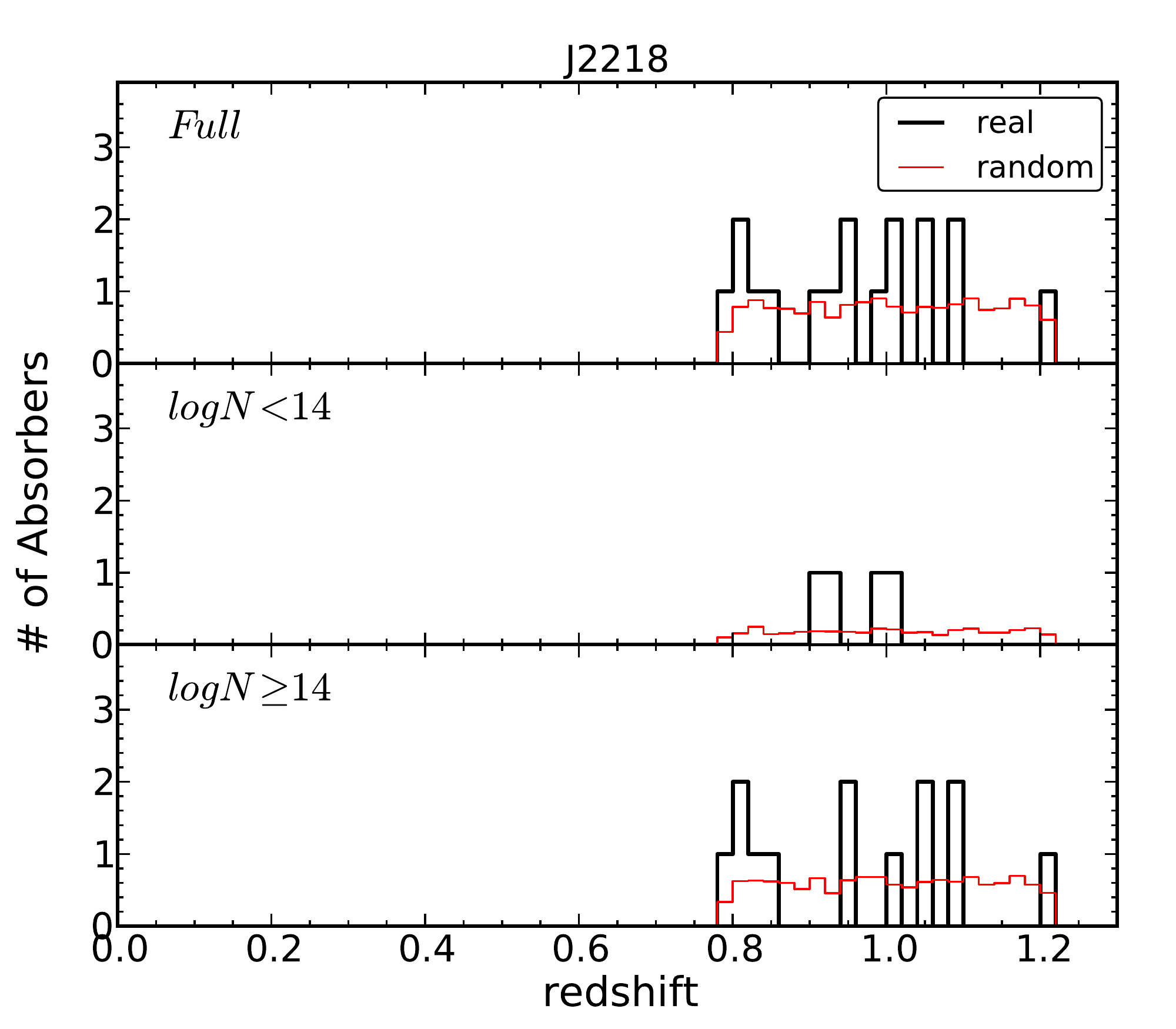}
  \end{minipage}

  \caption{Histograms of the \hi~absorption systems redshift
    distribution for our different fields ($0.02$ binning).  The black
    thick solid lines correspond to the real distributions whereas the
    red thin solid lines correspond to the normalized random
    expectation drawn from samples of $200 \times$ the real sample
    sizes. A full description of the random generator algorithm can be
    found in \Cref{random:HI}. Top panels show the full \hi~samples
    while the middle and bottom panels show subsamples based on
    \nhi~cuts.}\label{fig:zhist_abs}

\end{figure*}
 
\begin{figure*}
  \begin{minipage}{0.49\textwidth}
    \includegraphics[width=1\textwidth]{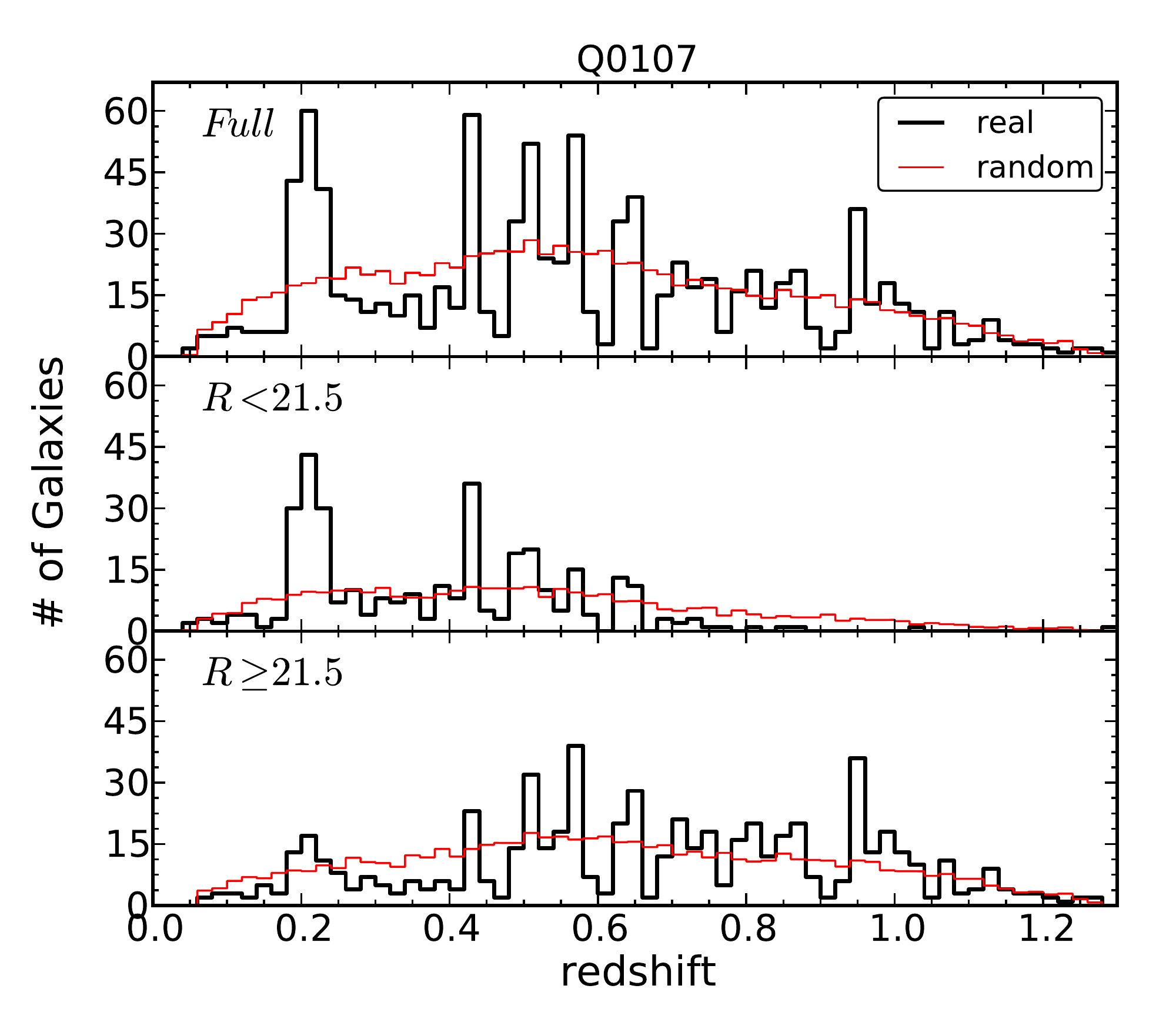}
  \end{minipage}
  \begin{minipage}{0.49\textwidth}
    \includegraphics[width=1\textwidth]{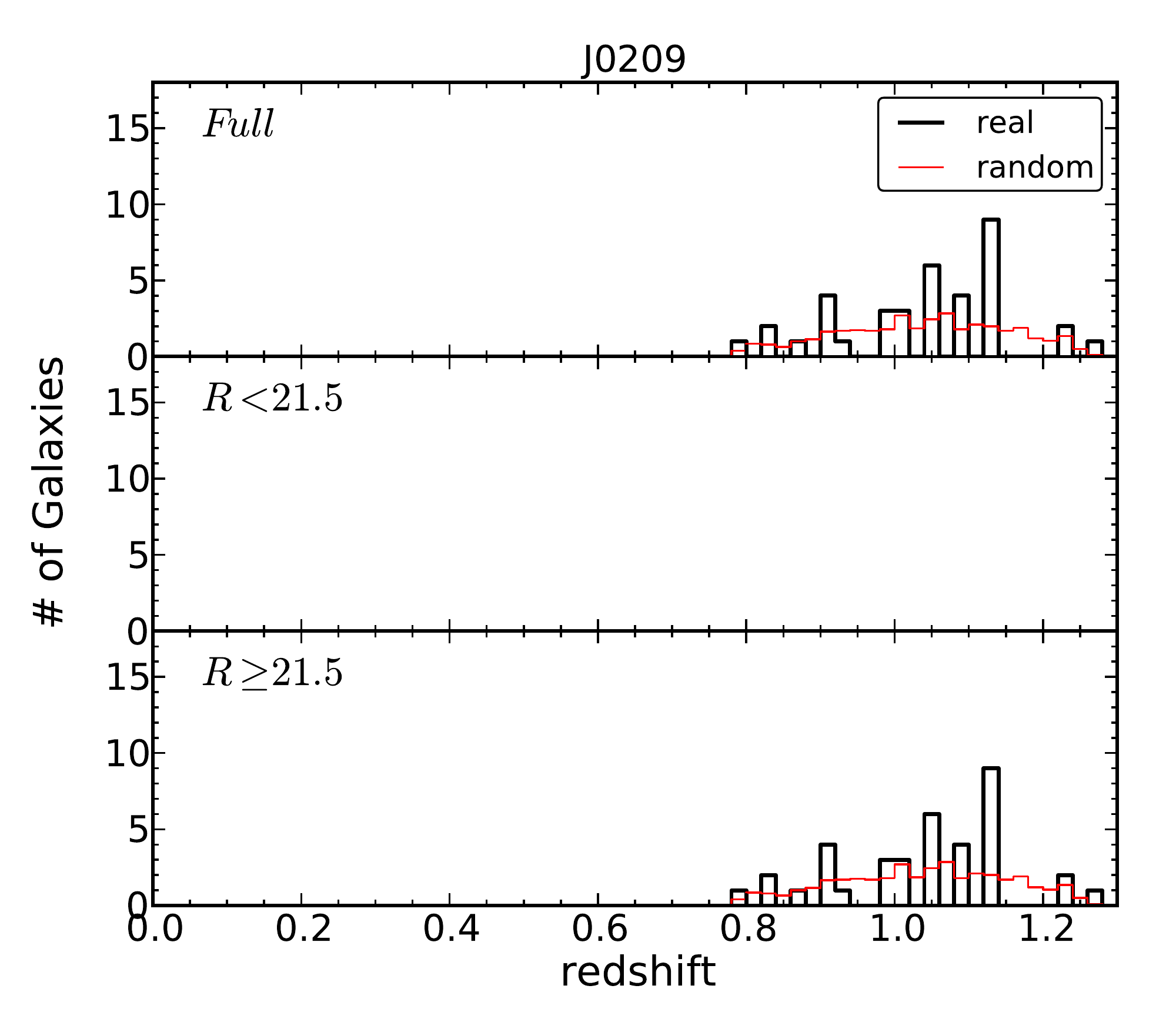}
  \end{minipage}
  \begin{minipage}{0.49\textwidth}
    \includegraphics[width=1\textwidth]{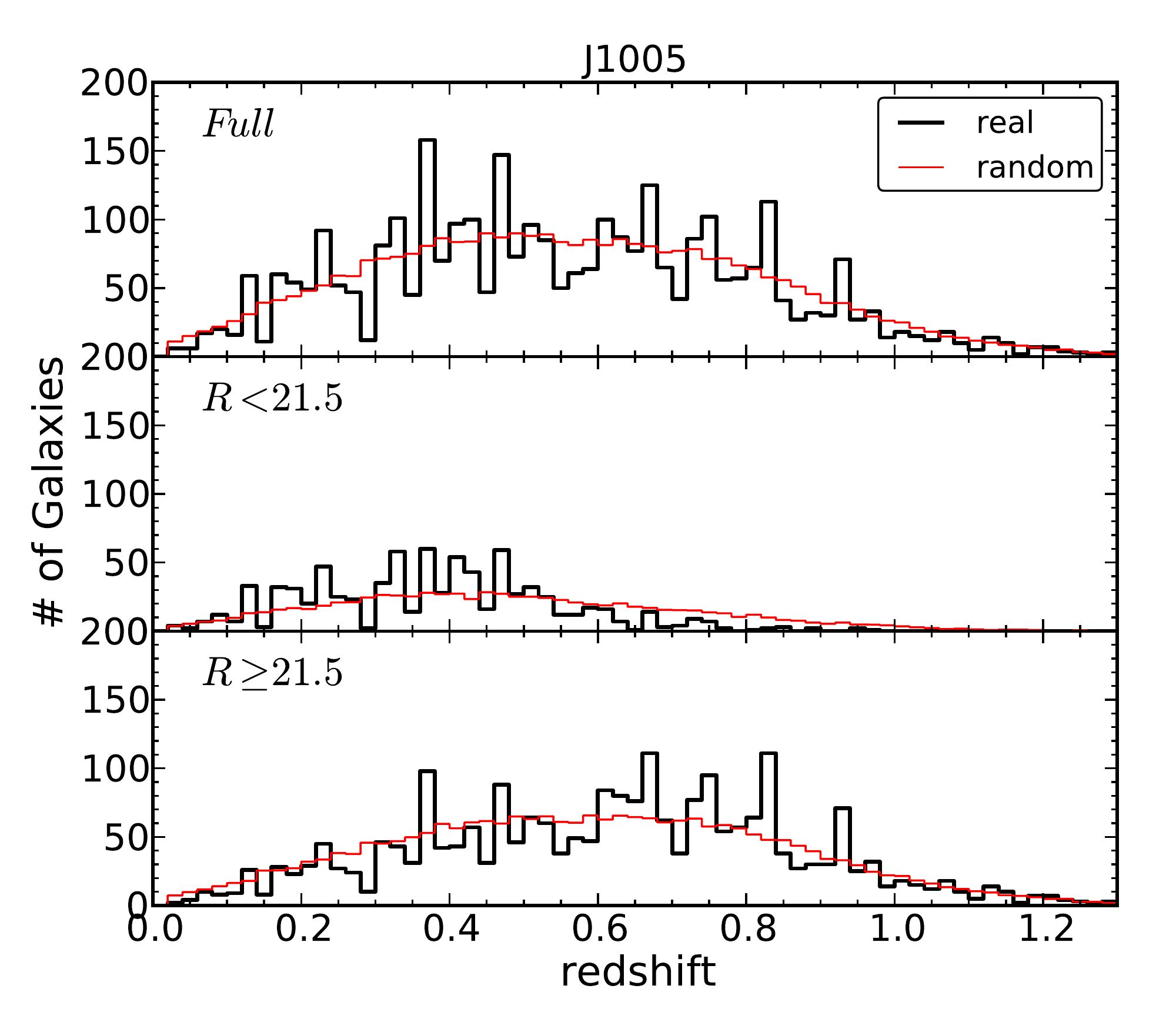}
  \end{minipage}
  \begin{minipage}{0.49\textwidth}
    \includegraphics[width=1\textwidth]{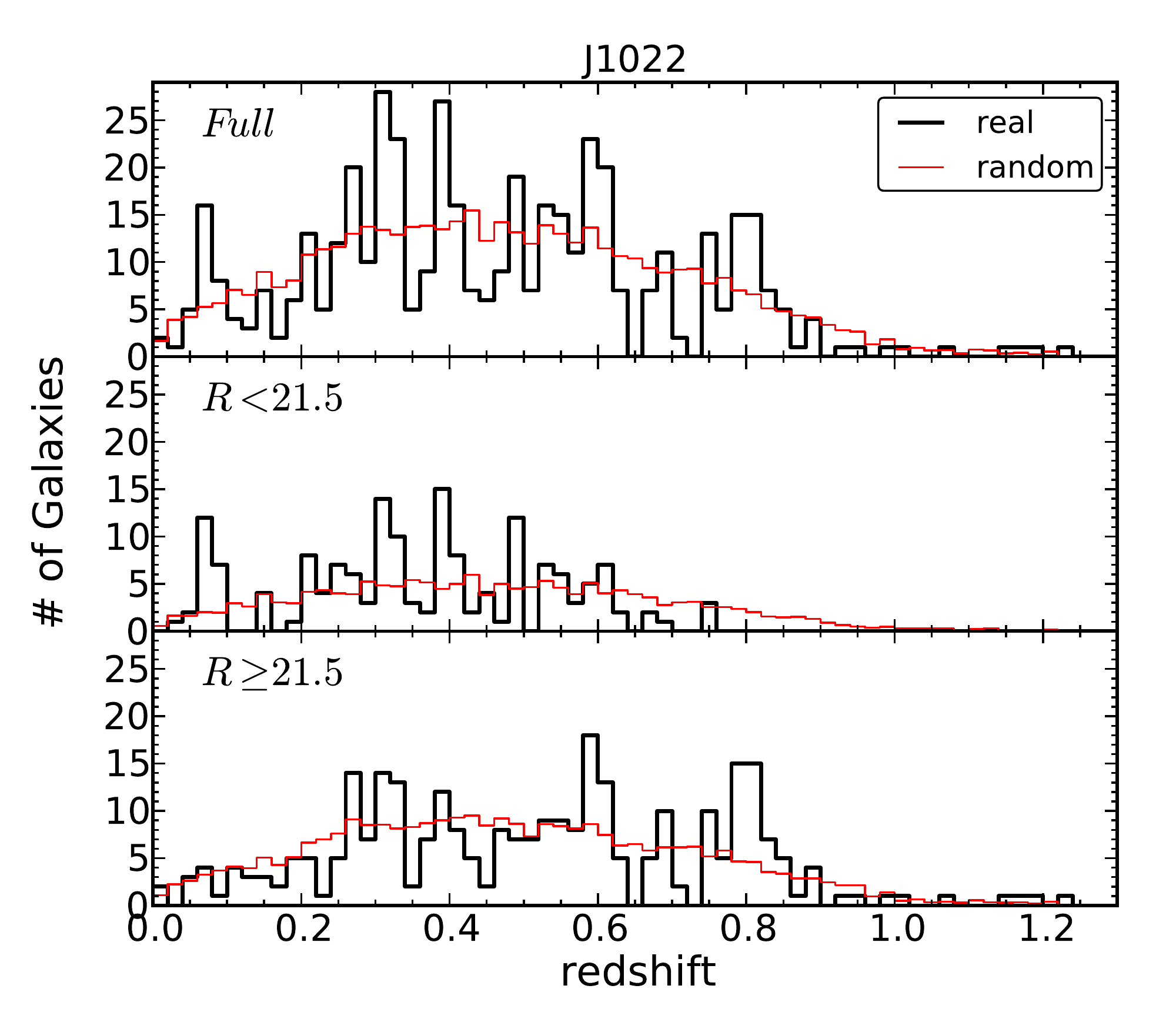}
  \end{minipage}
  \begin{minipage}{0.49\textwidth}
    \includegraphics[width=1\textwidth]{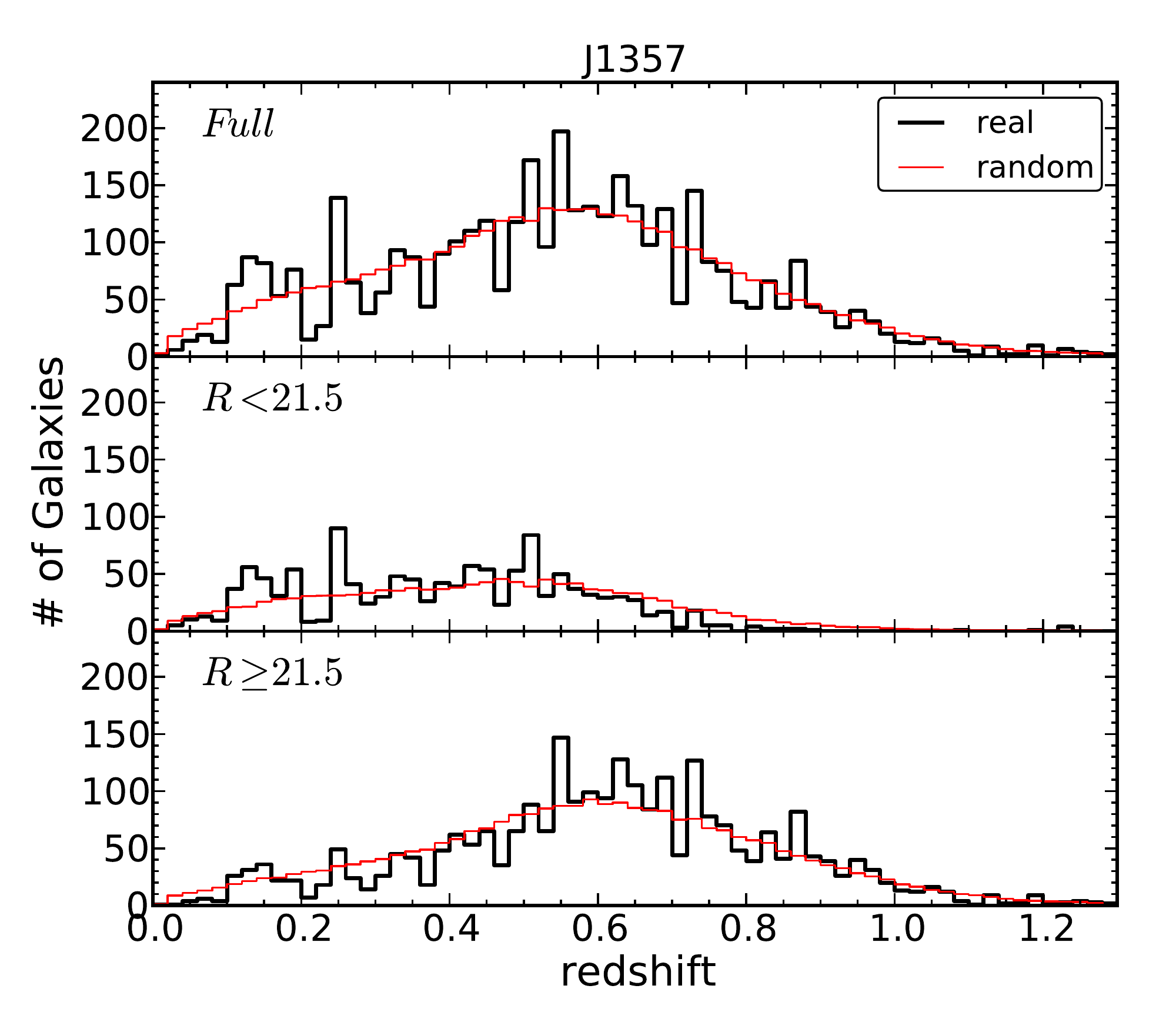}
  \end{minipage}
  \begin{minipage}{0.49\textwidth}
    \includegraphics[width=1\textwidth]{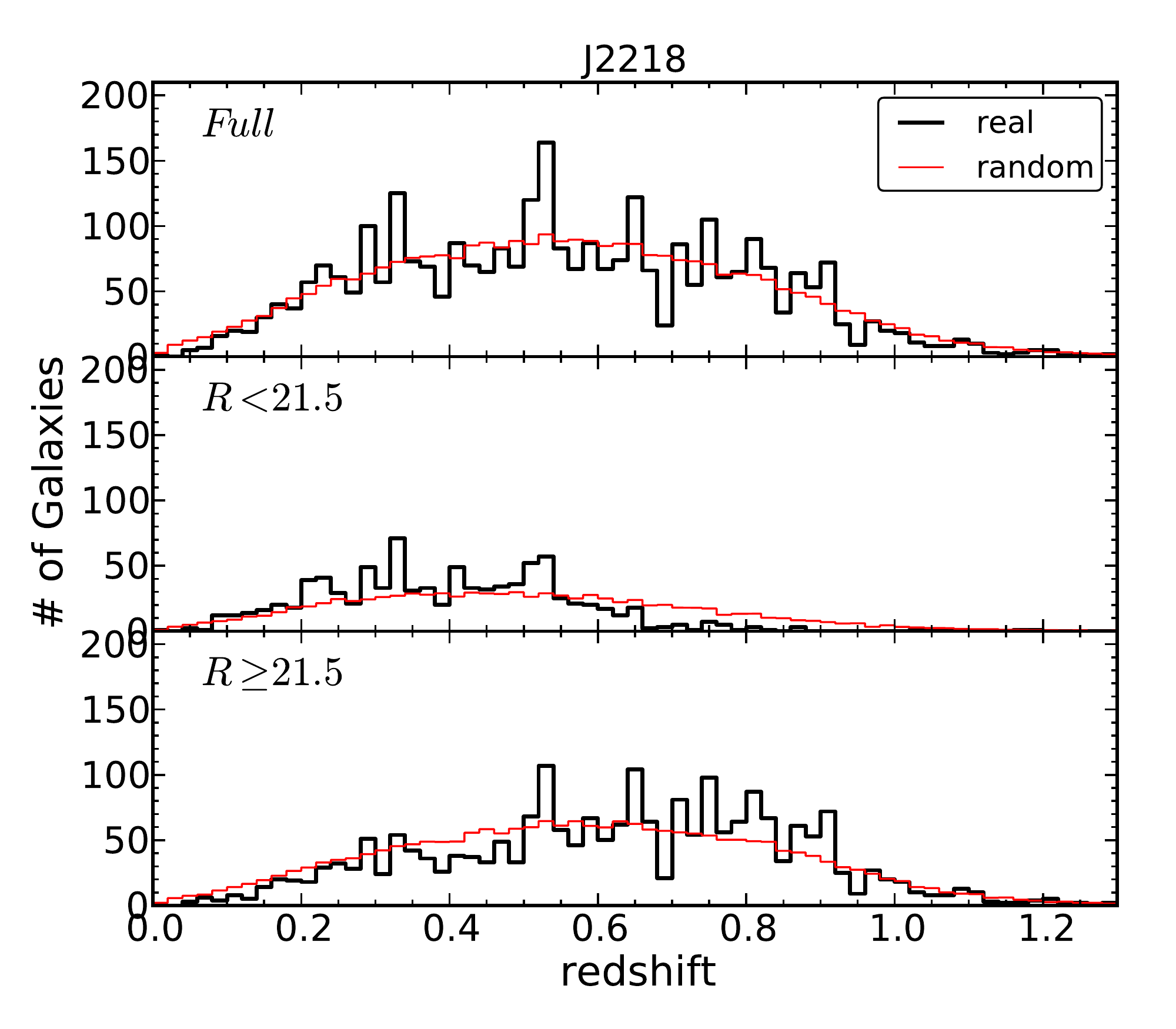}
  \end{minipage}
    
  \caption{Histograms of the galaxy redshift distribution for our
    different fields ($0.02$ binning).  The black thick solid lines
    correspond to the real distributions whereas the red thin solid
    lines correspond to the normalized random expectation drawn from
    samples of $20 \times$ the real sample sizes. A full description of
    the random generator algorithm can be found in
    \Cref{random:gal}. Top panels show the full galaxy samples while
    the middle and bottom panels show subsamples based on
    \rband~magnitude cuts.}\label{fig:zhist_gal}

\end{figure*}

\begin{figure*}
  \begin{minipage}{0.49\textwidth}
    \includegraphics[width=1\textwidth]{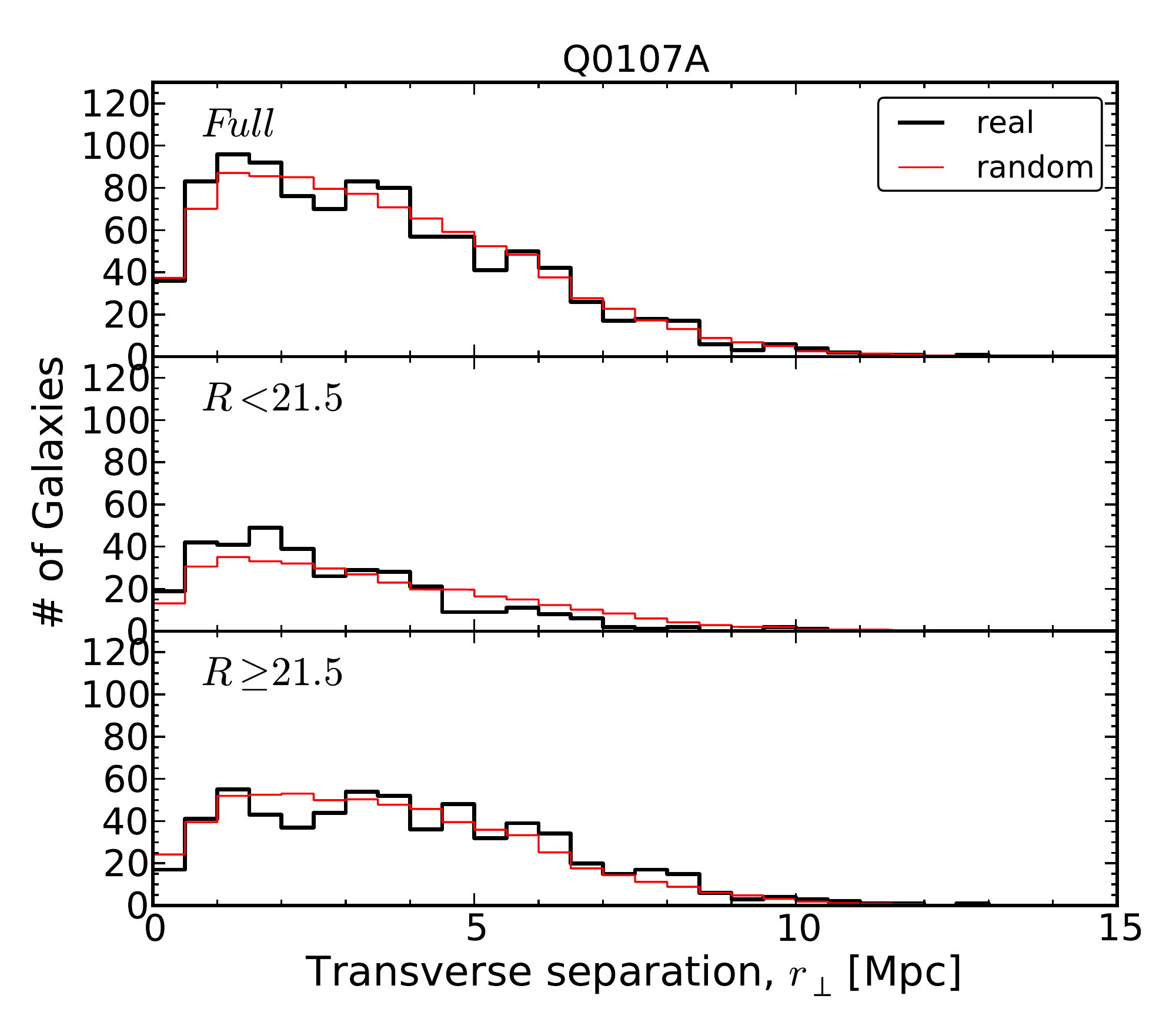}
  \end{minipage}
  \begin{minipage}{0.49\textwidth}
    \includegraphics[width=1\textwidth]{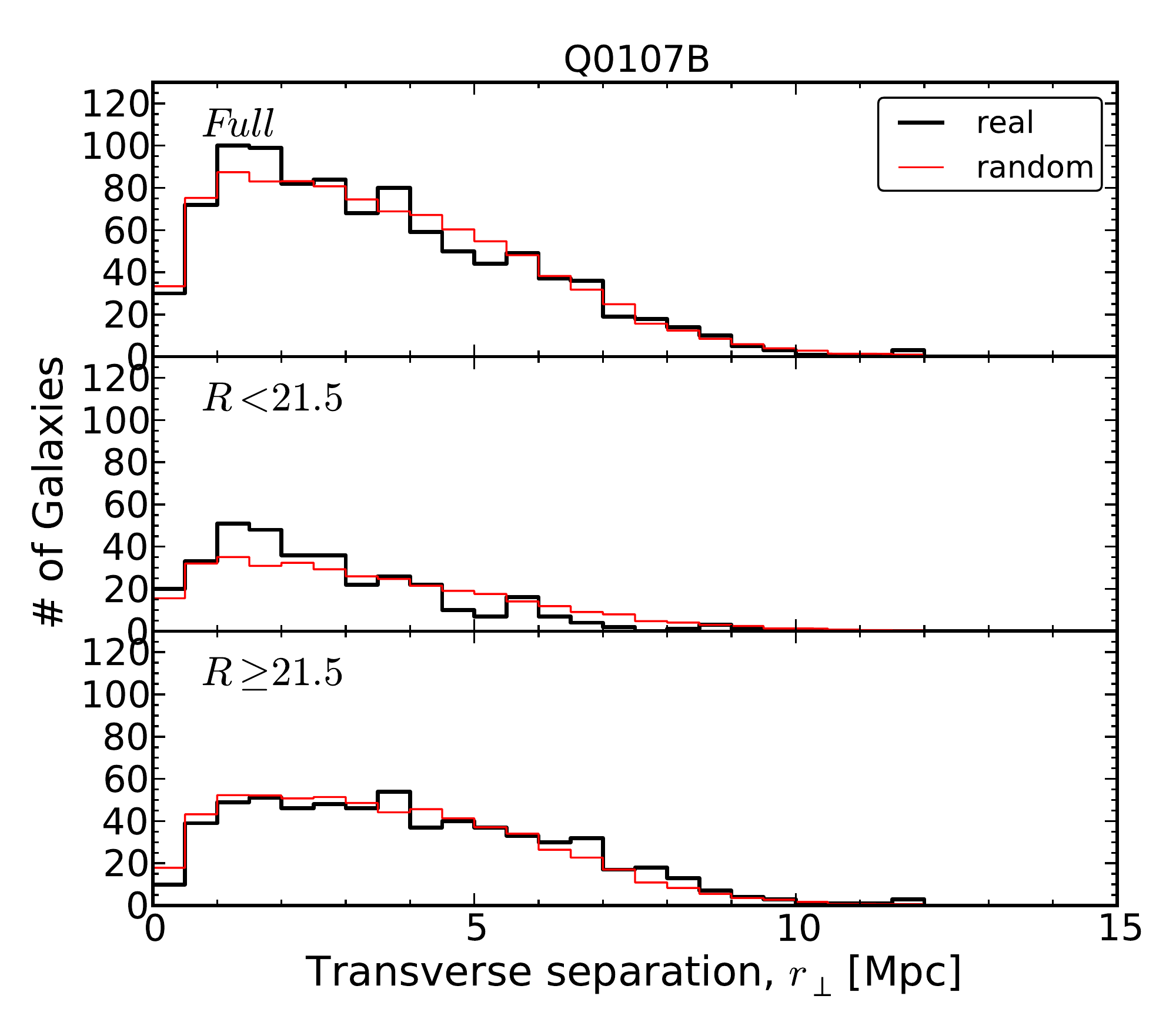}
  \end{minipage}
  \begin{minipage}{0.49\textwidth}
    \includegraphics[width=1\textwidth]{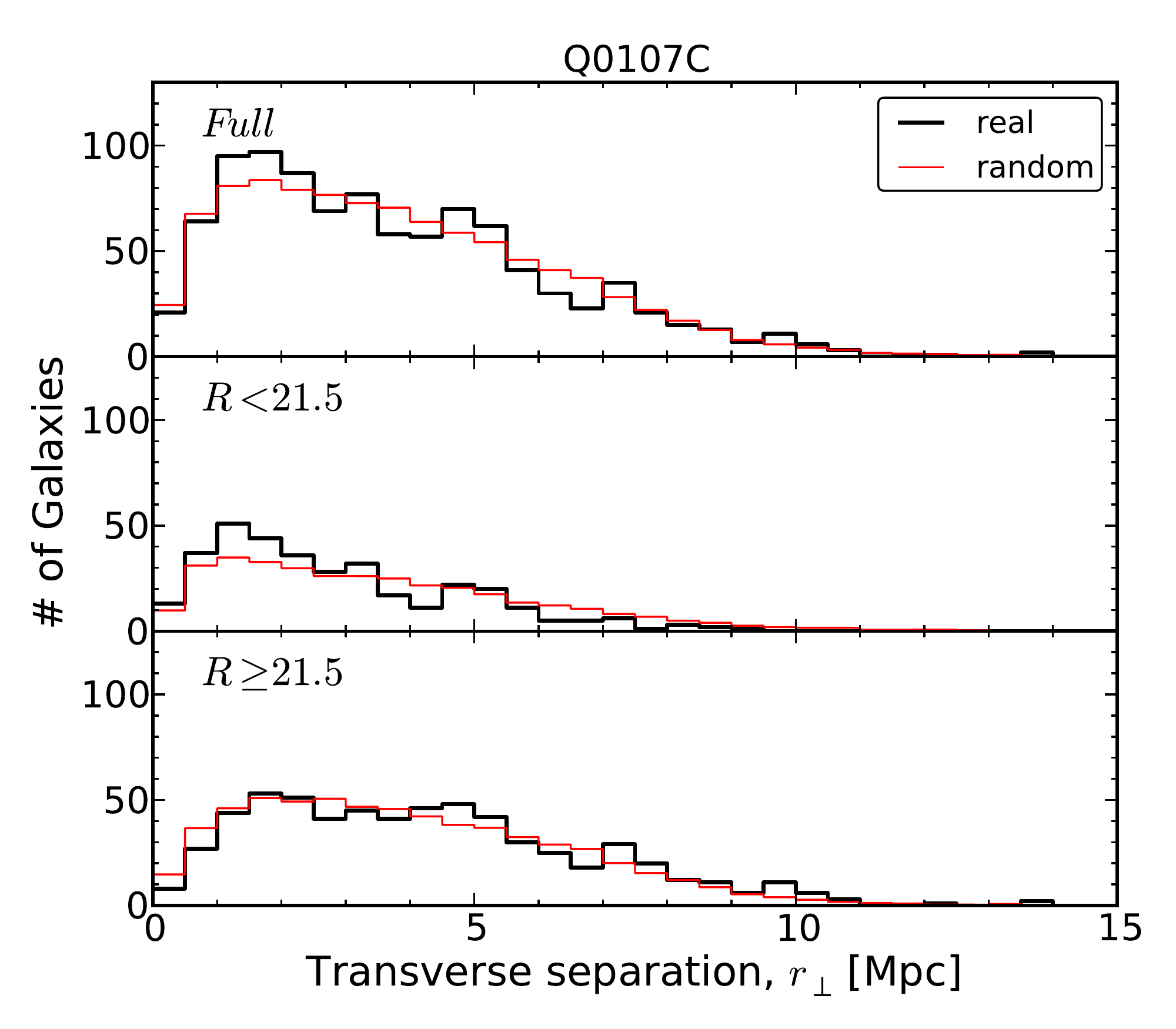}
  \end{minipage}
  \begin{minipage}{0.49\textwidth}
    \includegraphics[width=1\textwidth]{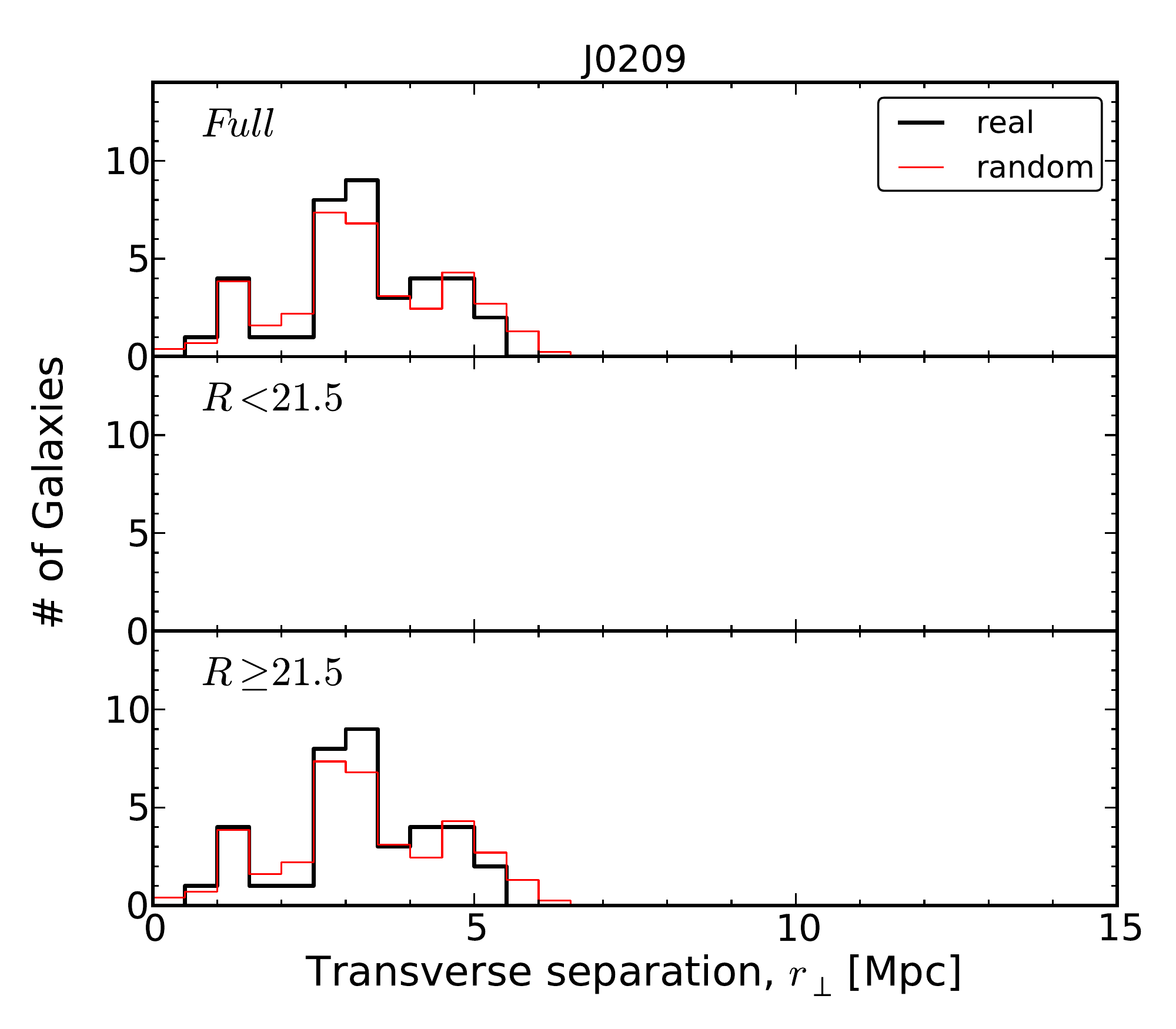}
  \end{minipage}
  \begin{minipage}{0.49\textwidth}
    \includegraphics[width=1\textwidth]{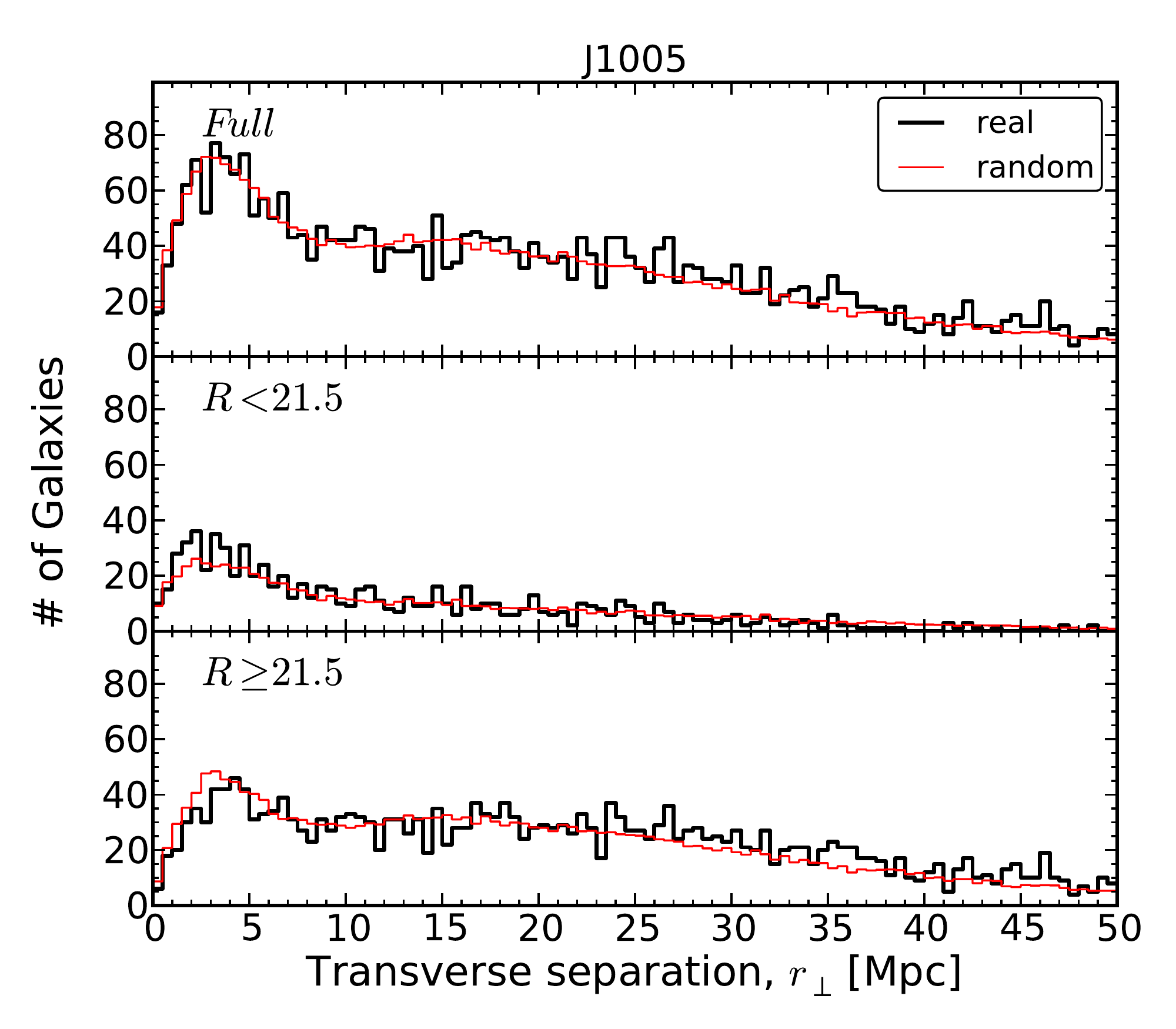}
  \end{minipage}
  \begin{minipage}{0.49\textwidth}
    \includegraphics[width=1\textwidth]{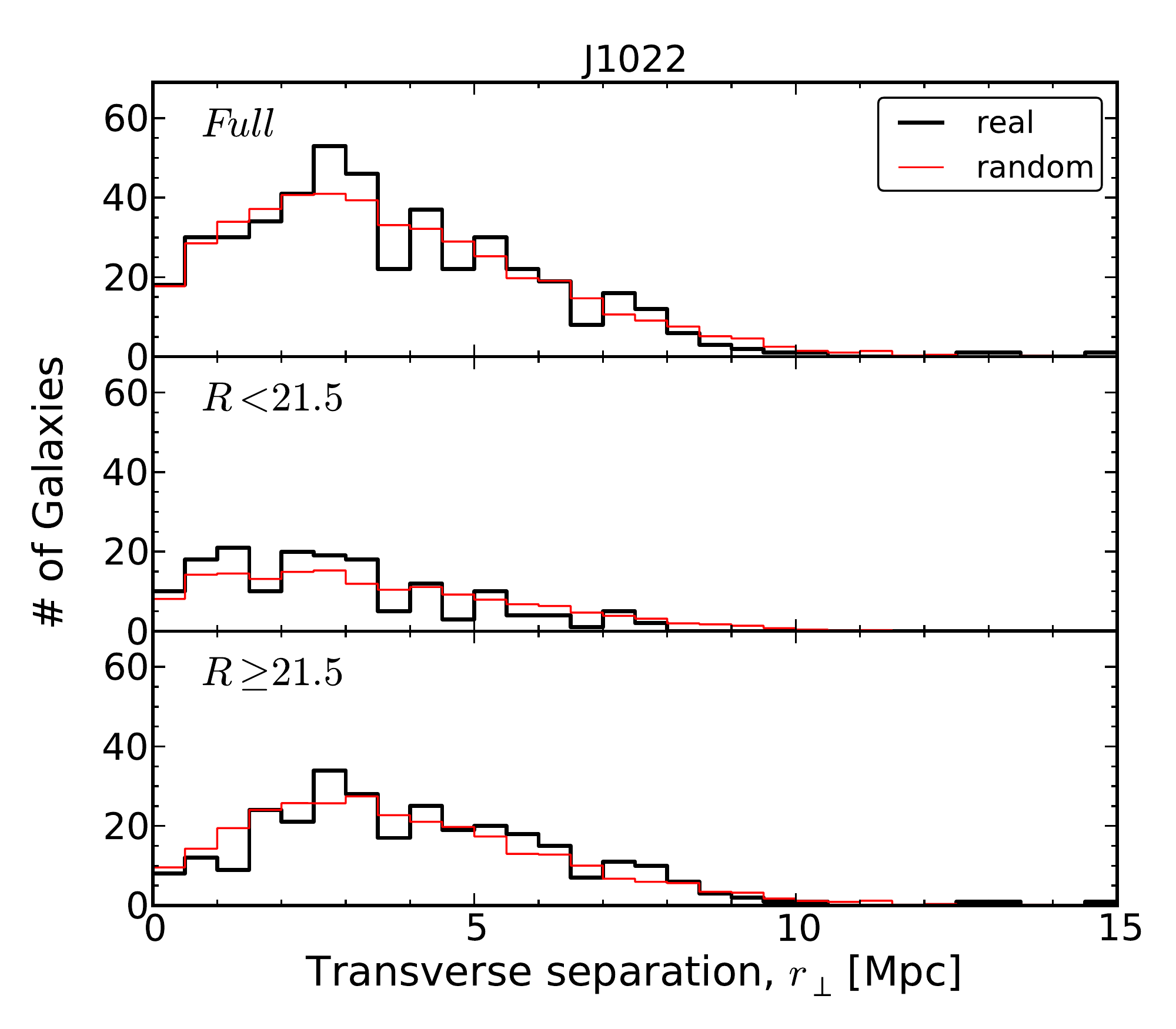}
  \end{minipage}
  \caption{Histograms of the galaxy transverse separation distribution
    for our different fields ($0.5$ \mpc binning).  The black thick
    solid lines correspond to the real distributions whereas the red
    thin solid lines correspond to the normalized random expectation
    drawn from samples of $20 \times$ the real sample sizes. A full
    description of the random generator algorithm can be found in
    \Cref{random:gal}. Top panels show the full galaxy samples while
    the middle and bottom panels show subsamples based on
    \rband~magnitude cuts.}\label{fig:yzhist_gal}

\end{figure*}

\begin{figure*}
  \begin{minipage}{0.49\textwidth}
    \includegraphics[width=1\textwidth]{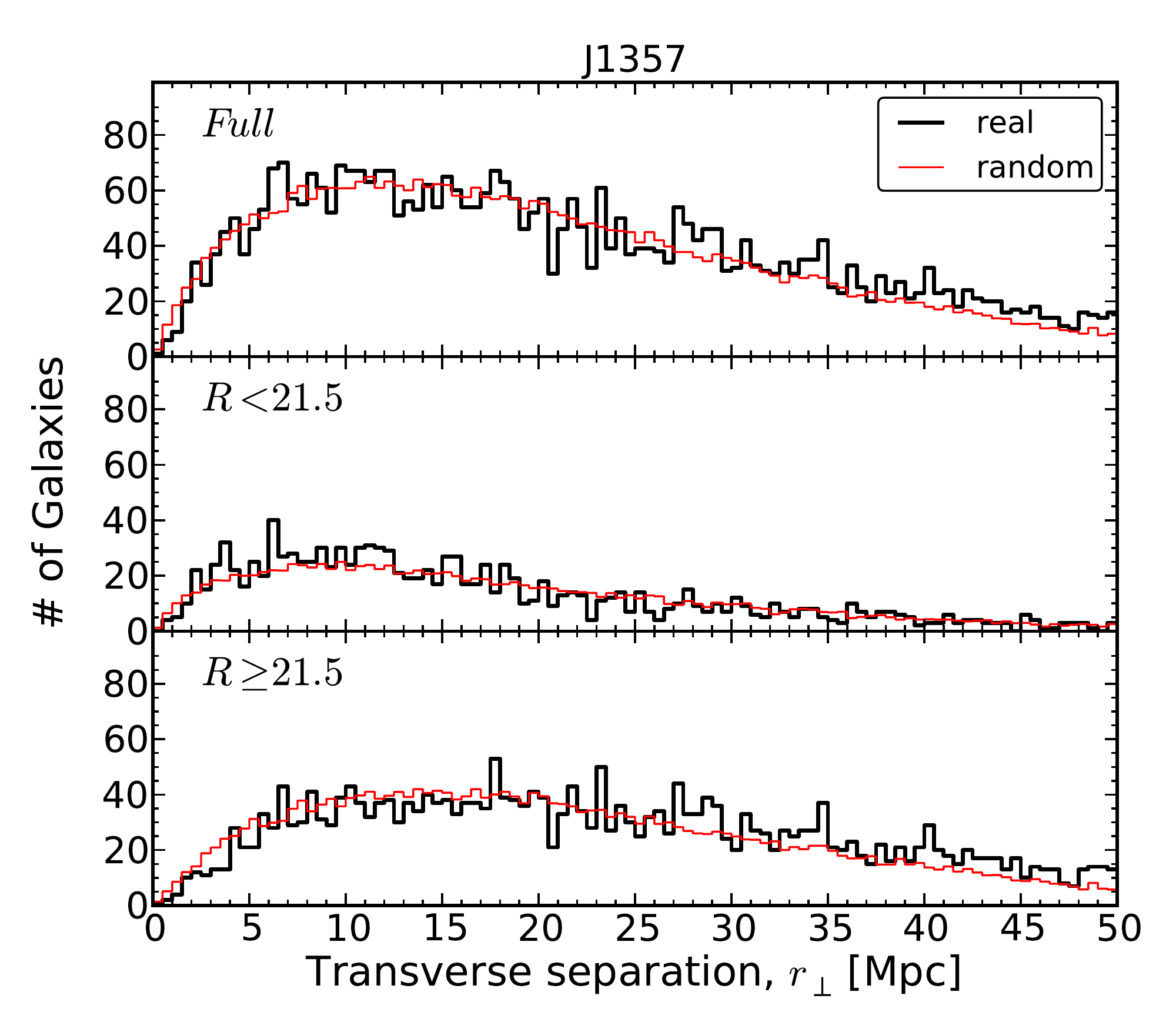}
  \end{minipage}
  \begin{minipage}{0.49\textwidth}
    \includegraphics[width=1\textwidth]{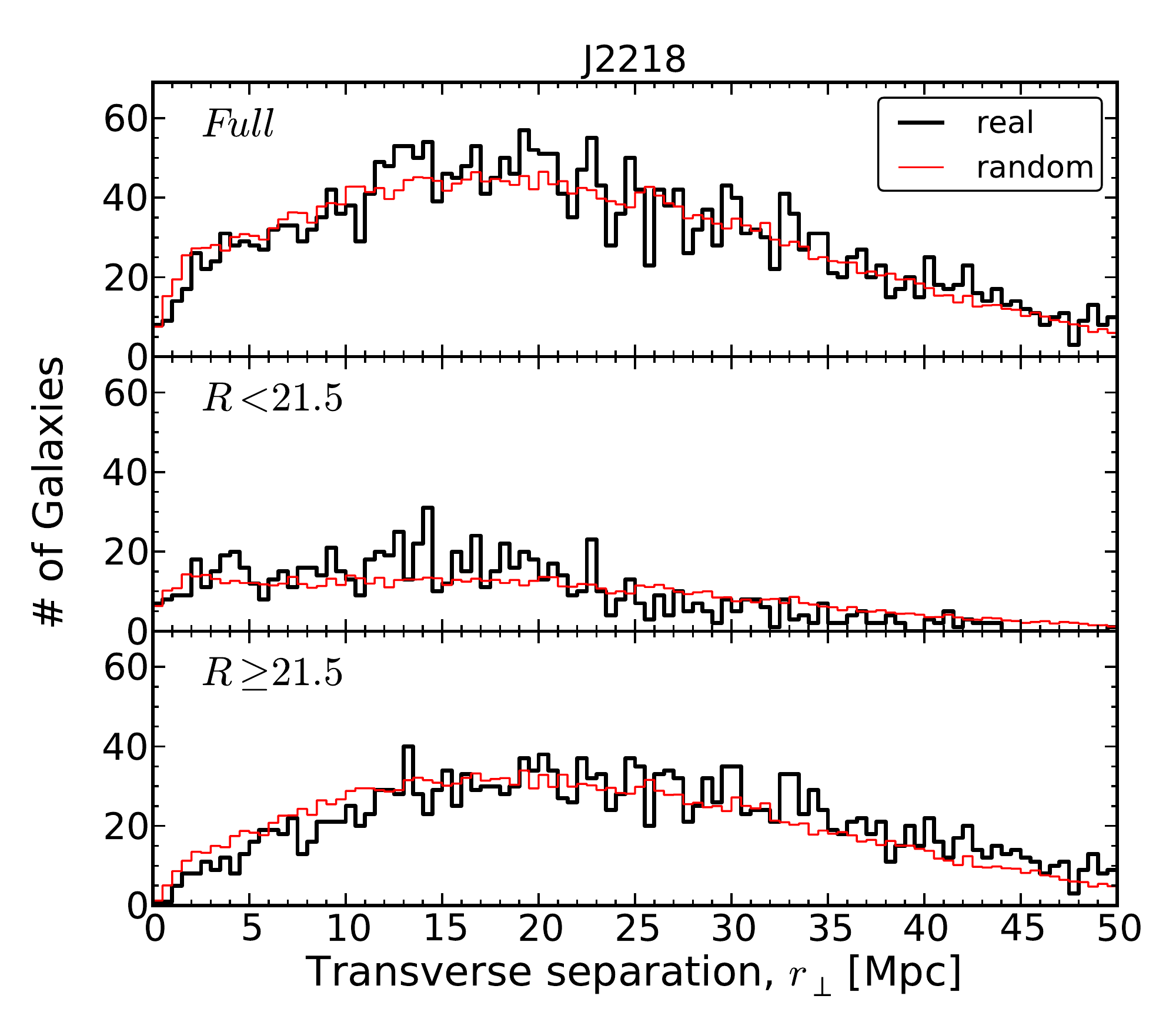}
  \end{minipage}
    
  \contcaption{}
\end{figure*}

\subsubsection{Random absorption lines}\label{random:HI}

We created random samples for individual observations made with a given
instrument and/or instrument setting (i.e. resolution, wavelength
coverage, etc). This means that we treat the two channels of \ac{cos}
(\ac{fuv} and \ac{nuv}) independently for the creation of the random
samples, and also for \ac{fos}. For a given absorption system with (RA,
DEC, \zabs, \nhi, \bhi) we create \nabsrand~random ones, varying the
redshift but preserving the rest of its parameters.

The random redshifts were chosen based on the properties of the
spectrum in which the original absorption system was observed. We first
estimated the minimum rest-frame equivalent width of a transition that
could have been observed in the spectrum at a redshift $z$. For
unresolved features, the minimum equivalent width for a line to be
detected at wavelength $\lambda$ is

\begin{equation}
W_{\rm min}(\lambda) \approx sl \ \frac{\rm FWHM}{\langle S/N
  \rangle_{\lambda}} \ \rm{,}
\label{eq:Wmin1}
\end{equation}

\noindent where $sl$ is the significance level of the detection in
standard deviation units, \ac{fwhm} is the `full-width at half maximum'
of the line spread function (LSF) of the spectrograph in \AA, and
$\langle S/N \rangle_{\lambda}$ is the average signal-to-noise per
resolution element. Transforming $\lambda$ coordinates to redshift
coordinates for a given rest-frame transition at $\lambda_0$
(i.e. $\lambda \ \rightarrow \ z=\frac{\lambda}{\lambda_0} - 1$), and
assuming a constant spectral resolution $R \equiv \frac{\lambda}{\rm
  FWHM}$, the rest-frame minimum equivalent width is then given by

\begin{equation}
W_{\rm r, min}(z) \approx sl \frac{\lambda_0}{ R\langle S/N
  \rangle_{z}} \ \rm{.}
\label{eq:Wmin2}
\end{equation}

\noindent Finally, for a given absorber with equivalent width, $W_{\rm
  HI}^{\rm obs}$, we compare it with $W_{\rm r, min}(z)$ and place
\nabsrand~random absorbers uniformly at redshifts where the condition
$W_{\rm HI}^{\rm obs} \ge W_{\rm r, min}(z)$ is satisfied. We masked
out spectral regions over a velocity window of $\pm200$ \kms~around the
position where strong Galactic absorption could have been detected
(namely: \ion{C}{2}, \ion{N}{5}, \ion{O}{1}, \ion{Si}{2}, \ion{P}{3},
\ion{S}{2} and \ion{Fe}{2}) {\it before} the random redshifts are
assigned.

Even though we have direct measurements of the equivalent widths for
the real absorption systems, we do not use them directly in order to
avoid confusion from blended systems. We use instead the approximation
given by \citet[][see his equation 9.27]{Draine2011} to convert the
inferred \nhi~and \bhi~to a $W_{\rm HI}^{\rm obs}$. Note that passing
from $W_{\rm HI} \rightarrow (N_{\rm HI},b_{\rm HI})$ is not always
robust because of the flat part of the curve-of-growth, but passing
from $(N_{\rm HI},b_{\rm HI}) \rightarrow W_{\rm HI}$ is.

We mainly based our search of \hi~absorption systems on the
\lya~transition (for which $\lambda_0 = 1215.67$~\AA), but in some
cases we extended it to \lyb~in spectral regions with no
\lya~coverage. For the \lyb~detected systems, we applied the same
method described above but changing the transition parameters
accordingly.

\Cref{fig:zhist_abs} presents the redshift distribution of real (black
lines) and random (red lines) absorbers in each of our independent
fields using \nabsrand$=200$.

\subsubsection{Random galaxies}\label{random:gal}

The random galaxies were created for each field and instrument
independently. This means that we treat different galaxy surveys
independently for the creation of the random samples, even when the
galaxy surveys come from the same field. For a given observed galaxy
with (RA, DEC, \zgal, magnitude, spectral type, etc.) we create
\ngalrand~random ones, varying the redshift, but preserving the rest of
its parameters. This approach ensures the selection function is well
matched by the random galaxies.

The random redshifts (\zrandgal) were chosen based on the observed
redshift distribution. We made sure that our randoms resembled the
observed galaxy distribution independently of the observed magnitude of
the galaxies. To do so, we selected multiple subsamples of galaxies at
different magnitude bins, whose empirical redshift distributions are
used as proxies for the redshift selection function. We used magnitude
bins of size $1$, shifted by $0.5$ magnitudes, ranging from $15$ to
$25$. For the brighter and fainter ends of the subsamples we increased
the magnitude bin sizes to ensure a minimum of $20$ galaxies. For each
magnitude subsample, we computed histograms using redshift bins of
$\Delta z=0.01$ (arbitrary), which were then smoothed with a Gaussian
filter of standard deviation $\sigma=0.1$ (roughly corresponding to a
co-moving scale of $\approx 300$ \mpc~at redshift $z = 0.5$). This
large smoothing length is important to get rid of the \ac{lss} spikes
and valleys present in the real redshift distributions. The final
redshift probability distribution of a given magnitude bin is obtained
by cubic spline interpolation over the smoothed histograms. Thus, for a
given galaxy with observed magnitude $m$, we placed \ngalrand~randoms
according to the spline fit associated with the subsample of galaxies
centred on the closest magnitude bin to $m$. We also imposed the
redshifts of the random galaxies to lie between $z_{\rm min}<z_{\rm
  gal}^{\rm random}<z_{\rm max}$, where $z_{\rm min}$ and $z_{\rm max}$
are the minimum and maximum galaxy redshifts of the real sample.

\Cref{fig:zhist_gal} presents the redshift distribution of real (black
lines) and random (red lines) galaxies in each of our independent
fields using \ngalrand$=20$. Similarly, \Cref{fig:yzhist_gal} presents
the distribution in transverse separations of real (black lines) and
random (red lines) galaxies with respect to their respective \ac{qso}
\ac{los}.

\subsection{Projected correlations along the line-of-sight}\label{correlation:1d}

A useful quantity to compute from the two-dimensional correlation
functions is the projected correlation function along the \ac{los},

\begin{equation}
\Xi (r_{\perp}) = 2 \int_0^\infty \! \xi(r_{\perp},r_{\parallel}) \,
    {\rm d} r_{\parallel} \ \rm{,}
\label{eq:projected}
\end{equation}

\noindent as it will be insensitive to redshift distortions, at least
for the transverse separations involved in this work
\citep{Davis1983}. Therefore, one can find a relation between the
`real-space' correlation function (distortion free),
$\xi(r=\sqrt{r_{\parallel}^2+r_{\perp}^2})$, and $\Xi (r_{\perp})$, as

\begin{equation}\begin{split}
\Xi (r_{\perp}) &= 2 \int_0^\infty \! \xi(r) \, {\rm d} r_{\parallel}
\\ &= 2 \int_{r_{\perp}}^\infty \! \xi(r)\frac{r \, {\rm d}
  r}{\sqrt{r^2-r_{\perp}^2}} \ \rm{,}
\label{eq:davis}
\end{split}\end{equation}

\noindent which leads to $\xi(r)$ being given by the inverse Abel
transform,

\begin{equation}
\xi(r) = - \frac{1}{\pi} \int_{r}^\infty \! \frac{{\rm
    d}\Xi(r_{\perp})}{{\rm d}r_{\perp}} \frac{{\rm d}
  r_{\perp}}{\sqrt{r_{\perp}^2-r^2}} \ \rm{.}
\label{eq:Abel}
\end{equation}

\citet{Davis1983} showed that when $\xi(r)$ is described by a power-law
of the form,

\begin{equation}
  \xi(r) = \left(\frac{r}{r_0}\right)^{-\gamma} \ \rm{,}
  \label{eq:rscf}
\end{equation}

\noindent then \Cref{eq:davis} yields to

\begin{equation}
  \Xi (r_{\perp}) = A(r_0,\gamma) r_{\perp}^{1-\gamma} \ \rm{,}
  \label{eq:fit}
\end{equation}

\noindent where $A(r_0,\gamma) = r_0^{\gamma}\Gamma(1/2)\Gamma[(\gamma
  - 1)/2] / \Gamma(\gamma/2)$ and $\Gamma$ is the Gamma
function. Therefore, $r_0$ and $\gamma$ of $\xi(r)$ can be obtained
directly from a power-law fit to $\Xi(r_{\perp})$, using
\Cref{eq:fit}. Note that this method is only valid for $\gamma > 1$.

In practice, we will use $r_{\parallel}^{\rm max}=20$~\mpc as the
integration limit in \Cref{eq:projected}. A larger integration limit
will increase the `shot noise' while not adding much correlation
power. As long as the vast majority of correlated pairs are included in
the integration limit (which is the case), this approach will suffice
\citep[e.g.][]{Davis1983, Ryan-Weber2006}. In order to further reduce
the `shot noise', we summed all the cross-pairs along the \ac{los},
e.g. $D_{\rm a}D_{\rm g}(r_{\perp}) = \sum_i D_{\rm a}D_{\rm
  g}(r_{\perp},r_{\parallel,i})$ (and so on for the others), and then
computed the \citeauthor*{Landy1993} estimators, $\xi^{\rm
  LS}(r_{\perp})$, using these collapsed cross-pairs,

\begin{equation}
\Xi (r_{\perp}) = 2 r_{\parallel}^{\rm max} \xi^{\rm LS}(r_{\perp})
\ \rm{.}
\label{eq:projected2}
\end{equation}

\noindent This approach is justified given the cylindrical geometry of
our survey, for which the `random-random' pairs (denominator of the LS
estimator) is almost constant along the $r_{\parallel}$-axis for the
scales involved in this study (e.g. see right panel of
\Cref{fig:counts_ag}). We compared the absolute values of $\Xi$ from
our adopted approach with that of a direct integration (as in
\Cref{eq:projected} using $r_{\parallel}^{\rm max}=20$~\mpc as the
integration limit). We obtained differences of $\lesssim 5\%$ in the
correlation amplitudes, indicating that our approach is appropriate.


\subsection{Relations between auto- and cross-correlations}\label{dependence}

We use the Cauchy–-Schwarz inequality,

\begin{equation}
\xi_{\rm ag}^2 \le \xi_{\rm gg}\xi_{\rm aa} \rm{,}
\label{eq:inequality}
\end{equation}

\noindent as the main tool to address the connection between \hi~and
galaxies. The equality only holds when the density fluctuations that
give rise to \hi~absorption systems and galaxies are linearly
dependent. However, in the most general case, the product of the
auto-correlation functions does not necessarily equal $\xi_{\rm
  ag}^2$. If we do assume that both \hi~absorption systems and galaxies
trace the same underlying dark matter density distribution
\citep[e.g.][]{Ryan-Weber2006}, we have

\begin{equation}\begin{split}
\xi_{\rm gg} &= b_{\rm g}^2 \xi_{\rm DM} \\ 
\xi_{\rm aa} &= b_{\rm a}^2 \xi_{\rm DM} \\ 
\xi_{\rm ag} &= b_{\rm a}b_{\rm b} \xi_{\rm DM} \rm{,}
\label{eq:biases}
\end{split}\end{equation}

\noindent where $\xi_{\rm DM}$ is the dark matter auto-correlation
function (assumed positive) and $b_{\rm g}$ and $b_{\rm a}$ are the
galaxy and \hi~`absolute biases' (also positives), respectively. If
these biases are independent of the scale (i.e. linear biases), then
the equality of \Cref{eq:inequality} holds. If that is the case, one
can use the ratio between the correlation functions to infer the dark
matter halo masses of one population relative to the other
\citep[e.g.][]{Mo1993,Ryan-Weber2006}. On the other hand, if $\xi_{\rm
  ag}^2 < \xi_{\rm gg}\xi_{\rm aa}$ we can no longer assume such a
simplistic model. In such a case, the observed difference with respect
to $\xi_{\rm ag}^2 = \xi_{\rm gg}\xi_{\rm aa}$ can be used to: (i) get
insights on the baryonic physics affecting \hi~absorption systems
and/or galaxies, assuming that the standard cosmological paradigm is
correct; or (ii) put constraints on the current cosmological paradigm,
assuming that the baryonic physics is fully understood. In this paper
we will focus on the former.\\

\citet{Adelberger2003} showed a third possibility: $\xi_{\rm ag}^2$
exceeding $\xi_{\rm gg}\xi_{\rm aa}$ for correlation functions measured
from discrete and volume limited samples. In the hypothetical case of
an \hi--galaxy one-to-one correspondence, then $\xi_{\rm gg} = \xi_{\rm
  aa}$, but $\xi_{\rm ag}$ will appear higher at the very small scales
because in the case of auto-correlations we exclude the correlation of
an object with itself, whereas in $\xi_{\rm ag}$ that correlation is
present \citep[][see their appendix A]{Adelberger2003}. Such a
behaviour between auto- and cross-correlations will indicate that the
two populations of objects are indeed the same physical entities. The
geometry of our survey might not be suitable for testing this idea, as
we are only mapping \hi~absorption systems along single \ac{los} for
which the completeness level of galaxies close to these absorbers is
low. Still, we will bear this result in mind for the interpretation of
our results.

\begin{figure*}
  \begin{minipage}{1.05\textwidth}
    \centering
    \includegraphics[width=\textwidth]{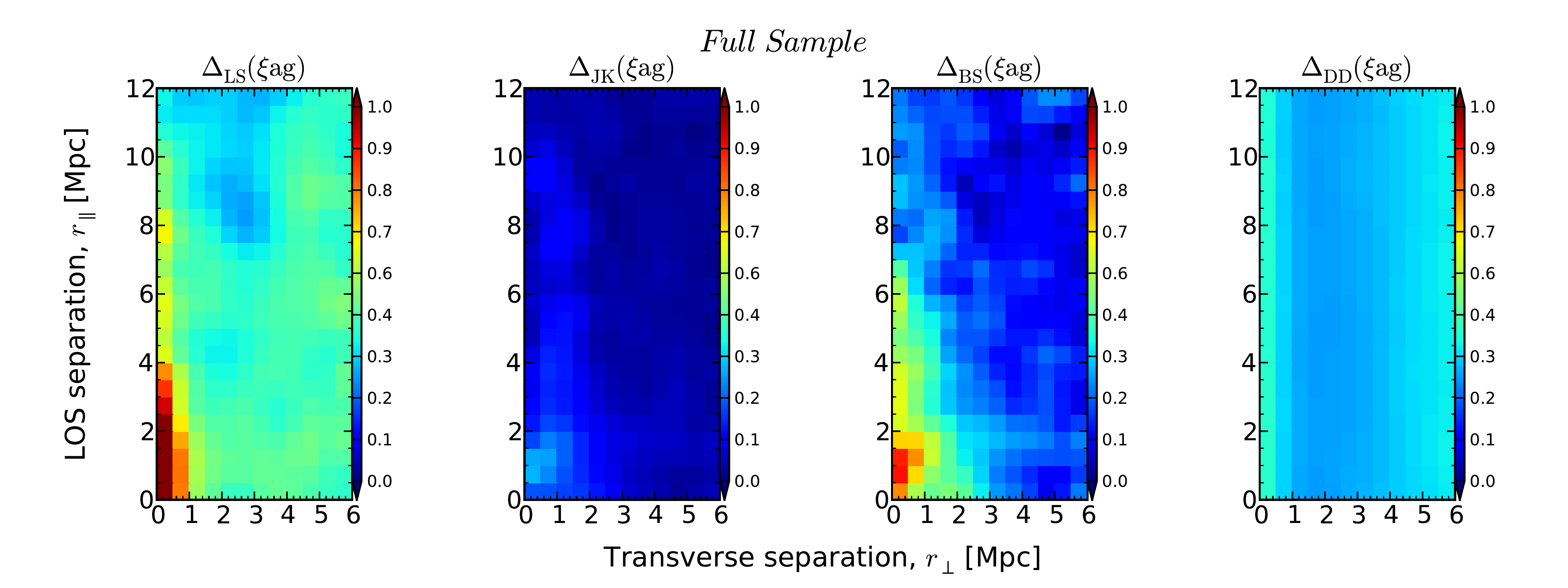}
  \end{minipage}
  
  \caption{Uncertainty estimations (square root of variances) of the
    \hi-galaxy cross-correlation, $\Delta(\xi^{\rm LS}_{\rm ag})$,
    measured from our `Full Sample' as a function of separations both
    along the line-of-sight ($r_{\parallel}$; $y$-axes) and transverse
    to the line-of-sight ($r_{\perp}$; $x$-axes). From left to right:
    uncertainty from the \citet{Landy1993} analytical approximation,
    $\Delta_{\rm LS}$ (\cref{eq:var_LS}); uncertainty from a
    `jackknife' resampling, $\Delta_{\rm JK}$ (\cref{eq:var_JK});
    uncertainty from a `bootstrap' resampling, $\Delta_{\rm BS}$
    (\cref{eq:var_BS}); and the commonly used Poissonian uncertainty,
    $\Delta_{\rm DD}$ (\cref{eq:var_DD}).}\label{fig:errors}

\end{figure*}

\subsection{Uncertainty estimation}\label{uncertainty}
When dealing with cross-correlations, it is important to realize that
the statistical uncertainties will be dominated by those of the {\it
  smallest} sample. If we consider a sample composed of a single object
and another sample composed of $100$ objects, the number of cross-pairs
is $100$, but none of these pairs are truly independent as they all
share a common object. Therefore, assuming Poissonian uncertainty for
the number of pairs \citep[as commonly done in the literature;
  e.g.][]{Chen2009} is not optimal, as it will underestimate the true
uncertainty. For {\it correlated} distributions, none of the pairs are
independent because the number of systems at a given scale will depend
on the number of systems at all other scales, and deviations from the
Poissonian expectation will be more important at the scales where the
correlation signal is large. Indeed, \citet{Landy1993} showed that the
variance of $\xi^{\rm LS}$ can be approximated by (in our notation),

\begin{equation}
\Delta^2_{\rm LS}(\xi^{\rm LS}) \approx \frac{(1+\xi^{\rm
    LS})^2}{n_{\rm DD}(RR/n_{\rm RR})} \approx \frac{(1+\xi^{\rm
    LS})^3}{DD}\ \rm{.}
\label{eq:var_LS}
\end{equation}

\noindent This variance is greater than the commonly used

\begin{equation}
\Delta^2_{\rm DD}(\xi^{\rm LS}) = \frac{1+\xi^{\rm LS}}{DD} \ \rm{,}
\label{eq:var_DD}
\end{equation}

\noindent by a factor of $\sim (1+\xi)^2$, and so we caution the use of
the latter as it might still underpredict the real uncertainty.\\

In order to test whether the uncertainty given by \Cref{eq:var_LS} is
reasonable for our survey, we also computed the `jackknife' and
`bootstrap' variances. The `jackknife' variance is computed as

\begin{equation}
\Delta^2_{\rm JK}(\xi) = \frac{1}{N_{\rm f}(N_{\rm
    f}-1)}\sum\limits_i^{N_{\rm f}} (\xi^{*}_i - \bar{\xi}^{*})^2
\ \rm{,}
\label{eq:var_JK}
\end{equation}

\noindent where $\xi^{*}_i$ is the $i$-th `pseudo-value' of the
correlation function, $\xi^{*}_i \equiv N_{\rm f}\xi - (N_{\rm
  f}-1)\xi_{-i}$, with $\xi_{-i}$ being the value of the correlation
function measured when the $i$-th field is removed from the sample, and
$\bar{\xi}^{*}$ is the mean of the `pseudo-values'. The `bootstrap'
variance is computed by creating $N_{\rm bs}=500$ sets of $N_{\rm f}$
fields, randomly chosen (with repetition) from the set of real
fields,\footnote{Note that for $6$ fields, the total number of possible
  combinations is $\frac{(6+6-1)!}{6!(6-1)!}=462$.} so

\begin{equation}
\Delta^2_{\rm BS}(\xi) = \frac{1}{N_{\rm bs}}\sum\limits_i^{N_{\rm bs}}
(\xi_i - \bar{\xi})^2 \ \rm{,}
\label{eq:var_BS}
\end{equation}

\noindent where $\xi_{i}$ is the correlation measured from the $i$-th
random set, and $\bar{\xi}$ is the mean of these `bootstrap'
measurements. Uncertainties for the projected correlations, $\Xi$, and
the ratio $(\xi_{\rm ag})^2/(\xi_{\rm gg}\xi_{\rm aa})$, were
calculated analogously.

As an example, \Cref{fig:errors} shows these $4$ uncertainty
estimations (square root of the variances) for our measurements of
$\xi_{\rm ag}^{\rm LS}(r_{\perp},r_{\parallel})$. From left to right:
$\Delta_{\rm LS}$, $\Delta_{\rm JK}$, $\Delta_{\rm BS}$ and
$\Delta_{\rm DD}$. All these uncertainty estimations are within $\sim
1$ order of magnitude consistent with each other, but systematic trends
are present. $\Delta_{\rm LS}$ and $\Delta_{\rm BS}$ give the largest
uncertainties while $\Delta_{\rm JK}$ and $\Delta_{\rm DD}$ give the
smallest. We also observe that $\Delta_{\rm LS}$, $\Delta_{\rm JK}$ and
$\Delta_{\rm BS}$ peak at the smallest scales (where the correlation
amplitudes are greater) while $\Delta_{\rm DD}$ does not. Similar
behaviors are observed for the uncertainties associated to our
$\xi_{\rm gg}^{\rm LS}(r_{\perp},r_{\parallel})$ measurements (not
shown).

\begin{figure}
  \includegraphics[width=0.45\textwidth]{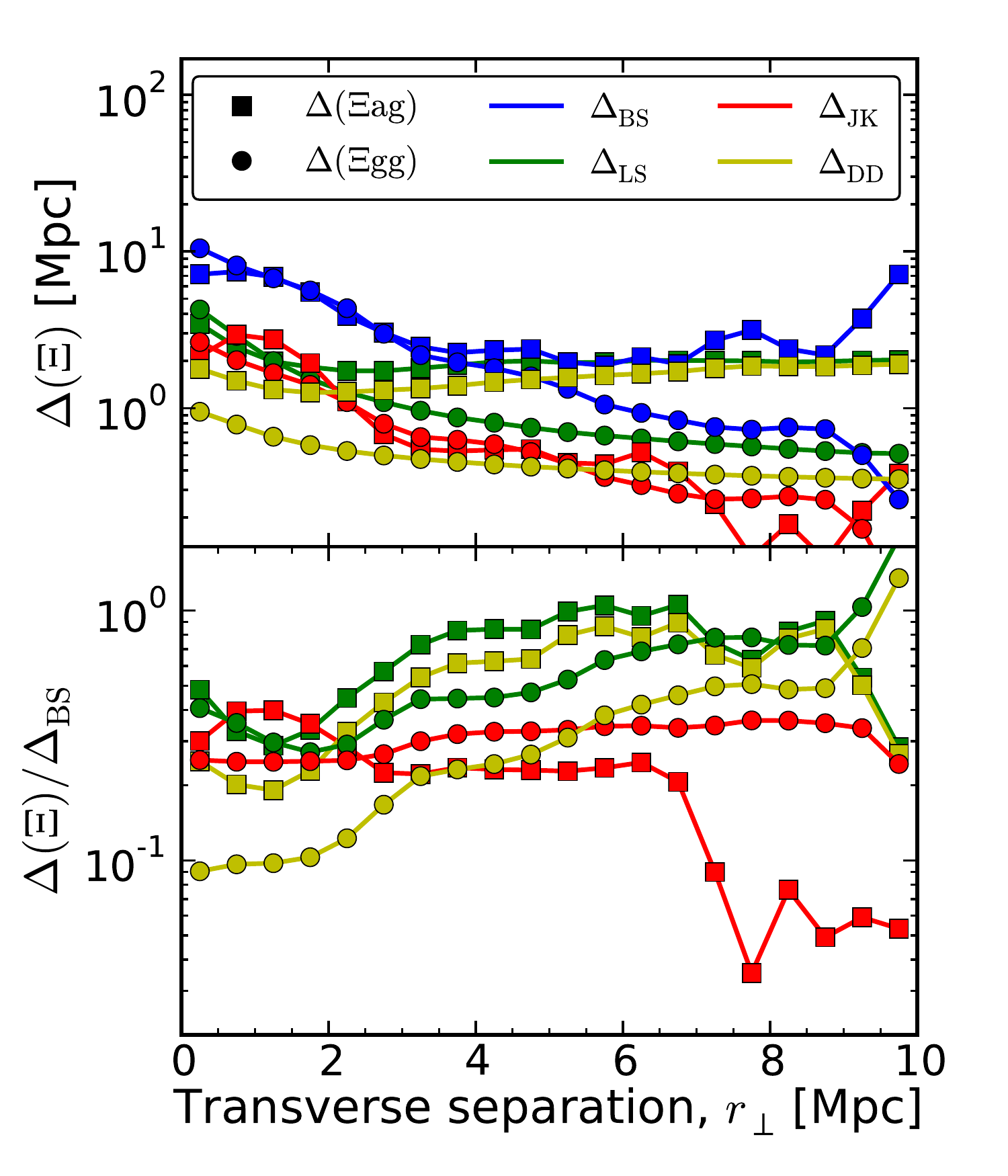}
  \caption{Uncertainty estimations (square root of variances) of the
    projected \hi-galaxy cross-correlation ($\Xi_{\rm ag}$; squares)
    and galaxy auto-correlation ($\Xi_{\rm gg}$; circles) measured from
    our full sample as a function of separations transverse to the
    line-of-sight ($r_{\perp}$). The top panel shows the uncertainty
    from the \citet{Landy1993} analytical approximation, $\Delta_{\rm
      LS}$ (\Cref{eq:var_LS}; green lines); uncertainty from a
    `jackknife' resampling, $\Delta_{\rm JK}$ (\Cref{eq:var_JK}; red
    lines); uncertainty from a `bootstrap' resampling, $\Delta_{\rm
      BS}$ (\Cref{eq:var_BS}; blue lines); and the commonly used
    Poissonian uncertainty, $\Delta_{\rm DD}$ (\Cref{eq:var_DD}; yellow
    lines). The bottom panel shows the ratio between these
    uncertainties and $\Delta_{\rm BS}$.}\label{fig:errors_T}

\end{figure}

\Cref{fig:errors_T} shows these $4$ uncertainty estimations for our
measurements of the projected correlations $\Xi(r_{\perp})$, for both
\hi--galaxy (squares) and galaxy--galaxy (circles): $\Delta_{\rm LS}$
(green lines), $\Delta_{\rm JK}$ (red lines), $\Delta_{\rm BS}$ (blue
lines) and $\Delta_{\rm DD}$ (yellow lines). The top panel shows the
absolute values for these different uncertainties, while the bottom
panel shows the ratio of a given uncertainty estimation and
$\Delta_{\rm BS}$. As before, we observe systematic trends, but all
uncertainties are consistent within $\sim 1$ order of magnitude of each
other. In contrast to the two-dimensional uncertainties, $\Delta_{\rm
  BS}$ is the largest in this case. Focusing on the smallest scales
(where the correlation amplitudes are greater) we see that $\Delta_{\rm
  JK}$ and $\Delta_{\rm LS}$ are in closer agreement to $\Delta_{\rm
  BS}$ than $\Delta_{\rm DD}$.

\begin{table*}
  \begin{minipage}{0.63\textwidth}
    \centering
    \caption{Summary of the `Full Sample' used for the
      cross-correlation analysis, as a function of
      $r_{\perp}$.}\label{tab:full_sample}
    \begin{tabular}{@{}lcccccc@{}}
      \hline & $< 0.5$ \mpc &  $< 1$ \mpc  & $< 2$ \mpc& $< 10$ \mpc & $<50$ \mpc & Total\\ 
             & (1)          & (2)          & (3)       & (4)         & (5)        &  (6) \\
      \hline
      
      Galaxies   &141    & 466  & 1354  & 6871      & 19509   & 17509\\
      \ \ \SF    &105    & 339  & 997   & 4756      & 9963    & 8293\\
      \ \ \nSF   &24     & 66   & 193   & 779       & 2011    & 1743\\
      \hi        & \dots & \dots& \dots & \dots     & \dots   & 654\\
      \ \ \strong& \dots & \dots& \dots & \dots     & \dots   & 165\\
      \ \ \weak  & \dots & \dots& \dots & \dots     & \dots   & 489\\

    \hline                                     
    \end{tabular}
  \end{minipage}
  \begin{minipage}{0.63\textwidth}
    (1): Number of galaxies at transverse distances $r_{\perp}< 0.5$
    \mpc from a \ac{qso} \ac{los}. (2): Number of galaxies at
    transverse distances $r_{\perp}< 1$ \mpc from a \ac{qso}
    \ac{los}. (3): Number of galaxies at transverse distances
    $r_{\perp}< 2$ \mpc from a \ac{qso} \ac{los}.  (4) Number of
    galaxies at transverse distances $r_{\perp}< 10$ \mpc from a
    \ac{qso} \ac{los}.  (5) Number of galaxies at distances $r_{\perp}<
    50$ \mpc from a \ac{qso} \ac{los}. (6) Total number of galaxies and
    \hi absorption systems in the `Full Sample'. Note that the vast
    majority of galaxies in the triple \ac{qso} field Q0107 have been
    counted three times in columns (1), (2), (3), (4) and (5).
  \end{minipage}
\end{table*}

These results suggest that $\Delta_{\rm LS}$ is preferable over
$\Delta_{\rm DD}$ and even over $\Delta_{\rm JK}$ (at least when the
number of independent fields is small, like in our case). A more in
depth study of the error estimation for auto-
\citep[e.g.][]{Norberg2009} and cross-correlations is beyond the scope
of this paper.

For the results of this paper, we will adopt uncertainties given by
$\Delta_{\rm BS}$. As has been shown, $\Delta_{\rm BS}$ gives, in
general, the most conservative uncertainty estimation at all scales. An
exception to this rule was found for $\xi_{\rm
  aa}(r_{\perp},r_{\parallel})$ and $\Xi_{\rm aa}(r_{\perp})$, in which
$\Delta_{\rm LS} > \Delta_{\rm BS}$. This is due to the combination of
the special survey geometry in which $\xi_{\rm aa}$ is measured. Thus,
for such a sample we adopted $\Delta_{\rm LS}$ as the uncertainty.

\subsection{Calibration between galaxy and \hi absorbers redshift frames}\label{calibration}

Before computing the final two-point correlation functions, we
calibrated the redshift frames between our \hi absorption systems and
galaxies, using the idea presented by \citet{Rakic2011}: that in an
isotropic Universe, the mean \hi absorption profile around galaxies
should be symmetric. Thus, we measured the \hi--galaxy cross-correlation
using $r_{\parallel,ij} \equiv X_i- X_j$ instead of $r_{\parallel,ij}
\equiv |X_i- X_j|$, and applied a constant redshift shift to all our
galaxies such that the cross-correlation appears symmetric with respect
to the $r_{\parallel} = 0 $ axis at the scales involved in this
analysis. This redshift shift corresponded to $+0.0002$ (smaller than
the galaxy redshift uncertainty). Note that this shift has not been
added to the redshifts reported in Tables \ref{tab:gals_Q0107} to
\ref{tab:gals_J2218}. The final two-point correlation functions were
still calculated using $r_{\parallel,ij} \equiv |X_i- X_j|$ in order to
reduce the `shot noise'.



\section{Results}\label{results}

In this section we present the results of the two point correlation
analysis, following the formalisms described in \Cref{analysis}. We
used the \hi and galaxy samples described in Sections \ref{data:IGM},
\ref{data:galaxy}, \ref{abs:samples} and \ref{gal:samples}, but
excluding: (i) \hi and galaxies falling in their respective `c'
categories (see Sections \ref{gal:reliability} and
\ref{abs:reliability}); (ii) \hi and galaxies at $z<0.01$ and at
$z>1.3$; (iii) \hi systems at redshifts within $5000$ \kms of the
redshift of the \ac{qso} in which the absorption line was observed; and
(iv) galaxies at projected distances greater than $50$ \mpc from the
centre of their closest field. We will refer to this sample as the
`Full Sample', which comprises: $654$ \hi~absorption systems, of which,
$165$ are classified as \strong and $489$ as \weak (see \Cref{abs:logn}
for definitions); and $17509$ galaxies, of which, $8293$ are classified
as \SF and $1743$ as \nSF (see \Cref{gal:sclass} for definitions).

\Cref{tab:full_sample} summarizes relevant information regarding our
`Full Sample'. The following results were computed with random samples
$200 \times$ and $20 \times$ larger than the real \hi~and galaxy
samples, respectively. Even though we have galaxies up to $50$ \mpc
from the \ac{qso} \ac{los}, we will focus only on clustering at scales
$r_{\perp}<10$ \mpc, as at larger scales our results get considerably
noisier. Galaxies at $r_{\perp}>10$\mpc are still used for the
galaxy--galaxy auto-correlation though. In the case of the \hi--\hi
auto-correlation, we only focus on scales $r_{\perp}<2$ \mpc, as we
have no data sampling larger transverse scales.

\subsection{Two-dimensional correlations}\label{results:2d}

\begin{figure*}
  \begin{minipage}{\textwidth}
    \flushleft
    \includegraphics[width=\textwidth]{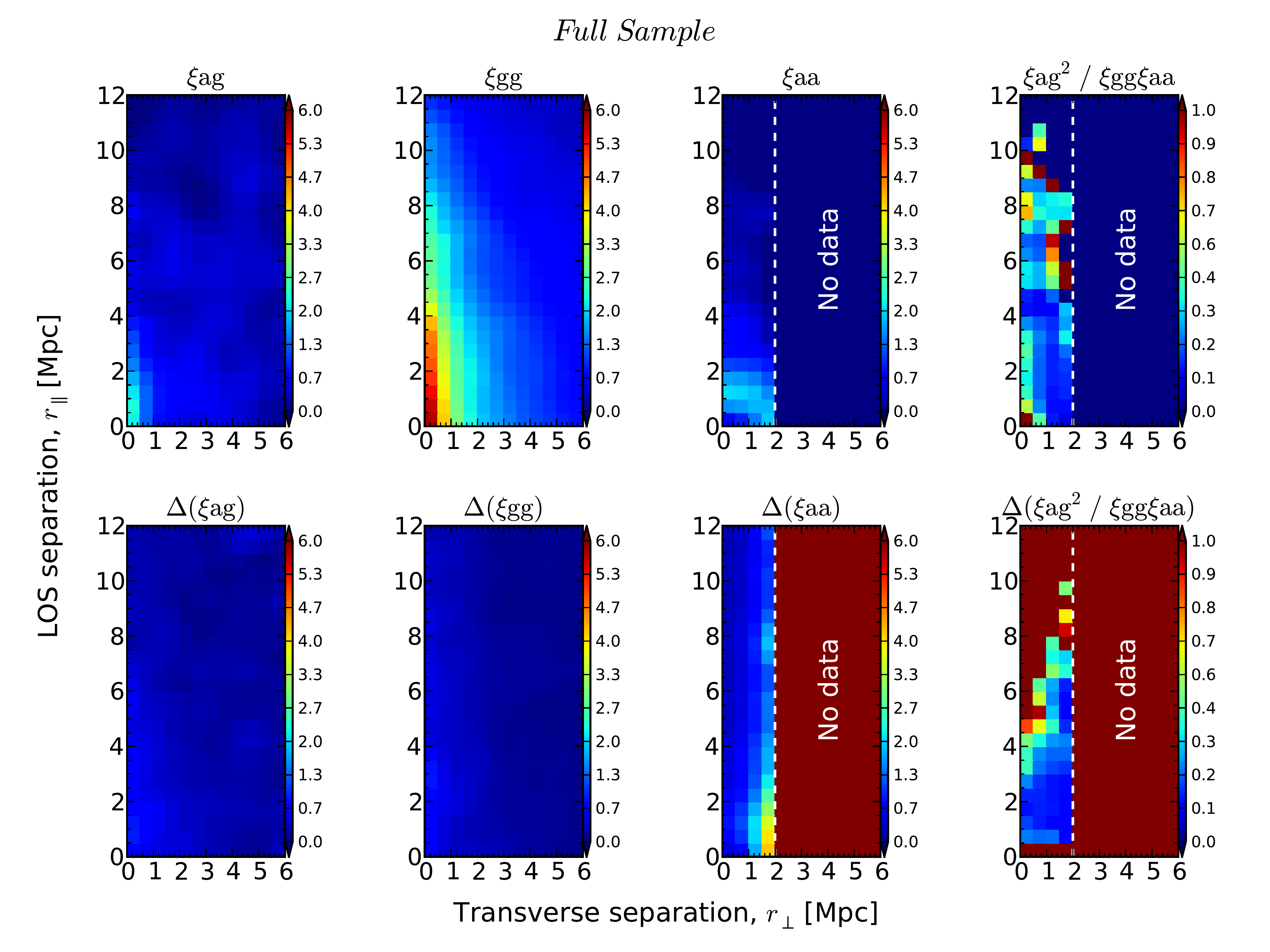}
  \end{minipage}
  \caption{Two-dimensional correlation functions for galaxies and
    \hi~absorption system (top panels) and their respective
    uncertainties (bottom panels), as a function of separations along
    ($r_{\parallel}$; $y$-axes) and transverse to the line-of-sight
    ($r_{\perp}$; $x$-axes). From left to right: the
    galaxy-\hi~cross-correlation ($\xi_{\rm ag}^{\rm LS}$;
    \cref{eq:LSag}), the galaxy-galaxy~auto-correlation ($\xi_{\rm
      gg}^{\rm LS}$; \cref{eq:LSgg}), the \hi-\hi~auto-correlation
    ($\xi_{\rm aa}^{\rm LS}$; \cref{eq:LSaa}) and the ratio, $(\xi^{\rm
      LS}_{\rm ag})^2/(\xi^{\rm LS}_{\rm gg}\xi^{\rm LS}_{\rm
      aa})$. Note that our data are not suitable for measuring the
    $\xi^{\rm LS}_{\rm aa}$ and $(\xi^{\rm LS}_{\rm ag})^2/(\xi^{\rm
      LS}_{\rm gg}\xi^{\rm LS}_{\rm aa})$ at scales $r_{\perp}> 2$
    Mpc. The correlation functions in this figure were calculated using
    an arbitrary binning of $0.5$ \mpc~with cross-pairs counts smoothed
    with a Gaussian filter of standard deviation of $0.5$ \mpc~along
    both directions. See \Cref{correlation:2d,results:2d} for further
    details.} \label{fig:xi_all}

\end{figure*}

\begin{figure*}
  \begin{minipage}{\textwidth}
    \flushleft
    \includegraphics[width=\textwidth]{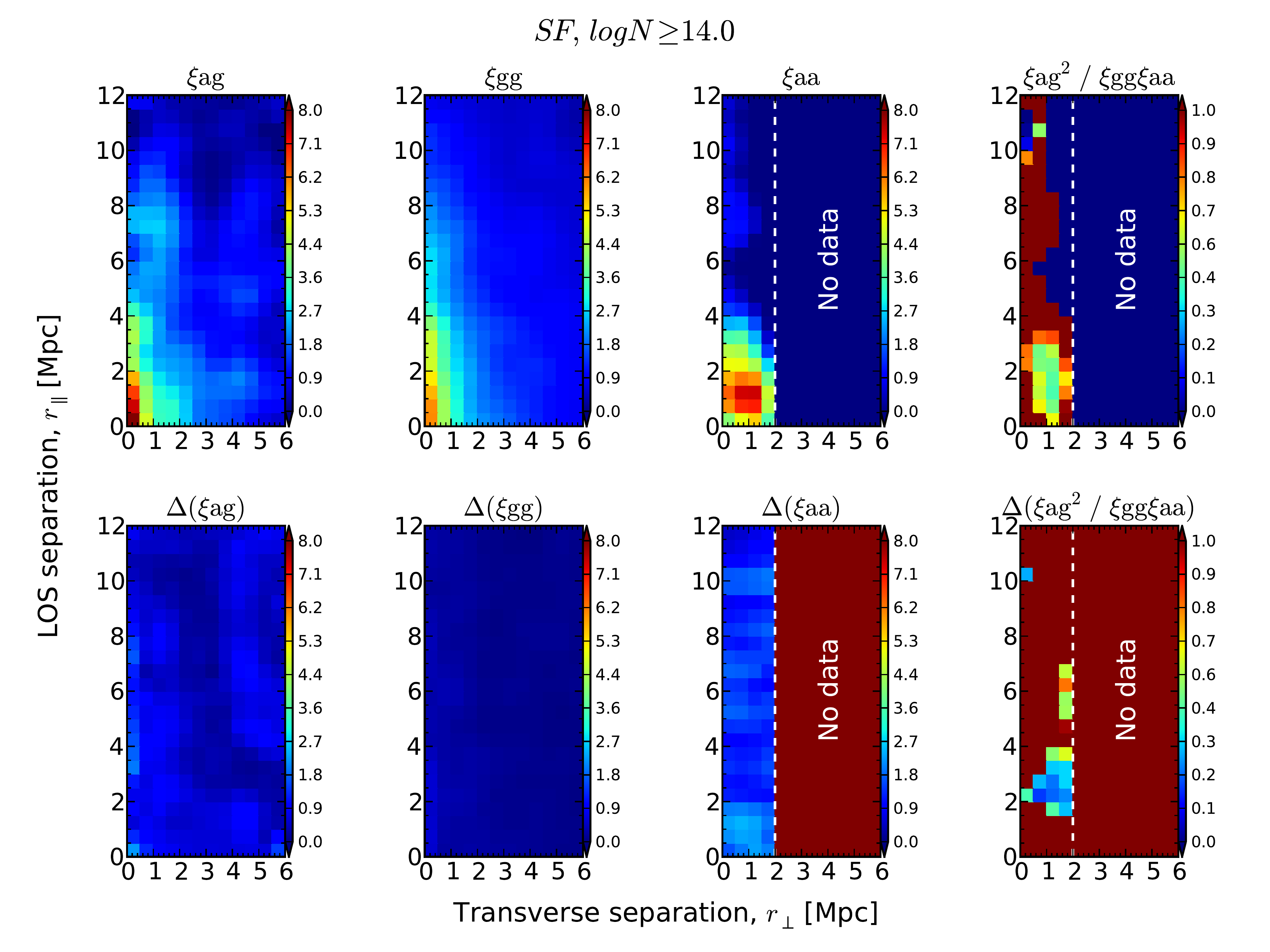}
  \end{minipage}
  \caption{Same as Figure \ref{fig:xi_all} but for \SF~galaxies and
    \hi~absorption systems with $N_{\rm HI} \ge 10^{14}$ \cm
    (\strong).} \label{fig:xi_em_st}

\end{figure*}

 \begin{figure*}
  \begin{minipage}{\textwidth}
    \flushleft
    \includegraphics[width=\textwidth]{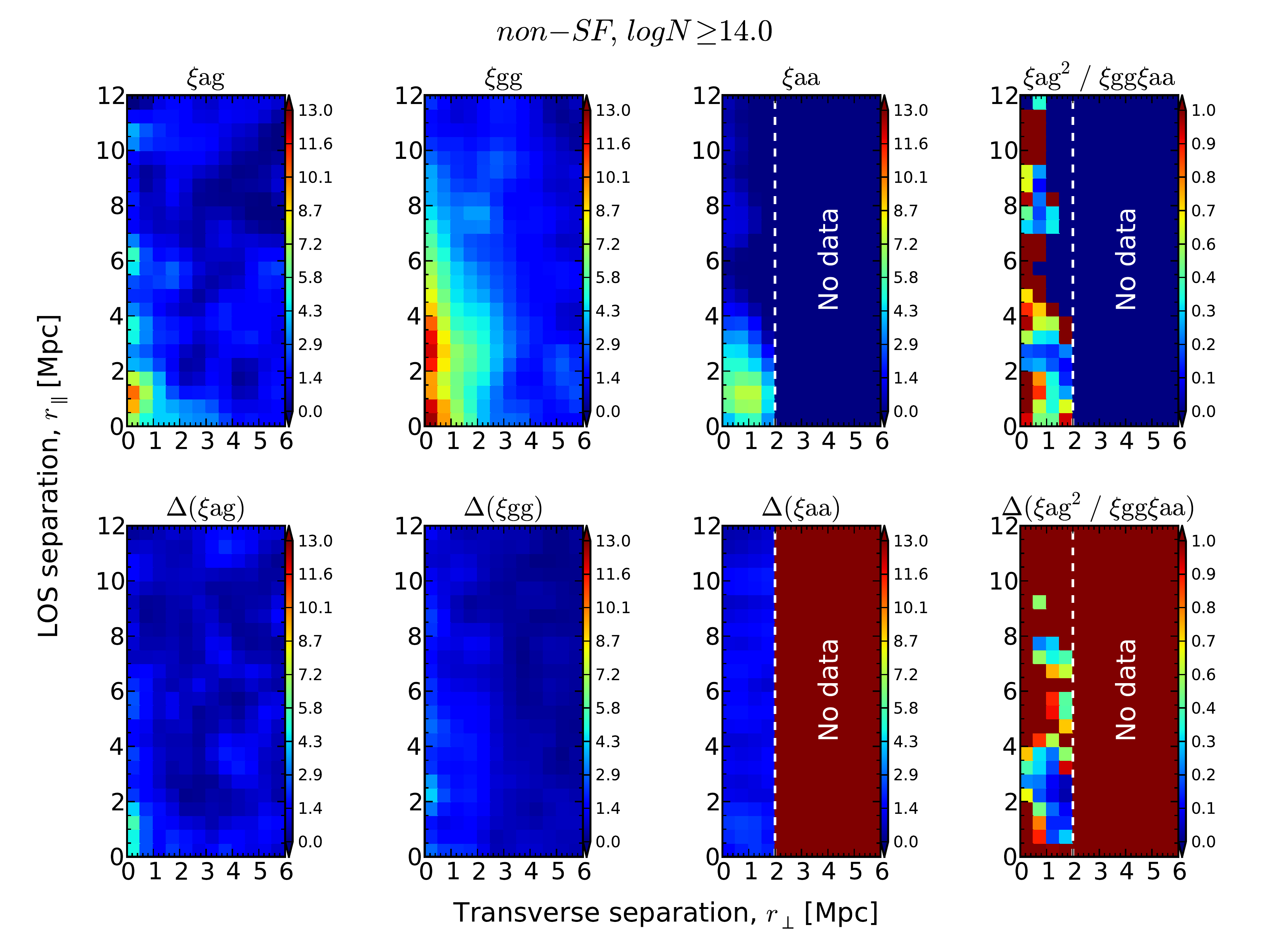}
  \end{minipage}
  \caption{Same as Figure \ref{fig:xi_all} but for \nSF~galaxies and
    \hi~absorption systems with $N_{\rm HI} \ge 10^{14}$
    \cm (\strong).} \label{fig:xi_ab_st}

\end{figure*}

\begin{figure*}
  \begin{minipage}{\textwidth}
    \flushleft
    \includegraphics[width=\textwidth]{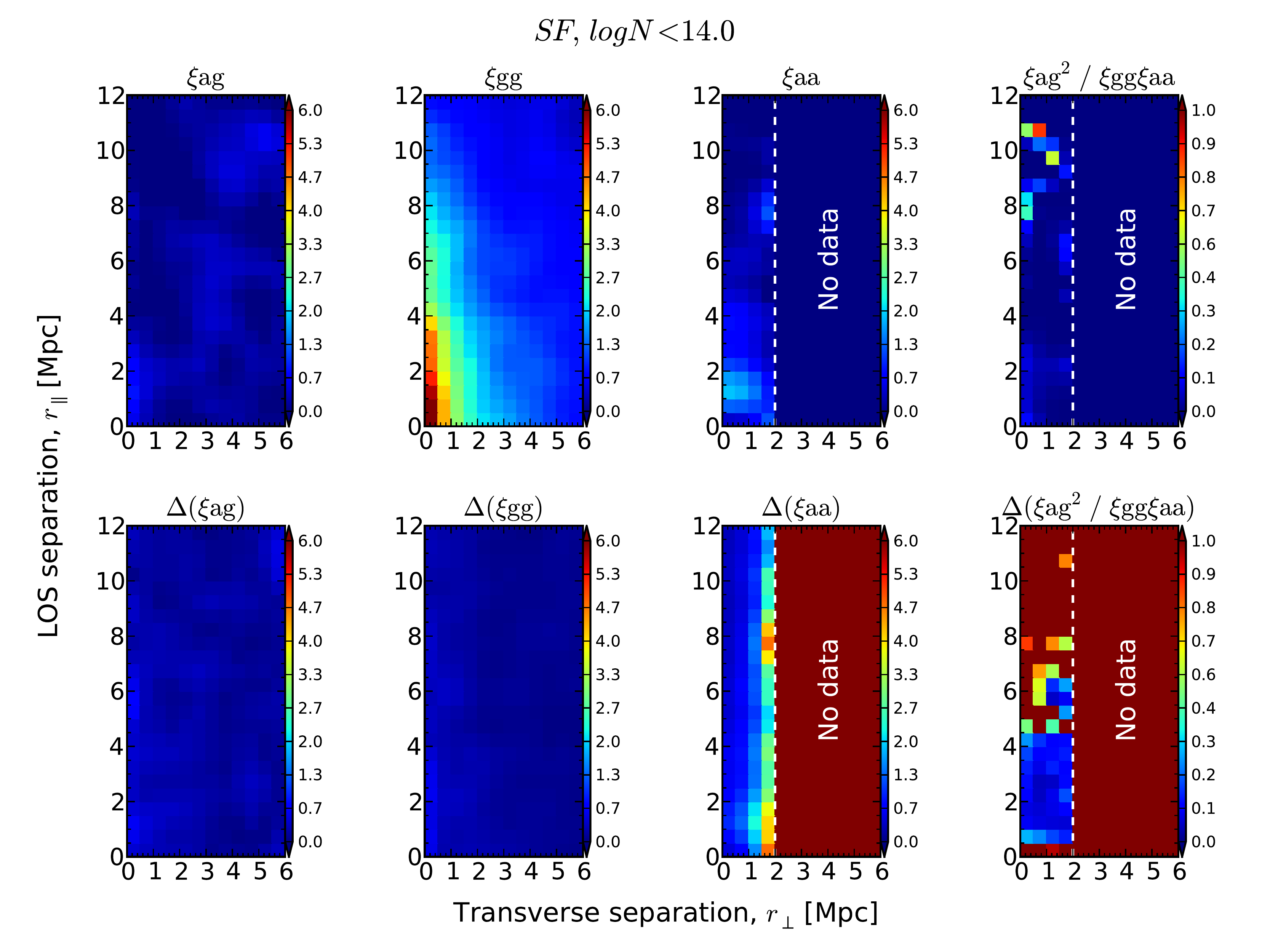}
  \end{minipage}
  \caption{Same as Figure \ref{fig:xi_all} but for \SF~galaxies and
    \hi~absorption systems with $N_{\rm HI}< 10^{14}$
    \cm (\weak).} \label{fig:xi_em_wk}

\end{figure*}

 \begin{figure*}
  \begin{minipage}{\textwidth}
    \flushleft
    \includegraphics[width=\textwidth]{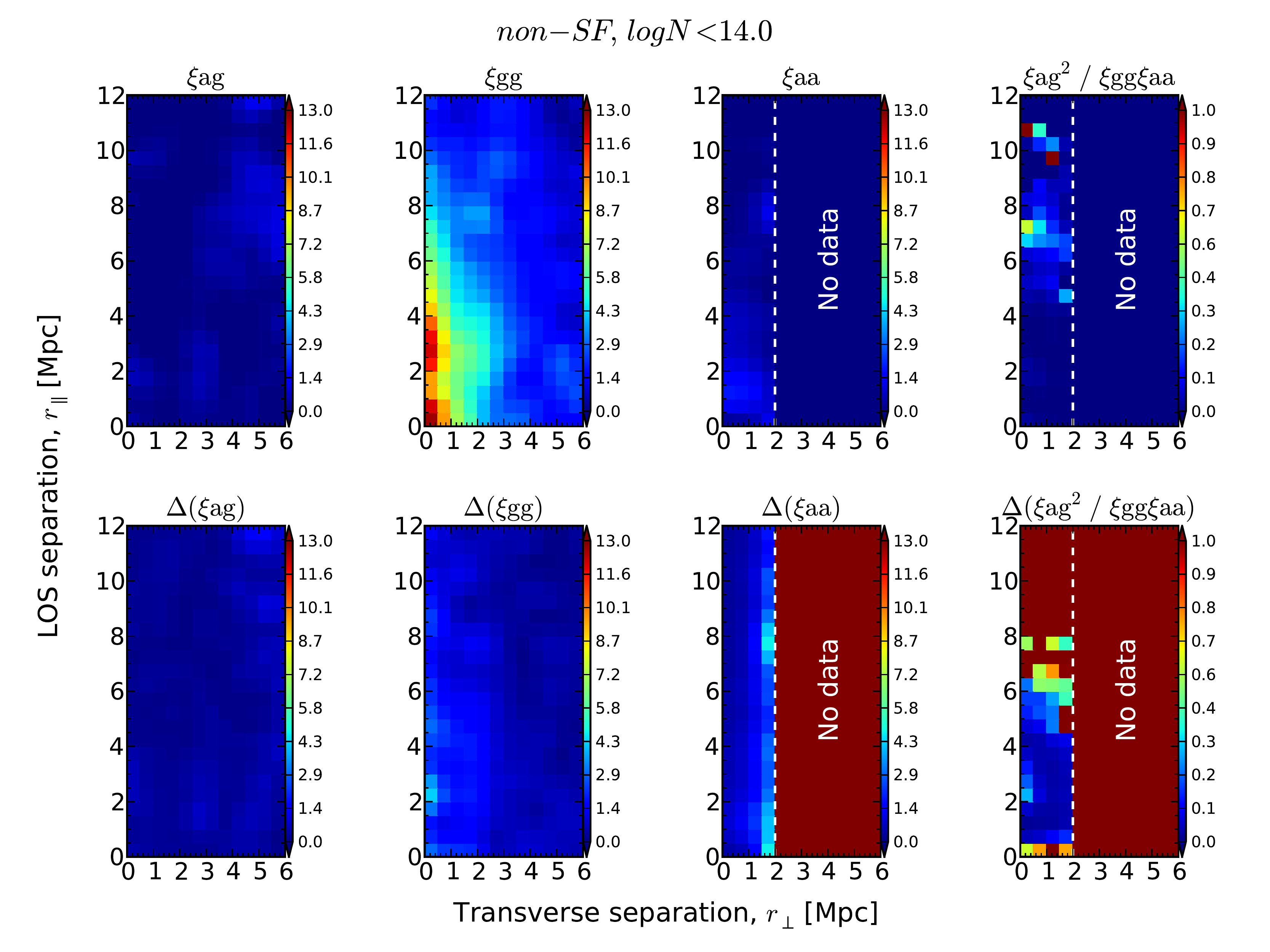}
  \end{minipage}
  \caption{Same as Figure \ref{fig:xi_all} but for \nSF~galaxies and
    \hi~absorption systems with $N_{\rm HI}< 10^{14}$
    \cm (\weak).} \label{fig:xi_ab_wk}

\end{figure*}

\subsubsection{Full Sample}
In \Cref{fig:xi_all} we show the two-dimensional correlation functions
(top panels) and their respective uncertainties (bottom panels) for our
`Full Sample' of \hi-absorption systems and galaxies. The first three
panels, from left to right, show the \hi--galaxy cross-correlation
($\xi_{\rm ag}^{\rm LS}$; \Cref{eq:LSag}), the
galaxy--galaxy~auto-correlation ($\xi_{\rm gg}^{\rm LS}$;
\Cref{eq:LSgg}) and the \hi--\hi~auto-correlation ($\xi_{\rm aa}^{\rm
  LS}$; \Cref{eq:LSaa}), respectively.  We see that the amplitudes of
$\xi_{\rm ag}$ and $\xi_{\rm aa}$ are comparable (within the
uncertainties), whereas the amplitude of $\xi_{\rm gg}$ is greater than
these two (see also \Cref{tab:xi_peaks}). Also, the fact that both
$\xi_{\rm gg}$ and $\xi_{\rm ag}$ peak at the smallest separations
confirms that the redshift frames for \hi~absorption systems and
galaxies are self consistent (by construction; see
\Cref{calibration}). The decrease in the $\xi_{\rm aa}$ signal at the
smallest $r_{\parallel}$ separations is because we cannot always
resolve two real absorption systems separated by less than the typical
width of an absorption feature. This width corresponds to $\sim 16$
\kms ($\sim 100$ \kms) for \ac{fuv} (\ac{nuv}) data, which in co-moving
distance correspond to $\sim 0.26$ \mpc ($\sim 1.6$ \mpc) at $z=0.5$.

Our sample of \hi absorption systems is not large enough to measure
$\xi_{\rm ag}$ or $\xi_{\rm aa}$ anisotropies at a high confidence
level. Still, we can obtain qualitative features by looking at the
corresponding `iso-correlation' contours. We observe deviations from an
isotropic signal in both $\xi_{\rm ag}$ and $\xi_{\rm gg}$. Apart from
a decrease of the $\xi_{\rm aa}$ signal at the smallest $r_{\parallel}$
separations, we do not see significant anisotropies in $\xi_{\rm
  aa}$. The typical uncertainty for our single galaxy redshift
determination, $\Delta z_{\rm gal} \approx 0.0006/\sqrt{2}$, is
equivalent to $\sim 1.7-1.4$ \mpc~at $z=0.1-0.5$, which corresponds to
an `anisotropy ratio' of $\sim 3:1$ for pixels of $0.5$ \mpc~each. If
the observed anisotropies are dominated by redshift uncertainties, we
should expect the $\xi_{\rm ag}$ contours to be consistent with this
ratio (neglecting the much smaller contribution from the \hi~redshift
uncertainty) and the $\xi_{\rm gg}$ one to be $\sim 4:1$ (greater by a
factor of $\sqrt{2}$). These expectations are consistent with what we
see in our `Full Sample' for the smallest scales, whereas for scales
$\gtrsim 4$ \mpc~the anisotropy looks somewhat reduced. We do not
detect compression along the \ac{los} at larger scales either
\citep[e.g.][]{Kaiser1987}. The only anisotropy observed can be fully
explained by galaxy redshift uncertainties.

The fourth panel of \Cref{fig:xi_all} shows the ratio, $(\xi^{\rm
  LS}_{\rm ag})^2/(\xi^{\rm LS}_{\rm gg}\xi^{\rm LS}_{\rm aa})$.  We
see that the majority of the bins at the smallest scales have values
$(\xi_{\rm ag})^2/(\xi_{\rm gg}\xi_{\rm aa}) < 1$. This result suggests
that, contrary to what is usually assumed, the population of
\hi~absorption systems (as a whole) and galaxies do not linearly trace
the same underlying dark matter distributions (see \Cref{dependence}).

In the following, we will split the \hi~absorber sample into \strong
($N_{\rm HI} \ge 10^{14}$ \cm) and \weak ($N_{\rm HI}< 10^{14}$ \cm),
and the galaxy sample into \SF~and \nSF. In this way we can isolate the
contribution of each sub-population of \hi and galaxies to the
correlation functions, and to the $(\xi_{\rm ag})^2/(\xi_{\rm
  gg}\xi_{\rm aa})$ ratio.

\subsubsection{`Strong' \hi systems and \SF galaxies}

\Cref{fig:xi_em_st} is analogous to \Cref{fig:xi_all} but for \strong
\hi~systems ($N_{\rm HI} \ge 10^{14}$ \cm) and \SF~galaxies. We see
that in this case the $\xi_{\rm ag}$, $\xi_{\rm gg}$ and $\xi_{\rm aa}$
are all comparable within the errors (see also
\Cref{tab:xi_peaks}). Anisotropy signals behave in the same way as for
our `Full Sample', i.e. are dominated by our galaxy redshift
uncertainty and with no detected compression along the \ac{los} at
large scales.

In this case, the ratio $(\xi_{\rm ag})^2/(\xi_{\rm gg}\xi_{\rm aa})$
is consistent with $1$, suggesting that \SF~galaxies and \nhi~$\ge
10^{14}$ \cm~systems do trace the same underlying dark matter
distribution. The comparable clustering amplitudes may also indicate
that they typically belong to dark matter haloes of similar masses. We
will address these points more quantitatively in \Cref{results:1d}

\subsubsection{`Strong' \hi systems and \nSF galaxies}

\Cref{fig:xi_ab_st} shows the correlation functions for \strong
\hi~systems ($N_{\rm HI} \ge 10^{14}$ \cm) and \nSF~galaxies. In this
case, $\xi_{\rm ag}$, $\xi_{\rm gg}$ and $\xi_{\rm aa}$ are all
comparable within the errors (see also \Cref{tab:xi_peaks}), but
$\xi_{\rm gg}$ appears systematically larger. As before, the anisotropy
is dominated by the galaxy redshift uncertainty and no (significant)
compression along the \ac{los} at large scales is detected.

Interestingly, there is a displacement in the $\xi_{\rm ag}$ peak
relative to the smallest bin. This signal also appears symmetric with
respect to the $r_{\parallel}=0$ axis, after computing $\xi_{\rm ag}$
using $r_{\parallel,ij} \equiv X_i- X_j$ (not plotted) instead of
$r_{\parallel,ij} \equiv |X_i- X_j|$. We also checked that the signal
remained using only `a' labelled \hi systems and galaxies. This suggests
that this feature might be real. A similar (although more uncertain)
feature was observed by \citet{Wilman2007}, from their observation of
the \hi-`absorption-line-dominated galaxy' cross-correlation (see their
figure 4).\footnote{We note that there is a small overlap between
  \citet{Wilman2007} sample and ours.}  \citet{Pierleoni2008} also
reported a similar signal from hydrodynamical simulations (see their
figure 7), although their samples of \hi and galaxies are not directly
comparable to our \strong and \nSF ones. A more detailed comparison
between our results and those from previous studies will be presented
in \Cref{discussion:comparison}.

The ratio $(\xi_{\rm ag})^2/(\xi_{\rm gg}\xi_{\rm aa})$ seems also
consistent with $1$, which suggests that \nSF~galaxies and \nhi~$\ge
10^{14}$ \cm~systems trace the same underlying dark matter distribution
linearly. 

Comparing Figures \ref{fig:xi_em_st} and \ref{fig:xi_ab_st} we see that
$\xi_{\rm gg}$ for \nSF~galaxies is larger than that of \SF~galaxies
(as has been shown by many authors). Given that $\xi_{\rm aa}$ is the
same in both cases, one would expect $\xi_{\rm ag}$ to be also larger
for \nSF than that of \SF galaxies. Although within the uncertainties
our results indicate that the $\xi_{\rm ag}$ amplitude is independent
of galaxy type, we do see a somewhat larger cross-correlation signal
for \nSF galaxies (see \Cref{tab:xi_peaks}). We will address these
points more quantitatively in \Cref{results:1d}.

\subsubsection{`Weak' \hi systems and galaxies}

\Cref{fig:xi_em_wk,fig:xi_ab_wk} show the two-dimensional correlation
functions for \weak \hi~absorption systems ($N_{\rm HI} < 10^{14}$ \cm)
and \SF~and \nSF~galaxies, respectively. These results are dramatically
different than those for \strong \hi systems and galaxies. In
particular, $\xi_{\rm ag}$ is significantly weaker than $\xi_{\rm gg}$
but also weaker than $\xi_{\rm aa}$, for both types of
galaxies. Consequently, the ratios $(\xi_{\rm ag})^2/(\xi_{\rm
  gg}\xi_{\rm aa})$ are both smaller than one. This is a very strong
indication (given the comparatively smaller uncertainties) that the
underlying baryonic matter distributions giving rise to \weak
\hi~absorption systems and galaxies are not linearly dependent. Given
that the signal in the $\xi_{\rm ag}$ is marginally consistent with
zero, we do not observe anisotropies either.\\

To summarize, \strong systems and galaxies are consistent with tracing
the same underlying dark matter distribution linearly, whereas \weak
systems are not. Therefore, the fact that $(\xi_{\rm ag})^2/(\xi_{\rm
  gg}\xi_{\rm aa}) <1$ in the `Full Sample' should be primarily driven
by the presence of \hi systems with $N_{\rm HI} < 10^{14}$ \cm. We also
note that the amplitude of $\xi_{\rm aa}$ is weaker for \weak systems
than that for \strong systems.

Because redshift uncertainties affect $\xi_{\rm ag}$, $\xi_{\rm gg}$
and $\xi_{\rm aa}$ in different ways, the interpretation of the
two-dimensional $(\xi_{\rm ag})^2/(\xi_{\rm gg}\xi_{\rm aa})$ is not
straightforward. In the following we present the results for the
projected correlation functions, which are not affected by velocity
distortions along the \ac{los}, and have smaller statistical
uncertainties.

\begin{table}
  \begin{minipage}{0.42\textwidth}
    \centering
    \caption{Strength of the two-dimensional correlations,
      $\xi(r_{\perp},r_{\parallel})$, at their
      peaks.\tablenotemark{a}}\label{tab:xi_peaks}
    \begin{tabular}{@{}lrrr@{}}
      \hline 
           & \multicolumn{1}{c}{$\xi_{\rm ag}^{\rm peak}$} &\multicolumn{1}{c}{$\xi_{\rm gg}^{\rm peak}$} & \multicolumn{1}{c}{$\xi_{\rm aa}^{\rm peak}$}\\
      \hline
      Full Sample &   2.3 $\pm$  0.9 &  5.7 $\pm$  0.7 &  2.1 $\pm$  0.9\\
      \ \ `Strong'--\SF  &   8.3 $\pm$  2.2 &  6.1 $\pm$  0.6 &  7.5 $\pm$  2.3\\
      \ \ `Strong'--\nSF & 10.3 $\pm$  5.6 & 12.6 $\pm$  3.0 &  7.5 $\pm$  2.3\\
      \ \ `Weak'--\SF    &   0.9 $\pm$  0.6 &  6.1 $\pm$  0.6 &  1.9 $\pm$  0.9 \\
      \ \ `Weak--\nSF    &   0.6 $\pm$  0.5 & 12.6 $\pm$  3.0 &  1.9 $\pm$  0.9 \\
      
      \hline                                     
    \end{tabular}
  \end{minipage}
  \begin{minipage}{0.42\textwidth}
    $^{\rm a}$ Note that peaks are not necessarly at the smallest scale
    bins (see Figures \ref{fig:xi_all} to \ref{fig:xi_ab_wk}). \\
  \end{minipage}
\end{table}

\subsection{Correlations projected along the line-of-sight}\label{results:1d}

\begin{figure*}
  \centering
  \includegraphics[width=0.7\textwidth]{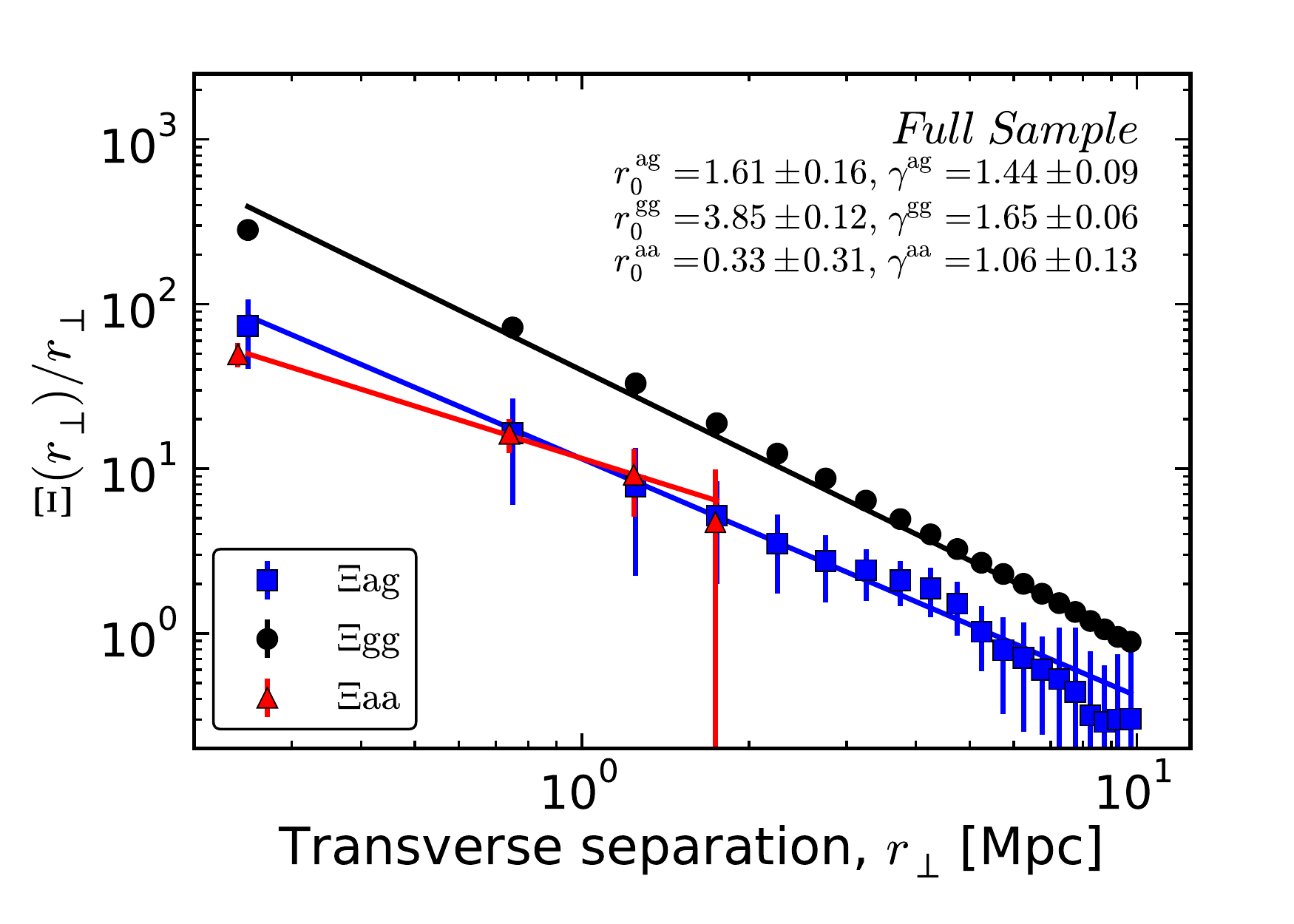}
  \caption{Projected along the line-of-sight correlation functions (see
    \cref{eq:projected}) divided by the transverse separation,
    $\Xi(r_{\perp})/r_{\perp}$, for our `Full Sample' of galaxies and
    \hi~absorption systems. Different symbols/colors show our different
    measurements: the blue squares correspond to the
    galaxy-\hi~cross-correlation ($\Xi_{\rm ag}$); the black circles to
    the galaxy-galaxy auto-correlation ($\Xi_{\rm gg}$); and the red
    triangles to the \hi-\hi~auto-correlation ($\Xi_{\rm aa}$; slightly
    shifted along the $x$-axis for the sake of clarity). The lines
    correspond to the best power-law fits (\Cref{eq:fit}) to the data,
    from a non-linear least squares analysis. The parameters $r_0$ and
    $\gamma$ correspond to those of the real-space correlation
    function, $\xi(r)$, when described as a power-law of the form
    presented in \Cref{eq:rscf}. Note that points and uncertainties are
    both correlated, and that uncertainties smaller than the symbols
    are not shown.}
\label{fig:xi_all_T}

\end{figure*}

\begin{figure*}
  \centering
  \includegraphics[width=1.01\textwidth]{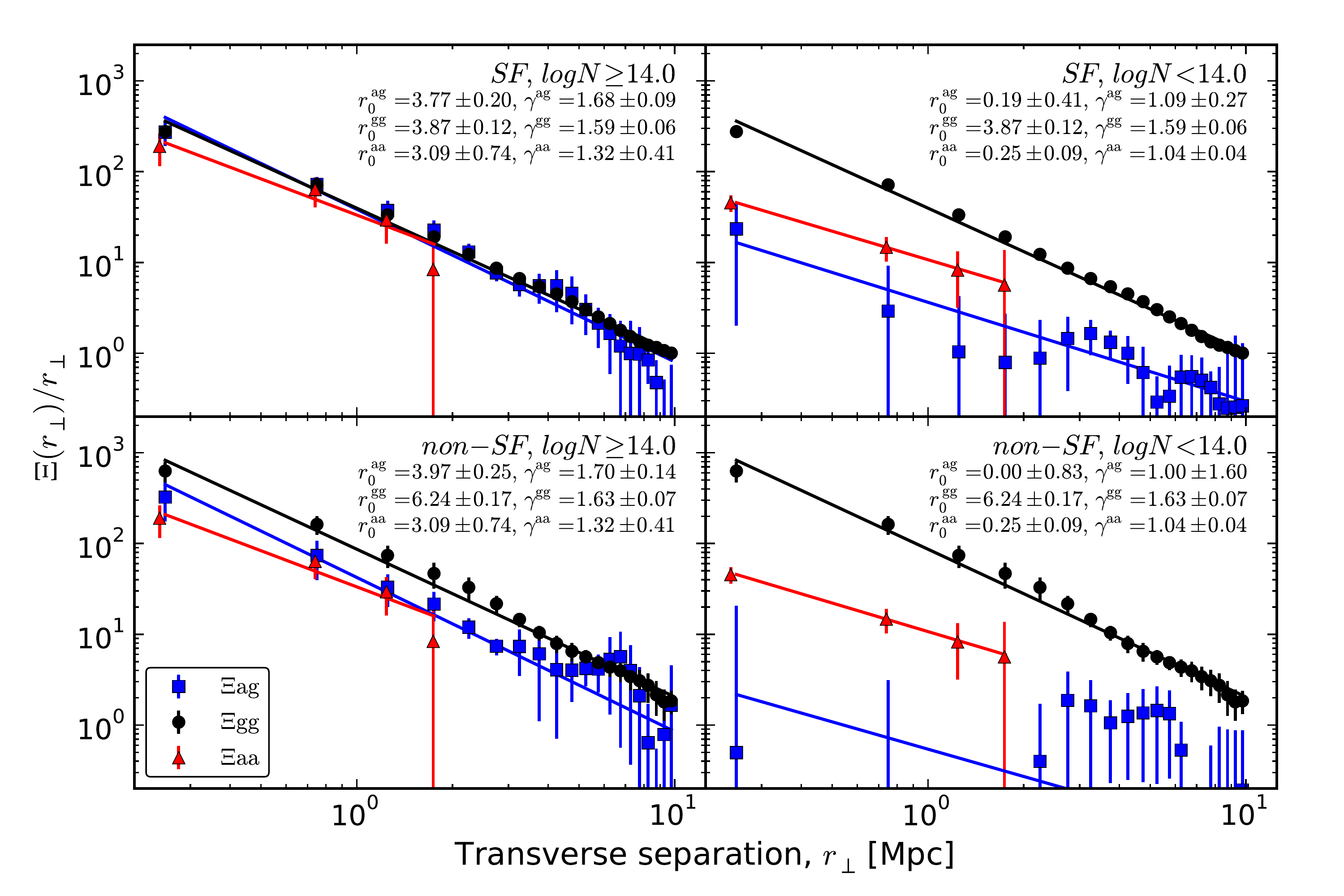}
  \caption{Same as Figure \ref{fig:xi_all_T} but for our different
    subsamples: \SF~galaxies and `strong' ($N_{\rm HI} \ge10^{14}$ \cm)
    \hi~absorption systems (top left); \SF~galaxies and `weak' ($N_{\rm
      HI} < 10^{14}$ \cm) \hi~absorption systems (top right);
    \nSF~galaxies and `strong' \hi~absorption systems (bottom left);
    and \nSF~galaxies and `weak' \hi~absorption systems (bottom
    right). Note that points and uncertainties are both correlated, and
    that uncertainties smaller than the symbols are not shown. }

\label{fig:xi_split_T}

\end{figure*}

\begin{table*}
  \begin{minipage}{0.7\textwidth}
    \centering
    \caption{Best fit parameters to the real-space correlation function
      assuming power-law of the form presented in
      \Cref{eq:rscf}.}.\label{tab:xi_params}
    \begin{tabular}{@{}lcccccccc@{}}
      \hline 
       & \multicolumn{2}{c}{$\xi_{\rm ag}(r)$} &&\multicolumn{2}{c}{$\xi_{\rm gg}(r)$} && \multicolumn{2}{c}{$\xi_{\rm aa}(r)$}\\ 
      \cline{2-3} \cline{5-6}\cline{8-9}
                                 & $r_0$ (\mpc)    & $\gamma$        &         & $r_0$ (\mpc)     & $\gamma$                &&$r_0$ (\mpc)    & $\gamma$                 \\
      \hline
      Full Sample        &   1.6 $\pm$  0.2  & 1.4 $\pm$ 0.1   &         &  3.9 $\pm$  0.1 &  1.7 $\pm$  0.1   && 0.3 $\pm$  0.3& 1.1 $\pm$  0.1\\
      \ \ `Strong'--\SF  &   3.8 $\pm$  0.2  & 1.7 $\pm$ 0.1   &         &  3.9 $\pm$  0.1 &  1.6 $\pm$  0.1   && 3.1 $\pm$  0.7& 1.3 $\pm$  0.4\\
      \ \ `Strong'--\nSF &   4.0 $\pm$  0.3  & 1.7 $\pm$ 0.1   &         &  6.2 $\pm$  0.2 &  1.6 $\pm$  0.1   && 3.1 $\pm$  0.7& 1.3 $\pm$  0.4\\
        \ \ `Weak'--\SF  &   0.2 $\pm$  0.4  & 1.1 $\pm$ 0.3   &         &  3.9 $\pm$  0.1 &  1.6 $\pm$  0.1   && 0.3 $\pm$  0.1& 1.0 $\pm$  0.1\\
       \ \ `Weak--\nSF   &   0.0 $\pm$  0.8  & 1.0 $\pm$ 1.6   &         &  6.2 $\pm$  0.2 &  1.6 $\pm$  0.1   && 0.3 $\pm$  0.1& 1.0 $\pm$  0.1\\
       
       \hline                                     
    \end{tabular}
    \end{minipage}
  \end{table*}

\subsubsection{Full Sample}

\Cref{fig:xi_all_T} shows the projected (along the \ac{los}; see
\Cref{eq:projected}) correlation functions divided by the transverse
separation, $\Xi(r_{\perp})/r_{\perp}$, for our `Full Sample' of
\hi~absorption systems and galaxies. Different symbols/colors show our
different measurements: the blue squares correspond to the \hi--galaxy
cross-correlation ($\Xi_{\rm ag}$), the black circles to the
galaxy--galaxy auto-correlation ($\Xi_{\rm gg}$), and the red triangles
to the \hi--\hi~auto-correlation ($\Xi_{\rm aa}$; slightly shifted
along the $x$-axis for the sake of clarity). The lines correspond to
the best power-law fits (\Cref{eq:fit}) to the data, from a non-linear
least squares analysis. The parameters $r_0$ and $\gamma$ correspond to
those of the real-space correlation function, $\xi(r)$, when described
as a power-law of the form presented in \Cref{eq:rscf}. Uncertainties
in these fits include the variances and covariances of both
parameters. From this figure, we see that a power-law fit is a good
description of the data, hence justifying the use of Equations
\ref{eq:rscf} and \ref{eq:fit}.\footnote{We note that there might be
  some tension in fitting $\Xi_{\rm gg}$ with a single power-law
  function. We did not explore more complicated fits in order to keep
  the analysis and further comparisons as simple as possible.}
\Cref{tab:xi_params} summarizes the best power-law fit parameters for
our different samples.

We find that $\xi_{\rm ag}(r)$ has a correlation length of $r_0^{\rm
  ag} = 1.6 \pm 0.2$ \mpc and slope $\gamma^{\rm ag}=1.4 \pm 0.1$,
whereas $\xi_{\rm gg}(r)$ and $\xi_{\rm aa}(r)$ have correlation
lengths of $r_0^{\rm gg} = 3.9 \pm 0.1$ \mpc and $r_0^{\rm aa} = 0.3
\pm 0.3$ \mpc, and slopes $\gamma^{\rm gg}=1.7 \pm 0.1$, $\gamma^{\rm
  aa}=1.1 \pm 0.1$, respectively. Thus, the clustering of \hi
absorption systems and galaxies is weaker than the clustering of
galaxies with themselves, and the clustering of \hi systems with
themselves is weaker still. We also see that the slopes are
inconsistent with each other at the $1\sigma$ confidence level (c.l.),
which is in tension with the assumption that these objects trace the
same underlying dark matter distribution linearly (see
\Cref{dependence}). Moreover, the slope of the $\xi_{\rm aa}(r)$ is
consistent with $\gamma=1$, indicating that this distribution is at the
limit in which the methodology adopted here is valid (see
\Cref{correlation:1d}).

As was the case for the two-dimensional results, in the following we
will split the \hi and galaxy samples into \weak and \strong, and \SF
and \nSF, respectively, in order to isolate different contributions
from these sub-populations into the observed correlations.

\subsubsection{`Strong' \hi systems and \SF galaxies}\label{results:1d:strong_sf}

The top left panel of \Cref{fig:xi_split_T} shows the projected
correlation functions for our \strong \hi systems ($N_{\rm HI} \ge
10^{14}$~\cm) and \SF galaxies. In this case, we find that the
$\xi_{\rm ag}(r)$ has a correlation length of $r_0^{\rm ag} = 3.8 \pm
0.2$~\mpc and slope $\gamma^{\rm ag}=1.7 \pm 0.1$, whereas $\xi_{\rm
  gg}(r)$ and $\xi_{\rm aa}(r)$ have correlation lengths of $r_0^{\rm
  gg} = 3.9 \pm 0.1$~\mpc and $r_0^{\rm aa} = 3.1 \pm 0.7$~\mpc, and
slopes $\gamma^{\rm gg}=1.6 \pm 0.1$, $\gamma^{\rm aa}=1.3 \pm 0.4$,
respectively (see also \Cref{tab:xi_params}). Thus, all have
correlation lengths and slopes agreeing with each other at the
$1\sigma$ c.l.. The fact that all have comparable correlation lengths
and slopes supports the hypothesis that \strong \hi systems and \SF
galaxies do trace the same underlying dark matter distribution.

\subsubsection{`Strong' \hi systems and \nSF galaxies}

The bottom left panel of \Cref{fig:xi_split_T} shows the projected
correlation functions for our \strong \hi systems ($N_{\rm HI} \ge
10^{14}$ \cm) and \nSF galaxies. In this case we find that \xiag has a
correlation length of $r_0^{\rm ag} = 4.0 \pm 0.3$ \mpc and slope
$\gamma^{\rm ag}=1.7 \pm 0.1$, whereas \xigg has a correlation length
of $r_0^{\rm gg} = 6.2 \pm 0.2$ \mpc and slope $\gamma^{\rm gg}=1.6 \pm
0.1$ (see also \Cref{tab:xi_params}). The parameters for \xiaa are the
same as in the previous case (see \Cref{results:1d:strong_sf}). The
fact that the slopes are all consistent supports the idea that \strong
\hi systems and \nSF galaxies also trace the same underlying dark
matter distribution. This is an expected result in view of what was
observed for the case of \strong \hi systems and \SF galaxies, and
because it is well known that \SF and \nSF do trace the same underlying
dark matter distribution. We also see that the galaxy--galaxy
auto-correlation is significantly larger than the \hi--galaxy
cross-correlation and the \hi--\hi auto-correlation. The most simple
explanation for such a difference is that the linear bias (see
\Cref{dependence}) of \nSF is greater than that of \SF galaxies. This
has been commonly interpreted as \nSF galaxies belonging, on average,
to more massive dark matter haloes than \SF galaxies. The fact that the
correlation length for the \strong \hi--galaxy cross-correlation is
(marginally) larger for \nSF than \SF galaxies is also expected because
the \hi population is the same in both cases. However, we will see in
\Cref{results:1d:ratio} that this length is smaller than what is
expected from the linear dependence hypothesis.

\subsubsection{`Weak' \hi systems and galaxies}

\begin{figure*}
  \begin{minipage}{0.49\textwidth}
    \centering
    \includegraphics[width=1\textwidth]{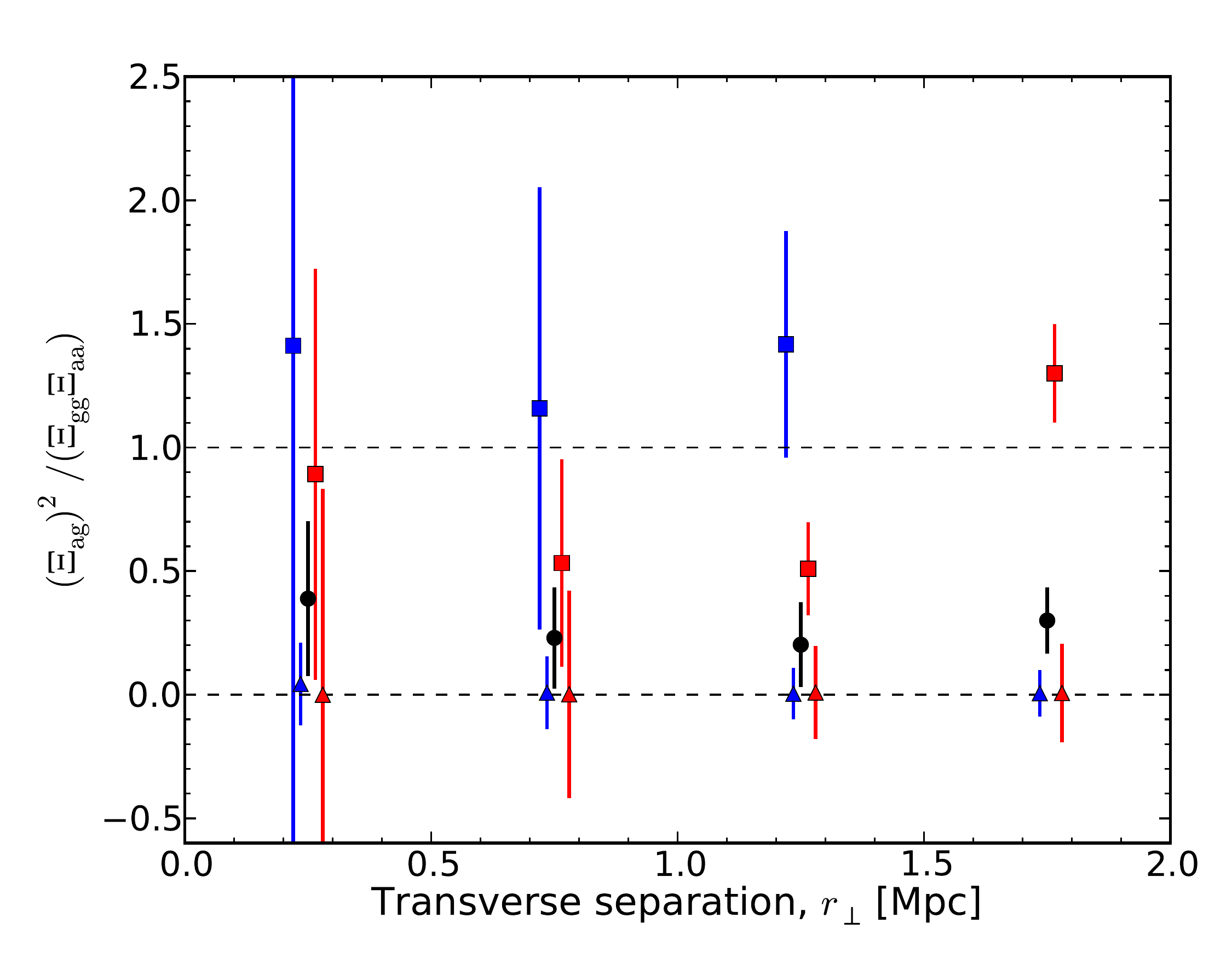}
  \end{minipage}
  \begin{minipage}{0.49\textwidth}
    \centering
    \includegraphics[width=1\textwidth]{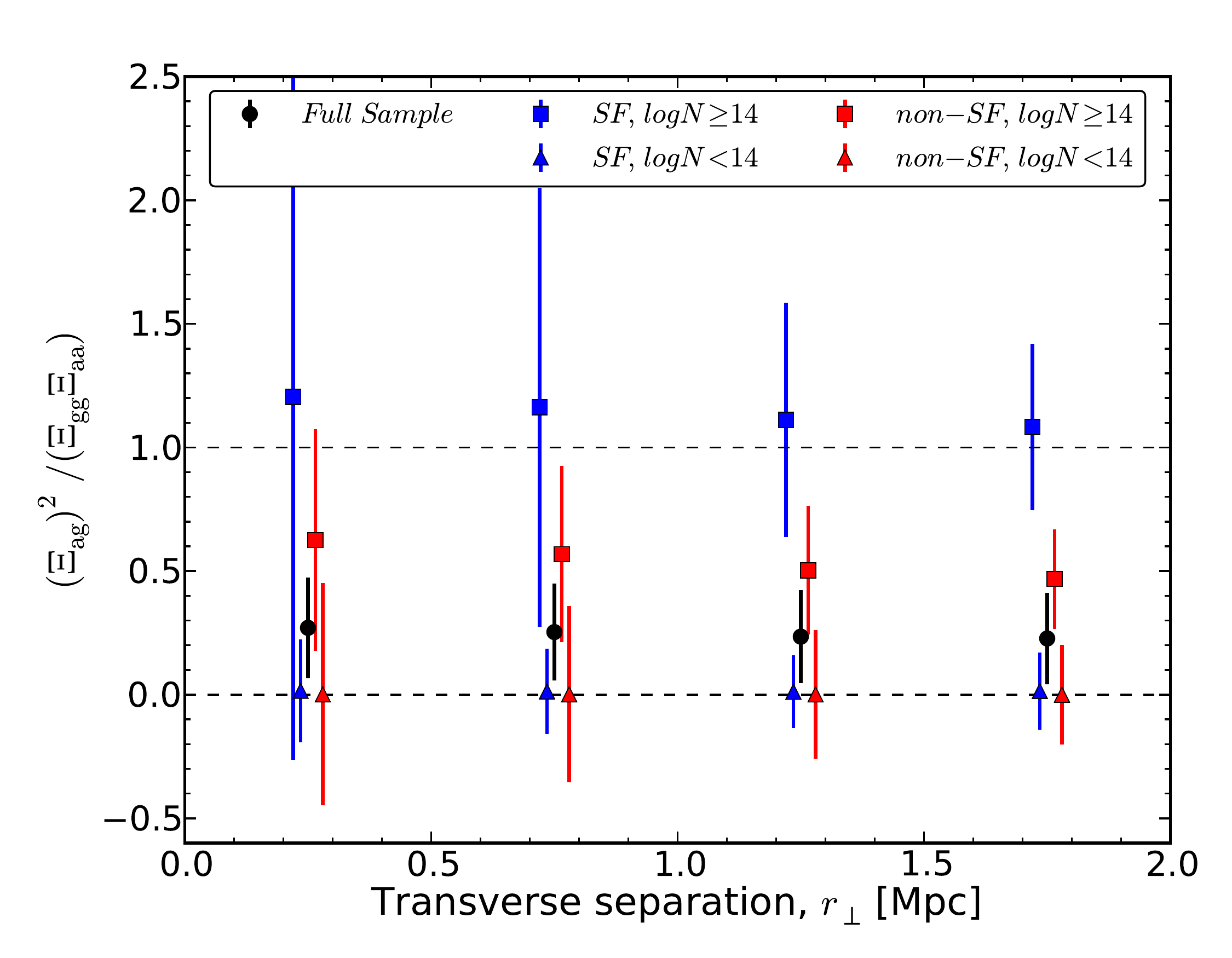}
  \end{minipage}
   
  \caption{The ratio $(\Xi_{\rm ag})^2/(\Xi_{\rm gg}\Xi_{\rm aa})$ as a
    function of transverse separation, $r_{\perp}$. Results from
    different samples of galaxies and \hi~absorption systems are shown
    by different colors/symbols. The black circles correspond to our
    `Full Sample'; blue and red symbols correspond to \SF~and
    \nSF~galaxies respectively; and squares and triangles correspond to
    `strong' ($N_{\rm HI} \ge 10^{14}$ \cm) and `weak' ($N_{\rm HI} <
    10^{14}$ \cm) \hi~absorption systems respectively. The left panel
    shows the results from our adopted Gaussian smoothing of $0.5$~\mpc
    standard deviation while the right panel shows it applying a
    Gaussian smoothing of $1$~\mpc standard deviation. The smoothings
    were applied to the cross-pairs only, {\it before} calculating the
    different $\Xi$ and the corresponding ratios (see
    \Cref{analysis}). Note that the fifth point associated to \strong
    \hi systems and \SF galaxies in the left panel is out of range and
    hence not shown. Uncertainties were obtained directly from the
    `bootstrap' resampling technique of our independent fields (see
    \Cref{uncertainty}). Note that points and uncertainties are both
    correlated. For further details see \Cref{results:1d:ratio}.}
  
\label{fig:projected_ratios}

\end{figure*}

The top right panel of \Cref{fig:xi_split_T} shows the projected
correlation functions for our \weak \hi systems ($N_{\rm HI} < 10^{14}$
\cm) and \SF galaxies. In this case we find that \xiag has a
correlation length of $r_0^{\rm ag} = 0.2 \pm 0.4$~\mpc and slope
$\gamma^{\rm ag}=1.1 \pm 0.3$, whereas \xiaa has a correlation length
of $r_0^{\rm aa} = 0.3 \pm 0.1$~\mpc and slope $\gamma^{\rm aa}=1.0 \pm
0.1$ (see also \Cref{tab:xi_params}). The parameters for \xigg are the
same as in \Cref{results:1d:strong_sf}. These results are dramatically
different from those involving \strong \hi systems. In particular for
the \hi--galaxy cross-correlation, not only is the power-law fit
questionable, but also the correlation length is smaller than both
galaxy--galaxy and \hi--\hi auto-correlations. Moreover, the
correlation length of the cross-correlation is consistent with $r_0=0$,
i.e., no correlation.

The results for \weak \hi systems and \nSF galaxies are even more
dramatic. The bottom right panel of \Cref{fig:xi_split_T} shows the
projected correlation functions for these samples. In this case we find
that \xiag has a correlation length of $r_0^{\rm ag} = 0.0 \pm 0.8$
\mpc and slope $\gamma^{\rm ag}=1.0 \pm 1.6$. Although consistent
within errors with the \weak \hi-\SF galaxy cross-correlation, this
correlation length is even smaller. This result goes in the opposite
direction to what would be expected in the case of linear dependency,
because the clustering of \nSF galaxies with themselves is stronger
than that of \SF. Therefore, these results are a strong indication that
\weak \hi systems and galaxies do not trace the same underlying dark
matter distribution linearly.

\subsubsection{Ratio $(\Xi_{\rm ag})^2/(\Xi_{\rm gg} \Xi_{\rm aa})$}\label{results:1d:ratio}

\Cref{fig:projected_ratios} shows the ratio $(\Xi_{\rm ag})^2/(\Xi_{\rm
  gg} \Xi_{\rm aa})$ for our different samples. The black circles
correspond to our `Full Sample'; blue and red symbols correspond to
\SF~and \nSF~galaxies respectively; and squares and triangles
correspond to `strong' ($N_{\rm HI} \ge 10^{14}$ \cm) and `weak'
($N_{\rm HI} < 10^{14}$ \cm) \hi~absorption systems respectively. The
left panel shows the results from our adopted Gaussian smoothing of
$0.5$~\mpc standard deviation. Given that the points are all
correlated, we expected this ratio to be roughly independent of the
scale, at least below $\lesssim 2$ \mpc. Thus, we attribute the large
variation seen in the left panel of \Cref{fig:projected_ratios} to
`shot noise' and repeated the calculation using a Gaussian smoothing of
$1$~\mpc standard deviation. The right panel of
\Cref{fig:projected_ratios} show the results from this last
calculation. We note that the smoothings were applied to the
cross-pairs only, {\it before} calculating the different $\Xi$ and the
corresponding ratios (see \Cref{analysis} for details), and that the
uncertainties were obtained directly from the `bootstrap' resampling of
our independent fields (see \Cref{uncertainty}).

These results are consistent with what we found for the two-dimensional
correlations. We see that the `Full Sample' have ratios inconsistent
with $1$. Taking the bin at $1.25$ \mpc as representative, we find that
$(\Xi_{\rm ag})^2/(\Xi_{\rm gg} \Xi_{\rm aa}) \approx 0.2 \pm 0.2$,
which gives a high confidence level (c.l; $> 3\sigma$) for ruling out
the hypothesis of linear dependency between the underlying matter
distribution giving rise to \hi and galaxies. The same is true for our
samples of \weak \hi systems and \SF galaxies, for which $(\Xi_{\rm
  ag})^2/(\Xi_{\rm gg} \Xi_{\rm aa}) \approx 0.0 \pm 0.2$. In the case
of \weak \hi systems and \nSF galaxies, we find $(\Xi_{\rm
  ag})^2/(\Xi_{\rm gg} \Xi_{\rm aa}) \approx 0.0 \pm 0.3$ which is also
inconsistent with $1$, but the significance is somewhat reduced. Apart
from the fact that \weak systems and galaxies have this ratio
inconsistent with $1$, it is also interesting to note that all are
consistent with $0$. This result supports the conclusion that many
\weak \hi systems {\it are not} correlated with galaxies on scales
$\lesssim 2$ \mpc.

On the other hand, in the case of \strong \hi systems and \SF galaxies,
this ratio is $(\Xi_{\rm ag})^2/(\Xi_{\rm gg} \Xi_{\rm aa}) \approx 1.1
\pm 0.6$. Thus, we find consistency with the linear dependency
hypothesis, although with large uncertainty. The ratio for \strong \hi
systems and \nSF is $(\Xi_{\rm ag})^2/(\Xi_{\rm gg} \Xi_{\rm aa})
\approx 0.5 \pm 0.3$, which is consistent with neither $1$ nor $0$ (at
least at the $2\sigma$ c.l.). Given the large uncertainty in this case,
no strong conclusion can be drawn. Still, if we believe this ratio to
be $<1$, it would mean that a fraction of \nSF galaxies would not be
correlated with \strong \hi systems either. This fraction can be
estimated from the actual value of $(\Xi_{\rm ag})^2/(\Xi_{\rm gg}
\Xi_{\rm aa})$ (e.g. see \Cref{discussion:interpretation:nSF}).

\begin{figure*}
  \centering
  \includegraphics[width=1.01\textwidth]{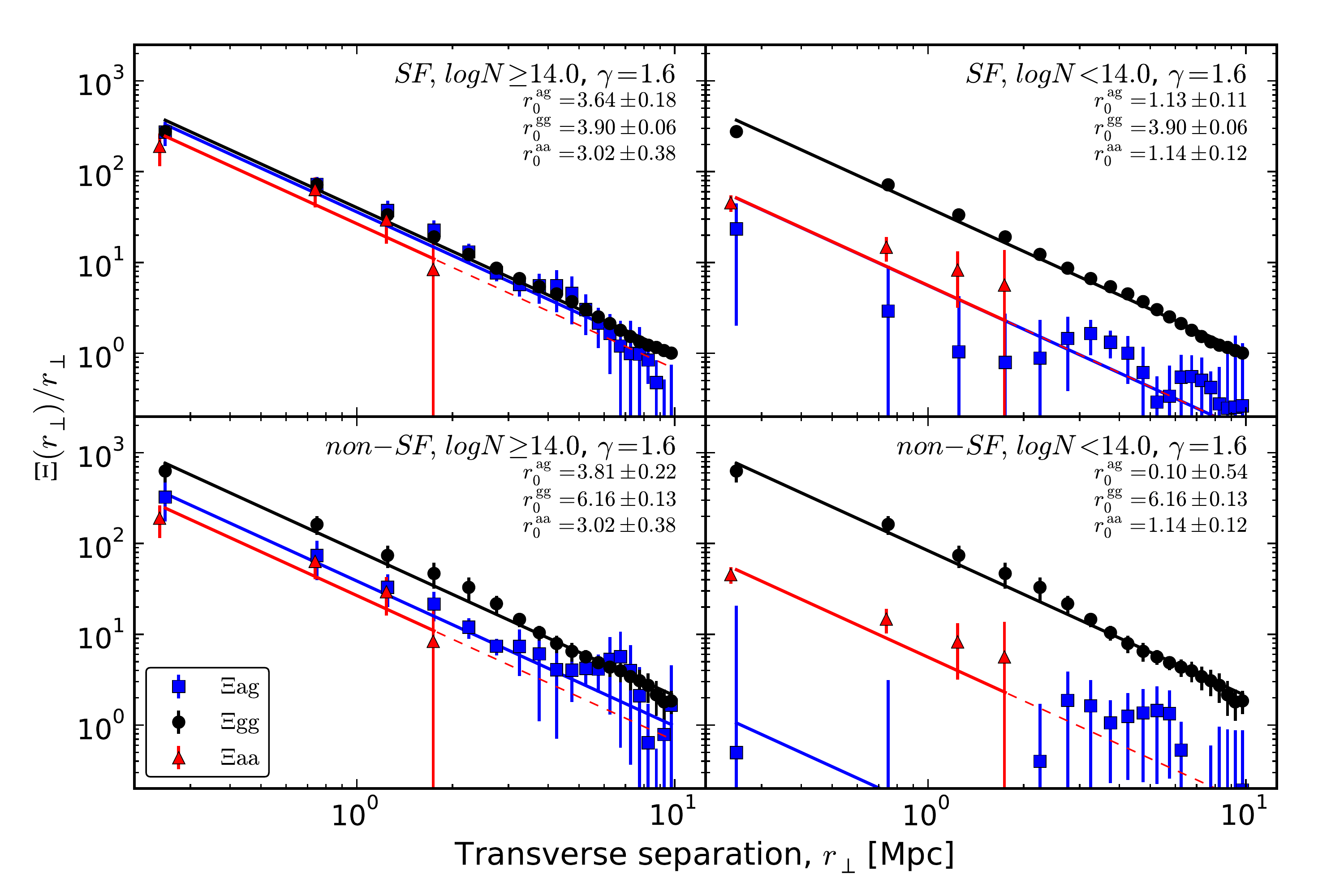}
  \caption{Same as Figure \ref{fig:xi_split_T} but using a fixed slope,
    $\gamma=1.6$, for the power-law fits. These fits are for
    illustrative purposes only (see \Cref{results:1d:gamma} for further
    details). Note that points and uncertainties are both correlated,
    and that uncertainties smaller than the symbols are not shown.}
\label{fig:xi_split_T2}

\end{figure*}

\subsubsection{Results assuming a fixed slope $\gamma = 1.6$}\label{results:1d:gamma}

As mentioned in \Cref{dependence}, if we assume that \hi and galaxies
do trace the same underlying dark matter distribution linearly, then we
can use the different correlation lengths to obtain the relative linear
biases between populations \citep[e.g.][]{Mo1993,Ryan-Weber2006}. For
this method to work, we require the slopes of the correlation functions
to be the same. Even though we have shown that this assumption is not
always valid (at a $>3\sigma$ c.l.), in this section we fix the slope
of the real-space correlations and repeat the analysis. We do this for
illustrative purposes, so these results should not be taken as
conclusive.

\Cref{fig:xi_split_T2} is the same as \Cref{fig:xi_split_T}, but using
a fixed slope of $\gamma=1.6$. Judging from the plots, the fits work
reasonably well for the galaxy--galaxy auto-correlations and the
\hi--galaxy cross-correlations for the \strong \hi systems, but they
fail to represent the \hi--\hi auto-correlations and the \weak
\hi--galaxy cross-correlations. These are expected results given what
we observed in the previous analysis.

The top left panel of \Cref{fig:xi_split_T2} shows the results for our
samples of \strong \hi systems and \SF galaxies. We see that the \xiag
has a correlation length of $r_0^{\rm ag} = 3.6 \pm 0.2$~\mpc, whereas
\xigg and \xiaa have correlation lengths of $r_0^{\rm gg} = 3.9 \pm
0.1$~\mpc and $r_0^{\rm aa} = 3.0 \pm 0.4$~\mpc respectively. All these
correlation lengths are consistent with each other within the
uncertainties, indicating that \strong \hi systems and \SF galaxies
trace the same underlying dark matter distribution linearly. In fact,
the ratio $(\xi_{\rm ag})^2/(\xi_{\rm gg}\xi_{\rm aa}) \approx 1.1 \pm
0.2$. From \Cref{eq:biases} we have that the relative linear biases
should be,

\begin{equation}
\begin{split}
  \left(\frac{b_{\rm g}}{b_{\rm a}}\right) = \left(\frac{r_0^{\rm
      gg}}{r_0^{\rm aa}}\right)^{\frac{\gamma}{2}} =
  \left(\frac{r_0^{\rm gg}}{r_0^{\rm ag}}\right)^{\gamma} \rm{,}\\
\label{eq:biases2}
\end{split}
\end{equation}

\noindent where $b_{\rm g}$ and $b_{\rm a}$ are the \SF and \strong \hi
biases respectively. Replacing the correlation lengths, we get that
$b_{\rm g}:b_{\rm a} \sim 1.1-1.2$, which implies these objects belong
to dark matter haloes of similar masses.

The bottom left panel of \Cref{fig:xi_split_T2} shows the results for
our samples of \strong \hi systems and \nSF galaxies. We see that \xiag
has a correlation length of $r_0^{\rm ag} = 3.8 \pm 0.2$ \mpc, whereas
\xigg has a correlation length of $r_0^{\rm gg} = 6.2 \pm 0.1$
\mpc. The correlation length for \xiaa is the same as before. In
contrast to the \SF case, the correlation length of \nSF galaxies with
themselves is significantly larger ($>3\sigma$ c.l.). In this case, the
ratio $(\xi_{\rm ag})^2/(\xi_{\rm gg}\xi_{\rm aa}) \approx 0.8 \pm
0.1$. Consequently, \Cref{eq:biases2} is at the limit of its
validity. Applying this equation, we find that $b_{\rm g}:b_{\rm a}
\sim 1.8-2.2$.

The top right panel of \Cref{fig:xi_split_T2} shows the results for our
samples of \weak \hi systems and \SF galaxies. We see that \xiag has a
correlation length of $r_0^{\rm ag} = 1.1 \pm 0.1$ \mpc, and \xiaa also
has $r_0^{\rm aa} = 1.1 \pm 0.1$ \mpc. The correlation length for \xigg
is the same as previously mentioned (two paragraphs above). In this
case, the ratio $(\xi_{\rm ag})^2/(\xi_{\rm gg}\xi_{\rm aa}) \approx
0.3 \pm 0.1$. Consequently, \Cref{eq:biases2} should not hold. Still,
if we apply this equation anyway, we find that $b_{\rm g}:b_{\rm a}
\sim 2.6-7.6$.

The bottom right panel of \Cref{fig:xi_split_T2} shows the results for
our samples of \weak \hi systems and \nSF galaxies. We see that \xiag
has a correlation length of $r_0^{\rm ag} = 0.1 \pm 0.5$ \mpc. The
parameters for \xiaa and \xigg are the same as previously mentioned
(one and two paragraphs above respectively). In this case, the ratio
$(\xi_{\rm ag})^2/(\xi_{\rm gg}\xi_{\rm aa}) \approx 0.0 \pm
0.1$. Consequently, \Cref{eq:biases2} should not hold either. Still, if
we apply this equation, we find that $b_{\rm g}:b_{\rm a} \sim 4-700$.

\subsection{Consistency checks}
In order to check whether our results are robust, we have repeated the
analysis using only \hi systems and galaxies in their respective `a'
categories (i.e., best quality; see Sections \ref{gal:reliability} and
\ref{abs:reliability}). We found qualitative agreement with all our
previous results, but a systematic increase in the correlation
amplitudes by $\lesssim 10\%$ (with larger statistical uncertainties)
was observed. Such a difference is expected due to the presence of
random contamination in our `Full Sample' (e.g. catastrophic failures,
missidentification of \hi systems, etc.). Still, within the
uncertainties, the results from both analyses are fully consistent.

We also checked the effect of the Gaussian smoothing by repeating the
analysis without smoothing at all (but still using the same linear
grid). As expected, the new results for $r_0$ and $\gamma$ had
increased statistical uncertainties but were all consistent with our
previously reported values. We note that the slopes obtained from this
new analysis were systematically larger by $\sim 10\%$ in most of the
cases, but a $\sim 30\%$ increase was found for $\gamma^{\rm aa}$ and
$\gamma^{\rm ag}$ in samples involving \weak \hi systems.

\section{Discussion}\label{discussion}

\subsection{Comparison with previous results}\label{discussion:comparison} 

In this section we compare our results with those published in other
recent studies considering the \hi--galaxy two-point correlation
function at $z\lesssim1$.

\subsubsection{Comparison with \citet{Ryan-Weber2006} results ($z \sim 0$)}

\citet{Ryan-Weber2006} measured the \hi--galaxy cross-correlation at $z
< 0.04$ using \hi data from the literature
\citep{Impey1999,Penton2000,Bowen2002,Penton2004,Williger2006} and
galaxy data from the \ac{hipass} \citep{Doyle2005}. Their total sample
comprised $129$ \hi absorption systems with $10^{12.5} \lesssim N_{\rm
  HI} \lesssim 10^{15}$ \cm, from $27$ \ac{qso} \ac{los}; and $5317$
gas-rich galaxies.

Our results are in contrast with theirs. First, they found a strong
`finger-of-god' signal in the two-dimensional \hi--galaxy
cross-correlation, extending up to $\sim 10$ \hmpc (see their figure
3), corresponding to an `anisotropy ratio' of $\sim 10:1$. This
anisotropy signal is also larger than they observed for the
galaxy--galaxy auto-correlation, see their figure 2), meaning that it
can not be explained by the galaxy redshift uncertainties. This result
is in contrast to ours in that we do not see such a significant
`finger-of-god' signal, and the only anisotropy that we observe is
consistent with being due to the galaxy redshift uncertainty.

Another difference between our results and theirs is the correlation
length of the real-space correlations. They found $r_0^{\rm ag} = 7.2
\pm 1.4$ \hmpc (which in our adopted cosmology corresponds to $r_0^{\rm
  ag} \approx 10.3 \pm 2.0$ \hhmpc) imposing $\gamma^{\rm ag}$ to be
equal to that of the \xigg, $\gamma^{\rm ag} \equiv \gamma^{\rm gg}=1.9
\pm 0.3$. Although the slope is marginally consistent with what we find
(see \Cref{results:1d}), the correlation length is more than $3\sigma$
c.l. larger than any of our values.  If we set the slope of our
correlations to be $\gamma=1.9$, we do not find consistency either
(also note that a power-law fit for such a slope is not a good
representation of our data). \citet{Ryan-Weber2006} used this result to
rule out `mini-haloes' for the confinement of \hi absorption
systems. In view of our new results, we consider that this conclusion
must be revisited (see \Cref{discussion:interpretation:masses}).

Another intriguing result from \citet{Ryan-Weber2006} is the fact that
the amplitude of \xiag is greater than that of \xigg. They found a
\xigg correlation length of $r_0^{\rm gg} = 3.5 \pm 0.7$ \hmpc
($\approx 5.0 \pm 1.0$ \hhmpc), which is somewhat larger but marginally
consistent with our findings. In order to explain the larger $r_0^{\rm
  ag}$ value with respect to $r_0^{\rm gg}$, \xiaa should be greater
than both \xigg and \xiag. This hypothesis is difficult to understand
within the current cosmological paradigm, and in fact, it is not
supported by our results on the \hi--\hi auto-correlation either.

We note that the surveys have important differences, in particular
regarding the galaxy samples. \ac{hipass} selected galaxies based on
\hi emission, i.e. containing significant amounts of neutral gas. It
also includes low-surface brightness galaxies that might be lacking in
ours. The clustering of these galaxies is expected to be lower than
that of brighter galaxies in our sample though, which goes in the
opposite direction of what is needed to reconcile our results with
those of \citet{Ryan-Weber2006}. The much lower redshift range in their
sample might also have an impact on the clustering, as structures are
more collapsed. This might help to increase the correlation lengths,
but it should not make the \xiag amplitude greater than \xigg by
itself. Another possibility is that this might be a case in which the
ratio $(\xi_{\rm ag})^2/(\xi_{\rm gg}\xi_{\rm aa})>1$, meaning that the
\hi and galaxies observed actually correspond to the same physical
objects (see \Cref{dependence}). Such an effect should be most
noticeable at the smallest scales, but this is not supported by their
results. Indeed, there is a flattening in their reported $\Xi_{\rm
  ag}(r_{\perp})/r_{\perp}$ at $\lesssim 1$ \hmpc (see their figure 5)
which makes the \hi--galaxy cross-correlation consistent with the
galaxy--galaxy auto-correlation at these scales.\footnote{Note that a
  flattening in $\Xi(r_{\perp})/r_{\perp}$ means that $\Xi(r_{\perp})
  \propto r_{\perp}$.}

\subsubsection{Comparison with \citet{Wilman2007} results ($z \lesssim 1$)}

\citet{Wilman2007} measured the \hi--galaxy cross-correlation at
$z\lesssim 1$ using data from \citet{Morris2006}. Their total sample
comprised $381$ \hi absorption systems with $10^{13} \lesssim N_{\rm
  HI} \lesssim 10^{19}$ \cm, from $16$ \ac{qso} \ac{los}; and $685$
galaxies at $\lesssim 2 $ \hmpc from the \ac{qso} \ac{los}, of which,
$225$ were classified as `absorption-line-dominated' and $406$ as
`emission-line-dominated'.

We find qualitative agreement with their observational results in the
following sense: (i) no strong `finger-of-god' effect is seen in the
observed \hi--galaxy cross-correlation; (ii) the larger the \nhi, the
stronger the clustering with galaxies; and (iii) no evidence that
`emission-line-dominated' galaxies cluster more strongly with \hi
systems than `absorption-line-dominated' ones \citep[in contrast to
  what was reported by][see below]{Chen2009}.

\citet{Wilman2007} also performed a comparison with a cosmological
hydrodynamical simulation. In contrast to their observational results,
they did find a strong `finger-of-god' effect \citep[similar to that
  found by][]{Ryan-Weber2006} in their simulated data (see their figure
6). This prediction is in not supported by our observations, as we do
not detect such a strong anisotropy feature.


\subsubsection{Comparison with \citet{Pierleoni2008} results (simulations)}\label{discussion:comparison:pierleoni}

\citet{Pierleoni2008} investigated the observational results from
\citet{Ryan-Weber2006} and \citet{Wilman2007} in the context of a
cosmological hydrodynamical simulation. They selected samples of
simulated \hi absorption systems and galaxies, trying to match those
from \citet{Ryan-Weber2006}.

Contrary to the \citet{Ryan-Weber2006} observational results \citep[and
  the prediction from][]{Wilman2007}, they did not find a strong
`finger-of-god' signal in the mock \hi--galaxy cross-correlation (see
their figure 7), which agrees with our observational result. In
contrast, they find a compression along the \ac{los} at scales $\gtrsim
4$ \hmpc, similar to the expectation from the `Kaiser effect'
\citep{Kaiser1987}. We did not detect such a feature but note that our
survey was not designed to do so.

They also found that the peak in the two-dimensional \hi--galaxy
cross-correlation was offset along the \ac{los} by about $\sim 1$
\hmpc. A similar signal was observed in our sample of \strong \hi
systems and \nSF galaxies (see \Cref{fig:xi_ab_st}), but these two
results are not directly comparable. Indeed, we do not observe such a
feature in our `Full Sample'. Still, we caution the reader that a $\sim
1$ \hmpc displacement in the \ac{los} direction is comparable to our
galaxy redshift uncertainty ($\sim 1.4-1.7$ \hhmpc), and so such a
signal might get easily diluted.

Another qualitative agreement between our results and those from
\citet{Pierleoni2008} is that the amplitude of the \hi--galaxy
cross-correlation is smaller than that of the galaxy--galaxy
auto-correlation, and that the \hi--\hi auto-correlation is smaller
still (see their figures 3 and 9). Quantitatively, they found that
\xiag and \xigg have correlation lengths of $r_0^{\rm ag} = 1.4 \pm 0.1
$ \hmpc ($\approx 2.0 \pm 0.1$ \hhmpc) and $r_0^{\rm gg} = 3.1 \pm 0.2
$ \hmpc ($\approx 4.4 \pm 0.2$ \hhmpc), and slopes $\gamma^{\rm ag} =
1.29 \pm 0.03$ and $\gamma^{\rm gg} = 1.46 \pm 0.03$
respectively. These values are marginally consistent with our findings
(see \Cref{tab:xi_params}). Moreover, they predict a flattening of
\xiaa at scales $\lesssim 1$ \hmpc, which is also consistent with our
observations.

Finally, we also find agreement in the sense that the amplitude of the
\hi--galaxy cross-correlation significantly increases for high column
density absorbers, but little variation is observed for different
galaxy samples selected by mass (see their figure 4). Even though we do
not have direct measurements of galaxy masses in our galaxy samples,
the significantly larger auto-correlation amplitude of \nSF galaxies
with respect to \SF suggests that, on average, \nSF galaxies typically
belong to more massive dark matter haloes than \SF galaxies (see also
\Cref{discussion:interpretation:masses}).

\subsubsection{Comparison with \citet{Chen2009} results ($z \lesssim 0.5$)}\label{discussion:comparison:chen}

\citet{Chen2009} measured the \hi--galaxy cross-correlation at $z
\lesssim 0.5$ from their own \hi and galaxy survey \citep[including
  data from][]{Chen2005}. Their total sample comprised $195$ \hi
absorption systems with $10^{12.5} \lesssim N_{\rm HI} \lesssim
10^{16}$ \cm, from $3$ \ac{qso} \ac{los}; and $670$ galaxies at
$\lesssim 4 $ \hmpc from the \ac{qso} \ac{los}, of which $222$ are
classified as `absorption-line-dominated' and $448$ as
`emission-line-dominated'.

In this case, we find both agreements and disagreements. Our results
agree with theirs in the sense that the clustering of \strong
\hi~systems ($N_{\rm HI} \ge 10^{14}$ \cm) with galaxies is stronger
than that of \weak \hi systems and galaxies (see their figure 13), and
that \strong \hi systems and `emission-line-dominated' galaxies have
comparable clustering amplitudes. However, our results disagree with
their claim that \strong \hi systems cluster more strongly with
`emission-line-dominated' than with `absorption-line-dominated' (see
their figure 13). In fact, our findings are consistent with the
amplitude of the \hi--galaxy cross-correlation being independent of
spectral type (within the statistical uncertainties). Moreover, we find
that the \hi--galaxy cross-correlation for \nSF galaxies is
systematically stronger than that of \SF galaxies, which is the
opposite to what \citet{Chen2009} found.

Quantitatively, they reported a $\sim 6 \times$ smaller clustering
amplitude between \strong \hi absorption systems and \SF galaxies than
that of \nSF galaxies with themselves, whereas we find this difference
to be a factor of $\sim 2$ only. We note that their quoted statistical
errors are Poissonian, which underestimate the true uncertainties. The
Poissonian uncertainty in our survey is typically $\sim 1$ order of
magnitude smaller than our adopted `bootstrap' one (see
\Cref{uncertainty}). Thus, there is still room for their results to
agree with ours after taking this fact into account. There is also the
possibility that sample/cosmic variance is significantly affecting
their results. We note that one of the three \ac{qso} \ac{los} used by
them passes at $\sim 2$ \mpc from the Virgo Cluster. Even though this
single cluster is not likely to explain the discrepancy, any sightline
passing through it is also probing an unusually high overdensity in the
local Universe (which extends beyond the Virgo Cluster itself).

\subsubsection{Comparison with \citet{Shone2010} results ($z\sim1$)}

\citet{Shone2010} measured the \hi--galaxy cross-correlation at $0.7
\lesssim z \lesssim 1.5$ from their own \hi and galaxy survey. Their
total sample comprised $586$ \hi absorption systems with $10^{13.2}
\lesssim N_{\rm HI} \lesssim 10^{17}$ \cm, from $2$ \ac{qso} \ac{los};
and $193$ galaxies at $\lesssim 4 $ \hhmpc from the \ac{qso} \ac{los}
($196$ absorber--galaxy pairs used).

They found the peak in the two-dimensional \hi--galaxy cross-correlation
to be $\xi_{\rm ag}^{\rm peak} = 1.9 \pm 0.6$ (although displaced from
the smallest separation bin by $\sim 5$ \hhmpc along the \ac{los}; see
their figure 12), whereas $\xi_{\rm gg}^{\rm peak} = 10.7 \pm 1.4$ for
the galaxy--galaxy auto-correlation (see their figure 13). Our results
agree with theirs qualitatively in the sense that the clustering of \hi
and galaxies is weaker than that of the galaxies with themselves.

\subsubsection{Summary}
In summary, we have found agreements and disagreements with previously
published results. We consider the majority of the discrepancies to be
driven by the inherent difficulty of addressing uncertainties in this
type of analysis, which often lead to an underestimation of the errors.

\subsection{Interpretation of the results}\label{discussion:interpretation}

In this section we provide our preferred interpretation of our
observational results.

\subsubsection{Probabilistic interpretation (model independent)}
The clustering analysis provides an essentially model independent
statistic. The amplitude of the two-point correlation function
corresponds to the probability excess of finding a pair compared to the
Poissonian expectation. Thus, our results point towards the following
conclusions:

\begin{itemize}

\item The probability of finding a \SF galaxy at a distance $\lesssim
  5$ \mpc from another \SF galaxy, is $\sim 2 \times$ smaller than
  that of finding a \nSF galaxy at that same distance from another \nSF
  galaxy.\smallskip

\item The probability of finding a \hi absorption system with $N_{\rm
  HI} \ge 10^{14}$ \cm at a distance $\lesssim 5$ \mpc from a \SF
  galaxy, is approximately the same as that of finding a \SF galaxy at
  that same distance from another \SF galaxy.\smallskip

\item The probability of finding a \hi absorption system with $N_{\rm
  HI} < 10^{14}$ \cm at a distance $\lesssim 5$ \mpc from a \SF
  galaxy, is $\sim 10 \times$ smaller than that of finding a \SF galaxy
  at that same distance from another \SF galaxy.\smallskip

\item The probability of finding a \hi absorption system with $N_{\rm
  HI} \ge 10^{14}$ \cm at a distance $\lesssim 5$ \mpc from a \nSF
  galaxy, is $\sim 2 \times$ smaller than that of finding a \nSF galaxy
  at that same distance from another \nSF galaxy.\smallskip

\item The probability of finding a \hi absorption system with $N_{\rm
  HI} < 10^{14}$ \cm at a distance $\lesssim 5$ \mpc from a \nSF
  galaxy, is $\gtrsim 100 \times$ smaller than that of finding a \nSF
  galaxy at that same distance from another \nSF galaxy.\smallskip

\item The probability of finding a \hi absorption system with $N_{\rm
  HI} < 10^{14}$ \cm at a distance $\lesssim 2$ \mpc from another
  $N_{\rm HI} < 10^{14}$ \cm system is $\sim 4 \times$ smaller than
  that of finding a $N_{\rm HI} \ge 10^{14}$ \cm system at that same
  distance from another $N_{\rm HI} \ge 10^{14}$ \cm system.

\end{itemize}

Any physical model aiming to explain the connection between \hi
absorption systems and galaxies at $z\lesssim 1$ will need to take
these constraints into account.

\subsubsection{Velocity dispersion between \hi and galaxies}

We find that the two-dimensional \hi--galaxy cross-correlations do not
show detectable velocity distortions along the \ac{los} larger than
those expected from the galaxy redshift uncertainties. As mentioned,
the typical uncertainty for our single galaxy redshift determination is
$\Delta z_{\rm gal} \approx 0.0006/\sqrt{2}$, which is equivalent to
rest-frame velocity differences of $ \Delta v \sim 120-60$ \kms at
$z=0.1-1$, respectively. Any velocity dispersion between \hi systems
and galaxies greater than, or of the order of, this value, would have
been noticeable in the two-dimensional \hi--galaxy cross-correlation
signals. Therefore, we conclude that the bulk of \hi systems on
$\sim$~Mpc scales, have little velocity dispersion ($\lesssim 120$
\kms) with respect to the bulk of galaxies. Hence, no strong galaxy
outflow or inflow signal is detected in our data.

We emphasize that our results are based on \hi only. Given that \hi
does not exclusively trace gas originating in galaxy outflows or
inflows, we do not necessarily expect to find the same signatures as
those traced by metals. Moreover, our results are dominated by scales
somewhat larger than those typically associated to the \ac{cgm}, in
which the outflow or inflow signal is expected to be maximized. In view
of these considerations, it is not surprising that no strong outflow or
inflow signal is detected in our data.

We also emphasize that the cross-correlation analysis provides an
averaged statistical result; individual galaxies having strong \hi
inflows/outflows might still be present, but our results indicate that
these do not dominate the cross-correlation signal at $z\lesssim 1$.

\subsubsection{Spatial distribution of \hi and galaxies}\label{discussion:distribution}

The absolute and relative clustering amplitudes of our different
populations of \hi and galaxies can be used to give us an idea of their
spatial distribution. Our conclusions on this are as follows:

\begin{itemize}
\item The fact that \strong \hi systems and \SF galaxies have similar
  amplitudes and slopes for the auto- and cross-correlation, indicates
  that these are distributed roughly in the same locations.\smallskip

\item The fact that the auto-correlation of \nSF has also the same
  slope but a larger amplitude, indicates that there are sub-locations
  (within those where galaxies and \strong \hi systems reside) with a
  higher density of \nSF galaxies than \SF galaxies and/or \strong \hi
  systems. This interpretation also explains the fact that the ratio
  $(\xi_{\rm ag})^2/\xi_{\rm gg} \xi_{\rm aa}$ for \strong \hi systems
  and \nSF galaxies is consistent with neither $1$ nor $0$ (see
  \Cref{discussion:interpretation:nSF}).\smallskip

\item The suggestion that the self-clustering of \weak systems is not
  zero, and the fact that \weak \hi systems and galaxies have a ratio
  $(\xi_{\rm ag})^2/\xi_{\rm gg} \xi_{\rm aa} \approx 0$, indicate that
  these are not distributed in the same locations. Therefore, there are
  locations containing \weak \hi systems but roughly devoid of \strong
  \hi systems and galaxies of any kind.
\end{itemize}

This picture fits well with the recent results presented in
\citet{Tejos2012}, from their study of the distribution of \hi
absorption systems within and around galaxy voids at $z\lesssim
0.1$. They showed that galaxy voids are not empty, and in fact contain
about $\sim 20-40\%$ of \hi absorption line systems with $N_{\rm HI}
\gtrsim 10^{12.5}$ \cm. The remaining $\sim 60-80\%$ were found at the
edges of galaxy voids, hence sharing locations with galaxies.

Even though it seems natural to identify our \weak systems with those
systems found in galaxy voids, not all \weak systems need to be
unassociated with galaxies. Despite the fact that \citet{Tejos2012}
reported a (tentative) difference in the column density distributions
between \hi absorbers within and around galaxy voids (at the $\sim
2\sigma$ c.l.), they did not find sharp \nhi transitions between their
samples. The most important difference came from the presence of
`extremely weak' \hi systems, $N_{\rm HI} \lesssim 10^{13}$ \cm, that
were present within galaxy voids but not outside (see their figures 2
and 3). Such a low column density is at the limit of our current
completeness (see \Cref{abs:completeness}) and so we are not able to
give confident results on the clustering of these `extremely weak' \hi
systems either with themselves or with galaxies. Restricting the column
density range to $10^{13}\le N_{\rm HI} < 10^{14}$ \cm, there are
$19/50 \sim 40\%$ systems within galaxy voids in the \citet{Tejos2012}
sample. In the following we will estimate the fraction of \weak systems
that could still be associated with galaxies in our current sample.\\

It is straightforward to show that the two-point correlation function
between two populations, $a$ and $b$, each one composed by
sub-populations $a_i$ where $i \in \{0,1,...,N_{a}\}$, and $b_j$ where
$j \in \{0,1,...,N_{b}\}$, respectively, is

\begin{equation}
\xi_{ab} = \sum_{i}^{N_a}\sum_{j}^{N_b} f_i f_j \xi_{a_ib_j} \ \rm{,}
\label{eq:multicomponent}
\end{equation}

\noindent where $\xi_{a_ib_j}$ is the cross-correlation between the
$a_i$ and $b_j$ sub-populations (assumed positive), and $f_i$ and $f_j$
are the fractions of $a_i$ and $b_j$ objects over the total samples $a$
and $b$, respectively. Thus, if we think of \weak absorbers being
composed of two kinds of populations, we have,

\begin{equation}
  \xi_{aa}^{\rm weak} = f_{a_1}^2 \xi_{a_1a_1} + f_{a_2}^2 \xi_{a_2a_2} +
  2f_{a_1}f_{a_2} \xi_{a_1a_2} \ \rm{.}
\end{equation}

\noindent If we consider a scenario in which one of these populations
clusters in the same way as \strong \hi systems ($\xi_{a_1a_1} \equiv
\xi_{aa}^{\rm strong}$) and the other is completely random
($\xi_{a_2a_2} = \xi_{a_1a_2} \equiv 0 $), then,

\begin{equation}
\xi_{aa}^{\rm weak} = f_{a_1}^2 \xi_{aa}^{\rm strong} \ \rm{,}
\end{equation}

\noindent From this, we can estimate the fraction of \weak systems that
could be clustered like \strong ones as $f_{a_1} = \sqrt{\xi_{aa}^{\rm
    weak}/\xi_{aa}^{\rm strong}} \sim 0.5$. We note that the assumption
that one of the sub-populations has $\xi_{a_2a_2} \equiv 0$ might be
unrealistic, because $\xi_{aa}^{\rm weak}$ and $\xi_{aa}^{\rm strong}$
have marginally different slopes, and a random component does not
change the slope but only the amplitude of the correlation
function. Also, if both populations lie exclusively in different
locations, the cross-correlation should be $\xi_{a_1a_2} < 0$, which
makes $\xi_{a_1a_2} \equiv 0$ unrealistic too. These two effects go in
opposite directions for the final fraction estimation however, which
might in the end compensate each other. With this caveat in mind, this
rough estimation seems consistent with what \citet{Tejos2012} found for
systems in the range $10^{13} \lesssim N_{\rm HI} \lesssim 10^{14}$ \cm
($\sim 60\%$; see above).

\subsubsection{\hi and non-star-forming
  galaxies}\label{discussion:interpretation:nSF}

Our results point towards \strong \hi systems and \nSF galaxies having
a ratio $(\Xi_{\rm ag})^2/(\Xi_{\rm gg} \Xi_{\rm aa}) \approx 0.5 \pm
0.3$, which is consistent with neither $1$ nor $0$ at the $\sim 1\sigma$
c.l.. In order to explain this result we will consider the presence of
two types of \nSF galaxies: one type ($g_1$) that correlates linearly
with \strong \hi absorbers, and another type ($g_2$) that does
not. Thus,

\begin{equation}
\begin{split}
\frac{(\xi_{ag_1})^2}{(\xi_{aa})(\xi_{g_1g_1})} & \equiv 1 \ \rm{,}\\
\frac{(\xi_{ag_2})^2}{(\xi_{aa})(\xi_{g_2g_2})} & \equiv 0 \ \rm{.}\\
\end{split}
\end{equation}

\noindent Let $f_{g_1}$ and $f_{g_2}$ be the fraction of \nSF galaxies
of type $g_1$ and $g_2$, respectively, such that $f_{g_1}+f_{g_2}=1$.
Then, from \Cref{eq:multicomponent} we have,

\begin{equation}
\begin{split}
\xi_{ag} & = f_{g_1}\xi_{ag_1} + f_{g_2}\xi_{ag_2} \\
        &  = f_{g_1}\xi_{ag_1}  \ \rm{,}\\
\end{split}
\end{equation}

\noindent because $\xi_{ag_2}=0$. Similarly,

\begin{equation}
\begin{split}
\xi_{gg} &= f_{g_1}^2\xi_{g_1g_1} + f_{g_2}^2\xi_{g_2g_2}+2f_{g_1}f_{g_2}\xi_{g_1g_2} \\
         &= f_{g_1}^2\xi_{g_1g_1} + f_{g_2}^2\xi_{g_2g_2} \ \rm{,}\
\end{split}
\end{equation}

\noindent because $\xi_{g_1g_2}\approx 0$ also. Our observational
results indicate that,
\begin{equation}
\begin{split}
\frac{(\xi_{ag})^2}{(\xi_{aa})(\xi_{gg})} = \alpha \ \rm{,}\\
\end{split}
\end{equation}

\noindent with $0 < \alpha < 1$. Combining these relations we find the
following quadratic equation for $f_{g_1}$,

\begin{equation}
(1-\alpha - \alpha\beta) f_{g_1}^2 + 2\alpha\beta f_{g_1} - \alpha
  \beta = 0 \ \rm{.}
\label{eq:fg1}
\end{equation}

\noindent where $\beta \equiv \xi_{g_2g_2}/\xi_{g_1g_1}$. Solving
\Cref{eq:fg1} for a positive solution smaller than $1$, gives us our
estimation of the required fraction of \nSF galaxies that are
correlated with \strong \hi systems linearly, for the given
$(\xi_{ag})^2/(\xi_{aa})(\xi_{gg})$ and $\xi_{g_2g_2}/\xi_{g_1g_1}$
ratios.

Our proposed scenario aims to approximate what might be the case for
galaxy clusters, which contain an important fraction of \nSF galaxies
but whose diffuse \ac{igm} or \ac{cgm} can get destroyed by baryonic
physics \citep[e.g.][]{Morris1993,Lopez2008,Padilla2009,Yoon2012}. In
such a case, $\xi_{g_2g_2}/\xi_{g_1g_1} \gg 1$ because galaxy clusters
represent the most massive dark matter haloes. Measurements and
predictions for the auto-correlation of galaxy clusters point towards
correlation lengths of $r_0^{\rm cc}\sim 20-30$ \mpc
\citep[e.g.][]{Colberg2000, Estrada2009,Hong2012}, which would imply a
$\xi_{g_2g_2}/\xi_{g_1g_1} \sim 10 \pm 5$ (assuming a slope of
$\gamma=1.6$). Using this value together with
$(\xi_{ag})^2/(\xi_{aa})(\xi_{gg})=0.5\pm 0.3$, we find the fraction
$f_{g_1} \approx 0.75 \pm 0.15$ and consequently $f_{g_2} \approx
0.25 \pm 0.15$.\footnote{Note that the functional form of the
  solution of \Cref{eq:fg1} gives relatively well constrained results,
  even for $(\xi_{ag})^2/(\xi_{aa})(\xi_{gg})$ and
  $\xi_{g_2g_2}/\xi_{g_1g_1}$ ratios with large uncertainties (as in
  our case).}

Therefore, our results suggest that an important fraction of \nSF
galaxies ($\sim 60-90\%$) trace the same underlying dark matter
distribution as \strong \hi systems and \SF galaxies at scales
$\lesssim 2$ \mpc. This is in contrast with what can be inferred from
the results reported by \citet{Chen2009}, in which \strong \hi systems
cluster more weakly with \nSF than \SF galaxies (see
\Cref{discussion:comparison:chen}). In such a case,
$(\xi_{ag})^2/(\xi_{aa})(\xi_{gg})\approx 0$,\footnote{Otherwise
  $\xi_{\rm gg}^{\rm non-SF}<\xi_{\rm gg}^{\rm SF}$, which is in
  contradiction with their observations.} implying a fraction close to
$f_{g_1} \sim 0$.

Our simple interpretation agrees quite well with the recent
observational results presented by \citet{Thom2012}. These authors
found that $11/15 \sim 70\%$ of their sample of
`non-star-forming-galaxies' at low-$z$, have \hi absorption with
rest-frame equivalent widths, $W_r > 0.3$~\AA\ (equivalent to $\gtrsim
10^{14}$~\cm), within $300$~\kms from their systemic redshifts, and at
impact parameters $\lesssim 200$~\kpc (see their figures 2 and 3). By
definition, these \hi systems should be associated with the \ac{cgm} of
these galaxies. However, because of incompleteness in the galaxy
surveys, it is not certain that this gas is purely associated to these
`non-star-forming-galaxies' \citep[less luminous
  `star-forming-galaxies' could have been missed by their target
  selection; e.g.][]{Stocke2013}. Still, both \citet{Thom2012} and our
results point towards the conclusion that a significant fraction of
`non-star-forming-galaxies' share locations with \strong \hi systems at
scales $\lesssim 2$~\mpc. Thus, our results indicate that the `cold
gas' (traced by \strong \hi) around \nSF galaxies could be the rule
rather than the exception.

\begin{figure}
  \includegraphics[width=0.49\textwidth]{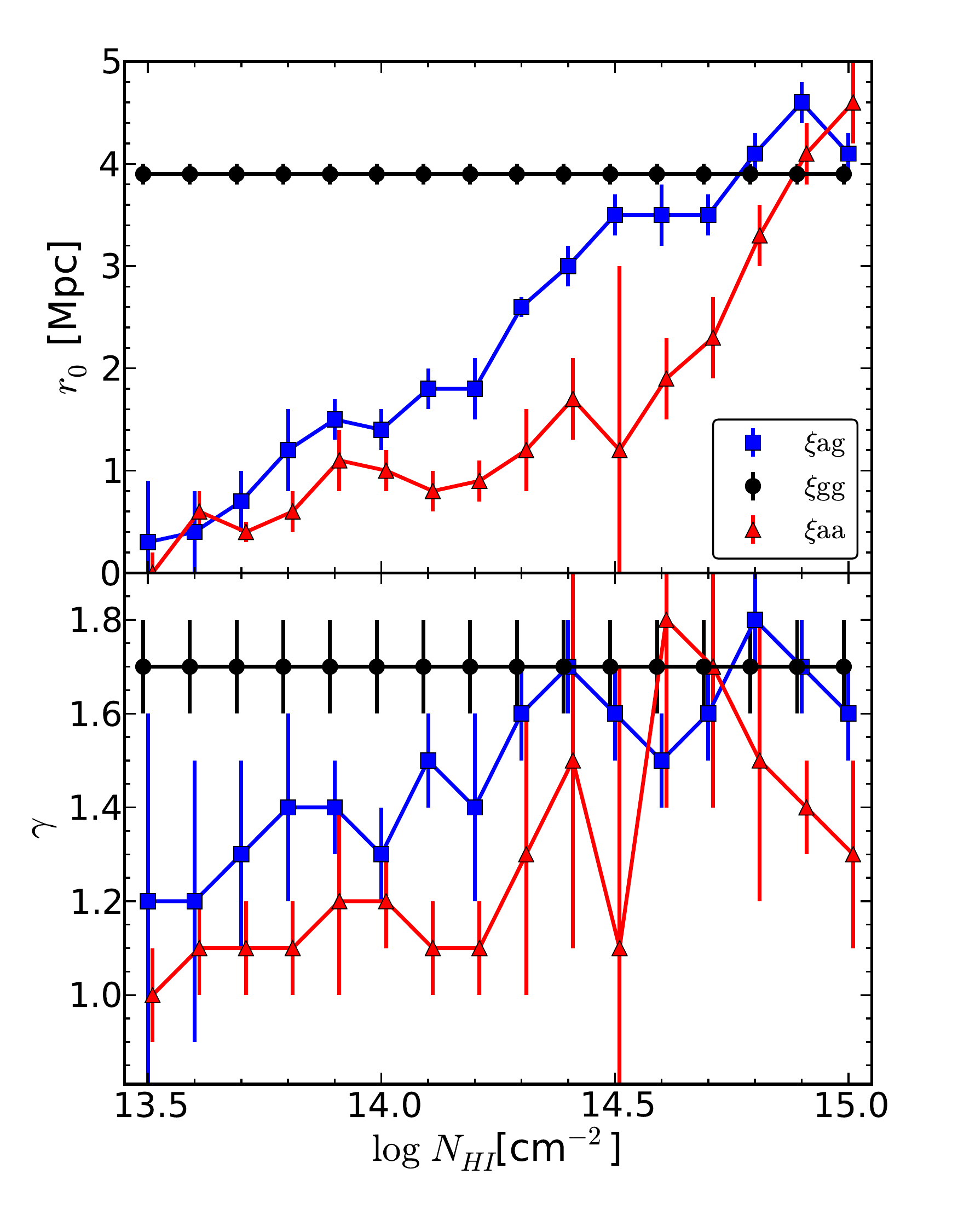}
  \caption{Correlation lengths (top panel) and slopes (bottom panel)
    from the best power-law fits of the `real-space' correlation
    functions of the form presented in \Cref{eq:rscf}, as a function of
    \hi column density bins of $1$ dex width each, from which the
    correlation functions were measured. Different symbols/colors show
    our different measurements: the blue squares correspond to the
    galaxy-\hi~cross-correlation ($\xi_{\rm ag}$); the black circles to
    the galaxy-galaxy auto-correlation ($\xi_{\rm gg}$; slightly
    shifted along the $x$-axis for the sake of clarity); and the red
    triangles to the \hi-\hi~auto-correlation ($\xi_{\rm aa}$; slightly
    shifted along the $x$-axis for the sake of clarity). Note that
    points associated with the galaxy auto-correlation are independent
    of \hi column density. Note that points and uncertainties are both
    correlated.}\label{fig:logn_limits}

\end{figure}

\subsubsection{Column density limit}\label{discussion:logn}

Our choice of a $10^{14}$ \cm limit was somewhat arbitrary (see
\Cref{abs:logn}). As mentioned, when we increase the limit for dividing
\strong versus \weak systems from $10^{14}$ to $\sim 10^{15-16}$ \cm,
we get larger cross-correlation amplitudes and slopes (although with
larger uncertainties due to the reduced number of systems above such
limits in our sample) for \strong compared to those from \weak
systems. Similarly, when we decrease the limit from $10^{14}$ to
$\sim10^{13}$ \cm, we observe a decrease in the cross-correlation
amplitudes and slopes of \strong systems. In this section we explore
more on this issue.

In order to quantify the \hi-galaxy cross-correlation dependence on \hi
column density, we have repeated the analysis using subsamples of the
\hi absorption systems based on \nhi limits, together with all galaxies
in our `Full Sample'. We used $16$ \nhi bins of $1$ dex width each,
shifted by $0.1$ dex, starting from $[10^{13},10^{14}]$ through
$[10^{14.5},10^{15.5}]$~\cm. \Cref{fig:logn_limits} shows the
correlation lengths (top panel) and slopes (bottom panel) from the best
power-law fits of the `real-space' correlation functions of the form
presented in \Cref{eq:rscf}, for each of those \nhi bins. Different
symbols/colors show our different measurements: the blue squares
correspond to the galaxy-\hi~cross-correlation ($\xi_{\rm ag}$); the
black circles to the galaxy-galaxy auto-correlation ($\xi_{\rm gg}$;
slightly shifted along the $x$-axis for the sake of clarity); and the
red triangles to the \hi-\hi~auto-correlation ($\xi_{\rm aa}$; slightly
shifted along the $x$-axis for the sake of clarity). Note that points
associated with the galaxy auto-correlation are independent of \hi
column density, and that points and uncertainties are both
correlated. As expected, we see an overall monotonic increase in the
correlation length and slopes with increasing \nhi. Such a behaviour
can be explained by assuming that the fraction of \hi systems that are
not correlated with galaxies decreases with an increase in the minimum
column density limit. Any change in the amplitude of the correlation
functions can be understood as a change in the `linear bias' and/or the
fraction of `random contamination' present. Changes in the slope of the
correlations (like the one we have marginally observed in this work)
would require the addition of baryonic physics, assuming a fixed
underlying dark matter slope. We also observe that the $\xi_{\rm ag}$
and $\xi_{\rm gg}$ have comparable amplitudes (within $2\sigma$) for
column density bins centred at $10^{14.5}$ \cm and above, which
corresponds to $N_{\rm HI} \gtrsim 10^{14}$ \cm. As mentioned, this is
one of the reasons that motivated our adopted limit of $10^{14}$ \cm
for splitting our \hi sample (see \Cref{abs:logn}).

These results show that there is a dramatic change in the \hi-galaxy
cross-correlation signal, where the correlation length changes from
being consistent with $0$~\mpc at \nhi~$\in [10^{13},10^{14}]$ \cm, to
being consistent with the $\xi_{\rm gg}$ value of $\sim 4$~\mpc at
\nhi~$\in [10^{14},10^{15}]$ \cm. The slope of the \hi--galaxy
cross-correlation also follows the same trend. This is an important
change occurring in about one order of magnitude column density
range. Given the $1$ dex binning used for this analysis (needed to
reduce the statistical uncertainties), we cannot rule out an even
sharper transition occurring within the $\sim 10^{13}-10^{14}$ \cm
range with our current data.

Recent theoretical results also suggest that a $10^{14}$ \cm limit
might have a physical meaning. \citet{Dave2010} used a cosmological
hydrodynamical simulation to study the properties of \hi absorption
systems from $z=2$ to $z=0$. They found an interesting bimodality in
the distribution of $\log N_{\rm HI}$ per unit path length at $\langle
z \rangle \approx 0.25$, where $N_{\rm HI}<10^{14}$~\cm systems are
dominated by the diffuse \ac{igm} and $N_{\rm HI}>10^{14}$~\cm are
dominated by the condensed \ac{igm} associated with galaxy halos (see
their figure 10). This theoretical result is supported by our
observations of the similar clustering amplitudes of all $\xi_{\rm
  ag}$, $\xi_{\rm aa}$ and $\xi_{\rm gg}$ albeit at a somewhat larger
limit of $N_{\rm HI} \sim 10^{14.5}$ \cm (see \Cref{fig:logn_limits}).

According to the results from \citet{Dave2010}, the diffuse \ac{igm}
approximately follows,

\begin{equation}
\frac{\rho}{\bar{\rho}} \approx 50 \left( \frac{N_{\rm HI}}{10^{14} \,
  {\rm cm}^{-2}} \right)^{0.74} 10^{-0.37z} \rm{,}
\label{eq:dave}
\end{equation}

\noindent where $\rho/\bar{\rho}$ is the local baryonic density in
units of the cosmic mean (see their equation 3 and figure 9). This
gives us an idea of the overdensities involved \citep[see also][for a
  similar relationship from analytical arguments]{Schaye2001}. A change
of one order of magnitude in column density corresponds to a factor of
$\sim 5$ (directly proportional) in $\rho/\bar{\rho}$, whereas a change
of one unit redshift corresponds to a factor of $\sim 2$ (inversely
proportional) in $\rho/\bar{\rho}$. Thus, a limit of $10^{14}$~\cm
would correspond to overdensities of $\sim 50\times$ and $\sim
25\times$ the cosmic mean at $z=0$ and $z=1$, respectively. Similarly,
limits of $10^{13}$ and $10^{15}$~\cm would correspond to $\sim 5
\times$ more and less than those values, respectively. We emphasize
that there are large scatters involved in this relation: roughly one
order of magnitude in overdensity for a fixed \hi column density, and
roughly half an order of magnitude in \hi column density for a fixed
overdensity. Such a scatter would likely end up diluting any sharper
\nhi transition.

\begin{figure*}
  \centering
  \includegraphics[width=1.01\textwidth]{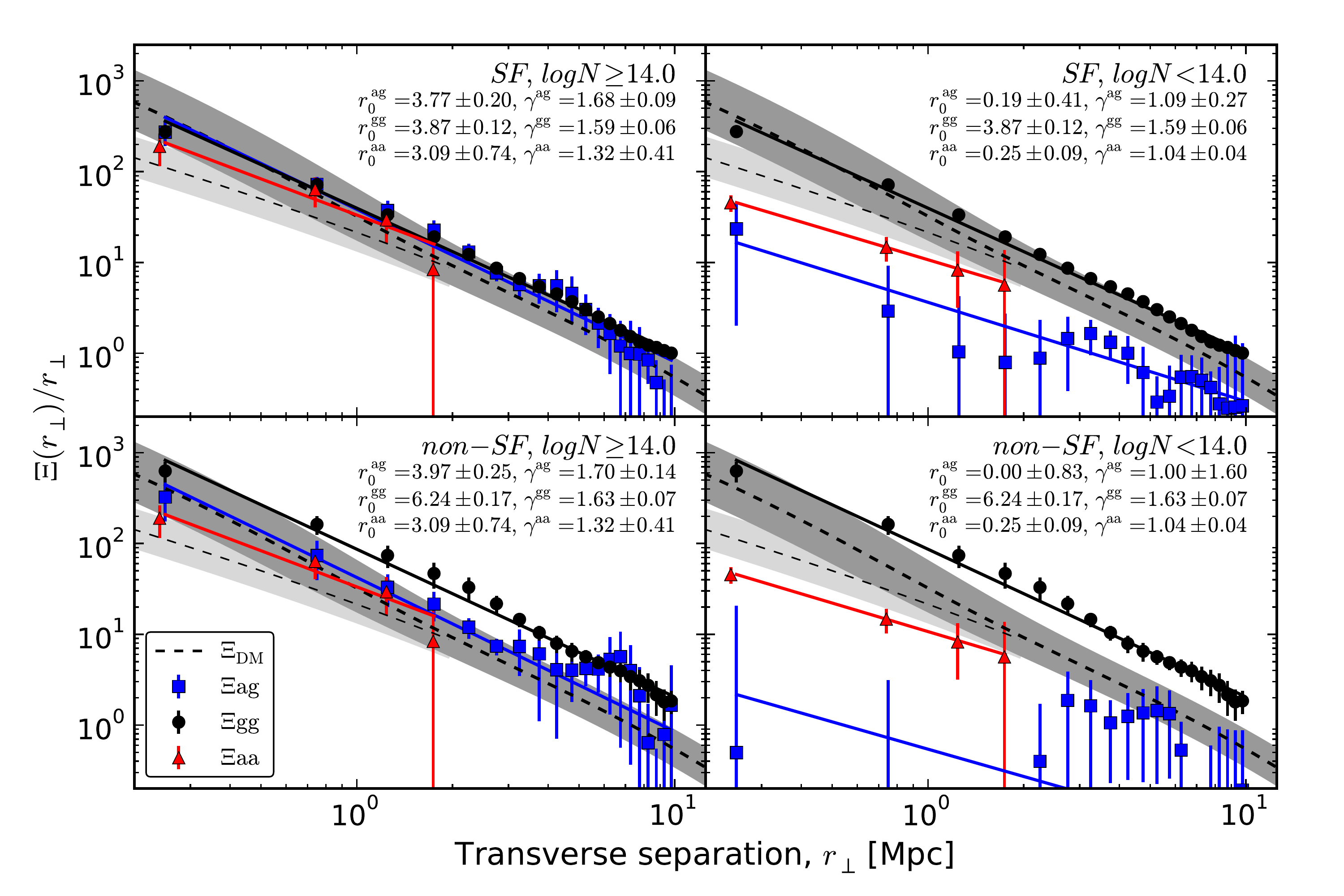}
  \caption{Same as Figure \ref{fig:xi_split_T} with the prediction for
    the dark matter clustering at $z \lesssim 1$ (dashed line). The
    shaded regions enclose the expected dark matter clustering between
    redshift $z=1$ (lower envelope) and $z=0$ (upper envelope) while
    the dashed lines themselves correspond to the expectation at
    $z=0.5$. These predictions were obtained from the dark matter power
    spectrum provided by \camb \citep{Lewis2000}, with (thick dashed
    lines and dark shaded regions) and without (thin dashed lines and
    light shaded regions) using the non-linear corrections of
    \citet{Smith2003}, for our adopted cosmological parameters and
    $\sigma_{8}=0.8$. See \Cref{discussion:interpretation:masses} for
    further details.}
\label{fig:xi_split_T3}

\end{figure*}

\subsubsection{Dark matter halo masses hosting \hi systems and galaxies}\label{discussion:interpretation:masses}

It is common practice to compare the observed clustering amplitudes of
extragalactic objects (e.g. galaxies, galaxy clusters, \ac{igm}
absorbers, etc.) with that of the expected theoretical (cold) dark
matter in a given cosmological framework, in order to infer a typical
dark matter halo mass for the confinement of such objects
\citep[e.g.][]{Mo1993,Ryan-Weber2006}. This method is model dependent,
and it is only applicable over narrow cosmological epochs (narrow
redshift ranges).

Our sample is composed of objects at $0\lesssim z \lesssim 1$, which
corresponds to about half of the history of the Universe. Thus, a
direct link between the clustering amplitudes reported in this paper
with a single dark matter halo mass is not meaningful. Still, simple
reasoning leads to the conclusion that the typical dark matter haloes
for the confinement of \hi systems and galaxies, should follow the same
trends as the amplitudes of their correlation functions. Therefore, the
most massive ones should correspond to \nSF galaxies, followed by \SF
galaxies, \strong \hi systems (both comparable) and \weak \hi systems,
in that same order.

\Cref{fig:xi_split_T3} is the same as \Cref{fig:xi_split_T} but
including the prediction for the dark matter clustering at $z \lesssim
1$ (thick dashed line). The shaded regions enclose the expected dark
matter clustering between redshift $z=1$ (lower envelope) and $z=0$
(upper envelope), while the dashed lines correspond to the expectation
at $z=0.5$. These predictions were obtained from the dark matter power
spectrum\footnote{Note that the power spectrum is the Fourier transform
  of the correlation function (and viceversa).}  provided by
\camb\footnote{\url{www.camb.info}} \citep{Lewis2000}, with (thick
dashed lines and dark shaded regions) and without (thin dashed lines
and light shaded regions) the non-linear corrections of
\citet{Smith2003}, for our adopted cosmological parameters and
$\sigma_{8}=0.8$. We see that the shape of the correlations for \strong
\hi systems and galaxies are approximately consistent with that of the
predicted dark matter in the non-linear regime. Their somewhat larger
amplitudes hint towards `absolute biases' $b\gtrsim1$. On the other
hand, the shape of the \weak \hi is marginally in disagreement with
that of the dark matter expectation in the non-linear regime. In this
case, the lower amplitude compared to that of the dark matter hints
towards an `absolute bias' $b<1$. We note that for the case of \weak
systems, a linear approximation for the dark matter clustering
\citep[i.e. neglecting the correction of][]{Smith2003}, gives a
somewhat better match in terms of slopes, although still with
amplitudes marginally above our observed ones. If a significant
fraction of \weak \hi systems reside in underdense regions (i.e. within
galaxy voids), a linear evolution should be expected even at
$z\approx0$. We speculate that \hi systems within galaxy voids are
still evolving in the linear regime, even at scales $\lesssim 2$ \mpc.

In view of these results, we revisit the claim by
\citet{Ryan-Weber2006} that \hi absorption systems with $\lesssim
10^{15}$ \cm reside preferentially in dark matter halo of masses $M\sim
10^{13.6}-10^{14.5}$ $h^{-1}_{100}$M$_{\sun}$, analogous to those of
massive galaxy groups. Given the significantly lower clustering
amplitude of our full sample of \hi systems compared to that of
galaxies, we conclude that \hi absorption systems are preferentially
found in dark matter haloes of masses smaller than those populated by
galaxies. At most, \SF galaxies and \strong \hi systems are typically
found in dark matter haloes of similar masses. Moreover, a significant
fraction of \weak \hi systems might reside in underdense regions with
`absolute biases' $b<1$.

\subsubsection{Three types of relationships between \hi and galaxies}

We have reported a significant ($>3\sigma$ c.l.) rejection of the
hypothesis that \hi absorption systems and galaxies (as a whole) trace
the same underlying dark matter distribution linearly (see
\Cref{results}).  We have found that this is mostly driven by \hi
absorption systems with column densities $N_{\rm HI}<10^{14}$ \cm
(\weak systems), which show little (consistent with $0$) correlation
with galaxies. On the other hand, \hi systems with $N_{\rm HI} \ge
10^{14}$ \cm (\strong systems) are consistent with such an
hypothesis. Thus, this indicates the presence of, at least, two types
of relationships between \hi and galaxies: (i) linear correlation, and
(ii) no correlation.

A third type of relationship comes from the fact that at small enough
scales, \hi systems and galaxies are a different manifestation of the
same physical object; a galaxy is also a very strong \hi absorption
system and, depending on the galaxy definition, the other way around
also applies. Our survey was not designed for studying scales $\lesssim
0.5$ \mpc, and so it is not surprising that we do not observe a
characteristic signal of a one-to-one association (see
\Cref{dependence}). Thus, we can not neglect the fact that this
relationship exists and should be included in our
interpretation. Still, the contribution of this one-to-one correlation
between \hi absorption systems and luminous galaxies to the total
fraction of \hi systems at $10^{13} \lesssim N_{\rm HI} \lesssim
10^{17}$ \cm is quite low.



This picture fits well with what was presented early by
\citet{Mo1994a}, and is in contrast to the commonly adopted
interpretation presented by \citet{Lanzetta1995} which claims that the
majority of low-$z$ \hi systems belong to the extended haloes of
luminous galaxies.

\subsection{Prospects and future work}\label{future}
In this section we will enumerate some of the projects that are
directly linked to our current study, but that we have not performed
here either because of lack of observational data or limited time. We
aim to address them in the near future.

\subsubsection{Comparison with simulations}

Even though many of our results are in good agreement with those
presented by \citet[][see
  \Cref{discussion:comparison:pierleoni}]{Pierleoni2008}, others have
not been properly compared with the predictions from simulations
yet. For instance, one of our key results is the fact that \weak \hi
systems and galaxies cluster more weakly than \weak \hi systems with
themselves, or than galaxies with themselves. As discussed in
\Cref{discussion:distribution}, this would imply that \weak \hi systems
and luminous galaxies do not trace the same underlying matter
distribution linearly. It is still to be seen if current cosmological
hydrodynamical simulations can reproduce this and {\it all} our
observational results.


\subsubsection{Cosmological evolution}

A complete picture of the relationship between the \ac{igm} and
galaxies requires understanding not only their statistical connection
at a given epoch, but also their cosmological evolution. Combining our
results with those from higher redshifts \citep[$z \sim 2-3$;
  e.g.][]{Adelberger2003,Adelberger2005,Crighton2011,Rudie2012,Rakic2012,Tummuangpak2013},
such an evolution can be studied. It is important to keep in mind that:
(i) galaxy samples in these high-$z$ studies are strongly biased
against `non-star-forming-galaxies', and (ii) the lower the redshift,
the higher the (average) overdensity traced by a fixed \nhi limit
\citep[e.g.][see \Cref{eq:dave}]{Schaye2001, Dave2010}. Thus, any
evolutionary analysis has to properly take into account such
differences.

We also note that because of observational limitations, the redshift
range between $z \sim 1$ and $2$ is currently unexplored for studies of
the \ac{igm}--galaxy connection. This is a very important cosmological
time, as it is when the star formation density starts to decline
\citep[e.g.][]{Hopkins2006}. We hope this will be covered in the near
future.

\subsubsection{Dependence on \hi Doppler parameter}

In our current analysis we have completely ignored the information
provided by the Doppler parameters of the \hi systems. Current
hydrodynamical simulations suggest that above a limit of $b_{\rm
  HI}\sim 50$ \kms, an important fraction of \hi lines trace the
\ac{whim} \citep[e.g.][see their figure 11]{Dave2010}. The \ac{whim} is
currently the best candidate to host the majority of the `missing
baryons' at low-$z$
\citep[e.g.][]{Cen1999,Dave2010,Tepper-Garcia2012,Shull2012}. However,
because of their expected large \bhi and low \nhi ($\lesssim 10^{13}$
\cm), its direct observation through \hi has been extremely
difficult. In fact, \hi can appear undetectable in such conditions
\citep{Savage2010}. Still, the \hi--galaxy cross-correlation could
provide an indirect way to observe the \ac{whim} by splitting the
samples by \bhi, and applying a similar reasoning as that presented in
\Cref{discussion:interpretation:nSF}.

\subsubsection{Cross-correlations for the \ac{cgm}}

Our current statistical results seem adequate for constraining the
\hi--galaxy connection on scales $\sim 0.5-10$ \mpc. An obvious
improvement would be to increase the galaxy completeness level at
scales $\lesssim 0.5$ \mpc. In this way the two-point correlation
function results can be directly linked to the studies of the \ac{cgm}
based on one-to-one absorber--galaxy associations
\citep[e.g.][]{Prochaska2011b,Tumlinson2011,Thom2012,Werk2013,Stocke2013}.
Correlations between metals and galaxies will also provide a useful
complement for such studies. Similarly, a better characterization of
the galaxies (e.g. stellar masses, specific star-formation rates,
morphology, etc.) in these samples will allow us to isolate their
relative contributions (and hence importance) to the observed
correlation amplitudes.

\subsubsection{`Extremely weak' \hi systems}

Our current data quality is not high enough to observe `extremely weak'
\hi systems ($N_{\rm HI} \lesssim 10^{13}$ \cm), but studying the
\hi--galaxy cross-correlation at such low column densities is clearly
worth exploring. There is strong observational evidence that the vast
majority of these absorbers reside within galaxy voids
\citep[e.g.][]{Manning2002,Tejos2012}. In such a case an {\it
  anti}-correlation between `extremely weak' \hi absorption systems and
galaxies should be expected, but this has not yet been observationally
confirmed (or refuted). There is also the interesting possibility that
these absorbers may represent a completely different type of \hi
absorption systems than those found co-existing with galaxies. If true,
such systems are good candidates for testing our current galaxy
formation paradigm \citep[e.g.][]{Manning2002,Manning2003}.

\section{Summary}\label{summary}

We have presented a new optical spectroscopic galaxy survey of $2143$
galaxies at $z\lesssim1$, around $3$ fields containing $5$ \acp{qso}
with \ac{hst} \ac{uv} spectroscopy.\footnote{Note that one of the
  fields has $3$ \acp{qso}.} These galaxies were observed by optical
\acl{mos} instruments such as \ac{deimos}, \ac{vimos} and \ac{gmos},
and were mostly selected based on magnitude limits ($R\sim 23-24$ mag;
no morphological criteria imposed). This selection also led to the
detection of $423$ stars and $22$ \ac{agn} within those fields. Out of
our new $2143$ galaxies, $1777$ have detectable star formation activity
within their past $\sim 1$ Gyr (referred to as \SF), while the remaining
$366$ have not (referred to as \nSF).

We have also presented a new spectroscopic survey of $669$ well
identified intervening \hi absorption line systems at $z \lesssim 1$,
observed in the spectra of $8$ \ac{qso} at $z \sim 1$. These systems
were detected in high-resolution \ac{uv} \ac{hst} spectroscopy from
\ac{cos} and \ac{fos}. Out of these $669$ \hi systems, $173$ have
column densities $10^{14} \le N_{\rm HI}\lesssim 10^{17}$ \cm (referred
to as \strong), while the remaining $496$ have $10^{13} \lesssim N_{\rm
  HI} < 10^{14}$ \cm (referred to as \weak).

Combining these new data with previously published galaxy catalogs from
the \ac{vvds} \citep{LeFevre2005,LeFevre2013}, \ac{gdds}
\citep{Abraham2004} and \citet{Morris2006} surveys, we have gathered a
sample of $17509$ galaxies with redshifts between $0.01<z<1.3$, and at
transverse separations $<50$ \mpc from their respective field centres;
and $654$ \hi~absorption systems at redshifts between $0.01<z<z_{\rm
  max}$, where $z_{\rm max}$ is the redshift corresponding to $5000$
\kms blueward of the redshift of their respective \acp{qso}. Out of
those $17509$ galaxies, $8293$ were classified as \SF and $1743$ as
\nSF; while out of those $654$ \hi systems, $165$ were classified as
\strong and $489$ as \weak.

Using these data, we have investigated the statistical connection
between the \acf{igm} and galaxies through a clustering analysis. This
dataset is the largest sample to date for such an analysis. We
presented observational results for the \hi--galaxy cross-correlation
and both the galaxy--galaxy and \hi--\hi auto-correlations at
$z\lesssim1$. The two-point correlation functions have been measured
both along and transverse to the \ac{los},
$\xi(r_{\perp},r_{\parallel})$, on a linear grid of $0.5$ \mpc in both
directions. We have measured the \hi--galaxy ($\xi_{\rm ag}$) and
galaxy--galaxy ($\xi_{\rm gg}$) correlations at transverse scales
$r_{\perp} \lesssim 10$ \mpc, and the \hi--\hi auto-correlation
($\xi_{\rm aa}$) at transverse scales $r_{\perp} \lesssim 2$ \mpc. We
have integrated these correlations along the \ac{los} up to $20$ \mpc,
and used the projected results to infer the shape of their
corresponding `real-space' correlation functions, $\xi(r)$, assuming
power-laws of the form $\xi(r)=(r/r_0)^{-\gamma}$. By comparing the
results from the \hi--galaxy cross-correlation with the \hi--\hi and
galaxy--galaxy auto-correlations, we have provided constraints on their
statistical connection, as a function of both \hi column density and
galaxy star-formation activity. We summarize our observational results
as follows:

\begin{itemize}
\item Two-dimensional correlations, $\xi(r_{\perp},r_{\parallel})$:
  \begin{itemize}
    
  \item {\it Full Sample:} the \hi--galaxy two-dimensional
    cross-correlation has comparable clustering amplitudes to those of
    the \hi--\hi auto-correlation, which are lower than those of the
    galaxy--galaxy auto-correlation. The peaks of these correlation
    functions were found to be $\xi_{\rm ag} = 2.3 \pm 0.9$, $\xi_{\rm
      aa} = 2.1 \pm 0.9$ and $\xi_{\rm gg} = 5.7 \pm 0.7$,
    respectively. \smallskip
 
  \item {\it `Strong' \hi systems and \SF galaxies:} the \hi--galaxy,
    \hi--\hi and galaxy--galaxy two-dimensional correlations all have
    comparable amplitudes. The peaks of these correlation functions
    were found to be $\xi_{\rm ag} = 8.3 \pm 2.2$, $\xi_{\rm aa} = 7.5
    \pm 2.3$ and $\xi_{\rm gg} = 6.1 \pm 0.6$, respectively. \smallskip

  \item {\it `Strong' \hi systems and \nSF galaxies:} the \hi--galaxy
    two-dimensional cross-correlation has comparable clustering
    amplitudes than those of the galaxy--galaxy auto-correlation, which
    are marginally higher than those of the \hi--\hi
    auto-correlation. The peaks of the correlation functions were found
    to be $\xi_{\rm ag} = 10.3 \pm 5.6$, $\xi_{\rm gg} = 12.6 \pm 3.0$
    and $\xi_{\rm aa} = 7.5 \pm 2.3$, respectively. \smallskip
     
  \item {\it `Weak' \hi systems and \SF galaxies:} the \hi--galaxy
    two-dimensional cross-correlation has much lower amplitudes than
    those of the galaxy--galaxy and \hi--\hi auto-correlations. The
    \hi-\hi auto-correlation has also lower amplitudes than those of
    the galaxy--galaxy auto-correlation. The peaks of the correlation
    functions were found to be $\xi_{\rm ag} = 0.9 \pm 0.6$, $\xi_{\rm
      gg} = 6.1 \pm 0.6$ and $\xi_{\rm aa} = 1.9 \pm 0.9$,
    respectively. \smallskip

  \item {\it `Weak' \hi systems and \nSF galaxies:} the \hi--galaxy
    two-dimensional cross-correlation has much lower amplitudes than
    those of the galaxy--galaxy and \hi--\hi auto-correlations. The
    \hi-\hi auto-correlation has also lower amplitudes than those of
    the galaxy--galaxy auto-correlation. The peaks of the correlation
    functions were found to be $\xi_{\rm ag} = 0.6 \pm 0.5$, $\xi_{\rm
      gg} = 12.6 \pm 3.0$ and $\xi_{\rm aa} = 1.9 \pm 0.9$,
    respectively.\smallskip
  \end{itemize}

\item Real space correlations, $\xi(r)\equiv(r/r_0)^{-\gamma}$:
  
  \begin{itemize}
  \item {\it Full Sample:} the \hi--galaxy cross-correlation has
    comparable clustering amplitudes than those of the \hi--\hi
    auto-correlation, which are lower than those of the galaxy--galaxy
    auto-correlation. The correlation lengths and slopes are found to
    be $r_0^{\rm ag} = 1.6 \pm 0.2$ \mpc and $\gamma^{\rm
      ag}=1.4\pm0.1$, $r_0^{\rm aa} = 0.3 \pm 0.3$ \mpc and
    $\gamma^{\rm aa}=1.1\pm0.1$, and $r_0^{\rm gg} = 3.9 \pm 0.1$ \mpc
    and $\gamma^{\rm gg}=1.7\pm0.1$, respectively.\smallskip
 
  \item {\it `Strong' \hi systems and \SF galaxies:} the \hi--galaxy,
    \hi--\hi and galaxy--galaxy correlations have all comparable
    amplitudes. The correlation lengths and slopes are found to be
    $r_0^{\rm ag} = 3.8 \pm 0.2$ \mpc and $\gamma^{\rm ag}=1.7\pm0.1$,
    $r_0^{\rm aa} = 3.1 \pm 0.7$ \mpc and $\gamma^{\rm aa}=1.3\pm0.4$,
    and $r_0^{\rm gg} = 3.9 \pm 0.1$ \mpc and $\gamma^{\rm
      gg}=1.6\pm0.1$, respectively.\smallskip

  \item {\it `Strong' \hi systems and \nSF galaxies:} the \hi--galaxy
    cross-correlation has comparable clustering amplitudes than those
    of the galaxy--galaxy auto-correlation, which are higher than those
    of the \hi--\hi auto-correlation.  The correlation lengths and
    slopes found to be $r_0^{\rm ag} = 4.0 \pm 0.3$ \mpc and
    $\gamma^{\rm ag}=1.7\pm0.1$, $r_0^{\rm gg} = 6.2 \pm 0.2$ \mpc and
    $\gamma^{\rm gg}=1.6\pm0.1$, and $r_0^{\rm aa} = 3.1 \pm 0.7$ \mpc
    and $\gamma^{\rm aa}=1.3\pm0.4$, respectively.\smallskip
     
  \item {\it `Weak' \hi systems and \SF galaxies:} the \hi--galaxy
    cross-correlation has much lower amplitudes than those of the
    galaxy--galaxy and \hi--\hi auto-correlations. The \hi-\hi
    auto-correlation has also lower amplitudes than those of the
    galaxy--galaxy auto-correlation. The correlation lengths and slopes
    are found to be $r_0^{\rm ag} = 0.2 \pm 0.4$ \mpc and $\gamma^{\rm
      ag}=1.1\pm0.3$, $r_0^{\rm gg} = 3.9 \pm 0.1$ \mpc and
    $\gamma^{\rm gg}=1.6\pm0.1$, and $r_0^{\rm aa} = 0.3 \pm 0.1$ \mpc
    and $\gamma^{\rm aa}=1.0\pm0.1$, respectively. We note however that
    a power-law fit for \hi--galaxy cross-correlation might not be a
    good description of the observations.\smallskip
  
  \item {\it `Weak' \hi systems and \SF galaxies:} the \hi--galaxy
    cross-correlation has much lower amplitudes than those of the
    galaxy--galaxy and \hi--\hi auto-correlations. The \hi-\hi
    auto-correlation has also lower amplitudes than those of the
    galaxy--galaxy auto-correlation. The correlation lengths and slopes
    are found to be $r_0^{\rm ag} = 0.0 \pm 0.8$ \mpc and $\gamma^{\rm
      ag}=1.0\pm1.6$, $r_0^{\rm gg} = 6.2 \pm 0.2$ \mpc and
    $\gamma^{\rm gg}=1.6\pm0.1$, and $r_0^{\rm aa} = 0.3 \pm 0.1$ \mpc
    and $\gamma^{\rm aa}=1.0\pm0.1$, respectively. We note, however,
    that a power-law fit for the real-space \hi--galaxy
    cross-correlation might not be a good description of the
    observations.\smallskip

\end{itemize}

\item Amplitudes:
  \begin{itemize}
  \item {\it \hi--galaxy cross-correlations:} The \hi--galaxy
    cross-correlation amplitudes are systematically higher for \strong
    systems than for \weak systems, and are also higher for \nSF
    galaxies than for \SF galaxies, with a much stronger dependence on
    \hi column density than galaxy star-formation activity. This is
    true for both the two-dimensional and the real-space correlations
    (see numbers above).\smallskip
    
  \item {\it Galaxy auto-correlations:} The galaxy--galaxy
    auto-correlation amplitudes are systematically higher for \nSF
    galaxies than for \SF galaxies. This is true for both the
    two-dimensional and the real-space correlations (see numbers
    above).\smallskip

  \item {\it \hi auto-correlations:} The \hi--\hi auto-correlation
    amplitudes are systematically higher for \strong systems than for
    \weak systems. This is true for both the two-dimensional and
    real-space correlations (see numbers above).\smallskip
  \end{itemize}
  
\item Velocity distortions:
  \begin{itemize}
  
  \item The two-dimensional \hi--galaxy cross-correlations do not show
    significant velocity distortions along the \ac{los}, apart from
    those expected by the galaxy redshift uncertainties. \smallskip

  \item The peak in the two-dimensional \hi--galaxy cross-correlation
    for \strong systems and \nSF galaxies appears shifted by $\sim 1$
    \mpc along the \ac{los} from $0$, and there is marginal evidence
    (not significant) that this might be a real feature.\smallskip

  \end{itemize}
  
\item Two-dimensional ratios, $(\xi_{\rm ag})^2/(\xi_{\rm gg}\xi_{\rm
  aa})$ on scales $< 2$ \mpc:
  \begin{itemize}
  \item {\it Full Sample:} the ratio $(\xi_{\rm ag})^2/(\xi_{\rm
    gg}\xi_{\rm aa})$ appears marginally inconsistent with
    $1$.\smallskip

  \item {\it `Strong' \hi systems and galaxies:} the ratio $(\xi_{\rm
    ag})^2/(\xi_{\rm gg}\xi_{\rm aa})$ appears roughly consistent
    (large uncertainties) with $1$, irrespective of the galaxy
    star-formation activity.\smallskip
    
  \item {\it `Weak' \hi systems and galaxies:} the ratio $(\xi_{\rm
    ag})^2/(\xi_{\rm gg}\xi_{\rm aa})$ appears inconsistent with $1$,
    irrespective of the galaxy star-formation activity.\smallskip

  \end{itemize}

\item Projected along the \ac{los} ratios, $(\Xi_{\rm ag})^2/(\Xi_{\rm
  gg}\Xi_{\rm aa})$ on scales $< 2$ \mpc:
  \begin{itemize}
    
  \item {\it Full Sample:} we find $(\Xi_{\rm ag})^2/(\Xi_{\rm
    gg}\Xi_{\rm aa}) \approx 0.2\pm 0.2$. This rules out the hypothesis
    that \hi systems and galaxies (as a whole) trace the same
    underlying dark matter distribution linearly, at a high statistical
    significance ($>3\sigma$ c.l.).\smallskip
    
  \item {\it `Strong' \hi systems and \SF galaxies:} we find $(\Xi_{\rm
    ag})^2/(\Xi_{\rm gg}\Xi_{\rm aa}) \approx 1.1\pm 0.6$. This is
    consistent (large uncertainties) with the hypothesis that \strong
    \hi systems and \SF galaxies trace the same underlying dark matter
    distribution linearly.\smallskip

  \item {\it `Strong' \hi systems and \nSF galaxies:} we find
    $(\Xi_{\rm ag})^2/(\Xi_{\rm gg}\Xi_{\rm aa}) \approx 0.5 \pm
    0.3$. This marginally rules out the hypothesis that \strong \hi
    systems and \nSF galaxies trace the same underlying dark matter
    distribution linearly (only at the $\sim 2\sigma$ c.l.).\smallskip

  \item {\it `Weak' \hi systems and \SF galaxies:} we find $(\Xi_{\rm
    ag})^2/(\Xi_{\rm gg}\Xi_{\rm aa}) \approx 0.0\pm 0.2$. This rules
    out the hypothesis that \weak \hi systems and \SF galaxies trace
    the same underlying dark matter distribution linearly, at a high
    statistical significance ($>3\sigma$ c.l.).\smallskip

  \item {\it `Weak' \hi systems and \nSF galaxies:} we find $(\Xi_{\rm
    ag})^2/(\Xi_{\rm gg}\Xi_{\rm aa}) \approx 0.0\pm 0.4$. This
    marginally rules out the hypothesis that \weak \hi systems and \nSF
    galaxies trace the same underlying dark matter distribution
    linearly (only at the $\sim 2\sigma$ c.l.).\smallskip
  \end{itemize}
  
\item `Absolute biases':
  \begin{itemize}
    \item {\it `Strong' \hi systems and galaxies:} their `absolute
      biases' are consistent with $b\gtrsim 1$.\smallskip

    \item {\it `Weak' \hi systems:} their `absolute biases' are
      consistent with $b<1$.\\

  \end{itemize}
\end{itemize}

Our interpretation of these results has led us to the following
conclusions:

\begin{itemize}

\item The bulk of \hi systems on $\sim$~Mpc scales have little velocity
  dispersion ($\lesssim 120$ \kms) with respect to the bulk of
  galaxies. Hence, no strong galaxy outflow or inflow signal is
  detected in our data. \smallskip

\item The vast majority ($\sim 100\%$) of \strong \hi systems and \SF
  galaxies are distributed in the same locations. We have identified
  these locations with the `overdense large-scale structure'.\smallskip

\item A fraction of \nSF galaxies are distributed in roughly the same
  way as \strong \hi systems and \SF galaxies but there are
  sub-locations---within those where galaxies and \strong \hi systems
  reside---with a much higher density of \nSF galaxies than \strong \hi
  systems and/or \SF galaxies. We have identified such locations as
  galaxy clusters. We estimated that only a $25 \pm 15 \%$ of \nSF
  galaxies reside in galaxy clusters and that the remaining $75 \pm 15
  \%$ co-exist with \strong \hi and \SF at scales $\lesssim 2$~\mpc,
  following the same underlying dark matter distribution, i.e. the
  `overdense large-scale structure'.\smallskip

\item An important fraction of \weak systems could reside in locations
  devoid of galaxies of any kind. We have identified such locations as
  galaxy voids, i.e. the `underdense large-scale structure'. At a limit
  of $N_{\rm HI} \ge 10^{13}$~\cm, we have estimated that roughly $\sim
  50\%$ of \weak systems reside within galaxy voids. At lower \nhi
  limits this fraction is likely to increase.\smallskip

\item The vast majority ($\sim 100\%$) of \strong \hi absorption
  systems at low-$z$ reside in dark matter haloes of masses comparable
  to those hosting the galaxies in our sample.\smallskip

\item At least $\sim 50\%$ of \weak \hi absorption systems with $N_{\rm
  HI} \ge 10^{13}$~\cm reside in dark matter haloes less massive than
  those hosting \strong \hi systems and/or the galaxies in our
  sample. At lower \nhi limits this fraction is likely to
  increase.\smallskip

\item We speculate that \hi systems within galaxy voids at $z\lesssim
  1$ might be still evolving in the linear regime even at scales
  $\lesssim 2$~\mpc.\smallskip

\item We conclude that there are {\it at least} three types of
  relationship between \hi absorption systems and galaxies at low-$z$:
  (i) one-to-one physical association; (ii) association because they
  both follow the same underlying dark matter distribution; and (iii)
  no association at all.
\end{itemize}

\section*{Acknowledgments}

We thank the anonymous referee for helpful comments and suggestions
which improved the paper. We thank Pablo Arnalte-Mur, Rich Bielby, John
Lucey, Peder Norberg and Tom Shanks for helpful discussions. We thank
Pablo Arnalte-Mur for his help in obtaining the prediction for the dark
matter clustering at $z\lesssim1$. We thank Mark Swinbank for his help
in the reduction of \ac{gmos} data.

We thank contributors to SciPy\footnote{\url{http://www.scipy.org}},
Matplotlib\footnote{\url{http://www.matplotlib.sourceforge.net}}, and
the Python programming language\footnote{\url{http://www.python.org}};
the free and open-source community; and the NASA Astrophysics Data
System\footnote{\url{http://adswww.harvard.edu}} for software and
services.

N.T. acknowledges grant support by CONICYT, Chile (PFCHA/{\it Doctorado
  al Extranjero 1$^{\rm a}$ Convocatoria}, 72090883). S.L.M. was
partially supported by STFC Rolling Grant PP/C501568/1 Extragalactic
Astronomy and Cosmology at Durham 2008–2013. J.B. acknowledges support
from HST-GO-11585.03-A and HST-GO-12264.03-A
grants. E.R.W. acknowledges support by Australian Research Council
DP1095600.

This work was mainly based on observations made with the NASA/ESA
Hubble Space Telescope under programs GO 12264 and GO 11585, obtained
at the Space Telescope Science Institute, which is operated by the
Association of Universities for Research in Astronomy, Inc., under NASA
contract NAS 5-26555; and on observations collected at the European
Southern Observatory, Chile, under programs 070.A-9007, 086.A-0970 and
087.A-0857. Some of the data presented herein were obtained at the
W.M. Keck Observatory, which is operated as a scientific partnership
among the California Institute of Technology, the University of
California and the NASA. The authors wish also to recognize and
acknowledge the very significant cultural role and reverence that the
summit of Mauna Kea has always had within the indigenous Hawaiian
community. This work was partially based on observations obtained at
the Gemini Observatory, which is operated by the Association of
Universities for Research in Astronomy, Inc., under a cooperative
agreement with the NSF on behalf of the Gemini partnership: the NSF
(United States), the National Research Council (Canada), CONICYT
(Chile), the Australian Research Council (Australia), Minist\'{e}rio da
Ci\^{e}ncia, Tecnologia e Inova\c{c}\~{a}o (Brazil) and Ministerio de
Ciencia, Tecnolog\'{i}a e Innovaci\'{o}n Productiva (Argentina).

\bibliographystyle{mn2e_internet.bst}
\bibliography{/home/ntejos/bib/IGM}

\clearpage
\appendix

\section[]{Data tables}\label{tables}


\begin{table*}
\centering
\begin{minipage}{0.98\textwidth}
\centering
\caption{\hi~absorption systems in QSO Q0107-025A.}\label{tab:HI_Q0107A}
\begin{tabular}{@{}cccc|cccc@{}}

\hline                            

$z$ & $\log{(N_{\rm HI}/{\rm cm}^{-2})}$ & $b_{\rm HI}$ (\kms) & label& 
$z$ & $\log{(N_{\rm HI}/{\rm cm}^{-2})}$ & $b_{\rm HI}$ (\kms) & label \\ 

(1)&(2)&(3)&(4)&(5)&(6)&(7)&(8) \\ 

\hline
0.003736 $\pm$ 0.000023 & 13.46 $\pm$ 0.09 & 30 $\pm$ 11 & b & 0.175444 $\pm$ 0.000016 & 13.73 $\pm$ 0.03 & 67 $\pm$ 7 & a \\ 
0.005063 $\pm$ 0.000030 & 13.34 $\pm$ 0.10 & 37 $\pm$ 13 & b & 0.177473 $\pm$ 0.000068 & 14.09 $\pm$ 0.34 & 30 $\pm$ 10 & a \\ 
0.006143 $\pm$ 0.002014 & 14.01 $\pm$ 9.45 & 47 $\pm$ 219 & c & 0.177697 $\pm$ 0.000035 & 14.63 $\pm$ 0.12 & 34 $\pm$ 5 & a \\ 
0.006291 $\pm$ 0.000409 & 15.14 $\pm$ 6.02 & 31 $\pm$ 205 & c & 0.181930 $\pm$ 0.000025 & 13.46 $\pm$ 0.05 & 58 $\pm$ 9 & a \\ 
0.006427 $\pm$ 0.007203 & 13.65 $\pm$ 55.41 & 34 $\pm$ 650 & c & 0.188633 $\pm$ 0.000021 & 12.90 $\pm$ 0.10 & 20 $\pm$ 9 & a \\ 
0.028710 $\pm$ 0.000009 & 13.49 $\pm$ 0.06 & 19 $\pm$ 4 & b & 0.188989 $\pm$ 0.000027 & 12.94 $\pm$ 0.10 & 30 $\pm$ 12 & a \\
\hline

    \end{tabular}
    
  \end{minipage}
\begin{minipage}{0.98\textwidth}
(1) and (5): \hi~redshift. 
(2) and (6): \hi~column density from Voigt profile fitting. 
(3) and (7): \hi~Doppler parameter from Voigt profile fitting.
(4) and (8): Confidence label: (a) `secure'; (b) `probable'; and (c) `uncertain' (see \Cref{abs:reliability} for definitions). 
See \Cref{abs:samples} for further details.
\end{minipage}
\end{table*}

\begin{table*}
\centering
\begin{minipage}{0.98\textwidth}
\centering
\contcaption{}
\begin{tabular}{@{}cccc|cccc@{}}

\hline                            
$z$ & $\log{(N_{\rm HI}/{\rm cm}^{-2})}$ & $b_{\rm HI}$ (\kms) & label& 
$z$ & $\log{(N_{\rm HI}/{\rm cm}^{-2})}$ & $b_{\rm HI}$ (\kms) & label \\ 

(1)&(2)&(3)&(4)&(5)&(6)&(7)&(8) \\ 

\hline
0.536282 $\pm$ 0.000237 & 15.17 $\pm$ 0.18 & 68 $\pm$ 24 & a& 0.728135 $\pm$ 0.000022 & 14.38 $\pm$ 0.09 & 23 $\pm$ 3 & a\\ 
0.536483 $\pm$ 0.000011 & 15.61 $\pm$ 0.08 & 34 $\pm$ 3 & a& 0.728846 $\pm$ 0.000028 & 14.41 $\pm$ 0.07 & 37 $\pm$ 4 & a\\ 
0.557532 $\pm$ 0.000015 & 14.81 $\pm$ 0.04 & 38 $\pm$ 2 & a& 0.786457 $\pm$ 0.000032 & 14.28 $\pm$ 0.02 & 110 $\pm$ 10 & a\\ 
0.579023 $\pm$ 0.000031 & 13.75 $\pm$ 0.26 & 6 $\pm$ 3 & a& 0.847815 $\pm$ 0.000042 & 13.85 $\pm$ 0.19 & 37 $\pm$ 30 & a\\ 
0.580353 $\pm$ 0.000043 & 14.00 $\pm$ 0.08 & 37 $\pm$ 8 & a& 0.876337 $\pm$ 0.000021 & 15.93 $\pm$ 0.10 & 19 $\pm$ 3 & a\\ 
0.640255 $\pm$ 0.000026 & 14.24 $\pm$ 0.08 & 28 $\pm$ 3 & a& 0.876569 $\pm$ 0.000125 & 15.30 $\pm$ 0.32 & 34 $\pm$ 8 & a\\ 
0.689857 $\pm$ 0.000029 & 13.97 $\pm$ 1.26 & 23 $\pm$ 43 & c& 0.889633 $\pm$ 0.000045 & 14.53 $\pm$ 57.89 & 13 $\pm$ 298 & c\\ 
0.718065 $\pm$ 0.000048 & 14.06 $\pm$ 0.10 & 33 $\pm$ 8 & a& 0.899105 $\pm$ 0.000156 & 14.20 $\pm$ 43.79 & 17 $\pm$ 512 & c\\ 
0.718936 $\pm$ 0.000006 & 15.63 $\pm$ 0.05 & 24 $\pm$ 1 & a& 0.906513 $\pm$ 0.000047 & 14.72 $\pm$ 2.55 & 44 $\pm$ 67 & c\\ 
0.719266 $\pm$ 0.000280 & 13.90 $\pm$ 0.23 & 105 $\pm$ 43 & a& 0.926727 $\pm$ 0.000056 & 14.32 $\pm$ 0.02 & 226 $\pm$ 15 & a\\ 

\hline

    \end{tabular}
    
  \end{minipage}
\begin{minipage}{0.98\textwidth}
(1) and (5): \hi~redshift. 
(2) and (6): \hi~column density from Voigt profile fitting. 
(3) and (7): \hi~Doppler parameter from Voigt profile fitting.
(4) and (8): Confidence label: (a) `secure'; (b) `probable'; and (c) `uncertain' (see \Cref{abs:reliability} for definitions). 
See \Cref{abs:samples} for further details.
\end{minipage}
\end{table*}

\begin{table*}
\centering
\begin{minipage}{0.98\textwidth}
\centering
\caption{\hi~absorption systems in QSO Q0107-025B.}\label{tab:HI_Q0107B}
\begin{tabular}{@{}cccc|cccc@{}}

\hline                            

$z$ & $\log{(N_{\rm HI}/{\rm cm}^{-2})}$ & $b_{\rm HI}$ (\kms) & label& 
$z$ & $\log{(N_{\rm HI}/{\rm cm}^{-2})}$ & $b_{\rm HI}$ (\kms) & label \\ 

(1)&(2)&(3)&(4)&(5)&(6)&(7)&(8) \\ 

\hline
0.006340 $\pm$ 0.000004 & 14.30 $\pm$ 0.08 & 31 $\pm$ 3 & b & 0.436121 $\pm$ 0.000024 & 13.74 $\pm$ 0.04 & 62 $\pm$ 7 & a \\ 
0.013367 $\pm$ 0.000010 & 13.35 $\pm$ 0.05 & 22 $\pm$ 5 & b & 0.467456 $\pm$ 0.000013 & 14.04 $\pm$ 0.03 & 44 $\pm$ 3 & a \\ 
0.041753 $\pm$ 0.000012 & 13.51 $\pm$ 0.04 & 42 $\pm$ 6 & b & 0.499367 $\pm$ 0.000009 & 15.00 $\pm$ 0.08 & 30 $\pm$ 3 & a \\ 
0.053188 $\pm$ 0.000011 & 13.64 $\pm$ 0.04 & 33 $\pm$ 5 & b & 0.499540 $\pm$ 0.000069 & 14.72 $\pm$ 0.17 & 57 $\pm$ 7 & a \\ 
0.060677 $\pm$ 0.000013 & 13.32 $\pm$ 0.06 & 28 $\pm$ 6 & b & 0.512259 $\pm$ 0.000097 & 13.64 $\pm$ 0.54 & 22 $\pm$ 58 & c \\ 
0.063119 $\pm$ 0.000008 & 13.47 $\pm$ 0.04 & 24 $\pm$ 3 & b & 0.517071 $\pm$ 0.000011 & 14.65 $\pm$ 0.02 & 44 $\pm$ 3 & a \\
\hline

    \end{tabular}
    
  \end{minipage}
\begin{minipage}{0.98\textwidth}
(1) and (5): \hi~redshift. 
(2) and (6): \hi~column density from Voigt profile fitting. 
(3) and (7): \hi~Doppler parameter from Voigt profile fitting.
(4) and (8): Confidence label: (a) `secure'; (b) `probable'; and (c) `uncertain' (see \Cref{abs:reliability} for definitions). 
See \Cref{abs:samples} for further details.
\end{minipage}
\end{table*}

\begin{table*}
\centering
\begin{minipage}{0.98\textwidth}
\centering
\caption{\hi~absorption systems in QSO Q0107-0232 .}\label{tab:HI_Q0107C}
\begin{tabular}{@{}cccc|cccc@{}}

\hline                            

$z$ & $\log{(N_{\rm HI}/{\rm cm}^{-2})}$ & $b_{\rm HI}$ (\kms) & label& 
$z$ & $\log{(N_{\rm HI}/{\rm cm}^{-2})}$ & $b_{\rm HI}$ (\kms) & label \\ 

(1)&(2)&(3)&(4)&(5)&(6)&(7)&(8) \\ 

\hline
0.198946 $\pm$ 0.000014 & 14.73 $\pm$ 0.11 & 52 $\pm$ 5 & b & 0.425130 $\pm$ 0.000014 & 13.93 $\pm$ 0.04 & 37 $\pm$ 5 & a \\ 
0.203349 $\pm$ 0.000015 & 14.79 $\pm$ 0.15 & 42 $\pm$ 6 & b & 0.428308 $\pm$ 0.000005 & 16.01 $\pm$ 0.18 & 35 $\pm$ 2 & a \\ 
0.227096 $\pm$ 0.000001 & 14.00 $\pm$ 0.01 & 50 $\pm$ 1 & b & 0.436369 $\pm$ 0.000014 & 13.76 $\pm$ 0.06 & 23 $\pm$ 10 & a \\ 
0.244724 $\pm$ 0.000001 & 14.00 $\pm$ 0.01 & 50 $\pm$ 1 & b & 0.441843 $\pm$ 0.000056 & 14.26 $\pm$ 0.03 & 208 $\pm$ 17 & a \\ 
0.261403 $\pm$ 0.000017 & 14.02 $\pm$ 0.05 & 43 $\pm$ 6 & b & 0.444586 $\pm$ 0.000040 & 13.70 $\pm$ 0.05 & 80 $\pm$ 13 & a \\ 
0.266532 $\pm$ 0.000003 & 14.53 $\pm$ 0.07 & 49 $\pm$ 4 & b & 0.487116 $\pm$ 0.000219 & 14.08 $\pm$ 0.06 & 384 $\pm$ 76 & a \\
\hline

    \end{tabular}
    
  \end{minipage}
\begin{minipage}{0.98\textwidth}
(1) and (5): \hi~redshift. 
(2) and (6): \hi~column density from Voigt profile fitting. 
(3) and (7): \hi~Doppler parameter from Voigt profile fitting.
(4) and (8): Confidence label: (a) `secure'; (b) `probable'; and (c) `uncertain' (see \Cref{abs:reliability} for definitions). 
See \Cref{abs:samples} for further details.
\end{minipage}
\end{table*}

\begin{table*}
\centering
\begin{minipage}{0.98\textwidth}
\centering
\caption{\hi~absorption systems in QSO J020930.7-043826.}\label{tab:HI_J0209}
\begin{tabular}{@{}cccc|cccc@{}}

\hline                            

$z$ & $\log{(N_{\rm HI}/{\rm cm}^{-2})}$ & $b_{\rm HI}$ (\kms) & label& 
$z$ & $\log{(N_{\rm HI}/{\rm cm}^{-2})}$ & $b_{\rm HI}$ (\kms) & label \\ 

(1)&(2)&(3)&(4)&(5)&(6)&(7)&(8) \\ 

\hline
0.060451 $\pm$ 0.000004 & 16.08 $\pm$ 0.79 & 41 $\pm$ 8 & c & 0.216171 $\pm$ 0.000012 & 14.23 $\pm$ 0.09 & 26 $\pm$ 4 & b \\ 
0.061037 $\pm$ 0.000006 & 13.96 $\pm$ 0.05 & 26 $\pm$ 3 & b & 0.216436 $\pm$ 0.000101 & 13.26 $\pm$ 0.31 & 39 $\pm$ 26 & b \\ 
0.062140 $\pm$ 0.000008 & 13.45 $\pm$ 0.06 & 16 $\pm$ 4 & b & 0.226416 $\pm$ 0.000012 & 13.08 $\pm$ 0.06 & 15 $\pm$ 5 & b \\ 
0.066439 $\pm$ 0.000014 & 13.96 $\pm$ 0.14 & 21 $\pm$ 8 & b & 0.226677 $\pm$ 0.000021 & 12.88 $\pm$ 0.10 & 20 $\pm$ 9 & b \\ 
0.066676 $\pm$ 0.000016 & 14.07 $\pm$ 0.09 & 27 $\pm$ 6 & b & 0.227182 $\pm$ 0.000015 & 13.08 $\pm$ 0.06 & 23 $\pm$ 6 & b \\ 
0.068032 $\pm$ 0.000026 & 13.39 $\pm$ 0.17 & 16 $\pm$ 9 & b & 0.227719 $\pm$ 0.000018 & 13.11 $\pm$ 0.06 & 30 $\pm$ 7 & b \\
\hline

    \end{tabular}
    
  \end{minipage}
\begin{minipage}{0.98\textwidth}
(1) and (5): \hi~redshift. 
(2) and (6): \hi~column density from Voigt profile fitting. 
(3) and (7): \hi~Doppler parameter from Voigt profile fitting.
(4) and (8): Confidence label: (a) `secure'; (b) `probable'; and (c) `uncertain' (see \Cref{abs:reliability} for definitions). 
See \Cref{abs:samples} for further details.
\end{minipage}
\end{table*}

\begin{table*}
\centering
\begin{minipage}{0.98\textwidth}
\centering
\contcaption{}
\begin{tabular}{@{}cccc|cccc@{}}

\hline                            
$z$ & $\log{(N_{\rm HI}/{\rm cm}^{-2})}$ & $b_{\rm HI}$ (\kms) & label& 
$z$ & $\log{(N_{\rm HI}/{\rm cm}^{-2})}$ & $b_{\rm HI}$ (\kms) & label \\ 

(1)&(2)&(3)&(4)&(5)&(6)&(7)&(8) \\ 

\hline
0.337387 $\pm$ 0.000102 & 12.12 $\pm$ 0.67 & 8 $\pm$ 58 & c& 0.459695 $\pm$ 0.000013 & 13.49 $\pm$ 0.64 & 7 $\pm$ 7 & c\\ 
0.338262 $\pm$ 0.000021 & 14.06 $\pm$ 0.02 & 98 $\pm$ 7 & a& 0.522512 $\pm$ 0.000026 & 13.78 $\pm$ 0.11 & 19 $\pm$ 9 & a\\ 
0.340036 $\pm$ 0.000014 & 14.00 $\pm$ 0.13 & 11 $\pm$ 3 & a& 0.522971 $\pm$ 0.000010 & 14.67 $\pm$ 0.02 & 46 $\pm$ 3 & a\\ 
0.340189 $\pm$ 0.000010 & 16.14 $\pm$ 2.98 & 4 $\pm$ 3 & c& 0.534301 $\pm$ 0.000012 & 13.88 $\pm$ 0.09 & 9 $\pm$ 5 & a\\ 
0.340357 $\pm$ 0.000007 & 15.72 $\pm$ 4.71 & 5 $\pm$ 6 & c& 0.569362 $\pm$ 0.000013 & 14.36 $\pm$ 0.05 & 18 $\pm$ 4 & a\\ 
0.346358 $\pm$ 0.000017 & 13.10 $\pm$ 0.08 & 19 $\pm$ 7 & a& 0.587029 $\pm$ 0.000018 & 14.49 $\pm$ 0.04 & 38 $\pm$ 5 & a\\
\hline

    \end{tabular}
    
  \end{minipage}
\begin{minipage}{0.98\textwidth}
(1) and (5): \hi~redshift. 
(2) and (6): \hi~column density from Voigt profile fitting. 
(3) and (7): \hi~Doppler parameter from Voigt profile fitting.
(4) and (8): Confidence label: (a) `secure'; (b) `probable'; and (c) `uncertain' (see \Cref{abs:reliability} for definitions). 
See \Cref{abs:samples} for further details.
\end{minipage}
\end{table*}

\begin{table*}
\centering
\begin{minipage}{0.98\textwidth}
\centering
\caption{\hi~absorption systems in QSO J100535.24+013445.7.}\label{tab:HI_J1005}
\begin{tabular}{@{}cccc|cccc@{}}

\hline                            

$z$ & $\log{(N_{\rm HI}/{\rm cm}^{-2})}$ & $b_{\rm HI}$ (\kms) & label& 
$z$ & $\log{(N_{\rm HI}/{\rm cm}^{-2})}$ & $b_{\rm HI}$ (\kms) & label \\ 

(1)&(2)&(3)&(4)&(5)&(6)&(7)&(8) \\ 

\hline
0.004126 $\pm$ 0.001790 & 13.87 $\pm$ 6.19 & 63 $\pm$ 209 & c & 0.185639 $\pm$ 0.000013 & 13.35 $\pm$ 0.06 & 24 $\pm$ 6 & a \\ 
0.004283 $\pm$ 0.000106 & 14.55 $\pm$ 1.30 & 46 $\pm$ 30 & c & 0.185822 $\pm$ 0.000021 & 12.77 $\pm$ 0.20 & 10 $\pm$ 12 & a \\ 
0.005998 $\pm$ 0.000037 & 13.01 $\pm$ 0.14 & 31 $\pm$ 17 & b & 0.185980 $\pm$ 0.000023 & 12.60 $\pm$ 0.16 & 10 $\pm$ 13 & a \\ 
0.017673 $\pm$ 0.000020 & 13.21 $\pm$ 0.09 & 25 $\pm$ 10 & b & 0.186191 $\pm$ 0.000021 & 12.56 $\pm$ 0.15 & 10 $\pm$ 11 & a \\ 
0.018720 $\pm$ 0.000008 & 13.53 $\pm$ 0.45 & 8 $\pm$ 6 & b & 0.186444 $\pm$ 0.000021 & 12.55 $\pm$ 0.15 & 10 $\pm$ 11 & a \\ 
0.020953 $\pm$ 0.000021 & 13.23 $\pm$ 0.08 & 29 $\pm$ 10 & b & 0.187977 $\pm$ 0.000011 & 13.21 $\pm$ 0.04 & 26 $\pm$ 4 & a \\
\hline

    \end{tabular}
    
  \end{minipage}
\begin{minipage}{0.98\textwidth}
(1) and (5): \hi~redshift. 
(2) and (6): \hi~column density from Voigt profile fitting. 
(3) and (7): \hi~Doppler parameter from Voigt profile fitting.
(4) and (8): Confidence label: (a) `secure'; (b) `probable'; and (c) `uncertain' (see \Cref{abs:reliability} for definitions). 
See \Cref{abs:samples} for further details.
\end{minipage}
\end{table*}

\begin{table*}
\centering
\begin{minipage}{0.98\textwidth}
\centering
\contcaption{}
\begin{tabular}{@{}cccc|cccc@{}}

\hline                            
$z$ & $\log{(N_{\rm HI}/{\rm cm}^{-2})}$ & $b_{\rm HI}$ (\kms) & label& 
$z$ & $\log{(N_{\rm HI}/{\rm cm}^{-2})}$ & $b_{\rm HI}$ (\kms) & label \\ 

(1)&(2)&(3)&(4)&(5)&(6)&(7)&(8) \\ 

\hline
0.352647 $\pm$ 0.000024 & 13.11 $\pm$ 0.08 & 28 $\pm$ 9 & a& 0.426592 $\pm$ 0.000014 & 13.15 $\pm$ 0.07 & 14 $\pm$ 6 & a\\ 
0.354692 $\pm$ 0.000022 & 13.37 $\pm$ 0.05 & 42 $\pm$ 7 & a& 0.426800 $\pm$ 0.000059 & 12.79 $\pm$ 0.45 & 8 $\pm$ 23 & c\\ 
0.362482 $\pm$ 0.000013 & 13.22 $\pm$ 0.10 & 11 $\pm$ 6 & a& 0.426932 $\pm$ 0.000039 & 13.03 $\pm$ 0.26 & 10 $\pm$ 18 & a\\ 
0.363048 $\pm$ 0.000014 & 13.14 $\pm$ 0.18 & 8 $\pm$ 7 & a& 0.433141 $\pm$ 0.000009 & 13.53 $\pm$ 0.04 & 22 $\pm$ 3 & a\\ 
0.363363 $\pm$ 0.000034 & 12.70 $\pm$ 0.22 & 8 $\pm$ 18 & a& 0.433472 $\pm$ 0.000018 & 13.13 $\pm$ 0.08 & 19 $\pm$ 7 & a\\ 
0.363567 $\pm$ 0.000018 & 13.18 $\pm$ 0.09 & 15 $\pm$ 8 & a& 0.433969 $\pm$ 0.000012 & 13.78 $\pm$ 0.03 & 41 $\pm$ 4 & a\\
\hline

    \end{tabular}
    
  \end{minipage}
\begin{minipage}{0.98\textwidth}
(1) and (5): \hi~redshift. 
(2) and (6): \hi~column density from Voigt profile fitting. 
(3) and (7): \hi~Doppler parameter from Voigt profile fitting.
(4) and (8): Confidence label: (a) `secure'; (b) `probable'; and (c) `uncertain' (see \Cref{abs:reliability} for definitions). 
See \Cref{abs:samples} for further details.
\end{minipage}
\end{table*}

\begin{table*}
\centering
\begin{minipage}{0.98\textwidth}
\centering
\caption{\hi~absorption systems in QSO J102218.99+013218.8.}\label{tab:HI_J1022}
\begin{tabular}{@{}cccc|cccc@{}}

\hline                            

$z$ & $\log{(N_{\rm HI}/{\rm cm}^{-2})}$ & $b_{\rm HI}$ (\kms) & label& 
$z$ & $\log{(N_{\rm HI}/{\rm cm}^{-2})}$ & $b_{\rm HI}$ (\kms) & label \\ 

(1)&(2)&(3)&(4)&(5)&(6)&(7)&(8) \\ 

\hline
0.050692 $\pm$ 0.000028 & 13.54 $\pm$ 0.08 & 46 $\pm$ 12 & b & 0.279424 $\pm$ 0.000045 & 13.99 $\pm$ 0.80 & 23 $\pm$ 17 & c \\ 
0.056892 $\pm$ 0.000017 & 17.24 $\pm$ 0.49 & 13 $\pm$ 2 & b & 0.279616 $\pm$ 0.000930 & 13.43 $\pm$ 3.51 & 43 $\pm$ 278 & c \\ 
0.057127 $\pm$ 0.000205 & 13.49 $\pm$ 0.61 & 50 $\pm$ 55 & c & 0.279935 $\pm$ 0.000149 & 13.13 $\pm$ 1.03 & 25 $\pm$ 32 & c \\ 
0.058275 $\pm$ 0.000052 & 13.43 $\pm$ 0.14 & 49 $\pm$ 24 & b & 0.290005 $\pm$ 0.000018 & 13.48 $\pm$ 0.07 & 24 $\pm$ 6 & a \\ 
0.058626 $\pm$ 0.000014 & 13.82 $\pm$ 0.11 & 20 $\pm$ 6 & b & 0.293121 $\pm$ 0.000016 & 13.68 $\pm$ 0.05 & 32 $\pm$ 6 & a \\ 
0.072400 $\pm$ 0.000009 & 14.09 $\pm$ 0.06 & 33 $\pm$ 4 & b & 0.293685 $\pm$ 0.000038 & 13.45 $\pm$ 0.15 & 24 $\pm$ 13 & a \\
\hline

    \end{tabular}
    
  \end{minipage}
\begin{minipage}{0.98\textwidth}
(1) and (5): \hi~redshift. 
(2) and (6): \hi~column density from Voigt profile fitting. 
(3) and (7): \hi~Doppler parameter from Voigt profile fitting.
(4) and (8): Confidence label: (a) `secure'; (b) `probable'; and (c) `uncertain' (see \Cref{abs:reliability} for definitions). 
See \Cref{abs:samples} for further details.
\end{minipage}
\end{table*}

\begin{table*}
\centering
\begin{minipage}{0.98\textwidth}
\centering
\caption{\hi~absorption systems in QSO J135726.27+043541.4.}\label{tab:HI_J1357}
\begin{tabular}{@{}cccc|cccc@{}}

\hline                            

$z$ & $\log{(N_{\rm HI}/{\rm cm}^{-2})}$ & $b_{\rm HI}$ (\kms) & label& 
$z$ & $\log{(N_{\rm HI}/{\rm cm}^{-2})}$ & $b_{\rm HI}$ (\kms) & label \\ 

(1)&(2)&(3)&(4)&(5)&(6)&(7)&(8) \\ 

\hline
0.017131 $\pm$ 0.000257 & 13.91 $\pm$ 0.73 & 61 $\pm$ 42 & c & 0.105164 $\pm$ 0.000009 & 13.77 $\pm$ 0.03 & 39 $\pm$ 4 & b \\ 
0.017248 $\pm$ 0.000014 & 14.08 $\pm$ 0.34 & 24 $\pm$ 15 & b & 0.105844 $\pm$ 0.000017 & 13.68 $\pm$ 0.04 & 64 $\pm$ 7 & b \\ 
0.017598 $\pm$ 0.000126 & 13.40 $\pm$ 0.81 & 50 $\pm$ 103 & c & 0.112071 $\pm$ 0.000023 & 13.23 $\pm$ 0.09 & 31 $\pm$ 10 & a \\ 
0.018668 $\pm$ 0.000011 & 13.37 $\pm$ 0.07 & 16 $\pm$ 6 & b & 0.112358 $\pm$ 0.000006 & 14.03 $\pm$ 0.15 & 18 $\pm$ 3 & a \\ 
0.018958 $\pm$ 0.000060 & 13.24 $\pm$ 0.26 & 33 $\pm$ 27 & b & 0.118585 $\pm$ 0.000007 & 13.63 $\pm$ 0.04 & 24 $\pm$ 3 & a \\ 
0.019134 $\pm$ 0.000011 & 13.77 $\pm$ 0.15 & 14 $\pm$ 5 & b & 0.120039 $\pm$ 0.000301 & 13.30 $\pm$ 1.22 & 39 $\pm$ 51 & c \\
\hline

    \end{tabular}
    
  \end{minipage}
\begin{minipage}{0.98\textwidth}
(1) and (5): \hi~redshift. 
(2) and (6): \hi~column density from Voigt profile fitting. 
(3) and (7): \hi~Doppler parameter from Voigt profile fitting.
(4) and (8): Confidence label: (a) `secure'; (b) `probable'; and (c) `uncertain' (see \Cref{abs:reliability} for definitions). 
See \Cref{abs:samples} for further details.
\end{minipage}
\end{table*}

\begin{table*}
\centering
\begin{minipage}{0.98\textwidth}
\centering
\contcaption{}
\begin{tabular}{@{}cccc|cccc@{}}

\hline                            
$z$ & $\log{(N_{\rm HI}/{\rm cm}^{-2})}$ & $b_{\rm HI}$ (\kms) & label& 
$z$ & $\log{(N_{\rm HI}/{\rm cm}^{-2})}$ & $b_{\rm HI}$ (\kms) & label \\ 

(1)&(2)&(3)&(4)&(5)&(6)&(7)&(8) \\ 

\hline
0.359820 $\pm$ 0.000020 & 13.52 $\pm$ 0.05 & 35 $\pm$ 6 & a& 0.640542 $\pm$ 0.000013 & 15.01 $\pm$ 0.06 & 36 $\pm$ 4 & a\\ 
0.361673 $\pm$ 0.000015 & 13.55 $\pm$ 0.05 & 28 $\pm$ 5 & a& 0.683699 $\pm$ 0.000009 & 15.14 $\pm$ 0.04 & 25 $\pm$ 2 & a\\ 
0.362677 $\pm$ 0.000024 & 13.59 $\pm$ 0.05 & 48 $\pm$ 8 & a& 0.817042 $\pm$ 0.000009 & 15.57 $\pm$ 0.03 & 31 $\pm$ 2 & a\\ 
0.379567 $\pm$ 0.000026 & 13.83 $\pm$ 0.05 & 60 $\pm$ 8 & a& 0.818074 $\pm$ 0.000042 & 14.53 $\pm$ 0.11 & 27 $\pm$ 11 & a\\ 
0.392532 $\pm$ 0.000031 & 13.45 $\pm$ 0.12 & 20 $\pm$ 9 & a& 0.819486 $\pm$ 0.000006 & 16.08 $\pm$ 0.02 & 36 $\pm$ 1 & a\\ 
0.392760 $\pm$ 0.000039 & 13.24 $\pm$ 0.21 & 18 $\pm$ 16 & a& 0.820566 $\pm$ 0.000014 & 15.00 $\pm$ 0.01 & 21 $\pm$ 3 & a\\
\hline

    \end{tabular}
    
  \end{minipage}
\begin{minipage}{0.98\textwidth}
(1) and (5): \hi~redshift. 
(2) and (6): \hi~column density from Voigt profile fitting. 
(3) and (7): \hi~Doppler parameter from Voigt profile fitting.
(4) and (8): Confidence label: (a) `secure'; (b) `probable'; and (c) `uncertain' (see \Cref{abs:reliability} for definitions). 
See \Cref{abs:samples} for further details.
\end{minipage}
\end{table*}

\begin{table*}
\centering
\begin{minipage}{0.98\textwidth}
\centering
\caption{\hi~absorption systems in QSO J221806.67+005223.6.}\label{tab:HI_J2218}
\begin{tabular}{@{}cccc|cccc@{}}

\hline                            

$z$ & $\log{(N_{\rm HI}/{\rm cm}^{-2})}$ & $b_{\rm HI}$ (\kms) & label& 
$z$ & $\log{(N_{\rm HI}/{\rm cm}^{-2})}$ & $b_{\rm HI}$ (\kms) & label \\ 

(1)&(2)&(3)&(4)&(5)&(6)&(7)&(8) \\ 

\hline
0.788733 $\pm$ 0.000049 & 14.22 $\pm$ 0.02 & 142 $\pm$ 13 & b & 0.948039 $\pm$ 0.000071 & 14.30 $\pm$ 1.07 & 36 $\pm$ 60 & c \\ 
0.801500 $\pm$ 0.000096 & 14.21 $\pm$ 0.06 & 111 $\pm$ 26 & b & 0.999144 $\pm$ 0.000091 & 13.86 $\pm$ 0.04 & 112 $\pm$ 23 & b \\ 
0.809249 $\pm$ 0.000097 & 14.28 $\pm$ 0.04 & 183 $\pm$ 24 & b & 1.012488 $\pm$ 0.000108 & 13.81 $\pm$ 0.06 & 87 $\pm$ 28 & b \\ 
0.839592 $\pm$ 0.000086 & 14.19 $\pm$ 0.13 & 61 $\pm$ 30 & b & 1.015784 $\pm$ 0.000103 & 14.08 $\pm$ 0.04 & 148 $\pm$ 24 & b \\ 
0.841663 $\pm$ 0.000058 & 15.05 $\pm$ 0.61 & 94 $\pm$ 38 & c & 1.048160 $\pm$ 0.000030 & 14.84 $\pm$ 0.08 & 71 $\pm$ 5 & a \\ 
0.844030 $\pm$ 0.000089 & 14.12 $\pm$ 0.09 & 72 $\pm$ 28 & b & 1.051686 $\pm$ 0.000080 & 14.04 $\pm$ 0.04 & 121 $\pm$ 18 & a \\ 
0.878483 $\pm$ 0.000117 & 15.58 $\pm$ 2.87 & 96 $\pm$ 99 & c & 1.083814 $\pm$ 0.000088 & 14.99 $\pm$ 15.67 & 26 $\pm$ 151 & c \\ 
0.886827 $\pm$ 0.006759 & 14.18 $\pm$ 21.99 & 90 $\pm$ 390 & c & 1.084660 $\pm$ 0.000347 & 14.11 $\pm$ 0.14 & 190 $\pm$ 38 & a \\ 
0.887010 $\pm$ 0.003798 & 14.19 $\pm$ 21.18 & 65 $\pm$ 989 & c & 1.093294 $\pm$ 0.000056 & 14.28 $\pm$ 1.64 & 27 $\pm$ 39 & c \\ 
0.919839 $\pm$ 0.000787 & 13.68 $\pm$ 0.42 & 147 $\pm$ 143 & b & 1.098911 $\pm$ 0.000068 & 14.00 $\pm$ 0.35 & 33 $\pm$ 27 & a \\ 
0.921107 $\pm$ 0.000234 & 13.73 $\pm$ 0.36 & 77 $\pm$ 45 & b & 1.130129 $\pm$ 0.000042 & 15.23 $\pm$ 0.56 & 24 $\pm$ 6 & c \\ 
0.944475 $\pm$ 0.000473 & 14.00 $\pm$ 0.13 & 315 $\pm$ 106 & b & 1.213798 $\pm$ 0.000032 & 14.51 $\pm$ 0.30 & 39 $\pm$ 9 & a \\ 
0.947890 $\pm$ 0.000220 & 14.18 $\pm$ 0.13 & 191 $\pm$ 62 & b & 1.217410 $\pm$ 0.000024 & 15.88 $\pm$ 0.59 & 70 $\pm$ 12 & c \\ 

\hline

    \end{tabular}
    
  \end{minipage}
\begin{minipage}{0.98\textwidth}
(1) and (5): \hi~redshift. 
(2) and (6): \hi~column density from Voigt profile fitting. 
(3) and (7): \hi~Doppler parameter from Voigt profile fitting.
(4) and (8): Confidence label: (a) `secure'; (b) `probable'; and (c) `uncertain' (see \Cref{abs:reliability} for definitions). 
See \Cref{abs:samples} for further details.
\end{minipage}
\end{table*}


\begin{table*}
\begin{minipage}{0.75\textwidth}
\centering
\caption{Spectroscopic catalog of objects in the Q0107 field.}\label{tab:gals_Q0107}
\begin{tabular}{@{}cccccccc@{}}
\hline                            
R.A.         & Dec.        & $z$ & $z$ label& Spec. Type & $R$   & CLASS\_STAR & Instrument \\ 

(degrees)    & (degrees)   &     &          &            & (mag) &             &            \\ 

(1)          & (2)         & (3) &  (4)     & (5)        & (6)   & (7)         & (8)        \\ 

\hline

17.38011 & -2.45953 & \dots & c & none	 & 22.04 $\pm$ 0.02 & 0.89 & VIMOS\\ 
17.38029 & -2.44843 & \dots & c & none	 & 21.58 $\pm$ 0.01 & 0.91 & VIMOS\\ 
17.38067 & -2.39631 & \dots & c & none	 & 22.86 $\pm$ 0.03 & 0.85 & VIMOS\\ 
17.38092 & -2.29300 & \dots & c & none	 & 22.76 $\pm$ 0.06 & 0.01 & VIMOS\\ 
17.38147 & -2.45457 & \dots & c & none	 & 20.60 $\pm$ 0.01 & 0.03 & VIMOS\\ 
17.38153 & -2.28402 & 0.8206 & a & SF	 & 22.62 $\pm$ 0.03 & 0.22 & VIMOS\\ 
17.38383 & -2.30767 & \dots & c & none	 & 22.85 $\pm$ 0.05 & 0.87 & VIMOS\\ 
17.38384 & -2.31244 & 0.2070 & a & SF	 & 21.49 $\pm$ 0.01 & 0.91 & VIMOS\\ 
17.38433 & -2.42912 & 0.5758 & a & SF	 & 21.85 $\pm$ 0.02 & 0.98 & VIMOS\\ 
17.38459 & -2.38049 & 0.5658 & a & SF	 & 21.46 $\pm$ 0.01 & 0.11 & VIMOS\\ 
17.38593 & -2.42506 & \dots & c & none	 & 21.88 $\pm$ 0.03 & 0.72 & VIMOS\\ 
17.38661 & -2.27211 & 0.1908 & a & non-SF	 & 18.48 $\pm$ 0.01 & 0.62 & VIMOS\\ 
17.38672 & -2.43483 & 0.2604 & a & SF	 & 22.25 $\pm$ 0.04 & 0.04 & VIMOS\\ 
17.38769 & -2.39048 & 0.1898 & a & non-SF	 & 18.92 $\pm$ 0.01 & 0.04 & VIMOS\\ 
17.38899 & -2.38348 & 0.4298 & a & non-SF	 & 19.57 $\pm$ 0.01 & 0.04 & VIMOS\\ 
17.38948 & -2.46353 & \dots & c & none	 & 22.78 $\pm$ 0.06 & 0.15 & VIMOS\\ 
17.38948 & -2.28029 & \dots & c & none	 & 22.40 $\pm$ 0.02 & 0.12 & VIMOS\\ 
17.39174 & -2.23779 & 0.8750 & b & SF	 & 21.92 $\pm$ 0.02 & 0.06 & VIMOS\\ 
17.39238 & -2.32387 & 0.3228 & a & SF	 & 19.35 $\pm$ 0.01 & 0.03 & VIMOS\\ 
17.39346 & -2.26905 & 0.5678 & b & SF	 & 22.00 $\pm$ 0.02 & 0.03 & VIMOS\\ 
17.39372 & -2.26188 & \dots & c & none	 & 23.22 $\pm$ 0.04 & 0.92 & VIMOS\\ 
17.39382 & -2.26352 & 0.1235 & a & SF	 & 20.76 $\pm$ 0.01 & 0.03 & VIMOS\\ 
17.39425 & -2.32939 & 0.1858 & a & SF	 & 20.60 $\pm$ 0.01 & 0.03 & VIMOS\\ 
17.39534 & -2.22252 & 0.4318 & b & SF	 & 21.52 $\pm$ 0.01 & 0.34 & VIMOS\\ 
17.39548 & -2.46720 & 0.4318 & a & SF	 & 21.51 $\pm$ 0.01 & 0.57 & VIMOS\\ 
17.39580 & -2.32021 & \dots & c & none	 & 22.84 $\pm$ 0.04 & 0.92 & VIMOS\\ 
17.39689 & -2.32676 & \dots & c & none	 & 22.78 $\pm$ 0.05 & 0.10 & VIMOS\\ 
17.39936 & -2.44477 & \dots & c & none	 & 21.03 $\pm$ 0.01 & 0.03 & VIMOS\\ 
17.40124 & -2.25178 & \dots & c & none	 & 22.86 $\pm$ 0.04 & 0.02 & VIMOS\\ 
17.40169 & -2.41860 & \dots & c & none	 & 22.95 $\pm$ 0.04 & 0.05 & VIMOS\\ 
17.40238 & -2.36791 & 0.7214 & b & SF	 & 22.92 $\pm$ 0.05 & 0.76 & VIMOS\\ 
17.40259 & -2.24309 & 0.5698 & a & SF	 & 22.66 $\pm$ 0.03 & 0.79 & VIMOS\\ 
17.40325 & -2.25935 & 0.0000 & b & star	 & 20.68 $\pm$ 0.01 & 0.03 & VIMOS\\ 
17.40331 & -2.27535 & 0.7548 & a & SF	 & 21.84 $\pm$ 0.02 & 0.03 & VIMOS\\ 
17.40371 & -2.25559 & \dots & c & none	 & 19.75 $\pm$ 0.01 & 0.03 & VIMOS\\ 
17.40562 & -2.22848 & \dots & c & none	 & 21.72 $\pm$ 0.01 & 0.25 & VIMOS\\ 
17.40621 & -2.39338 & 0.7564 & a & SF	 & 22.67 $\pm$ 0.04 & 0.77 & VIMOS\\ 
17.40647 & -2.44007 & 0.0000 & a & star	 & 18.79 $\pm$ 0.01 & 0.98 & VIMOS\\ 
17.40670 & -2.30130 & 0.5778 & b & SF	 & 21.37 $\pm$ 0.02 & 0.02 & VIMOS\\ 
17.40800 & -2.24577 & 0.5710 & a & SF	 & 20.94 $\pm$ 0.01 & 0.87 & VIMOS\\ 
17.40895 & -2.29587 & 0.4693 & a & non-SF	 & 21.61 $\pm$ 0.01 & 0.98 & VIMOS\\ 
17.40936 & -2.40779 & 0.5125 & b & SF	 & 21.93 $\pm$ 0.02 & 0.03 & VIMOS\\ 
17.40966 & -2.29878 & 0.4318 & b & SF	 & 21.90 $\pm$ 0.01 & 0.73 & VIMOS\\ 
17.41028 & -2.41474 & 0.0768 & a & SF	 & 17.72 $\pm$ 0.01 & 0.03 & VIMOS\\ 
17.41158 & -2.28715 & 0.0000 & a & star	 & 20.31 $\pm$ 0.01 & 0.98 & VIMOS\\ 

\hline
\end{tabular}
\end{minipage}
\begin{minipage}{0.75\textwidth}
{\bf Note.} Only a portion of this table is shown. The full table is available in the online version of the paper.
(1) Right ascension (J2000). 
(2) Declination (J2000). 
(3) Redshift. 
(4) Redshift label: secure (`a'), possible (`b'), no idea (`c'), undefined (`n'). 
(5) Spectral type: star-forming galaxy (\SF), non-star-forming (\nSF), star (`star'), active galactic nuclei (`AGN'), undefined (`none'). 
(6) \rband~magnitude (MAG\_AUTO) given by \sextractor; we note that these uncertainties might be underestimated by a factor of $\sim 3$. 
(7) CLASS\_STAR given by \sextractor. 
(8) Instrument. See \Cref{{gal:samples}} for further details. 
\end{minipage}
\end{table*}
 
\begin{table*}
\begin{minipage}{0.75\textwidth}
\centering
\caption{Spectroscopic catalog of objects in the J1005 field.}\label{tab:gals_J1005}
\begin{tabular}{@{}cccccccc@{}}
\hline                            
R.A.         & Dec.        & $z$ & $z$ label& Spec. Type & $R$   & CLASS\_STAR & Instrument \\ 

(degrees)    & (degrees)   &     &          &            & (mag) &             &            \\ 

(1)          & (2)         & (3) &  (4)     & (5)        & (6)   & (7)         & (8)        \\ 

\hline

151.20108 & 1.49272 & 0.0010 & a & star	 & 18.97 $\pm$ 0.01 & 0.98 & VIMOS\\ 
151.20265 & 1.61368 & 1.2043 & b & SF	 & 23.28 $\pm$ 0.05 & 0.66 & VIMOS\\ 
151.20276 & 1.43851 & 0.1284 & b & SF	 & 22.56 $\pm$ 0.03 & 0.15 & VIMOS\\ 
151.20418 & 1.44580 & 0.9792 & b & non-SF	 & \dots & \dots & VIMOS\\ 
151.20418 & 1.44879 & \dots & c & none	 & 23.00 $\pm$ 0.04 & 0.70 & VIMOS\\ 
151.20469 & 1.66112 & \dots & c & none	 & 22.13 $\pm$ 0.03 & 0.00 & VIMOS\\ 
151.20507 & 1.64756 & -0.0001 & a & star	 & 21.69 $\pm$ 0.01 & 0.98 & VIMOS\\ 
151.20593 & 1.57977 & 0.5020 & a & SF	 & 22.04 $\pm$ 0.02 & 0.57 & VIMOS\\ 
151.20654 & 1.46163 & \dots & c & none	 & 21.76 $\pm$ 0.01 & 0.98 & VIMOS\\ 
151.20737 & 1.52252 & 0.3741 & a & SF	 & 22.30 $\pm$ 0.02 & 0.90 & VIMOS\\ 
151.20745 & 1.44131 & \dots & c & none	 & 21.07 $\pm$ 0.01 & 0.01 & VIMOS\\ 
151.20786 & 1.59262 & 0.6756 & a & SF	 & 22.88 $\pm$ 0.05 & 0.00 & VIMOS\\ 
151.20786 & 1.65663 & 0.6171 & a & SF	 & 22.01 $\pm$ 0.03 & 0.02 & VIMOS\\ 
151.20807 & 1.51942 & 0.3758 & a & SF	 & 21.99 $\pm$ 0.02 & 0.07 & VIMOS\\ 
151.20824 & 1.61967 & \dots & c & none	 & 21.16 $\pm$ 0.01 & 0.98 & VIMOS\\ 
151.20876 & 1.65437 & 0.4140 & a & SF	 & 21.10 $\pm$ 0.01 & 0.02 & VIMOS\\ 
151.20898 & 1.60191 & -0.0003 & a & star	 & 22.30 $\pm$ 0.02 & 0.96 & VIMOS\\ 
151.20899 & 1.60833 & 0.1833 & a & SF	 & 20.15 $\pm$ 0.01 & 0.03 & VIMOS\\ 
151.21012 & 1.43022 & 0.0007 & a & star	 & 21.14 $\pm$ 0.01 & 0.98 & VIMOS\\ 
151.21039 & 1.49978 & 0.6186 & a & SF	 & 22.18 $\pm$ 0.02 & 0.52 & VIMOS\\ 
151.21094 & 1.48094 & 0.3369 & b & SF	 & 20.21 $\pm$ 0.01 & 0.03 & VIMOS\\ 
151.21137 & 1.56292 & 0.4217 & a & SF	 & \dots & \dots & VIMOS\\ 
151.21501 & 1.45867 & -0.0002 & a & star	 & 22.54 $\pm$ 0.02 & 0.95 & VIMOS\\ 
151.21624 & 1.66290 & 0.4349 & a & non-SF	 & 19.99 $\pm$ 0.01 & 0.03 & VIMOS\\ 
151.21690 & 1.46603 & 0.0004 & b & star	 & 22.21 $\pm$ 0.02 & 0.96 & VIMOS\\ 
151.21765 & 1.60560 & 0.3046 & b & SF	 & 20.63 $\pm$ 0.01 & 0.03 & VIMOS\\ 
151.21766 & 1.51071 & 0.2668 & a & SF	 & 21.46 $\pm$ 0.01 & 0.03 & VIMOS\\ 
151.21860 & 1.62702 & 0.3607 & a & non-SF	 & 20.61 $\pm$ 0.01 & 0.98 & VIMOS\\ 
151.21912 & 1.47104 & 0.2784 & a & SF	 & 21.62 $\pm$ 0.02 & 0.10 & VIMOS\\ 
151.21917 & 1.46904 & 0.8439 & a & SF	 & 21.37 $\pm$ 0.01 & 0.79 & VIMOS\\ 
151.22008 & 1.59780 & -0.0006 & b & star	 & 19.84 $\pm$ 0.01 & 0.98 & VIMOS\\ 
151.22048 & 1.59640 & 0.3799 & a & SF	 & 21.13 $\pm$ 0.01 & 0.37 & VIMOS\\ 
151.22182 & 1.63121 & 0.3408 & a & SF	 & 22.21 $\pm$ 0.02 & 0.44 & VIMOS\\ 
151.22212 & 1.66830 & 0.4357 & a & non-SF	 & 19.41 $\pm$ 0.01 & 0.03 & VIMOS\\ 
151.22541 & 1.62459 & 0.5973 & a & SF	 & 22.29 $\pm$ 0.03 & 0.15 & VIMOS\\ 
151.22703 & 1.57367 & 0.1773 & a & SF	 & 20.43 $\pm$ 0.01 & 0.04 & VIMOS\\ 
151.22793 & 1.50934 & \dots & c & none	 & 23.07 $\pm$ 0.05 & 0.00 & VIMOS\\ 
151.22802 & 1.64551 & 0.4308 & a & SF	 & 22.23 $\pm$ 0.02 & 0.15 & VIMOS\\ 
151.22852 & 1.47628 & 0.4130 & a & SF	 & 21.24 $\pm$ 0.01 & 0.97 & VIMOS\\ 
151.22882 & 1.48978 & 1.2499 & b & AGN	 & 21.68 $\pm$ 0.01 & 0.98 & VIMOS\\ 
151.23032 & 1.67427 & 0.4658 & a & SF	 & 21.16 $\pm$ 0.01 & 0.02 & VIMOS\\ 
151.23139 & 1.47889 & \dots & c & none	 & 20.40 $\pm$ 0.01 & 0.98 & VIMOS\\ 
151.23440 & 1.50143 & 0.0984 & a & SF	 & 21.45 $\pm$ 0.01 & 0.98 & VIMOS\\ 
151.23440 & 1.50273 & 0.9504 & b & SF	 & 23.16 $\pm$ 0.04 & 0.26 & VIMOS\\ 
151.23517 & 1.60390 & 0.6915 & a & SF	 & 22.23 $\pm$ 0.03 & 0.01 & VIMOS\\ 

\hline
\end{tabular}
\end{minipage}
\begin{minipage}{0.75\textwidth}
{\bf Note.} Only a portion of this table is shown. The full table is available in the online version of the paper.
(1) Right ascension (J2000). 
(2) Declination (J2000). 
(3) Redshift. 
(4) Redshift label: secure (`a'), possible (`b'), no idea (`c'), undefined (`n'). 
(5) Spectral type: star-forming galaxy (\SF), non-star-forming (\nSF), star (`star'), active galactic nuclei (`AGN'), undefined (`none'). 
(6) \rband~magnitude (MAG\_AUTO) given by \sextractor; we note that these uncertainties might be underestimated by a factor of $\sim 3$. 
(7) CLASS\_STAR given by \sextractor. 
(8) Instrument. See \Cref{{gal:samples}} for further details. 
\end{minipage}
\end{table*}

\begin{table*}
\begin{minipage}{0.75\textwidth}
\centering
\caption{Spectroscopic catalog of objects in the J1022 field.}\label{tab:gals_J1022}
\begin{tabular}{@{}cccccccc@{}}
\hline                            
R.A.         & Dec.        & $z$ & $z$ label& Spec. Type & $R$   & CLASS\_STAR & Instrument \\ 

(degrees)    & (degrees)   &     &          &            & (mag) &             &            \\ 

(1)          & (2)         & (3) &  (4)     & (5)        & (6)   & (7)         & (8)        \\ 

\hline

155.37715 & 1.39655 & 0.8507 & a & SF	 & 21.53 $\pm$ 0.02 & 0.00 & VIMOS\\ 
155.38284 & 1.41530 & 1.1483 & b & SF	 & 23.39 $\pm$ 0.06 & 0.00 & VIMOS\\ 
155.38420 & 1.40041 & 0.0000 & a & star	 & 20.79 $\pm$ 0.01 & 0.98 & VIMOS\\ 
155.38583 & 1.47086 & 0.6002 & a & non-SF	 & 21.43 $\pm$ 0.01 & 0.02 & VIMOS\\ 
155.38706 & 1.48588 & 0.5856 & b & non-SF	 & \dots & \dots & VIMOS\\ 
155.38832 & 1.46557 & 0.6024 & a & non-SF	 & 21.82 $\pm$ 0.02 & 0.13 & VIMOS\\ 
155.38857 & 1.40637 & -0.0001 & a & star	 & 21.86 $\pm$ 0.01 & 0.98 & VIMOS\\ 
155.38882 & 1.45225 & 0.5859 & a & SF	 & 21.92 $\pm$ 0.02 & 0.42 & VIMOS\\ 
155.38967 & 1.48727 & 0.5850 & a & SF	 & \dots & \dots & VIMOS\\ 
155.39031 & 1.45981 & 0.3872 & a & non-SF	 & 21.95 $\pm$ 0.02 & 0.19 & VIMOS\\ 
155.39137 & 1.44240 & 0.2793 & a & SF	 & 22.00 $\pm$ 0.02 & 0.03 & VIMOS\\ 
155.39313 & 1.37581 & 0.3280 & a & SF	 & 22.76 $\pm$ 0.03 & 0.05 & VIMOS\\ 
155.39317 & 1.38055 & 0.5095 & a & SF	 & 21.71 $\pm$ 0.01 & 0.14 & VIMOS\\ 
155.39435 & 1.43998 & \dots & c & none	 & 21.64 $\pm$ 0.01 & 0.98 & VIMOS\\ 
155.39496 & 1.44664 & 0.8356 & a & AGN	 & 20.79 $\pm$ 0.01 & 0.98 & VIMOS\\ 
155.39525 & 1.45809 & 0.5434 & a & SF	 & 22.62 $\pm$ 0.03 & 0.01 & VIMOS\\ 
155.39590 & 1.41016 & 0.3792 & a & SF	 & 21.85 $\pm$ 0.02 & 0.02 & VIMOS\\ 
155.39685 & 1.47371 & 0.3886 & a & non-SF	 & 20.79 $\pm$ 0.01 & 0.04 & VIMOS\\ 
155.39896 & 1.39150 & 0.3822 & a & SF	 & 20.52 $\pm$ 0.01 & 0.02 & VIMOS\\ 
155.40109 & 1.43295 & 1.1802 & a & SF	 & 22.39 $\pm$ 0.03 & 0.17 & VIMOS\\ 
155.40404 & 1.42086 & 0.0001 & a & star	 & 20.06 $\pm$ 0.01 & 0.98 & VIMOS\\ 
155.40579 & 1.46164 & 0.4315 & b & non-SF	 & 21.78 $\pm$ 0.01 & 0.40 & VIMOS\\ 
155.40968 & 1.48033 & 0.3387 & a & SF	 & 22.10 $\pm$ 0.03 & 0.00 & VIMOS\\ 
155.40993 & 1.38644 & 0.5379 & a & SF	 & 21.64 $\pm$ 0.01 & 0.63 & VIMOS\\ 
155.41473 & 1.42251 & 0.2704 & a & SF	 & 22.32 $\pm$ 0.02 & 0.96 & VIMOS\\ 
155.41755 & 1.48266 & \dots & c & none	 & 20.59 $\pm$ 0.01 & 0.02 & VIMOS\\ 
155.41755 & 1.48306 & 0.6919 & a & SF	 & 20.59 $\pm$ 0.01 & 0.02 & VIMOS\\ 
155.42122 & 1.41347 & 0.5490 & a & SF	 & 20.92 $\pm$ 0.01 & 0.03 & VIMOS\\ 
155.42184 & 1.45377 & 0.7426 & b & non-SF	 & 22.24 $\pm$ 0.03 & 0.00 & VIMOS\\ 
155.42204 & 1.50142 & 0.7131 & a & SF	 & 21.65 $\pm$ 0.02 & 0.00 & VIMOS\\ 
155.42233 & 1.47884 & -0.0001 & a & star	 & 20.88 $\pm$ 0.01 & 0.15 & VIMOS\\ 
155.42239 & 1.50883 & 0.2774 & a & SF	 & \dots & \dots & VIMOS\\ 
155.42247 & 1.46069 & 0.6690 & a & SF	 & 22.38 $\pm$ 0.03 & 0.00 & VIMOS\\ 
155.42297 & 1.49994 & 0.0001 & a & star	 & 21.24 $\pm$ 0.01 & 0.07 & VIMOS\\ 
155.42303 & 1.64718 & 0.0001 & a & star	 & 22.18 $\pm$ 0.02 & 0.98 & VIMOS\\ 
155.42307 & 1.44942 & 0.9721 & b & AGN	 & 21.17 $\pm$ 0.01 & 0.88 & VIMOS\\ 
155.42318 & 1.42447 & 0.0003 & a & star	 & 20.24 $\pm$ 0.01 & 0.94 & VIMOS\\ 
155.42378 & 1.43253 & \dots & c & none	 & \dots & \dots & VIMOS\\ 
155.42378 & 1.43483 & 0.3786 & a & SF	 & 22.80 $\pm$ 0.03 & 0.04 & VIMOS\\ 
155.42421 & 1.59549 & 0.2793 & b & SF	 & 21.56 $\pm$ 0.02 & 0.01 & VIMOS\\ 
155.42462 & 1.65111 & \dots & c & none	 & 22.36 $\pm$ 0.02 & 0.97 & VIMOS\\ 
155.42480 & 1.57184 & 0.3838 & a & non-SF	 & 19.88 $\pm$ 0.01 & 0.03 & VIMOS\\ 
155.42488 & 1.63062 & 0.2604 & a & SF	 & 23.17 $\pm$ 0.04 & 0.14 & VIMOS\\ 
155.42488 & 1.58274 & 0.3730 & a & SF	 & 22.37 $\pm$ 0.03 & 0.01 & VIMOS\\ 
155.42501 & 1.57854 & 0.2211 & a & SF	 & 21.24 $\pm$ 0.01 & 0.09 & VIMOS\\ 

\hline
\end{tabular}
\end{minipage}
\begin{minipage}{0.75\textwidth}
{\bf Note.} Only a portion of this table is shown. The full table is available in the online version of the paper.
(1) Right ascension (J2000). 
(2) Declination (J2000). 
(3) Redshift. 
(4) Redshift label: secure (`a'), possible (`b'), no idea (`c'), undefined (`n'). 
(5) Spectral type: star-forming galaxy (\SF), non-star-forming (\nSF), star (`star'), active galactic nuclei (`AGN'), undefined (`none'). 
(6) \rband~magnitude (MAG\_AUTO) given by \sextractor; we note that these uncertainties might be underestimated by a factor of $\sim 3$. 
(7) CLASS\_STAR given by \sextractor. 
(8) Instrument. See \Cref{{gal:samples}} for further details. 
\end{minipage}
\end{table*}
 
\begin{table*}
\begin{minipage}{0.75\textwidth}
\centering
\caption{Spectroscopic catalog of objects in the J2218 field.}\label{tab:gals_J2218}
\begin{tabular}{@{}cccccccc@{}}
\hline                            
R.A.         & Dec.        & $z$ & $z$ label& Spec. Type & $R$   & CLASS\_STAR & Instrument \\ 

(degrees)    & (degrees)   &     &          &            & (mag) &             &            \\ 

(1)          & (2)         & (3) &  (4)     & (5)        & (6)   & (7)         & (8)        \\ 

\hline

334.33420 & 0.88225 & \dots & c & none	 & 22.66 $\pm$ 0.06 & 0.76 & VIMOS\\ 
334.33427 & 0.87096 & 0.7139 & a & SF	 & 22.48 $\pm$ 0.06 & 0.06 & VIMOS\\ 
334.33532 & 0.76281 & -0.0007 & a & star	 & 21.45 $\pm$ 0.02 & 0.95 & VIMOS\\ 
334.33539 & 0.95526 & -0.0006 & a & star	 & 20.36 $\pm$ 0.01 & 0.98 & VIMOS\\ 
334.33540 & 0.87713 & -0.0001 & a & star	 & 21.14 $\pm$ 0.02 & 0.98 & VIMOS\\ 
334.33540 & 0.87816 & \dots & c & none	 & 21.14 $\pm$ 0.02 & 0.98 & VIMOS\\ 
334.33553 & 0.89967 & 0.2770 & a & SF	 & 19.98 $\pm$ 0.01 & 0.02 & VIMOS\\ 
334.33776 & 0.94074 & -0.0008 & a & star	 & 21.05 $\pm$ 0.02 & 0.98 & VIMOS\\ 
334.33805 & 0.75379 & 0.4266 & a & non-SF	 & 20.18 $\pm$ 0.01 & 0.06 & VIMOS\\ 
334.33840 & 0.89023 & -0.0003 & a & star	 & 20.48 $\pm$ 0.01 & 0.98 & VIMOS\\ 
334.33872 & 0.86856 & -0.0002 & a & star	 & 22.82 $\pm$ 0.06 & 0.73 & VIMOS\\ 
334.34238 & 0.85964 & 0.0000 & a & star	 & 21.57 $\pm$ 0.02 & 0.97 & VIMOS\\ 
334.34268 & 0.95394 & 0.3552 & a & SF	 & 20.23 $\pm$ 0.01 & 0.02 & VIMOS\\ 
334.34329 & 0.80460 & \dots & c & none	 & 22.33 $\pm$ 0.05 & 0.88 & VIMOS\\ 
334.34470 & 0.80777 & 0.5634 & b & SF	 & 22.03 $\pm$ 0.04 & 0.45 & VIMOS\\ 
334.34497 & 0.81192 & -0.0006 & a & star	 & 21.33 $\pm$ 0.02 & 0.98 & VIMOS\\ 
334.34497 & 0.81424 & \dots & c & none	 & 22.04 $\pm$ 0.04 & 0.16 & VIMOS\\ 
334.34521 & 0.80232 & 0.2780 & b & non-SF	 & 21.17 $\pm$ 0.02 & 0.24 & VIMOS\\ 
334.34639 & 0.94785 & \dots & c & none	 & 21.87 $\pm$ 0.04 & 0.16 & VIMOS\\ 
334.34675 & 0.72373 & \dots & c & none	 & 21.52 $\pm$ 0.04 & 0.00 & VIMOS\\ 
334.34679 & 0.70623 & \dots & c & none	 & 21.64 $\pm$ 0.03 & 0.97 & VIMOS\\ 
334.34808 & 0.86693 & \dots & c & none	 & 22.62 $\pm$ 0.07 & 0.09 & VIMOS\\ 
334.34811 & 0.72829 & \dots & c & none	 & 21.43 $\pm$ 0.03 & 0.01 & VIMOS\\ 
334.34837 & 0.90869 & \dots & c & none	 & 20.85 $\pm$ 0.01 & 0.98 & VIMOS\\ 
334.34899 & 0.74437 & \dots & c & none	 & 21.08 $\pm$ 0.02 & 0.98 & VIMOS\\ 
334.34957 & 0.71501 & 2.6775 & b & AGN	 & 20.71 $\pm$ 0.01 & 0.98 & VIMOS\\ 
334.35004 & 0.73447 & 0.5102 & a & SF	 & 21.49 $\pm$ 0.03 & 0.01 & VIMOS\\ 
334.35004 & 0.73733 & \dots & c & none	 & \dots & \dots & VIMOS\\ 
334.35029 & 0.77934 & \dots & c & none	 & 21.11 $\pm$ 0.02 & 0.01 & VIMOS\\ 
334.35062 & 0.93009 & \dots & c & none	 & 21.87 $\pm$ 0.05 & 0.00 & VIMOS\\ 
334.35116 & 0.89441 & 0.5197 & a & non-SF	 & 20.90 $\pm$ 0.02 & 0.04 & VIMOS\\ 
334.35170 & 0.76951 & 0.2519 & a & SF	 & 22.05 $\pm$ 0.05 & 0.01 & VIMOS\\ 
334.35180 & 0.79917 & 0.0001 & a & star	 & 22.98 $\pm$ 0.07 & 0.73 & VIMOS\\ 
334.35212 & 0.78120 & -0.0010 & a & star	 & 19.11 $\pm$ 0.01 & 0.90 & VIMOS\\ 
334.35303 & 0.96757 & -0.0010 & a & star	 & 20.74 $\pm$ 0.01 & 0.98 & VIMOS\\ 
334.35317 & 0.71804 & \dots & c & none	 & 20.84 $\pm$ 0.02 & 0.42 & VIMOS\\ 
334.35417 & 0.91667 & \dots & c & none	 & 20.37 $\pm$ 0.02 & 0.01 & VIMOS\\ 
334.35472 & 0.81685 & -0.0005 & a & star	 & 19.54 $\pm$ 0.01 & 0.98 & VIMOS\\ 
334.35482 & 0.85446 & \dots & c & none	 & 20.40 $\pm$ 0.01 & 0.98 & VIMOS\\ 
334.35491 & 0.78289 & 0.4486 & b & non-SF	 & 21.54 $\pm$ 0.03 & 0.34 & VIMOS\\ 
334.35506 & 0.74805 & \dots & c & none	 & 21.14 $\pm$ 0.02 & 0.97 & VIMOS\\ 
334.35588 & 0.90488 & \dots & c & none	 & 20.56 $\pm$ 0.01 & 0.98 & VIMOS\\ 
334.35620 & 0.75890 & -0.0002 & a & star	 & \dots & \dots & VIMOS\\ 
334.35620 & 0.76016 & \dots & c & none	 & 21.08 $\pm$ 0.02 & 0.98 & VIMOS\\ 
334.35643 & 0.89244 & 0.0004 & a & star	 & 21.08 $\pm$ 0.02 & 0.98 & VIMOS\\ 

\hline
\end{tabular}
\end{minipage}
\begin{minipage}{0.75\textwidth}
{\bf Note.} Only a portion of this table is shown. The full table is available in the online version of the paper.
(1) Right ascension (J2000). 
(2) Declination (J2000). 
(3) Redshift. 
(4) Redshift label: secure (`a'), possible (`b'), no idea (`c'), undefined (`n'). 
(5) Spectral type: star-forming galaxy (\SF), non-star-forming (\nSF), star (`star'), active galactic nuclei (`AGN'), undefined (`none'). 
(6) \rband~magnitude (MAG\_AUTO) given by \sextractor; we note that these uncertainties might be underestimated by a factor of $\sim 3$. 
(7) CLASS\_STAR given by \sextractor. 
(8) Instrument. See \Cref{{gal:samples}} for further details. 
\end{minipage}
\end{table*}

\bsp

\label{lastpage}
\end{document}